\documentclass[sn-aps]{sn-jnl}

\usepackage{graphicx}%
\usepackage{multirow}%
\usepackage{amsmath,amssymb,amsfonts}%
\usepackage{amsthm}%
\usepackage{mathrsfs}%
\usepackage[title]{appendix}%
\usepackage{xcolor}%
\usepackage{textcomp}%
\usepackage{manyfoot}%
\usepackage{booktabs}%
\usepackage{algorithm}%
\usepackage{algorithmicx}%
\usepackage{algpseudocode}%
\usepackage{listings}%

\usepackage{lipsum}
\usepackage{graphicx,color}
\usepackage{bm}
\usepackage{natbib}
\usepackage{placeins}
\setcitestyle{square}
\usepackage{braket}
\usepackage{makerobust}
\usepackage[separate-uncertainty = true]{siunitx}%

\usepackage{array}%

\theoremstyle{thmstyleone}%
\theoremstyle{thmstyletwo}%

\theoremstyle{thmstylethree}%

\newcommand{\bs}[1]{\boldsymbol{#1}}
\newcommand{\up}{{\uparrow}}
\newcommand{\dw}{{\downarrow}}
\newcommand{\pd}{{\phantom{\dagger}}}

\newcommand{\didv}{\mathrm{d}I/\mathrm{d}V}

\begin{document}

\title[Magnet-superconductor hybrid quantum systems]{Magnet-superconductor hybrid quantum systems: 
a materials platform for topological superconductivity
}


\author*[1]{\fnm{Roberto} \sur{Lo Conte}}\email{r.lo.conte@rug.nl}

\author[2]{\fnm{Jens} \sur{Wiebe}}

\author[3]{\fnm{Stephan} \sur{Rachel}}

\author[4]{\fnm{Dirk K.} \sur{Morr}}

\author[2]{\fnm{Roland} \sur{Wiesendanger}}

\affil[1]{\orgdiv{Zernike Institute for Advanced Materials}, \orgname{University of Groningen}, \orgaddress{\city{9747 AG Groningen}, \country{The Netherlands}}}

\affil[2]{\orgdiv{Department of Physics}, \orgname{University of Hamburg}, \orgaddress{\city{20355 Hamburg}, \country{Germany}}}

\affil[3]{\orgdiv{School of Physics}, \orgname{University of Melbourne}, \orgaddress{\city{3010 Parkville, VIC}, \country{Australia}}}

\affil[4]{\orgdiv{Department of Physics}, \orgname{University of Illinois at Chicago}, \orgaddress{\city{Chicago, IL 60607}, \country{USA}}}

\abstract{Magnet-superconductor hybrid (MSH) systems have recently emerged as one of the most significant developments in condensed matter physics. This has generated, in the last decade, a steadily rising interest in the understanding of their unique properties. They have been proposed as one of the most promising platforms for the establishment of topological superconductivity, which holds high potential for application in future quantum information technologies. Their emergent electronic properties stem from the exchange interaction between the magnetic moments and the superconducting condensate. Given the atomic-level origin of such interaction, it is of paramount importance to investigate new magnet-superconductor hybrids at the atomic scale. In this regard, scanning tunneling microscopy (STM) and spectroscopy (STS) is playing a crucial role in the race to unveil the fundamental origin of the unique properties of MSH systems, with the aim to discover new hybrid quantum materials capable of hosting topologically non-trivial unconventional superconducting phases. In particular, the combination of STM studies with tight-binding model calculations have represented, so far, the most successful approach to unveil and explain the emergent electronic properties of MSHs. The scope of this review is to offer a broad perspective on the field of MSHs from an atomic-level investigation point-of-view. The focus is on discussing the link between the magnetic ground state hosted by the hybrid system and the corresponding emergent superconducting phase. This is done for MSHs with both one-dimensional (atomic chains) and two-dimensional (atomic lattices and thin films) magnetic systems proximitized to conventional s-wave superconductors. We present a systematic categorization of the experimentally investigated systems with respect to defined experimentally accessible criteria to verify or falsify the presence of topological superconductivity and Majorana edge modes. The discussion will start with an introduction to the physics of Yu-Shiba-Rusinov bound states at magnetic impurities on superconducting surfaces. This will be used as a base for the discussion of magnetic atomic chains on superconductors, distinguishing between ferromagnetic, antiferromagnetic and non-collinear magnetic ground states. A similar approach will be used for the discussion of magnetic thin film islands on superconductors. Given the vast number of publications on the topic, we limit ourselves to discuss works which are most relevant to the search for topological superconductivity.}


\maketitle

\newpage
\tableofcontents
\newpage

\section{Introduction to magnet-superconductor hybrid systems}\label{intro}
Magnet-superconductor hybrid (MSH) systems have raised considerable interest in solid state physics in recent years driven by the search for novel topological phases of quantum matter and their exciting physical properties. A prominent example is topological superconductivity (TSC) in quasi-1D and 2D MSH systems with localized or propagating Majorana boundary modes based on three important ingredients: (i) superconductivity, (ii) breaking of time-reversal symmetry by the magnetic subsystem, and (iii) spin-orbit coupling (SOC). The latter is required in order to allow for $p$-wave pairing correlations. Alternatively such $p$-wave pairing correlations may also be induced instead of (ii) and (iii) by non-collinear spin-structures. Majorana modes are of fundamental interest because they represent quasiparticles being their own antiparticles~\cite{MajoranaNUOV-CIM1937}. Simultaneously, they offer great potential for practical applications, including the realization of robust Majorana qubits as basic building blocks for fault-tolerant topological quantum computing~\cite{KitaevANN-PHY2003,NayakREV-MOD-PHYS2008,AliceaNAT-PHYS2011}.

Historically, superconductivity and magnetism were believed to be mutually exclusive. Superconductors as ideal diamagnets repel magnetic fields. On the other hand, sufficiently strong, externally applied magnetic fields or a sufficiently high concentration of magnetic impurities within a superconducting material can destroy the superconducting phase. In the dilute limit, individual magnetic impurities in a superconductor break up Cooper pairs locally and create pairs of spin-polarized Yu-Shiba-Rusinov (YSR) states within the superconductor’s energy gap~\cite{YuACT-PHYS-SIN1965,ShibaPROG-THEO-PHYS1968,Rusinov1969}. If two such magnetic impurities come close to each other, these YSR states can hybridize and form bonding or anti-bonding states. If more YSR impurities interact with each other, e.g. in 1D chains or 2D lattices, then YSR bands can form within the gap of the superconductor. Depending on the materials’ properties, including the chemical potential and the strength of the exchange coupling, a topologically non-trivial superconducting phase might emerge with associated Majorana boundary modes, i.e., localized Majorana zero-energy bound states at both ends of quasi-1D spin chains~\cite{Nadj-PergePRB2013,BrauneckerPRL2013,PientkaPRB2013,KlinovajaPRL2013} or dispersing chiral Majorana edge modes at the periphery of quasi-2D 
lattices~\cite{Rachel2017,Bedow2020}. In the case of a YSR chain, $p$-wave pairing correlations can open a (topological) gap within the YSR band being characteristic for a 1D topological superconductor hosting Majorana zero modes at the chain’s ends.

To study the exciting physics of interacting YSR states and the emergence of Majorana modes under well-defined experimental conditions, i.e., excluding the effects of impurities and disorder, artificially fabricated 1D YSR chains~\cite{Kim2018,Schneider2020,Schneider2021b,Mier2021} or 2D YSR lattices~\cite{SoldiniNAT-PHYS2023} have become very attractive as atomic-scale engineered platforms for topological superconductivity and associated Majorana modes. The tailored design of quantum materials is particularly important in view of avoiding disorder-induced low-energy states which can mimic Majorana modes or having detrimental effects on the topological superconducting state. Scanning tunneling microscopy (STM) based methods of single-atom manipulation are ideally suited for fabricating largely disorder-free low-dimensional MSH systems with atomic-scale precision~\cite{ChoiREV-MOD-PHYS2019}. Subsequent scanning tunneling spectroscopy (STS) investigations performed in-situ allow one to probe the local structural, electronic, and spin-dependent (magnetic) properties with atomic-scale spatial resolution. For non-perfect MSH systems, prepared for instance by self-assembly processes~\cite{NadjPerge2014,Ruby2015,Pawlak2016,Feldman2017,Ruby2017,Jeon2017}, the type of disorder and its influence on the (local) electronic states can be investigated with atomic spatial resolution. 
The spin-polarized (SP-)STM method~\cite{WiesendangerREV-MOD-PHYS2009} additionally allows the characterization of the type of magnetic order in 1D spin chains and 2D spin arrays (dilute regime) or ultrathin magnetic films (non-dilute regime), being either collinear (e.g. ferro- or antiferromagnetic states)~\cite{NadjPerge2014,Schneider2021,BazarnikNAT-COMM2023,SoldiniNAT-PHYS2023} or non-collinear (e.g. spin spirals or magnetic skyrmions)~\cite{Kim2018,BrüningARXIV2024}, which has a decisive influence on the topological phase diagram and the possible emergence of topologically non-trivial phases in a particular type of low-dimensional MSH system~\cite{Schneider2021b,Küster2022}. In the case of collinear magnetic ground states, a significant SOC strength is required in order to achieve topologically non-trivial phases, while in the presence of non-collinear spin states, topological superconducting phases can emerge even in the absence of SOC. However, many non-collinear spin textures are stabilized by interfacial Dzyaloshinskii-Moriya (DM) interactions~\cite{vonBergmannJoP-COND-MATT2014,SteinbrecherNAT-COMM2018}, which are resulting from non-vanishing SOC, so that topologically non-trivial phases in MSH systems in the absence of SOC can only result in case of non-collinear spin textures stabilized by frustrated magnetic exchange interactions.

The significant advances in research on non-collinear spin textures in low-dimensional magnetic systems~\cite{vonBergmannJoP-COND-MATT2014}, as well as improvements in the preparation of ultra-clean superconductor surfaces and interfaces, within the past two decades have laid the foundations for numerous recent studies of nano-scale MSH systems and their exciting topologically superconducting states. In particular, bottom-up fabricated, defect-free MSH systems have become one of the most important and valuable platform for testing the various theoretical predictions associated with the possible emergence of the long-sought Majorana quasiparticles, while semiconductor–superconductor hybrid systems still exhibit a significant degree of disorder, even for epitaxially grown semiconducting nanowires, hampering the identification of Majorana modes~\cite{AhnPRM2021,WoodsPR-APPL2021}.\newline
Several review papers have been published in the last few years discussing the search for Majoranas in different materials platforms~\cite{JaeckNAT-REV-PHYS2021,YazdaniSCI2023}. In the present manuscript, we would like to offer a more detailed discussion of the efforts done on one specific type of materials platform, providing an overview of recent theoretical and experimental studies of low-dimensional magnet-superconductor hybrid systems, in particular in view of designing novel topological phases of quantum matter and their real-space investigation down to the atomic level. A more detailed discussion of all the kind of topological superconducting phases that are theoretically expected to be established in MSH systems can be found here~\cite{SteffensenPRR2022}. 
Finally, we would like to point out that other important review articles have been written which discuss MSHs systems concerning proximity-effects~\cite{Buzdin2005Proximity}, topological superconductivity~\cite{ZlotnikovJSNM2021}, odd triplet superconductivity~\cite{Bergeret2005Odd}, and other effects combining superconductivity and spintronics~\cite{Eschrig2015Spinpolarized,LinderNAT-PHY2015,MaggioraPR2024}.

\section{Theory of Yu-Shiba-Rusinov impurities, spin chains and two-dimensional systems}\label{theory}

In the theory part of this review, we discuss the emergence of TSC in the manifold variants of MSH systems, theoretically proposed in the past decade.

\subsection{Spin-chains on superconducting substrates}\label{theory-1D}

In this section, we review the basic theory ideas of topological superconductivity in MSH chains. We begin by discussing the Kitaev chain as the standard model of 1D TSC and its connection to Majorana qubits. After a short overview of YSR impurities, we discuss the most important MSH chains which were proposed to host TSC phases: helical and ferromagnetic Shiba chains. We also discuss antiferromagnetic chains where the emergence of TSC is less obvious. Finally we focus on the few but notable works where quantum many-body interactions were included in the otherwise free-fermionic models.

The simplest model featuring Majorana zero modes is the Kitaev chain\,\cite{kitaev01pu131}, a one-dimensional (1D) tight-binding chain of spinless fermions equipped with a nearest-neighbor superconducting pairing term,
\begin{equation}
\label{eq:ham-kitaev}
H_K = \sum_{i=1}^{N-1} \left( -t c_i^\dagger c_{i+1}^\pd + \frac{\mu}{2} c_i^\dagger c_i^\pd + \Delta_p c_i^\dagger c_{i+1}^\dagger + {\rm H.c.} \right)\ .
\end{equation}
Here $c_i^\dag$ ($c_i^\pd$) is a fermionic creation (annihilation) operator, $t$ denotes the hopping amplitude, $\mu$ the chemical potential, and $\Delta_p = |\Delta_p| e^{i\phi}$ the superconducting order parameter with complex phase $\phi$. For spinless fermions, the superconducting pairing must be anti-symmetric in momentum and corresponds thus to $p$-wave pairing. With other words, Kitaev's model is nothing else than a 1D spinless $p$-wave superconductor, the prototype of a topological superconductor. As pointed out by Kitaev, this simple model possesses two topologically distinct phases. Following \cite{kitaev01pu131}, one can formally introduce Majorana operators 
\begin{equation}
\label{eq:majorana-operators}
\gamma_{2j-1} = c_j^\pd + c_j^\dag, \quad \gamma_{2j} = -i \left( c_j^\pd - c_j^\dag \right), \quad j=1,\ldots, N,
\end{equation}
satisfying the relations
\begin{equation}
\gamma_m^\dag =\gamma_m^\pd, \quad \{ \gamma_m, \gamma_n \} = 2 \delta_{mn}, \quad m, n = 1, \ldots , 2N\ .
\end{equation}
One can absorb the U(1) phase $\phi$ in the definitions of the Majorana operators\,\cite{kitaev01pu131} such that we assume $\Delta_p = |\Delta_p|$ in the following. The existence of the two different phases is most obvious by considering two extreme limits. The trivial limit corresponds to $t=\Delta_p=0$ and $\mu<0$ leading to
\begin{equation}
H_K^{\rm triv} = -\mu\frac{i}{2}\sum_j \gamma_{2j-1} \gamma_{2j}\ .
\end{equation}
With Eq.\,\eqref{eq:majorana-operators} one can see that both the Majorana operators are from the same site $j$, thus $H_K^{\rm triv}$ describes a trivial system. In contrast, the topological limit corresponds to $t=\Delta_p>0$ and $\mu=0$, leading to
\begin{equation}
H_K^{\rm topo} = -i t \sum_j \gamma_{2j} \gamma_{2j+1}\ .
\end{equation}
While being similar to $H_K^{\rm triv}$, the crucial difference is that the two Majorana operators belong to different fermionic sites $j$ and $j+1$. That is, Majorana operators from different sites are paired together. Moreover, for open boundary conditions we notice that $\gamma_1$ and $\gamma_{2N}$ are not contained in $H_K^{\rm topo}$. 

One can introduce new fermionic creation and annihilaiton operators, {\it Bogoliubov quasiparticle} operators, which ``live'' on the bonds rather than sites,
\begin{equation}
\label{eq:d-operators}
d_j^\pd = \frac{1}{2}\left(\gamma_{2j} + i \gamma_{2j+1}\right), \quad d_j^\dag = \frac{1}{2}\left( \gamma_{2j} - i \gamma_{2j+1}\right)\ .
\end{equation}
In particular, we can introduce the highly non-local $d$-operator
\begin{equation}
d_0^\pd = \frac{1}{2} \left( \gamma_1 + i \gamma_{2N}\right)\ .
\end{equation}
The zero-energy many-particle ground state of $H_K$ be $\ket{0}$. It follows that $\ket{1} = d_0^\dag \ket{0}$ is a second, degenerate zero-energy ground state of $H_K$. As mentioned above, since $\gamma_1$ and $\gamma_{2N}$ do not enter the Hamiltonian $H_K^{\rm topo}$, one can write\,\cite{kitaev01pu131}
\begin{equation}
-i \gamma_0 \gamma_{2N} \ket{0} = \ket{0}, \quad -i \gamma_0 \gamma_{2N} \ket{1} = -\ket{1}\ .
\end{equation}
While the particle-number is not conserved in any mean-field superconductor, the fermion parity is. The two ground states can thus be distinguished by the fermion parity operator $P = \prod_j \left( -i \gamma_{2j-1} \gamma_{2j}\right)$, with $\ket{0}$ having even and $\ket{1}$ odd fermion parity. In both ground states, the Majorana bound state, i.e., the superposition of the Majorana zero modes (MZMs) on both chain ends, is present (and thus visible in the LDOS), but only for $\ket{1}$ it is occupied. By performing local measurements, e.g. $dI/dV$ with an STM, one should be able to observe the Majorana bound state as a zero-bias peak in the energy gap of the superconductor. Qubits constructed from Majorana zero modes are considered to be protected against local perturbations and decoherence, one of the main sources of qubit errors and thus a major obstacle to topological quantum computing\,\cite{KitaevANN-PHY2003}.

Kitaev's model does not appear to be very physical, as electrons are {\it spinful} particles, and $p$-wave superconductors are extremely rare in nature despite decades of focussed research\,\cite{kallin-16rpp054502,sato-17rpp076501,mackenzie-17npjqm40}. This limitation was overcome by proposals for engineering the Kitaev chain in physical systems\,\cite{lutchyn-10prl077001,oreg-10prl177002}. Today it is well-established that conventional electrons can realize an effective $p$-wave or topological superconductor by essentially fulfilling the following three points: (i) being in proximity to a conventional $s$-wave superconductor; (ii) breaking time-reversal symmetry; (iii) breaking SU(2) spin symmetry. This compact list of requirements can be more intuitively understood as follows. (i) When an electron fluid, e.g.\ a wire or chain, is brought in vicinity to a superconductor, the so-called proximity effect induces superconducting fluctuations also in the electron fluid, making it effectively superconducting. Note that this is not an instance of 1D superconductivity, as the hybrid system chain--substrate is effectively three-dimensional. Any superconductor with a hard gap can be used, in particular, the well-studied phonon-mediated superconductors with $s$-wave symmetry including elemental superconductors such as Pb and Nb. It was shown, however, that also unconventional superconductors would work as long as they provide a hard gap\,\cite{crawford-20prb174510}. (ii) Time-reversal breaking is typically induced by a Zeeman field which polarizes the electrons' spin. This can be realized either through an externally applied magnetic field or through the magnetic moments of certain atoms. We stress that also other sources of time-reversal symmetry breaking could be utilized. (iii) Despite the spin-polarization, the proximity-induced superconductivity would still not be topologically non-trivial. It is further important to break the SU(2) symmetry, allowing the electon's spin to rotate or change with momentum. As a consequence, when the Fermi energy lies in the minority band the Fermi surface is non-degenerate and the induced superconductivity would be topological. In an alternative perspective,  the SU(2) symmetry breaking causes a mixing of angular momentum channels, and when performing a unitary transformation, the new basis would appear to be that of a $p$-wave superconductor. SU(2) symmetry is usually broken by virtue of Rashba spin-orbit coupling. Any breaking of inversion symmetry can induce Rashba spin-orbit coupling, thus one might think of substrates which are non-centrosymmetric or systems which are hybrid or heterostructures causing the asymmetry. Even simpler, any surface (which can be thought of a heterostructure crystal--vacuum) would induce at least close to the surface some Rashba spin-orbit coupling. Spin-orbit coupling scales with the atomic number $Z$ (often as $Z^4$) and thus heavier elements involved should feature substantial Rashba spin-orbit coupling. We will discuss further below in Section \ref{theory-1D-NC} that also a non-constant magnetic field configuration would fulfill (iii); in such cases, requirements (ii) and (iii) can be fulfilled simultaneously. 
The mechanism (i)-(iii) can be straight-forwardly generalized to two spatial dimensions, as discussed in Section\,\ref{theory-2D}. Refs.\,\cite{lutchyn-10prl077001,oreg-10prl177002} envisioned the combination of semiconductor--superconductor heterostructures. A similar idea, suggested earlier, was that of the non-degenerate surface states of topological insulators to be proximitized with a superconductor\,\cite{fu-08prl096407}. The so-called Shiba chain is yet another way of effectively realizing topological superconductivity, as discussed in the following. In particular, the semiconductor-heterostructure model and the ferromagnetic Shiba chains discussed in Section\,\ref{theory-1D-FM} possess similar or, in the simplest cases, even identical Hamiltonians. 

To conclude these introductory considerations of the Kitaev chain, it is worth mentioning that an explicit connection between the spinful Shiba chain model, or its variants, and the Kitaev model\,\cite{kitaev01pu131} was demonstrated in the Appendix of \cite{choy-11prb195442} as well as in \cite{martin-12prb144505}. It was further shown that such a mapping from Shiba chain to effective spinless $p$-wave superconductor usually leads to long-ranged power-law hoppings and pairings\,\cite{PientkaPRB2013}, and hoppings might also collect complex phases. The effect on the phase diagram was studied in Ref.\,\cite{PientkaPRB2013}.

\subsubsection{Single Yu-Shiba-Rusinov impurities}\label{theory-0D}

A magnetic impurity in a conventional superconductor causes localized states within the energy gap of the superconductor, referred to as YSR states or Shiba states\,\cite{YuACT-PHYS-SIN1965,ShibaPROG-THEO-PHYS1968,Rusinov1969}. While Shiba impurities are discussed in the experimental part of this review in Section\,\ref{experiment_0D} and are important to understand the experimental Shiba chain systems, the theory of Shiba states is discussed in great length in the literature and an exhaustive review article is available\,\cite{Balatsky2006}. Thus we will restrict the discussion here to a minimum.

A magnetic impurity violates time-reversal invariance, leading to a suppression of superconductivity. When assuming classical spins, one obtains\,\cite{ShibaPROG-THEO-PHYS1968} the well-known expression for the YSR bound-state energy
\begin{equation}
\label{eq:YSR-energy}
    E_{\rm YSR} = \frac{1-(J S \pi N_0/2)^2}{1+(J S \pi N_0/2)^2}
\end{equation}
within the superconducting gap. Here $J$ is the magnetic moment, $S$ the spin and $N_0$ the DOS of the normal-state.  Due to particle-hole symmetry, there is an electron state with $E_{\rm YSR}$ as well as a hole state with $-E_{\rm YSR}$. While the energy position must be symmetric with respect to the middle of the gap, the corresponding spectral weight in the LDOS or $dI/dV$ is typically asymmetric. The two states are fully spin-polarized. The wavefunction of the bound state is localized near the impurity site with typical length scale $r_0 \sim \xi_0/\sqrt{1 - E_{\rm YSR}^2}$\,\cite{Balatsky2006} where $\xi_0$ is the superconducting coherence length.

\subsubsection{Helical spin chains}\label{theory-1D-NC}

A YSR  state discussed in Section\,\ref{theory-0D} can hybridize with another magnetic impurity state in its vicinity, and when sufficiently many impurities are involved they can form bands, usually referred to as YSR bands or Shiba bands.
1D arrangements of hybridized magnetic impurities were coined {\it Shiba chains}. The first Shiba chains discussed in the context of topological superconductivity exhibited helical or spiral spin arrangements as a special case of non-collinear spin order. Before discussing the properties and the chronological development  of helical Shiba chains in detail, credit should be given to early works where similar concepts and settings were proposed and studied. Particular excitement was caused by the prospect of not relying on spin-orbit coupling, thus most of these early papers emphasized the mechanism for topological superconductivity without the need of spin-orbit coupling.

Chains of magnetic nanoparticles with random orientation of their magnetic moments placed on a superconducting substrate were shown to lead to TSC without spin-orbit coupling\,\cite{choy-11prb195442}. Nanoparticles such as quantum dots as large as 10nm were suggested; the considered model is essentially that of a helical Shiba chain discussed below, albeit without systematic ordering of the magnetic moments.
Another early idea to realize TSC without spin-orbit coupling consisted of a special arrangement of micromagnets, leading to a magnetic field that rotates along the wire\,\cite{kjaergaard-12prb020503} and to an effectively form-invariant Hamiltonian of a helical Shiba chain.
Helical magnetic superconductors were proposed in Ref.\,\cite{martin-12prb144505} as a natural material platform to host MZMs.
Ref.\,\cite{schecter-16prb140503} showed that atoms with magnetic moments, usually aligned ferromagnetically, form a spin spiral when coupled to a three-dimensional superconductor. The resulting system thus realizes a {\it self-organized} topological superconductor, again without any need of spin-orbit coupling.
Comprehensive follow-up work with chains in different crystallographic directions on a two-dimensional superconducting substrate\,\cite{christensen-16prb144509} confirmed the early proposals and demonstrated the stability of the topological phases towards perturbations, including weak spin-orbit coupling.

Interacting electronic liquids provide another interesting mechanism to create an effective system with a magnetic spin spiral, when coupling to  magnetic moments and a superconducting substrate; unsurprisingly, the resulting system leads to 1D topological superconductivity\,\cite{BrauneckerPRL2013}. 
In another interesting paper it was argued that localized magnetic moments in contact with an $s$-wave superconductor order into a spin helix by means of RKKY interactions transmitted by electrons, constituting another platform proposal for hosting MZMs\,\cite{KlinovajaPRL2013}.

\begin{figure}[t!]
\centering
\includegraphics[width=\textwidth]{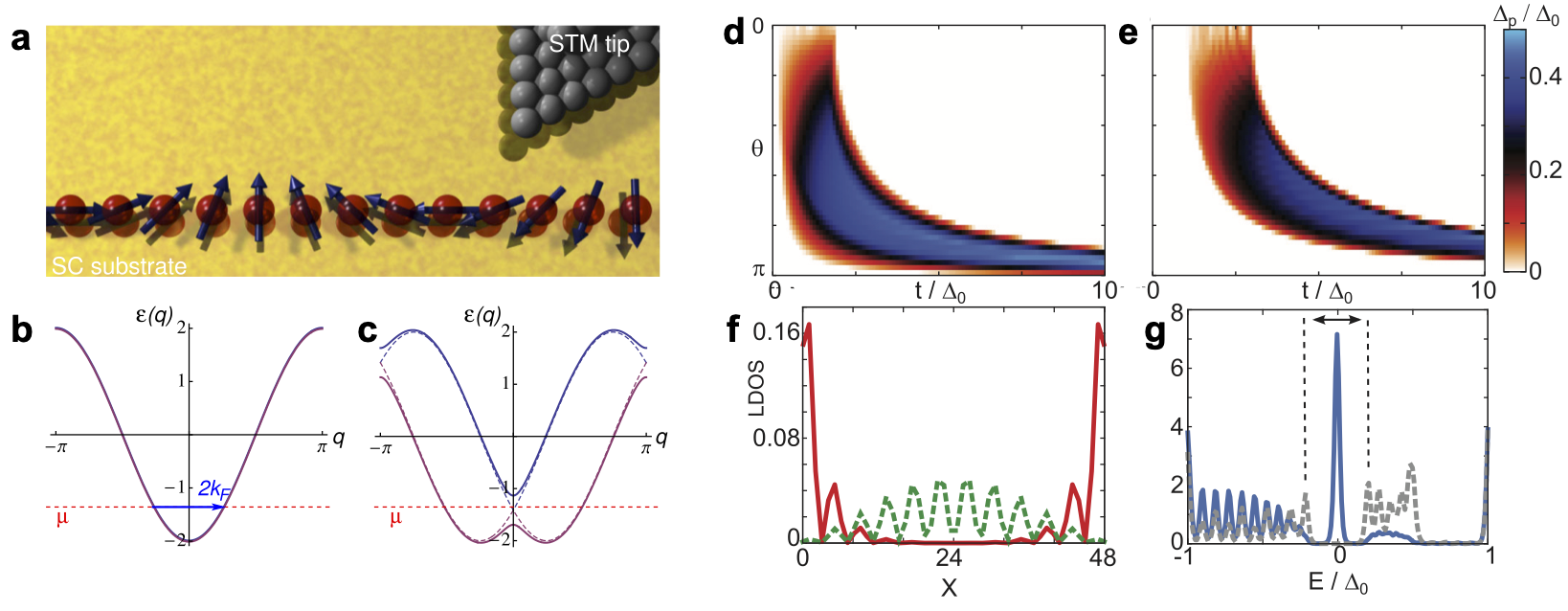}
\caption{\textbf{Helical Shiba chain}. \textbf{a} Chain of magnetic atoms forming a spin spiral on a superconducting substrate. \textbf{b} Normal-state spin-degenerate dispersion of free electrons in the absence of the magnetic moments. \textbf{c} Normal-state dispersion including magnetic moments: the two branches shift by momentum $G=2 k_F$, which is the natural wavevector of the spin spiral, and a gap at $q=0, \pi$ opens. [a-c taken from \cite{vazifeh-13prl206802}.] \textbf{d} Topological gap $\Delta_p$ (i.e., the effective minigap $\Delta_{\rm ind}$) of the helical Shiba chain as a function of spiral angle $\theta$ and hopping divided by the superconducting order parameter, $t/\Delta_0$.
Size of magnetic moments is $B/\Delta_0=3$ and \textbf{e} $B/\Delta_0=5$. \textbf{f} Example of LDOS vs.\ position $x$ for Majorana bound state (red solid) compared to the first excited state (green dashed). \textbf{g} Same as f but now for fixed position with varying energy: shown is the chain end (blue solid) and the middle of the chain (grey dashed).
The region indicated by the black arrow corresponds to the induced minigap $\Delta_{\rm ind} = \Delta_p$.
 [d-g taken from \cite{Nadj-PergePRB2013}.]}
\label{fig:th-helicalchain}
\end{figure}

The first paper explicitly mentioning multiple YSR states leading to YSR bands and topological superconductivity is Ref.\,\cite{Nadj-PergePRB2013}.
Here the idea of magnetic adatoms or impurities deposited on a superconducting substrate was explicated. The 1D array of magnetic atoms is embedded into a larger two-dimensional substrate; it is further assumed that the quantization axis of the magnetic atoms is rotating with position $x$ by the same angle $\theta$ between neighboring atoms\,\cite{Nadj-PergePRB2013}. This situation, later called {\it helical Shiba chain} is illustrated in Fig.\,\ref{fig:th-helicalchain}\,a. The generic Hamiltonian (only considering a 1D substrate) is given by
\begin{equation}
\label{eq:ham-helical-shiba}
H_{\rm S} = \sum_{i s} \left( t c_{i,s}^\dag c_{i+1,s}^\pd  - \frac{\mu}{2} c_{is}^\dag c_{is}^\pd +  {\rm H.c.}\right) + \sum_i \left( \Delta_0 c_{i,\up}^\dag c_{i,\dw}^\dag + {\rm H.c.}\right) + \sum_{i, s, s'} \left( \bs{B}_i \cdot \bs{\sigma} \right)_{ss'} c_{i,s}^\dag c_{i,s'}^\pd
\end{equation}
with the local superconducting $s$-wave order parameter, $\Delta_0 = \Delta_s={\rm const.}$; in principle, it should be treated self-consistently when coupled to the magnetic adatoms. $\bs{\sigma}$ denotes the vector of Pauli matrices, and $\bs{B}_i$ describes the magnetic moments of strength $|\bs{B}_i|=B$ which rotate with angle $\theta$ along the chain.

The dispersion relation of free particles is shown in Fig.\,\ref{fig:th-helicalchain}\,b. The spin spiral has two effects: it splits the previously spin-degenerate parabolas and opens a gap at $q=0$ and $\pi$. For topological superconductivity to emerge it is crucial that the Fermi energy or chemical potential is fine-tuned such that an odd number of bands cross the Fermi energy (see Fig.\,\ref{fig:th-helicalchain}\,c). Only when an odd number of crossings at the Fermi-energy is guaranteed, will the superconducting proximity effect induce the desired topological phase. We emphasize that further fine-tuning of the size of the magnetic moments of the adatoms (i.e., the size of $B$ in \eqref{eq:ham-helical-shiba}) is required for band-crossings to happen.

The phase diagrams depending on the ``helical angle'' $\theta$ and other parameters is carefully studied in Ref.\,\cite{Nadj-PergePRB2013}. In particular, it is emphasized that not every combination of $\theta$ and other parameters leads to the opening of a topological gap (see Fig.\,\ref{fig:th-helicalchain} d and e). The topological gap must be distinguished from the superconducting order parameter $\Delta_0$ which can be thought of as the bulk gap of the substrate superconductor. The topological gap, in contrast, is the induced minigap $\Delta_{\rm ind}$ which is responsible for the stability of the topological phase and the appearance of MZMs. We note that in the experimental systems, $\Delta_{\rm ind}$ is usually (much) smaller than the superconducting bulk gap. The required fine-tuning to stabilize a finite minigap suggests that not any magnetic chain with spin spiral will lead to topological superconductivity; instead an extensive search of various material combinations might be necessary.

In Fig.\,\ref{fig:th-helicalchain}\,a also the tip of an STM is shown which allows for local spectroscopy measurements. The topological phase can be nicely detected by local STM experiments: at the chain ends one expects a pronounced peak at zero energy in the $dI/dV$ data (or, similarly, in the LDOS), while there is no spectral weight in the middle of the chain; in contrast, higher-energy states do not have edge features but behave like ordinary bulk states (see Fig.\,\ref{fig:th-helicalchain}\,f and g). 

Around the same time, it was proposed that under generic conditions the pitch of the spin spiral that minimizes the free energy coincides with the one required to stabilize the topological phase. Such a proposal is intriguing as it indicates that the helical Shiba chain constitutes the case of a self-organized topological superconductor without the need for any parameter fine-tuning after all\,\cite{vazifeh-13prl206802}.
Shallow impurity states possess energies close to the band edges and tend to merge with the bulk states; topological superconductivity is less likely. In contrast, deep impurity states possess energies in the middle of the gap and are well separated from bulk states; in this regime, Shiba states start to overlap already for weak hybridization. A useful discussion of shallow and deep YSR impurities as well as YSR bands is given in Ref.\,\cite{PientkaPRB2013}. In addition, the authors show explicitly why the helical Shiba chain indeed leads to the physics of the Kitaev chain.
In Ref.\,\cite{kim-14prb060401} it was emphasized that the helical Shiba chain as an example of a magnet-superconductor hybrid system might not be stable under realistic conditions, i.e., upon inclusion of RKKY interactions between the magnetic adatoms as well as potential disorder. The authors argue further that by adding sufficiently strong Rashba spin-orbit coupling the persistance of the helical spin state can be guaranteed and, consequently, the existence of a topological phase.

As discussed later in the experimental part about non-collinear spin chains (Section\,\ref{experiment_1-1D-NC}), it is well-established that 
helical magnetic orders in spin chains can be realized in materials and even fabricated using bottom-up atomic chains\,\cite{SteinbrecherNAT-COMM2018}. In combination with a superconducting substrate, Fe/Re(0001) is so far the only experimental system for which a non-collinear spin order has been confirmed directly using
SP-STS measurements, and where signatures of topological superconductivity and Majorana modes were observed\,\cite{Kim2018,Schneider2020}, see Section\,\ref{experiment_1-1D-NC}. 

MSH chains with non-collinear spin spirals were more systematically studied in Ref.\,\cite{Beck2023b}. Fe chains on a Au monolayer on top of superconducting Nb(110) were build with the aim to induce stronger spin-orbit coupling. Ferromagnetic Fe chains measured with STS led to gapless YSR bands. In contrast, DFT calculations predicted that already a simple spiral configurations with 4-site unit cell should open a small minigap\,\cite{Beck2023b}.

\subsubsection{Ferromagnetic spin chains}\label{theory-1D-FM}

The pioneering experiments on self-assembled Fe chains on a superconducting Pb substrate, realizing a Shiba chain for first time, showed that this system features ferromagnetic order\,\cite{NadjPerge2014}, see Section~\ref{Sec:ExpMeth}. Despite not possessing a helical spin order, the presence of localized zero-energy states at the chain ends was reported together with the conclusion that the Fe/Pb(110) system realizes the topological phase of a Shiba chain. A simple explanation was given by earlier theory work\,\cite{braunecker-10prb045127} where the connection between helical and ferromagnetic order was pointed out. 
Performing a local gauge transformation from the helical order to a ferromagnetic alignment also transforms the kinetic terms in the Hamiltonian: the hopping acquires additional complex contributions which can be identified as anisotropic Rashba spin-orbit coupling; in addition, shifts in the chemical potential can be induced. These findings were later explicitly exercised in Refs.\,\cite{kjaergaard-12prb020503,martin-12prb144505,yang-16prb224505}. 

Remembering our original requirements (i)-(iii) at the beginning of Section\,\ref{theory-1D}, the ferromagnetic Shiba chain is in fact exactly matching these: while the helical Shiba chain combines (ii) and (iii) in its helical magnetic ordering, i.e., time-reversal breaking and SU(2) spin symmetry breaking, the ferromagnetic Shiba chain splits the requirements (ii) and (iii) into different contributions. The ferromagnetic alignment breaks time-reversal symmetry and Rasha spin-orbit coupling breaks the SU(2) symmetry. Two comments are in order: (a) The presence of Rashba spin-orbit coupling, isotropic or anisotropic, is crucial for stabilizing a topological phase and MZMs. At the surface there will always be inversion symmetry breaking and thus some Rashba spin-orbit coupling should be present, but it remains unclear if that is sufficient, or heavy elements are required to amplify the effect. (b) The ferromagnetic Shiba Hamiltonian now takes the form almost identical to the simple models of semiconductor-superconductor heterostructures\,\cite{lutchyn-10prl077001,oreg-10prl177002,brydon-15prb064505,hui-15sr8880} -- with the difference that the Zeeman term describes magnetic moments and not an externally applied field. Specifically, in the helical Shiba chain Hamiltonian Eq.\,\eqref{eq:ham-helical-shiba} the last term containing the helical magnetic field needs to be replaced by Rashba spin-orbit coupling and a constant ferromagnetic Zeeman term,
\begin{equation}
\label{eq:ham-FM-Shiba}
H_{\rm RSOC} + H_{\rm FM} = \sum_{i,s,s'} i \alpha c_{i+1,s}^\dag c_{i,s'}^\pd (\hat d_{ij} \times \bs{\sigma}_{ss'})\cdot \hat z  + {\rm H.c.} + \sum_{i,s,s'} \left(J\cdot \sigma_z\right)_{ss'} c_{i,s}^\dag c_{i,s'}^\pd
\end{equation}
where $\alpha$ is the amplitude of the Rashba spin-orbit coupling and $J$ is the strength of the ferromagnetic  moment. 
$\hat d_{ij}$ is a unit vector pointing from lattice site $i$ to lattice site $j$, $\hat d_{ij}= \bs{r}_j - \bs{r}_i$.
Here we assumed that the ferromagnetic moments with size $J$ are aligned out-of-plane which is crucial for the opening of a gap.

\begin{figure}[t!]
\centering
\includegraphics[width=\textwidth]{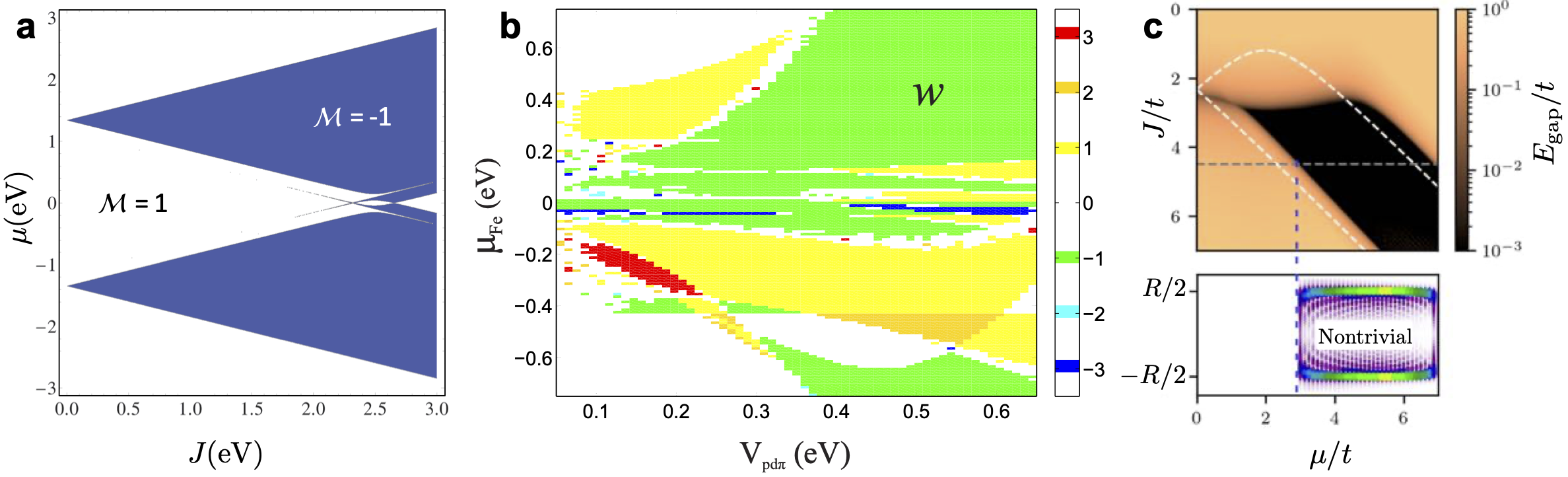}
\caption{\textbf{Ferromagnetic Shiba chain}. \textbf{a} Phase diagram of a straight Fe chain on superconducting substrate. The blue region indicates the topological superconducting phase. \textbf{b} Phase diagram showing the winding number for a linear Fe chain when an additional magnetic symmetry $M_T$ is present, see main text. [a-b taken from \cite{li-14prb235433}.] \textbf{c} Phase diagram of the simplest ferromagnetic Shiba chain: comparison of the 1D model (white dashed line) compared to a finite chain of length $R$ embedded into a larger two-dimensional substrate (black region indicates presence of zero-energy states). Zero-energy LDOS in chain direction is shown along the grey dashed line in the phase diagram. [c taken from \cite{crawford-20prb174510}.]}
\label{fig:th-FMchain}
\end{figure}

In the first paper where the ferromagnetic Shiba chain was studied in detail a Slater--Koster tight-binding model was derived applicable for transition-metals such as Fe\,\cite{li-14prb235433}.
Accounting for the $d$-electron character of the Fe atoms results in a multi-band model more involved than the minimal model in Eq.\,\ref{eq:ham-helical-shiba} and \ref{eq:ham-FM-Shiba}. It is shown that the atomic spin-orbit coupling for isolated Fe chains is not capable of opening gaps when superconductivity is induced via the proximity effect; Rashba spin-orbit coupling is found to be crucial for stabilizing topological superconductivity. An interesting finding for the regime of weak substrate coupling and long chains is that the proximity-induced superconducting gap on the chain is $\Delta_{\rm chain} \sim \Delta_s \alpha / J$, emphasizing the need of sizable spin-orbit coupling.
Properties of the Majorana zero modes for different chain-type geometries in the different regimes were investigated in detail\,\cite{li-14prb235433}. For instance, the topological phase diagram for straight transition-metal chains is shown in Fig.\,\ref{fig:th-FMchain}\,a ($\mathcal{M}=-1$ indicates the topological phase). It matches with the previously mentioned heuristic argument that the superconductivity is topological if the number of crossings of the normal-state bands is odd. For the physically realistic regime where the effective magnetic moment or {\it exchange splitting} $J$ is comparable to the bandwidth the system is almost always topological\,\cite{li-14prb235433}. 
Other studied geometries include a zigzag chain or chains embedded into the surface, which is chosen to be Pb(110) motivated by the  experiments\,\cite{NadjPerge2014}.

In addition, a variant of the Fe chain was studied with additional symmetry assumptions, where a magnetic symmetry $M_T$\,\cite{fang-14prl106401} (combined mirror and time-reversal symmetry) is present if the chain is straight, free of disorder and magnetic moments are fully in-plane. Then the symmetry classification\,\cite{Altland1997,Ryu2010,Kitaev2009} changes from class D (with the $\mathbb{Z}_2$-valued Kitaev number $\mathcal{M}$) to class BDI (with a $\mathbb{Z}$-valued winding number $w$) which offers the possibility to protect even multiple Majorana end states\,\cite{li-14prb235433}. The Majorana number $\nu$\,\cite{kitaev01pu131}, which defines the invariant $\mathcal{M} = (-1)^\nu$, is related to the winding number $w$ via $\nu = w\  {\rm mod}\ 2$.
The corresponding phase diagram is shown in Fig.\,\ref{fig:th-FMchain}\,b as a function of the chemical potential of the Fe adatoms and the hybridization $V_{pd\pi}$, the only two parameters not determined from ab initio calculations.

Subsequent studies confirmed the presence of topological superconducting phases for the ferromagnetic Shiba chain models\,\cite{hui-15sr8880,peng-15prl106801,dumitrescuc-15prb094505,sau-15prl127003,crawford-20prb174510}. In particular, it was shown how the phase diagram of the simple one-orbital ferromagnetic Shiba chain with length $R$, depending on magnetic moment $J/t$ and chemical potential $\mu/t$, is only altered in some regions when the chain is embedded into a two-dimensional substrate\,\cite{crawford-20prb174510}, see Fig.\,\ref{fig:th-FMchain}\,c.

With the experimental success also critical voices were raised. While the Shiba chain platform offers the general advantage of conducting local probe measurements thanks to the availability of STM techniques, it was criticized that the presence of a zero-energy peak at chain ends is not sufficient to prove the Majorana character\,\cite{dumitrescuc-15prb094505,sau-15prl127003} (see also Section~\ref{Sec:ExpMeth}). For instance, Ref.\,\cite{dumitrescuc-15prb094505} concludes that the experimentally observed small gap\,\cite{NadjPerge2014} in combination to the temperature the experiment is performed at, should lead to a spreading of spectral weight of the Majorana zero modes exceeding the spectral gap; with other word, one expects a rather weak and smeared out zero-energy peak, in contrast to the clearly localized MZMs observed in the Fe/Pb(110) experiments\,\cite{NadjPerge2014}. The combination of a small energy gap and comparatively large temperatures along with the issue of surprisingly short Majorana localization length were also mentioned in this commentary\,\cite{PLeeJournalClub}.

A theory answer to this challenge was provided in Ref.\,\cite{peng-15prl106801}: the authors show that the coupling of the superconductor to the magnetic adatoms leads to a renormalization of the Fermi velocity. Crucially, this directly affects the Majorana localization length and provides an answer as to why Majorana zero modes appear to be so localized in the first experiments (while one might naively expect this localization length to exceed the superconducting coherence length). A nice discussion in Ref.\,\cite{dassarma-15njp075001} confirms these findings about the Majorana localization, and points out why the same effect will be absent in the semiconductor-superconductor heterostructures due to the different transverse confinement sizes. For the semiconductor-based platform the Majorana localization length is predicted to be of more than a magnitude larger. Also Ref.\,\cite{theiler-19prb214504} emphasized the strong hybridization between MBS and in-gap YSR states.

\begin{figure}[t!]
\centering
\includegraphics[width=0.5\textwidth]{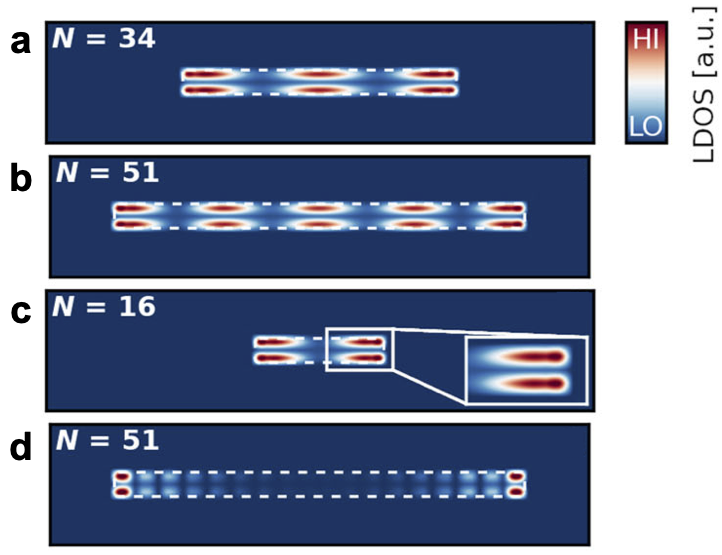}
\caption{\textbf{Zero-energy side feature states}. \textbf{a} Example for chain length $L=34$ with parameters chosen to describe Mn[001] on Nb(110). \textbf{b} Example for chain length $L=51$ with parameters chosen to describe Mn[001] on Nb(110). \textbf{c} Example for chain length $L=16$ with parameters chosen to describe Fe[001] on Nb(110). 
 \textbf{d} Example for chain length $L=51$ with parameters chosen to reveal a spatial distribution of spectral weight reminiscent of the double-eye feature observed in Fe/Pb(110)\,\cite{Feldman2017}. 
[a-d taken from \cite{Crawford2022}.]}
\label{fig:th-sidefeatures}
\end{figure}

Both Refs.\,\cite{li-14prb235433} and \cite{dumitrescuc-15prb094505} had already emphasized the complication by using transition metals as adatoms placed on superconducting substrates. The experiments on Mn chains on Nb(110) substrate lifted the complexity of material modeling to an even higher level. While experiments could show the existence of a topological gap\,\cite{Schneider2021b} (see Fig.\,\ref{fig:1DExpMethods}), the spatial distribution of the zero-energy  states within the minigap did not look like anything expected for a MBS: significant oscillatory spectral weight in the chain middle along with a reduction of spectral weight along the line of adatoms; instead, spectral weight was pushed to the sides (dubbed {\it side features} (see Fig.\,\ref{fig:1DExpMethods}). Only by combining ab initio calculations with extensive tight-binding BdG simulations\,\cite{Crawford2022}, zero-energy states in the topological superconducting phase could be identified which exhibit a spatial profile which is essentially identical to the experiments\,\cite{Schneider2021b,Crawford2022}. Within a simpler, effective four-orbital model it is possible to derive the same side feature states, see Fig.\,\ref{fig:th-sidefeatures}\,a and b. These findings were also applicable to the related hybrid structure consisting of Fe chains on a Nb(110) substrate\,\cite{Crawford2022}, see Fig.\,\ref{fig:th-sidefeatures}\,c. Most interestingly, by slightly increasing the superconducting order parameter, considered as a phenomenological parameter, the topological gap size can be slightly increased; as a consequence, the spectral weight of the associated Majorana modes is suppressed in the chain middle, and the resulting spatial profile appears to be very similar\,\cite{Crawford2022} to the previously observed double-eye feature in the Fe/Pb(110) system\,\cite{Feldman2017}, as shown in Fig.\,\ref{fig:th-sidefeatures}\,d. 

The importance of insights from ab initio calculations was already mentioned; there are more first-principle calculations for 3d transition metals Mn, Cr, Fe, Co on superconducting Pb\,\cite{kobialka-20prb205143}. In a different work, variants of such systems involving Nb-substrates were investigated in Ref.\,\cite{nyari-23prb134512}, leading to both ferromagnetic and helical Shiba chains, some of them predicted to be topological superconductors. Moreover, ab initio calculations for Fe chains on a Au monolayer on Nb(110) were investigated by combining DFT and BdG methods\,\cite{Beck2023b}. Magnetic ordering between Mn, Fe and Cr atoms on Nb(110) was studied using DFT in Ref.\,\cite{crawford-23prb075410}.

A few final comments about theoretical work on one-dimensional magnet-superconductor hybrid systems, i.e., Shiba chains are in order. The effect of disorder on the phase diagram of both helical and ferromagnetic Shiba chains was studied in Ref.\,\cite{weststrom-16prb104519}. Magnetic and potential impurities were also investigated in \cite{crawford-20prb174510}. The importance of spin-resolved measurements was explored in Ref.\,\cite{li-18prb125119}, as such measurements can distinguish trivial zero-energy states from MZMs. Topological superconductivity in ferromagnetic atom chains beyond the deep-impurity regime were investigated in Ref.\,\cite{poyhonen-16prb014517}. Instead of using conventional $s$-wave superconductors, Ref.\,\cite{crawford-20prb174510} proposed to deposit magnetic adatoms on unconventional superconductors (such as Fe-based extended $s$-wave superconductors with their high transition temperatures) which works equally well.

These combined efforts of advanced ab initio methods and involved model building demonstrate that a previously unexpected fine-tuning of various material parameters is essential to stabilize a TSC phase with a minigap large enough to localize Majorana modes.

\subsubsection{Antiferromagnetic spin chains}\label{theory-1D-AFM}

Compared to other Shiba chain systems, antiferromagnetic adatom chains on superconducting substrates have attracted little attention for a long time. The first mention of such alternating spin alignment was in the early helical Shiba chain paper\,\cite{Nadj-PergePRB2013} where ferromagnet and antiferromagnet are considered as special cases of the spin spiral ($\theta=0, \pi$). In fact, it is emphasized that from a topological perspective these two cases would support the largest parameter space for topological superconductivity. However, the spectral gap remains closed for both cases, which can be seen in Fig.\,\ref{fig:th-helicalchain}\,d and e. Two early works addressed antiferromagnetic Shiba chains\,\cite{heimes-14prb060507,heimes-15njp023051}. Either by adding a Zeeman field and supercurrents\,\cite{heimes-14prb060507} or including Rashba spin-orbit coupling\,\cite{heimes-15njp023051} it is possible to stabilize topological phases even for the antiferromagnetic alignment. From an experimental perspective, antiferromagnetic ordering is very well possible and material realizations are known (see Section\,\ref{experiment_1-1D-AFM}); also from a theory perspective, classical spin models can prefer antiparallel alignment, in particular, when Ising anisotropies are large\,\cite{heimes-15njp023051}. Performing more involved Monte Carlo simulations can also lead to parameter regimes with antiferromagnetic ordering\,\cite{neuhaus-22prb165415}. Going beyond classical spins, ab initio calculations showed that ferromagnetic vs.\ antiferromagnetic alignment can depend on the type of adatoms as well as the crystallographic chain directions\,\cite{crawford-23prb075410}, in line with recent experimental findings\,\cite{Schneider2021,Schneider2022,Schneider2023}.

A key difference between ferromagnetic and antiferromagnetic Shiba chains is the role of Rashba spin-orbit coupling. While a requirement for both systems, in the simplest one-orbital models, the Rashba amplitude enters the superconducting energy dispersion explicitly only for the antiferromagnetic case\,\cite{crawford-23prb075410}. As a consequence, some minimal Rashba spin-orbit coupling must be provided to stabilize the topological phase\,\cite{heimes-15njp023051}. In Ref.\,\cite{crawford-23prb075410} it is shown how the topological parameter space shrinks with shrinking Rashba spin-orbit coupling, in contrast to the ferromagnetic case. Extending from purely 1D one-orbital models to a chain embedded on a 2D substrate as well as describing a multi-orbital model (in an attempt to address chains on a Nb(110) substrate), these different models differ in their spectral and topological properties; they all have in common, however, that they support topological superconducting phases.

The Nb(110) superconducting substrate is a particular interesting case because for Mn, close-packed chains in [001] and [1$\bar{1}$0] directions were shown to be ferromagnetic\,\cite{Schneider2021b,Schneider2022} while those in [1$\bar{1}$1] direction are antiferromagnetic\,\cite{Schneider2023} (see Table\,\ref{table:spinchains}). Thinking about future braiding experiments involving T-junctions\,\cite{AliceaNAT-PHYS2011}, one might imagine that a material which orders in one chain direction ferromagnetically and in another one antiferrromagnetically, is topologically non-trivial for both chain directions. For this case it is interesting to see that the topological regime can persist even when the topological phase of a ferromagnetic chain segment is connected with the topological phase of an antiferromagnetic chain segment, forming a hybrid ferromagnetic-antiferromagnetic Shiba chain network\,\cite{crawford-23prb075410}. We note that antiferromagnetic ordering was also investigated in the context of the Rashba nanowires\,\cite{kobialka-21prb125110}.

\subsubsection{Quantum spin chains}\label{theory-1D-quantum}

\begin{figure}[h!]
\centering
\includegraphics[width=\textwidth]{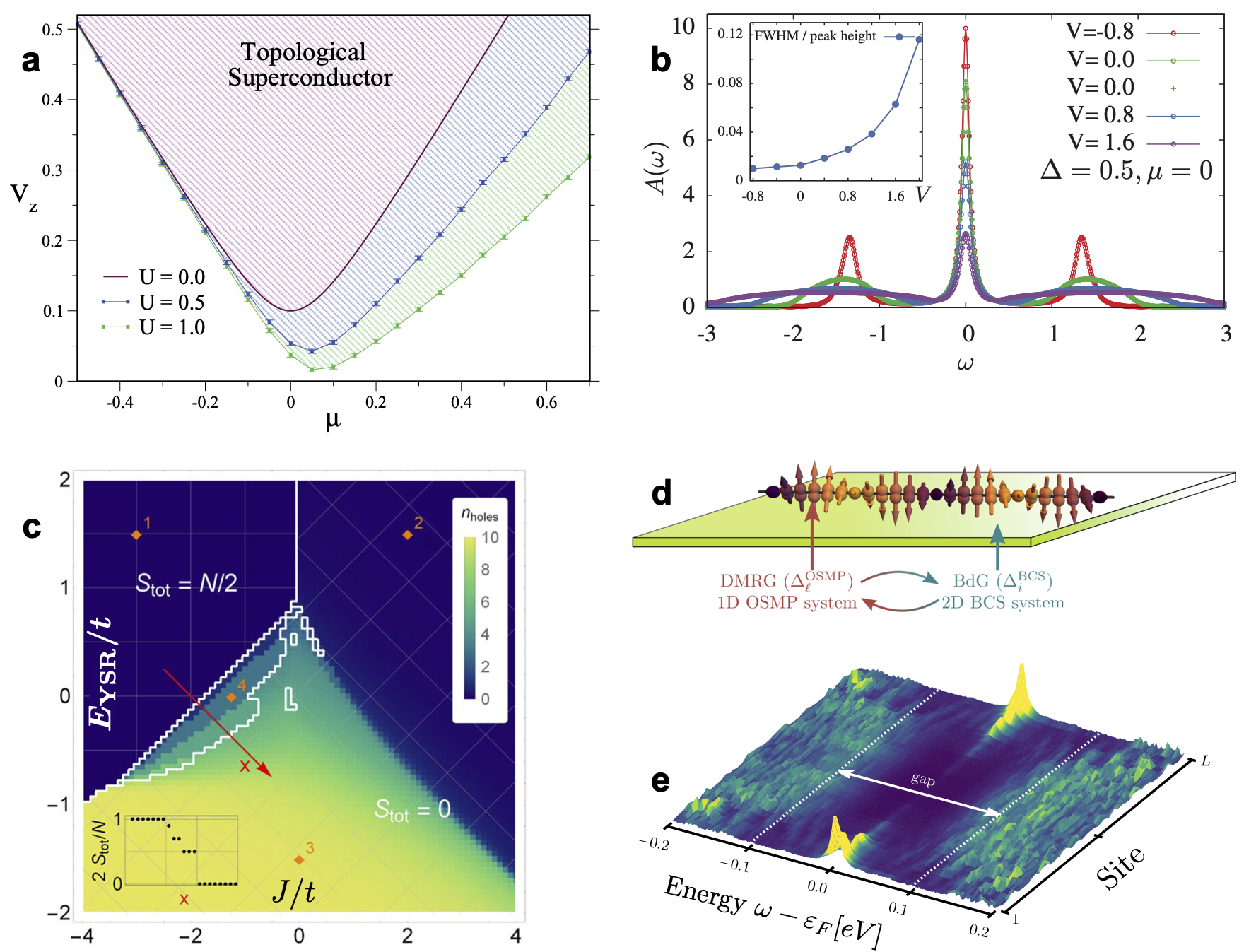}
\caption{\textbf{Quantum spin chains on superconducting substrates}. \textbf{a} Small but finite local Coulomb interactions $U$, added to a model similar to the ferromagnetic Shiba chain, broaden the topological phase. [a taken from\,\cite{stoudenmire-11prb014503}.]
\textbf{b} Single-particle spectral function $A(\omega)$ computed for various values of nearest-neighbor repulsion $V$ for the interacting Kitaev chain \eqref{eq:ham-kitaev} (parameters $\Delta_p/t=0.5$ and $\mu=0$), revealing a suppression of the peak of the MZM with increasing interaction strength. [b taken from\,\cite{thomale-13prb161103}.]
\textbf{c} Phase diagram of the effective $t-J$ model for $N=10$ adatoms including superconducting pairing studied in Ref.\,\cite{steiner-22prl036801}, depending on the YSR-energy $E_{\rm YSR}/t$ (which acts as an effective chemical potential) and the exchange coupling $J/t$. Number of holes $n_{\rm holes}$ and the total spin $S_{\rm tot}$ characterize the phase space. The small region framed by the white lines is a metallic ferromagnet; when adding spin-orbit coupling it turns into a TSC with MZMs [c taken from\,\cite{steiner-22prl036801}.]
\textbf{d} Illustration of hybrid method which combines density-matrix renormalization group to treat the spin-spin interactions quantum-mechanically, while a mean-field Bogoliubov--de Gennes formalism solves the 2D superconducting substrate.
\textbf{e} Emergence of zero-energy edge modes when the spins form a spin-spiral due to the interactions. [d and e taken from\,\cite{herbrych-21nc2955}.]
}
\label{fig:th-quantumchains}
\end{figure}

When considering clusters of magnetic impurities such as Shiba chains and lattices, it is often argued that the involved magnetic moments of the adatoms are large and behave mostly classically. That would allow for the convenient treatment of Shiba systems as free-fermion models, as done in the previous subsections of this Section\,\ref{theory}.
As we have seen in previous parts of this Review, there are works in the literature where RKKY interactions between the spins have been considered\,\cite{KlinovajaPRL2013}. In fact, it is common sense that it might not be justified to ignore the quantum mechanical character of the magnetic moments and interactions are generally present. The ``gap'' between such thoughts and the literature is of course explainable by the technical challenges to include quantum many-body interactions in a system which also features a superconducting proximity effect.

There are a few but notable works where spin-spin interactions or, more generally, Coulomb interactions between the electrons, are considered and treated quantum mechanically. The two first papers addressing this issue were published at the same time\,\cite{gangadharaiah-11prl036801,stoudenmire-11prb014503}; they applied Jordan-Wigner and renormalization group techniques or density-matrix renormalization group, bosonization and Hartree-Fock methods, respectively.
Ref.\,\cite{gangadharaiah-11prl036801} studied the effect of a nearest-neigbhor repulsion term for the spinless electrons of Kitaev's model\,\cite{kitaev01pu131}. While the interactions generally decrease the topological gap size rather quickly, it is shown that for weak interactions the gap renormalization is nonuniversal and that there are regimes supporting MZMs. Ref.\,\cite{stoudenmire-11prb014503} took a slightly different approach and considered the spinful model\,\cite{lutchyn-10prl077001,oreg-10prl177002}, originally proposed for the seminconductor nano-wires, and complemented it with a local Coulomb interaction term. As mentioned previously, the semiconductor nano-wire Hamiltonian is essentially that of a ferromagnetic Shiba chain, now with additional Hubbard interaction. Small but finite Hubbard interactions were found to not only stabilize but even broaden the topological phase\,\cite{stoudenmire-11prb014503}, see Fig.\,\ref{fig:th-quantumchains}\,a. By using other methods such as Hartree-Fock and bosonization, also other limits of the same model were thoroughly analyzed. It was further noticed that the presence of interactions allows to reduce the size of magnetic field, leading to the idea, that a topological superconductor model could be realized without any need of a magnetic field term (regardless of its origin, external field or magnetic moments). It was shown that a ferromagnetic Heisenberg interaction term accomplishes exactly that: topological superconductivity without a Zeeman field\,\cite{stoudenmire-11prb014503}. From the perspective of a ferromagnetic Shiba chain this is of course not surprising at all. What matters is the ferromagnetic alignment of the magnetic moments, i.e., the time-reversal symmetry breaking, and not the underlying mechanism. 

The role of interactions was further studied with regard to the zero-energy peak observable either in global or local transport experiments\,\cite{thomale-13prb161103}: in the presence of interactions, the single-particle quantity LDOS becomes ill-defined and is usually replaced by the single-particle spectral function. It was shown how nearest-neighbor interactions $V$ can broaden or sharpen the zero-energy peak (see Fig.\,\ref{fig:th-quantumchains} b), explicated for the interacting Kitaev chain using the density-matrix renormalization group.
Other notable work addressing the role of interactions in 1D topological superconducting systems are Refs.\,\cite{lutchyn-11prb214528,fidkowski-12prb245121}. Even the braiding dynamics of MZMs using exact diagonalization (specifically an $S$ gate) was studied in the presence of interactions\,\cite{sekania2017}. Only small attractive or repulsive interactions were found to play a stabilizing role for the braiding operation.

In a recent work, a chain of magnetic impurities were treated involving many-body interactions with one substrate site per impurity site\,\cite{steiner-22prl036801}. While Coulomb interactions might be the reasonable extension to an interacting model for the case of the semiconductor nanowires, for Shiba physics the inclusion of the Kondo term as well as additional Heisenberg-type RKKY interactions between the magnetic impurities is appropriate. 
By focussing on the physically relevant limit, the authors project to a subspace excluding doubly occupied sites and obtain effectively an extended 1D $t-J$ model which is solved with exact diagonalization for chain length $N=10$. The phase diagram (Fig.\,\ref{fig:th-quantumchains}\,c) is significantly altered compared to the classical case, highlighting the role of quantum magnetism in YSR chains. The authors further show that the metallic ferromagnet phase (small region surrounded by white lines in Fig.\,\ref{fig:th-quantumchains}\,c) turns into a topological superconductor upon inclusion of Rashba spin-orbit coupling and $z$-anisotropic spin exchange\,\cite{steiner-22prl036801}.
While this study is limited to very short chains (since exact diagonalization was used), the results are nevertheless concerning, in particular, when speculating about chains built from $4f$ adatoms  such as Gd\,\cite{Wang2023}.

Another interesting work considered so-called orbital-selective Mott insulators, where Mott-localized electrons in one orbital coexist with itinerant electrons in the remaining orbitals\,\cite{herbrych-21nc2955}. An appropriate model is the generalized Kondo-Heisenberg Hamiltonian (similar to what was studied in \cite{steiner-22prl036801}), also containing Hubbard interactions. The authors used a hybrid approach involving density-matrix renormalization group and Bogoliubov--de Gennes equations, as illustrated in Fig.\,\ref{fig:th-quantumchains}\,d, which allows to incorporate the superconducting substrate even as a 2D system. When interactions are strong enough, such an orbital-selective Mott insulator eventually forms an exotic magnetic order called block-spiral phase, which leads to a topological superconducting phase in the presence of proximity-induced superconductivity. The corresponding zero-energy Majorana states are shown in Fig.\,\ref{fig:th-quantumchains}\,e. While this system does not constitute a Shiba chain, we acknowledge that the physical mechanism is very similar. It might be an exciting avenue to proximitize strongly correlated electron systems with conventional superconductors in the search for Majorana modes.

\subsection{Two-dimensional MSH systems}\label{theory-2D}

The Hamiltonian for two dimensional (2D) MSH systems, consisting of a monolayer (or 2D islands) of magnetic adatoms placed on the surface of an $s$-wave superconductor is given by 
\begin{align}
\mathcal{H} =& \; -t_e \sum_{{\bf r}, {\bf r}', \sigma} c^\dagger_{{\bf r}, \sigma} c_{{\bf r}', \sigma} - \mu \sum_{{\bf r}, \sigma} c^\dagger_{{\bf r}, \sigma} c_{{\bf r}, \sigma} + \mathrm{i} \alpha \sum_{{\bf r}, {\bm \delta }, \sigma, \sigma^\prime} c^\dagger_{{\bf r}, \sigma}  \left({\bm \delta} \times \boldsymbol{\sigma} \right)^z_{\sigma, \sigma^\prime}  c_{{\bf r} + {\bm \delta}, \sigma^\prime} \nonumber \\
    & + \Delta \sum_{{\bf r}} \left( c^\dagger_{{\bf r}, \uparrow} c^\dagger_{{\bf r}, \downarrow} + c_{{\bf r}, \downarrow} c_{{\bf r}, \uparrow} \right) +  {\sum_{{\bf R} , \alpha, \beta}} c^\dagger_{{\bf R}, \sigma} \left[J {\bf S}_{\bf R} \cdot \boldsymbol{\sigma} \right]_{\sigma,\sigma^\prime} c_{{\bf R}, \sigma^\prime} \; .
    \label{eq:H}
\end{align}
Here, the operator $c^\dagger_{{\bf r}, \sigma}$ creates an electron with spin $\sigma$ at site ${\bf r}$, $t_e$ is the nearest-neighbor hopping amplitude on a 2D square lattice, $\mu$ is the chemical potential, $\alpha$ is the Rashba spin-orbit coupling between nearest-neighbor sites ${\bf r}$ and ${\bf r} +{\bm \delta}$, $\boldsymbol{\sigma}$ is the vector of Pauli matrices, and $\Delta$ is the $s$-wave superconducting order parameter. The last term in Eq.(\ref{eq:H}) describes the coupling between the magnetic adatoms with spin ${\bf S}_{\bf R}$ of magnitude $S$ at site ${\bf R}$ and the conduction electrons, with exchange coupling $J$. The magnetic structure of the MSH system, such as ferromagnetic, antiferromagnetic, etc. is encoded in the spatial form of ${\bf S}_{\bf R}$. Note that the lattice of magnetic adatoms can be the same as that of the underlying $s$-wave superconductor (in which case we refer to it as a dense lattice), or a sublattice thereof (also refered to as a sparse lattice), or it can be finite, describing a 2D island of magnetic adatoms. Due to the hard superconducting gap, which suppresses Kondo screening \cite{Balatsky2006,Heinrich2018}, we can consider the spins of the magnetic adatoms to be classical in nature.

\subsubsection{Two-dimensional ferromagnetic MSH systems}\label{theory-2D-FM}

\begin{figure}[H]
\centering
\includegraphics[width=0.75\textwidth]{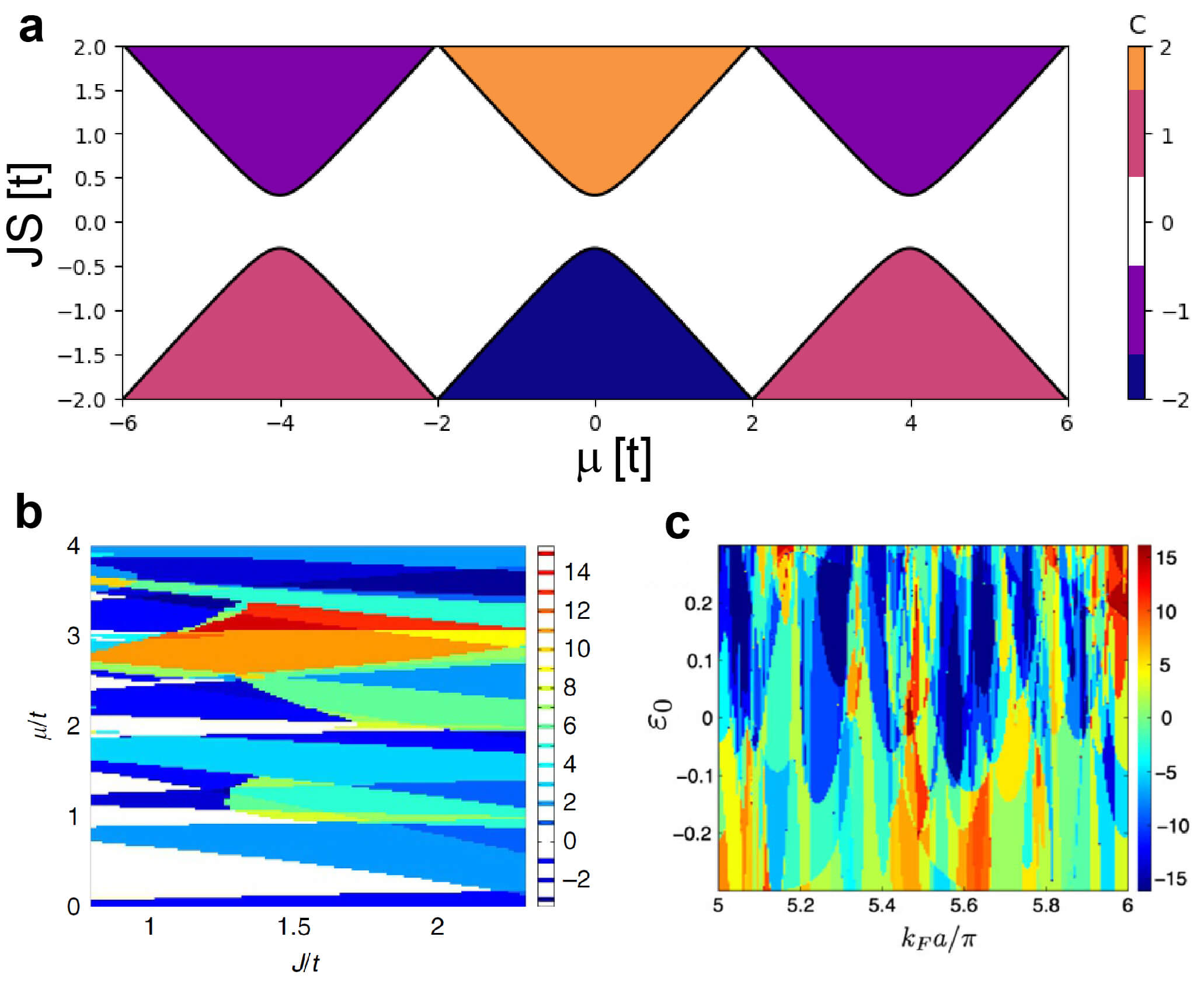}
\caption{Topological phase diagram of a ferromagnetic MSH system in the $(\mu,JS)$-plane for \textbf{a} a dense magnetic lattice, and {\bf b},{\bf c} sparse magnetic lattices. Panels {\bf b} and {\bf c} are adapted from Refs.\cite{Rontynen2015,Li2016}, respectively.}
\label{fig:FM_PD}
\end{figure}

We begin by considering MSH systems in which the monolayer of magnetic adatoms is ferromagnetically ordered. Since the time reversal symmetry is broken in such a system, it belongs to the Altland-Zirnbauer class D~\cite{Altland1997, Kitaev2009, Ryu2010}, exhibiting strong topological superconducting phases that are characterized by the Chern number $C$ as their topological invariant. The Chern number is computed via \cite{Kitaev2009, Ryu2010, Avron1983} 
\begin{align}
 C & =  \frac{1}{2 \pi  \mathrm{i}} \int_{\text{BZ}} d^2k \, \mathrm{Tr} \, ( P_{\bf{k}} [ \partial_{k_x} P_{\bf{k}}, \partial_{k_y} P_{\bf{k}} ] )  \nonumber \\
 P_{\bf{k}} & = \sum_{E_n({\bf{k}}) < 0} |\Psi_n({\bf{k}}) \rangle \langle \Psi_n({\bf{k}})|,
 \label{eq:C}
\end{align}
where $E_n({\bf{k}})$ and $|\Psi_n({\bf{k}}) \rangle$ are the eigenenergies and the eigenvectors of the Hamiltonian in Eq.~(\ref{eq:H}). The index $n$ and the trace both run over all degrees of freedom, such as the spin, magnetic sublattice, and Nambu degrees of freedom.  $P_{\bf{k}}$ is the projector onto the occupied states. For an MSH system with a square lattice symmetry, where every lattice site of the superconducting surface is covered by a magnetic adatom, the topological phase diagram contains only two topological phases with $C=\pm 1,\pm 2$ in the ($\mu,JS$)-plane, as shown in Fig.~\ref{fig:FM_PD}a. 
Note that a change in the magnetization direction as reflected in $JS \rightarrow -JS$ leads to a sign change in the topological invariant, which reflects a change in the chirality of the edge states \cite{Rachel2017}. 
Any phase transition in which the Chern number changes is accompanied by a gap closing at time-reversal invariant momentum points in the Brillouin zone, reflecting the bulk-boundary correspondence. The phase transition line between the trivial $C=0$ and (i) the $C=\pm 2$ phase is accompanied by gap closings at ${\bf k}=(0,\pi)$) (and symmetry related points), and is described by the relation
\begin{align}
    \mu = \pm \sqrt{(JS)^2-\Delta^2} \ , 
\end{align}
and (ii)  the $C=\pm 1$ phases is accompanied by gap closings at ${\bf k}=(\pm \pi,\pm \pi)$ for $\mu<0$ and at ${\bf k}=(0,0)$ for $\mu>0$, and described by the relation
\begin{align}
    \mu = \pm 4t \pm \sqrt{(JS)^2-\Delta^2} \ .
\end{align}
The phase diagram shows a richer variety of strong topological phases for sparser, but still ordered ferromagnetic coverings \cite{Rontynen2015,Li2016}, such as the $(2 \times 2)$ or $(3 \times 3)$-coverings \cite{Li2016}. Due to the increased unit cell in sparser systems that results in the back-folding of the electronic structure into the reduced BZ, and thus an increase in the number of bands, the corresponding topological phase diagram shows phases with higher Chern numbers, as shown in Figs.~\ref{fig:FM_PD}b and c \cite{Rontynen2015,Li2016}.

Due to the bulk-boundary correspondence, MSH systems possess edge modes at any interface where the Chern number changes by $\Delta C \not = 0$. Each interface possesses $|\Delta C|$ edge modes, which are chiral in nature, traverse the superconducting gap, and are localized at the edge \cite{Rontynen2015,Rachel2017}. These edge modes can be best studied for MSH systems possessing two parallel, straight edges separating regions with and without  magnetic adatoms (such an arrangement is referred to as a ribbon geometry). In such systems, the translational symmetry along the edge allows one to compute the electronic dispersion as a function of the momentum parallel to the edge. 
\begin{figure}[H]
\centering
\includegraphics[width=\textwidth]{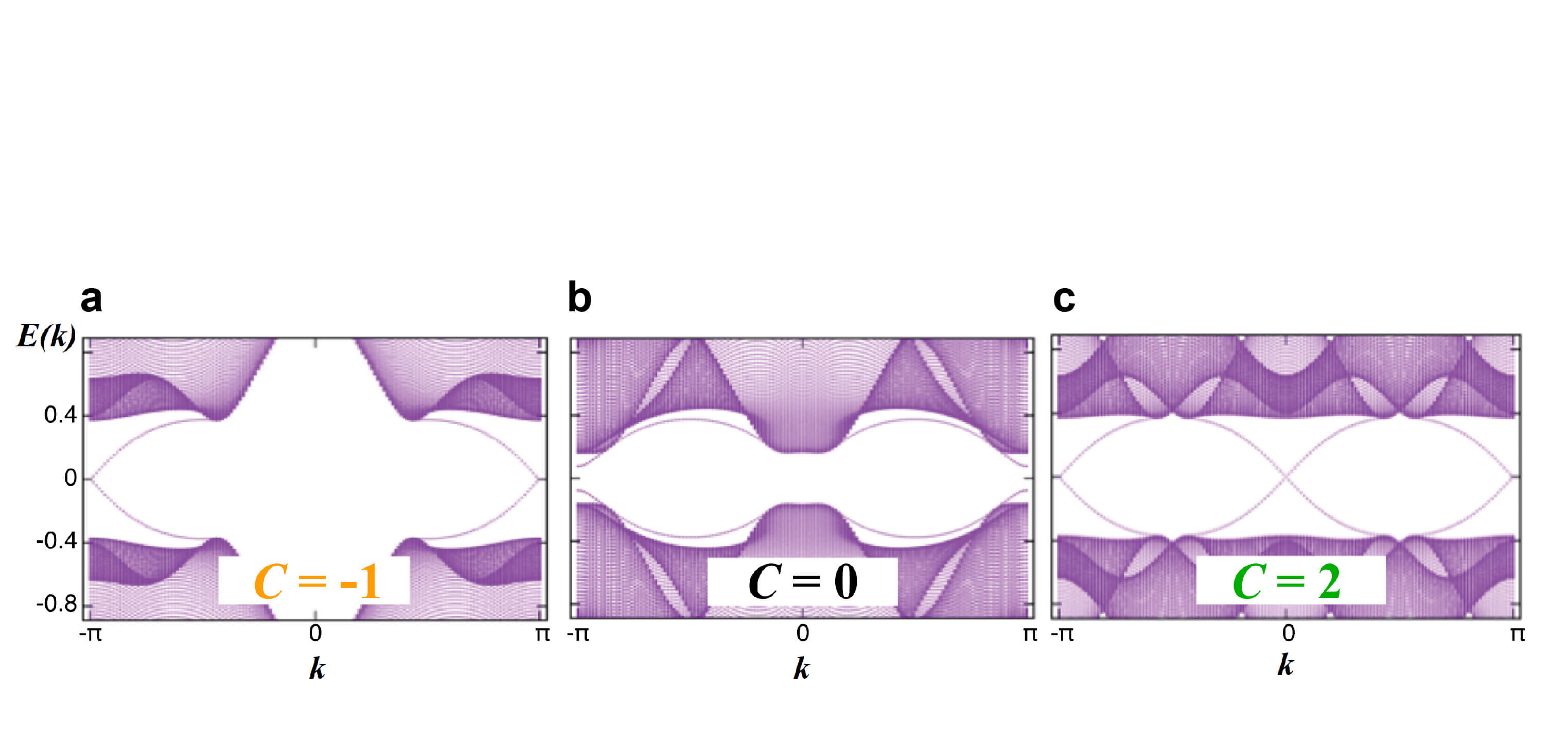}
\caption{Electronic dispersion of an MSH system in a ribbon geometry as a function of momentum along the edge. Here, the region covered by magnetic adatoms is in the \textbf{a} $C=-1$, \textbf{b} $C=0$, or \textbf{c} $C=+2$ phase. Figure adapted from Ref.~\cite{Rachel2017}.}
\label{fig:Edge}
\end{figure}
This is illustrated in Fig.~\ref{fig:Edge}, where the dispersion for an MSH systems in a ribbon geometry is shown for the three different cases where the region of the ribbon systems that is covered by magnetic adatoms is either in the \textbf{a} $C=-1$, \textbf{b} $C=0$, or \textbf{c} $C=+2$ phase (the region not covered by magnetic adatom is topologically trivial and thus in the $C=0$ phase). As expected, when the system is in the $C=-1$ ($C=+2$) phase, there are $|\Delta C| = 1$ ($|\Delta C| = 2$) Majorana edge modes per edge that traverse the superconducting gap and possess a linear dispersion around zero energy. In contrast, when the MSH system is in the $C=0$, no topological edge modes exist.

\begin{figure}[H]
\centering
\includegraphics[width=\textwidth]{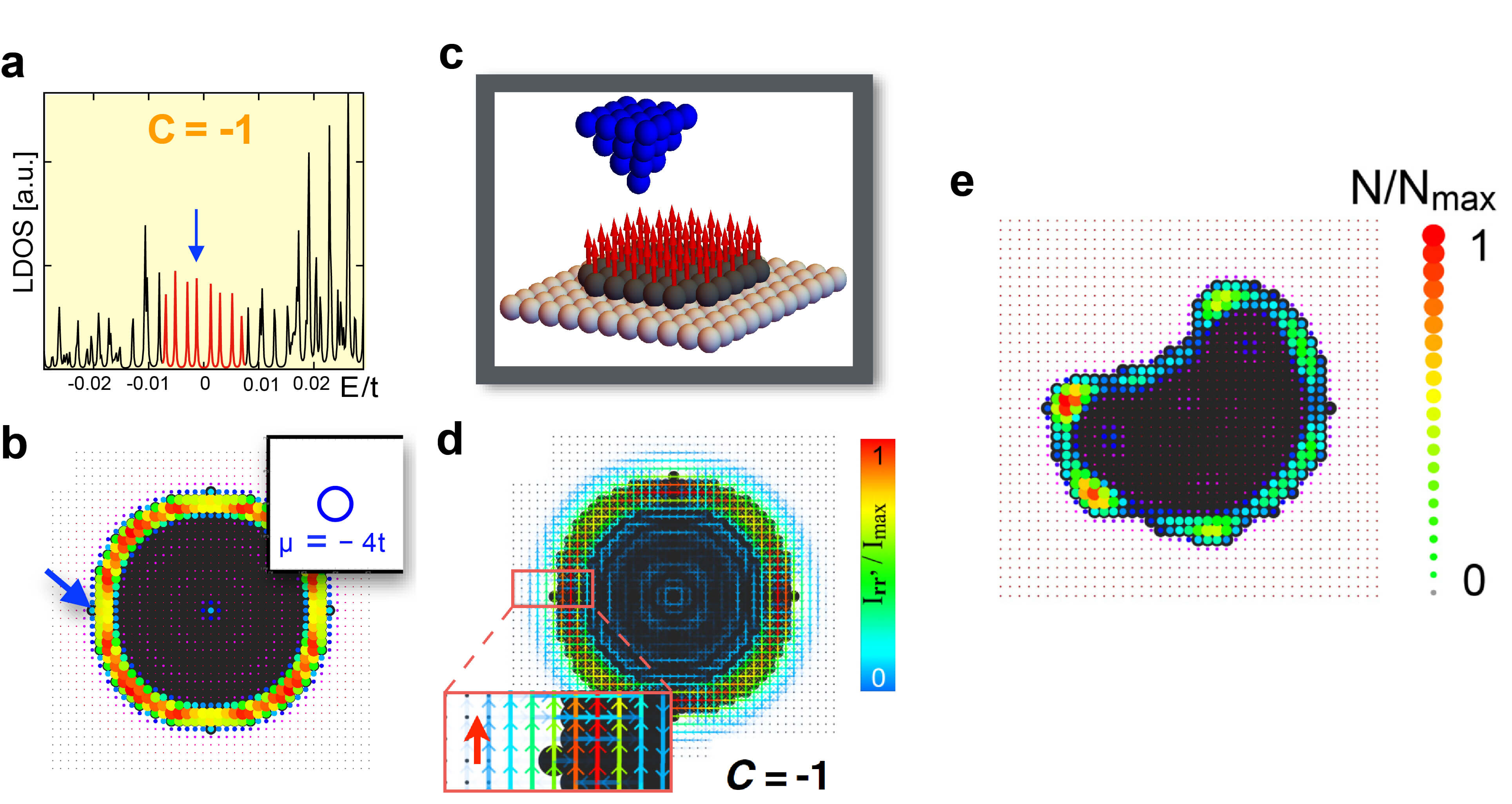}
\caption{\textbf{a} LDOS as a function of energy at an MSH island's edge (see panel \textbf{c}), exhibiting nearly equally spaced low energy states reflecting the existence of topologically protected Majorana edge modes. \textbf{b} LDOS of the lowest energy Majorana edge mode, with the spectral weight confined near the island's edge. \textbf{c} Schematic of electron tunneling from an STM tip into the Majorana edge modes of an MSH island. \textbf{d} Spatial profile of the supercurrent carried by Majorana edge modes, whose chirality reflects the sign of $C$. \textbf{e} LDOS of the lowest energy Majorana edge mode for a strongly disordered island, with the spectral weight still confined near the ferromagnetic island's edge. Figure adapted from Ref.\cite{Rachel2017}.}
\label{fig:Island}
\end{figure}

While ribbon geometries allow one to compute the momentum-dependent dispersion of Majorana edge modes, they are experimentally very difficult to realize. Experimentally more relevant are MSH islands that are often naturally occurring due to self-assembly, as in the case of Fe monolayer islands on Re(0001)\cite{Palacio-MoralesSCI-ADV2019}. 
The existence of edge modes for such finite size islands was investigated in Ref.\cite{Rachel2017} where it  was found that the edge mode consists of a series of discrete, and equally spaced energy levels with the energy spacing increasing with decreasing island size (see Fig.~\ref{fig:Island}a). As already discussed for the ribbon system, the low-energy Majorana modes (see Fig.~\ref{fig:Island}b) are localized at the edge of the island, with the localization length increasing with increasing energy of the Majorana modes. Moreover, it was shown that the differential tunneling conductance associated with the flow of charge from a scanning tunneling microscope (STM) tip into the Majorana modes (see Fig.~\ref{fig:Island}c) is quantized and proportional to $|C|$. These edge modes also carry a supercurrent whose chirality reflects the sign of the Chern number ${\rm sgn}(C)$ (see Fig.~\ref{fig:Island}d). Finally, all of these properties are topologically protected and thus robust against disorder in the form of the island edges, with the zero-energy Majorana mode still persisting even in the presence of significant spatial disorder (see Fig.~\ref{fig:Island}e).

\begin{figure}[H]
\centering
\includegraphics[width=0.5\textwidth]{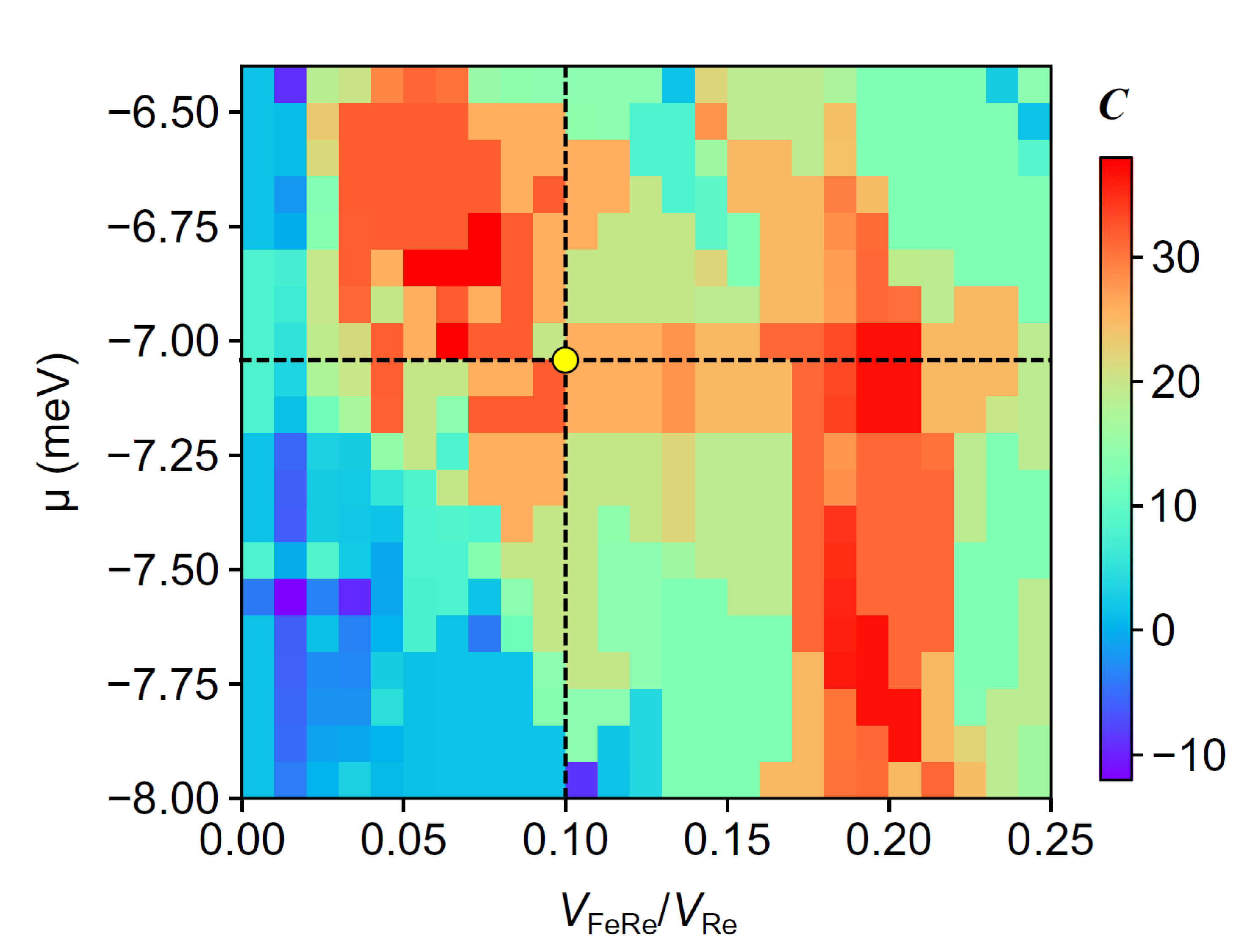}
\caption{Topological phase diagram for a 10-band model in the $(V_{FeRe}/V_{Re},\mu)$-plane, representing the Fe(ML)/Re(0001)-O(2x1) MSH system. Here, $V_{FeRe}$ is the hybridization between the Fe and Re layers, and $V_{Re}$ represents the hopping between the $d$-orbitals of the Re layer. Figure taken from Ref.\cite{Palacio-MoralesSCI-ADV2019}.}
\label{fig:FeRe_PD}
\end{figure}

The above results, which were based on the study of simple one-band models, identified the unique, and topologically protected features of MSH systems. In contrast, the experimental realization of MSH systems occurred in significantly more complex multi-band systems, such as in Pb(ML)/Co-Si/Si(111)~\cite{MenardNAT-COMM2017} or Fe(ML)/Re(0001)-O(2x1)~\cite{Palacio-MoralesSCI-ADV2019} heterostructures, raising the intriguing question of how the multi-band nature of these systems, and the presence of additional interactions, affect their topological phase diagram. To investigate this question, a 10-band model for the Fe(ML)/Re(0001)-O(2x1) MSH systems was developed~\cite{Palacio-MoralesSCI-ADV2019}, accounting for the 5 $d$-orbitals of Fe and Re and including further interactions, such as a \(\textbf{L}\cdot\textbf{S}\)-coupling. The topological phase diagram arising from this model (see Fig.~\ref{fig:FeRe_PD}) shows an extensive range in parameter space where topological phases can exist. The high value of the Chern number in this phase diagram arises from the the presence of a large number of bands in this 10-orbital system.   
The LDOS, computed using this multi-band model, reflects the existence of topologically protected Majorana edge modes for a finite-size magnetic island and is thus in very good agreement with the experimentally measured d\textit{I}/d\textit{V}, as shown in Figs.~\ref{fig:Fe-Re0001_structure} and \ref{fig:Fe-Re0001_specs}. This provides strong evidence for the existence of a topological superconducting phase in the Fe(ML)/Re(0001)-O(2x1) MSH system.

Another proposed approach to the quantum engineering of topological superconductivity was motivated by the ability to place a layer of the van der Waals material CrBr$_3$ on the surface of the $s$-wave superconductor NbSe$_2$ \cite{KezilebiekeNAT2020,KezilebiekeNANOLETT2022}. 

\begin{figure}[H]
\centering
\includegraphics[width=0.9\textwidth]{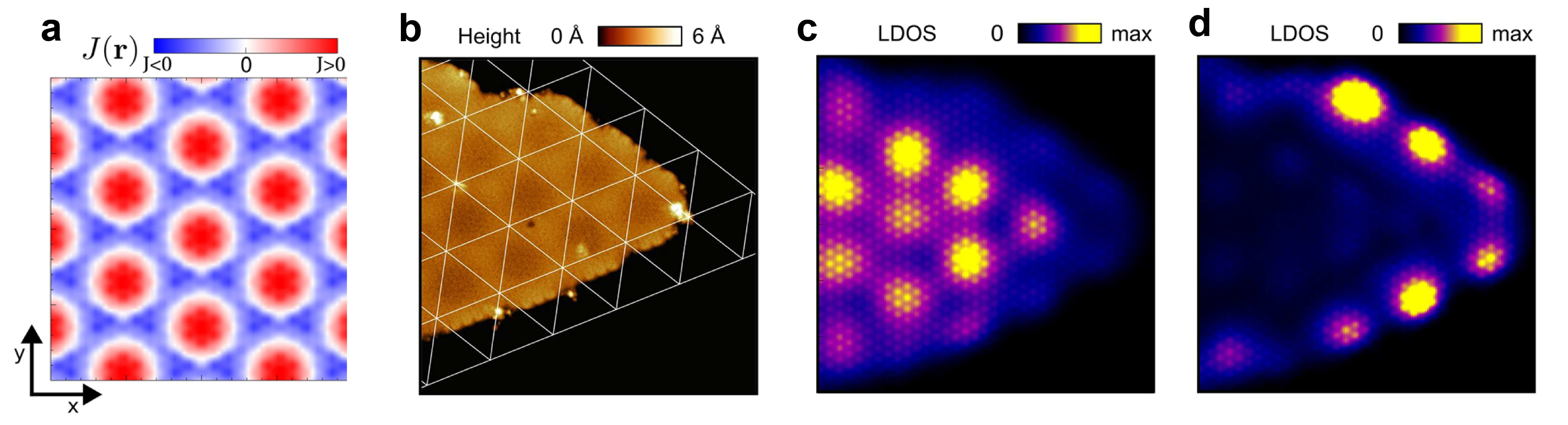}
\caption{{\bf a} Spatial Moir\'{e} pattern of the magnetic exchange $J$. {\bf b} Topography of an experimentally investigated island of CrBr$_3$ placed on the surface of NSe$_2$. Spatial form of the LDOS for an island similar to that shown in b and  c for an energy inside the YSR band, and d for zero energy. Figures adapted from Ref.\cite{KezilebiekeNANOLETT2022}.}
\label{fig:Moire}
\end{figure}

Specifically, it was argued \cite{KezilebiekeNANOLETT2022} that the lattice mismatch between CrBr$_3$ and the surface of NbSe$_2$ gives rise to a Moir\'{e}-type spatial pattern in the chemical potential and/or the local magnetic exchange (see Fig.~\ref{fig:Moire}a). This spatial variation, and the associated back-folding of bands, in turn extends the parameter range over which topological superconductivity can emerge. Moreover, considering a finite-size island of such an MSH system (see Fig.~\ref{fig:Moire}b), it was shown that the spatial pattern in the magnetic exchange (see Fig.~\ref{fig:Moire}a) is reflected in the spatial form of the LDOS of the YSR band which emerges close to the superconducting gap edge (see Fig.~\ref{fig:Moire}c). Interestingly enough, a similar pattern also emerges for the Majorana edge mode, as shown in Fig.~\ref{fig:Moire}d.

The effects of various types of disorder on the topological phase diagram of ferromagnetic MSH systems was investigated in Ref.\cite{Mascot2019}. To characterize the topological phase of the system in the presence of disorder, which breaks the translational invariance of the system, one can compute the Chern number in real space using \cite{Prodan2010,Prodan2017,Prodan2011} 
\begin{equation} \label{C real space}
C = \frac{1}{2\pi i} \mathrm{Tr} \left[ P [\delta_1 P, \delta_2 P] \right]
\end{equation}
\begin{equation}
\delta_i P = \sum_{m=-Q}^Q c_m e^{-2\pi i m \hat{x}_i / N} P e^{2\pi i m \hat{x}_i / N}
\end{equation}
where $P$ is the projector onto the occupied spectrum in real space, $N^2$ are the number of sites in the system, and $c_m$ are central finite difference coefficients for approximating the partial derivatives.
The coefficients for positive $m$ can be calculated by solving the following linear set of equations for $\vec{c} = (c_1, \dots, c_Q)$:
\begin{equation}
\hat{A} \vec{c} = \vec{b}, \; A_{ij} = 2j^{2i-1}, \; b_i = \delta_{i,1}, \; i,j \in \{1, \dots, Q \}
\end{equation}
while for negative $m$, we have $c_{-m} = -c_m$. To achieve a small error in the calculation of the Chern number, one usually takes the largest possible value of $Q$ given by $Q=N/2$. Moreover, important insight into the effects of disorder on the stability of a topological superconductor can be gained by considering the scaled Chern number density, defined as the partial trace over spin and Nambu space, and given by
\begin{equation} \label{C density}
C({\bf r}) = \frac{N^2}{2\pi i} \mathrm{Tr}_{\tau,\sigma} \left[ P [\delta_1 P, \delta_2 P] \right]_{\bf r,r}
\end{equation}
such that $C=\sum_{\bf r}[C({\bf r})]/N^2$.
\begin{figure}[H]
\centering
\includegraphics[width=0.9\textwidth]{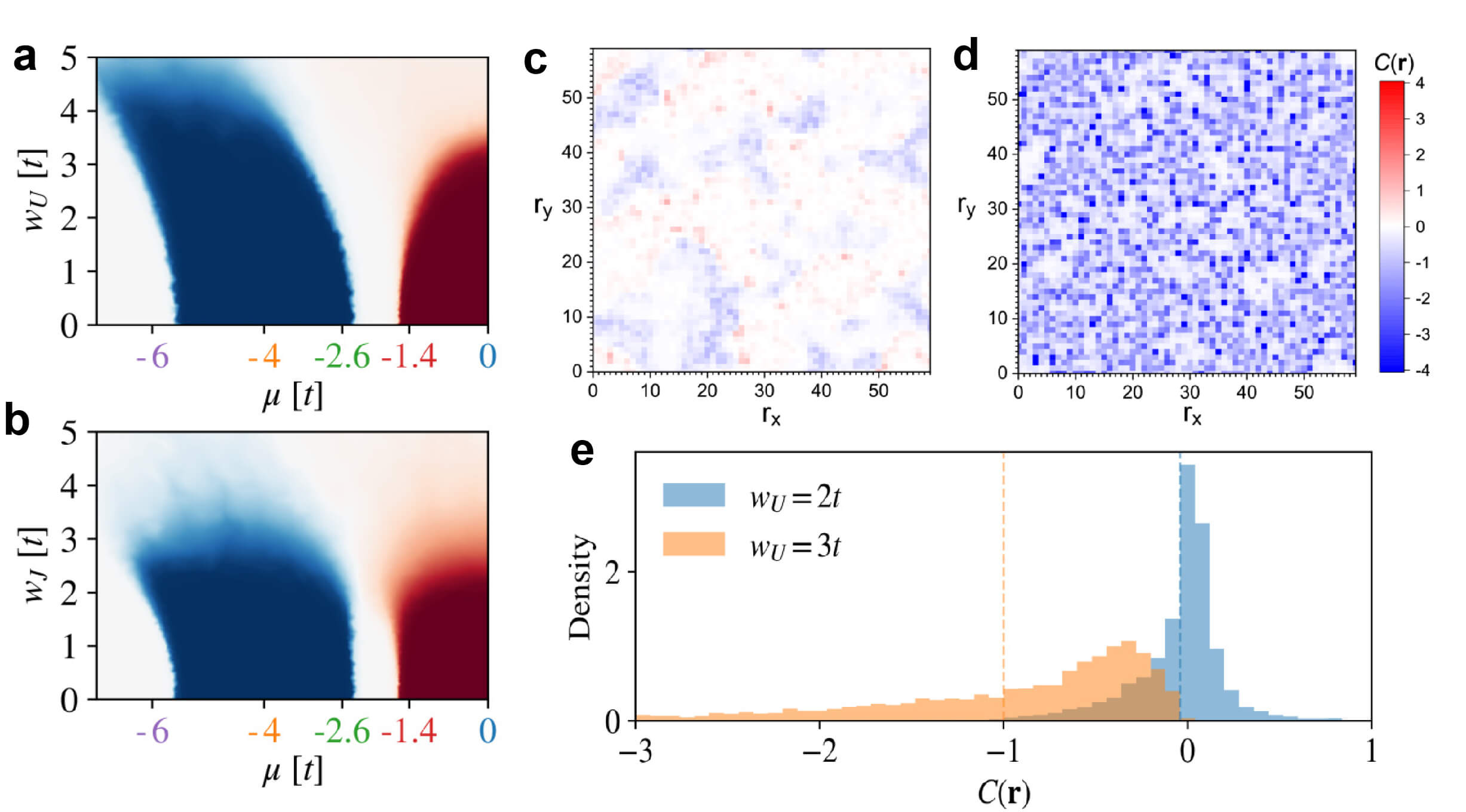}
\caption{Topological phase diagram for \textbf{a} potential disorder in the $(\mu,w_U)$-plane, and \textbf{b} magnetic disorder in the $(\mu,w_J)$-plane. Spatial plots of the Chern number density $C({\bf r})$ in the case of potential disorder for $\mu=-6t$ with \textbf{c} $w_U=2t$, and \textbf{d} $w_U=3t$.
\textbf{e} Distributions of the Chern number density $C({\bf r})$ for the two cases shown in \textbf{c} and \textbf{d}. The vertical dashed line in \textbf{c} show the mean values of the distribution, corresponding to the macroscopic Chern number. Figure adapted from Ref.\cite{Mascot2019}.}
\label{fig:MSH_disorder}
\end{figure}
Using this approach, the topological phase diagram was computed as a function of the strength of potential, $w_U$, and magnetic disorder, $w_J$, as shown in Figs.~\ref{fig:MSH_disorder}a and \ref{fig:MSH_disorder}b, respectively. This study showed that the topological superconducting phases are destroyed only for very large disorder strength which is comparable to the electronic bandwidth of the MSH system, thus demonstrating the robustness of topological superconducting phases against disorder effects. Note the reentrant behavior in the small $\mu$ region around $\mu=-6t$ (see Figs.~\ref{fig:MSH_disorder}a and b), where for small disorder strength the system is topologically trivial, while for larger disorder strength, the system undergoes a transition into a topological phase with $C=-2$. Moreover, it is also instructive to consider how disorder affects the local nature of the topological phase using the Chern number density, as shown in Figs.~\ref{fig:MSH_disorder}c and \ref{fig:MSH_disorder}d. Here, the spatial form of $C({\bf r})$ is shown for a weaker disorder distribution with $w_U = 2t$ (see Fig.~\ref{fig:MSH_disorder}c) and a stronger one with $w_U = 3t$ (Fig.~\ref{fig:MSH_disorder}d). It is apparent that with increasing disordere strength, the spatial variations in $C({\bf r})$ increase. However, despite the broad distribution of $C({\bf r})$ shown in Fig.~\ref{fig:MSH_disorder}e, the total Chern number is quantized within $0.00007\%$, demonstrating that the Chern number density is a suitable quantity to locally characterize the topological nature of the system. 

\begin{figure}[H]
\centering
\includegraphics[width=0.9\textwidth]{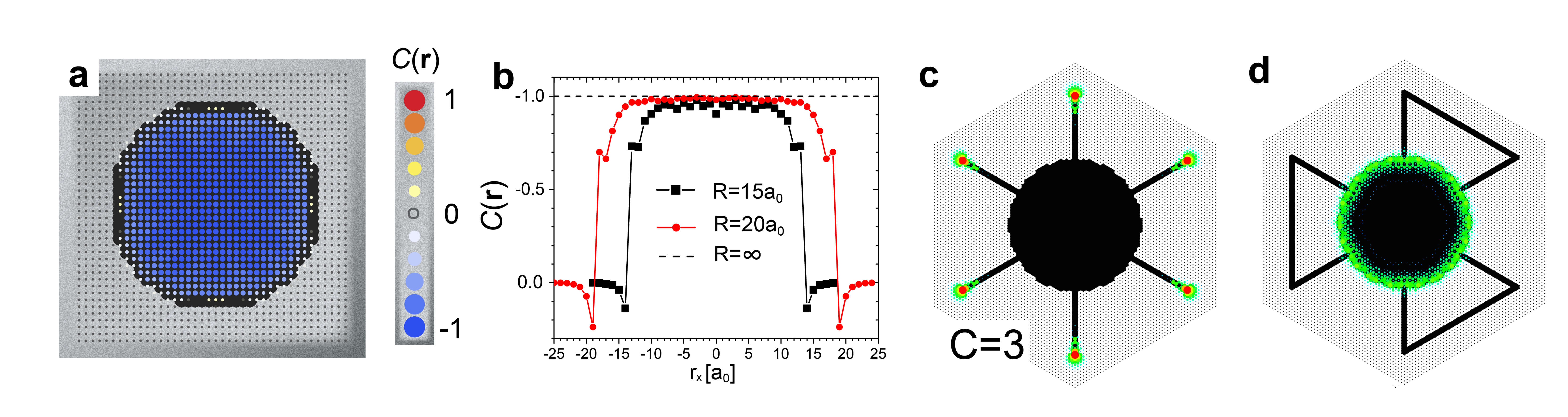}
\caption{ {\bf a} Spatial plot of the Chern number density $C({\bf r})$ for a finite size magnetic island. {\bf b} Linecut of $C({\bf r})$ through the center of two magnetic islands. Zero energy LDOS for {\bf c} a magnetic island with six topological chains attached, and {\bf d} an island with the ends of two neighboring chains connected, giving rise to a windmill-like structure and junctions with an even number of chains. Figure adapted from Ref.\cite{Mascot2019a}. }
\label{fig:Counting}
\end{figure}

The ability to compute the Chern number density in real space also allows one to characerize the topological nature of finite size magnetic islands, as was shown in Ref.\cite{Mascot2019a}.
In particular, if the parameters are chosen such that the 2D MSH system is in the topological $C=-1$ phase, then $C({\bf r})$ is non-zero inside a finite size island (see Fig.\ref{fig:Counting}a), but vanishes outside the island. Indeed, a linecut through the island (see Fig.\ref{fig:Counting}b) reveals that the Chern number density in the interior of the island is the same as that for a translationally invariant 2D MSH system.  These finite size islands also allow one to directly visualize and count the Chern number in real space. Consider for example a finite size island in the $C=3$ phase. Such an island possesses 3 Majorana edge modes, and hence 6 Majorana modes at zero energy. When one attaches 6 topological chains to such an island, for example, via atomic manipulation techniques, then the 6 MZMs are transferred to the ends of the chains, as shown in Fig.\ref{fig:Counting}c, and no zero-energy mode remains along the edge of the island. This result suggests a new real space approach to detecting the Chern number of a two-dimensional topological superconductor through atomic manipulation: if spectral weight for a zero-energy state remains located at the edge of the island when $N-1$ chains are attached, but vanishes for $N$ chains, then the Chern number of the 2D topological superconductors is given by $|C|=N/2$. These modes can be transferred back to the island by connecting the ends of these chains, as shown in Fig.\ref{fig:Counting}d.

Finally, we note that while the MSH systems discussed here are ones in which magnetic adatoms are placed on the surface of an $s$-wave superconductor, it was recently proposed that the surface of the iron-based superconductor FeSe$_{0.45}$Te$_{0.55}$ could be an intrinsic two-dimensional MSH system \cite{Mascot2022a,Wong2022,Xu2023a}. In this material, the full superconducting gap with $s_{\pm}$-symmetry, the Rashba-spin orbit interaction at the surface arising from the broken inversion symmetry, and the recently observed ferromagnetism \cite{Zaki2019,Li2021,McLaughlin2021} provide all of the fundamental building blocks for the emergence of topological superconductivity, similar to the MSH systems discussed above [see Eq.(\ref{eq:H})].

\subsubsection{Two-dimensional antiferromagnetic MSH systems}\label{theory-2D-AFM}

\begin{figure}[H]
\centering
\includegraphics[width=0.9\textwidth]{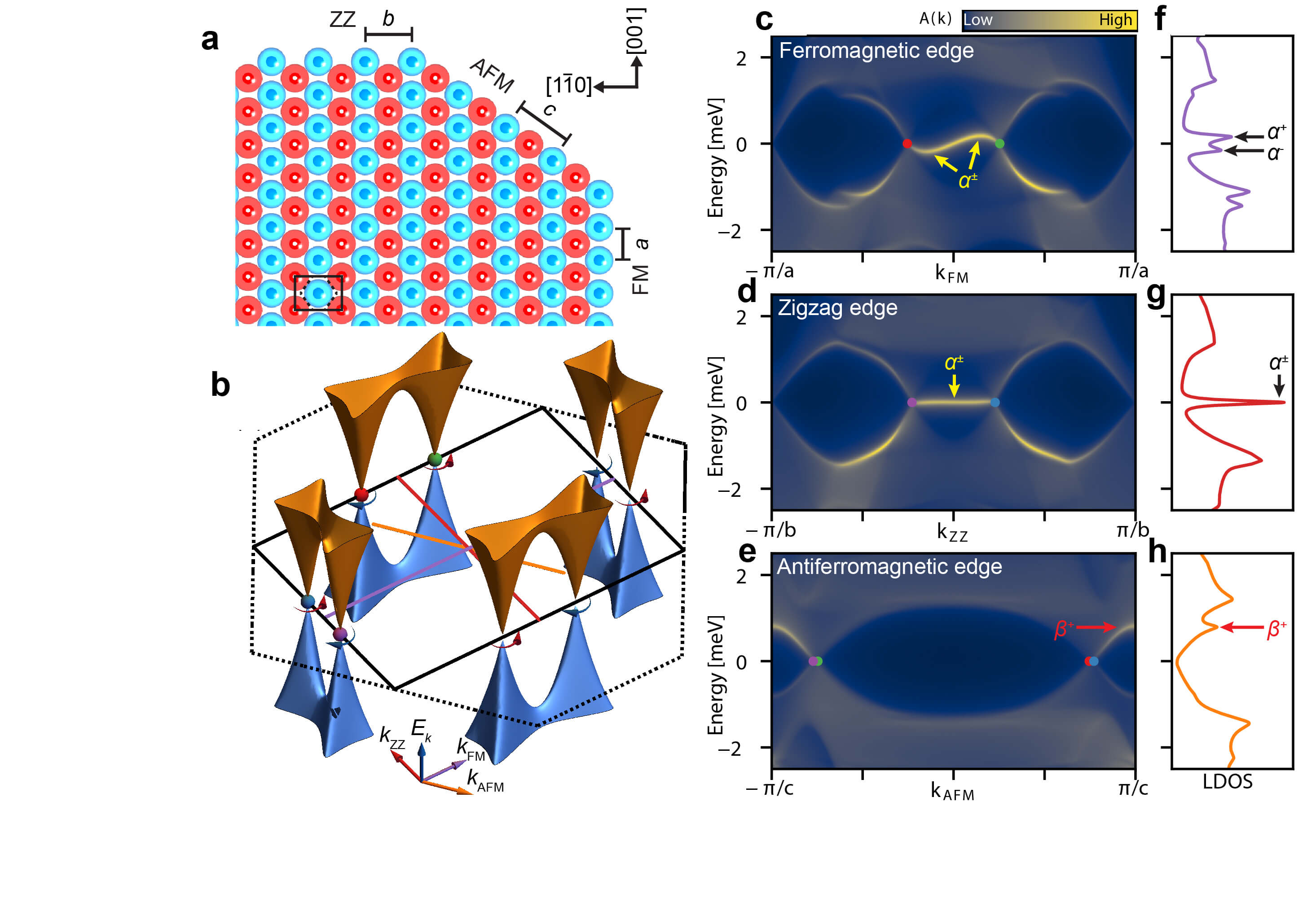}
\caption{\textbf{a} Real space structure of an antiferromagnetically ordered mono-atomic layer with the symmetry of the (110) surface of a body-centered cubic crystal. \textbf{b} Superconducting quasiparticle dispersion. The dispersion shows nodal points on the boundary of the magnetic Brillouin zone, shown as a black rectangle (the structural Brillouin zone is shown as dashed lines). The winding number associated with each nodal point is indicated by the curved arrows (+1 red, -1 blue). \textbf{c}-\textbf{e} Spectral function for each edge type and \textbf{f}-\textbf{h} corresponding LDOS at (\textbf{c},\textbf{f}) a FM edge, (\textbf{d},\textbf{g}) a ZZ edge, and (\textbf{e},\textbf{h}) an AFM edge. In addition to the edge modes, the spectral functions have contributions from the 2D bulk electronic structure, which originate from the projected bulk electronic band structure shown in \textbf{b} onto the respective momentum axes, as indicated by the colored lines in \textbf{b}. A dispersive edge band connects the projected nodal points at the FM edge and a flat edge band for the ZZ edge (yellow arrows), resulting in strong low-energy peaks in the LDOS (black arrows). For the AFM edge, a state corresponding to the van-Hove singularity connecting two neighboring cones appears and is indicated by red arrows. Figure adapted from Ref.~\cite{BazarnikNAT-COMM2023}.}
\label{fig:AFM_theory}
\end{figure}

The possibility to build MSH systems with antiferromagnetic (AFM) layers, such as in the Mn/Nb(110)~\cite{LoContePRB2022,BazarnikNAT-COMM2023} heterostructure, has raised the intriguing question of which new topological phases might arise from it. The antiferromagnetic structure on a surface-projected bcc lattice is shown in Fig.~\ref{fig:AFM_theory}a. 

The resulting electronic structure is shown in Fig.~\ref{fig:AFM_theory}b, exhibiting 8 nodal points that are located on the boundary of the magnetic Brillouin zone. To investigate the topological nature of this system, one computes the topological charge $q$ of these nodal points (which is allowed by the system's chiral symmetry). The non-zero value of the topological charge $q=\pm 1$ associated with these nodal points has two important consequences. First, their existence is topologically protected against disorder effects. Second, for any system with an edge, these topological nodal points are connected by edge modes, whose specific form depends on the real space direction of the edge, and the edge's magnetic structure. In particular, for the lattice structure shown in Fig.~\ref{fig:AFM_theory}a, three different edges are possible, which are referred to as the ferromagnetic (FM), zig-zag (ZZ), and antiferromagnetic (AFM) edges. A calculation of the momentum-resolved spectral function at theses three different edges, shown in Figs.~\ref{fig:AFM_theory}c-e, reveals the existence of a non-dispersive edge mode along the zig-zag edge and of a very weakly dispersing edge mode along the ferromagnetic edge. In contrast, for the antiferromagnetic edge, the nodal points are connected by a $\beta$-mode located close to the bulk bands. The existence of qualitatively different edge modes is a characteristic signature of the topological nodal-point superconducting phase and leads to characteristic signatures in the local density of states (LDOS) at the system's edges, as shown in Figs.~\ref{fig:AFM_theory}f-h.

\begin{figure}[H]
\centering
\includegraphics[width=\textwidth]{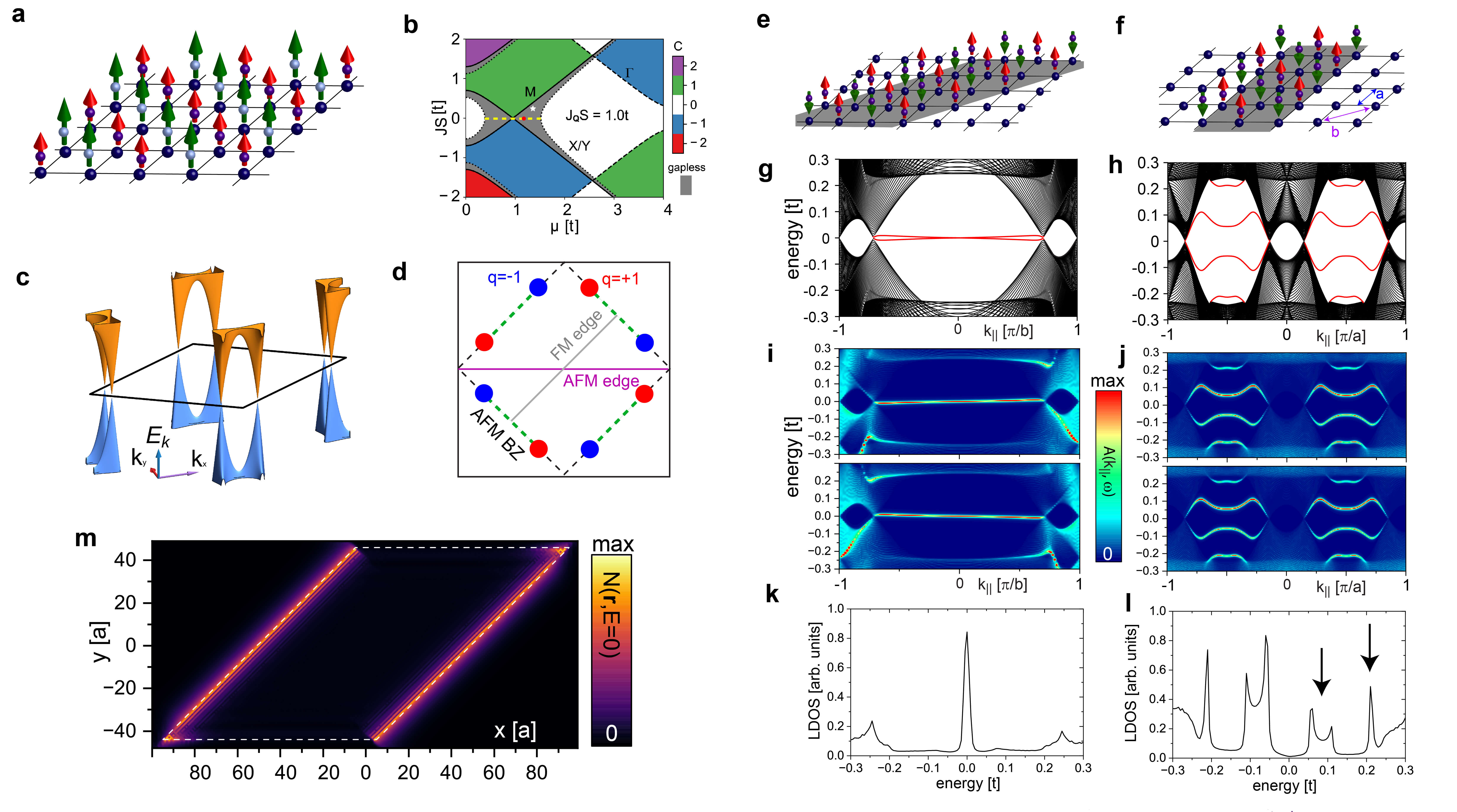}
\caption{\textbf{a} Schematic structure of the checkerboard magnetic structure with exchange couplings $J \pm J_{\bf Q}$ in the two sublattices. \textbf{b} Topological phase diagrams of the MSH system in the $(\mu,JS)$-plane for $J_{\bf Q} S=t$.  \textbf{c} Energy dispersion of the bulk system in the magnetic BZ. \textbf{d} Nodal points and edge modes (dashed green lines) projected onto momenta parallel to the FM and AFM edges. Schematic representation of \textbf{e} diagonal (FM) and \textbf{f} vertical (AFM) edges. Electronic structure of a ribbon with \textbf{g} FM and \textbf{h} AFM edges. \textbf{i}, \textbf{j} Spectral functions at the ribbon's left (upper panel) and right (lower panel) edges corresponding to \textbf{g} and \textbf{h}. \textbf{k}, \textbf{l} LDOS at the edges corresponding to \textbf{g} and \textbf{h}. \textbf{m} Zero-energy LDOS of a magnetic island (dashed white line) with FM and AFM edges. Figure adapted from Ref.\cite{Kieu2023}.}
\label{fig:Checkerboard}
\end{figure}

It was subsequently shown \cite{Kieu2023} that the AFM MSH system can be generalized to an MSH system with a checkerboard magnetic structure (see Fig.~\ref{fig:Checkerboard}a) with two sublattices of magnetic adatoms characterized by exchange couplings $J \pm J_{\bf Q}$. Note that for $J=0$, the system possesses an antiferromagnetic structure. The topological phase diagram in the $(\mu, JS)$-plane for $J_{\bf Q}S=t$ is shown in Fig.~\ref{fig:Checkerboard}b: in addition to strong topological superconducting phases for $JS \not = 0$, the system also possesses gapless regions, in particular around $JS = 0$. A plot of the electronic dispersion for an antiferromagnetic systems with $JS = 0$, as denoted by the red dot in Fig.~\ref{fig:Checkerboard}b shows the existence of 8 nodal points (see Fig.~\ref{fig:Checkerboard}c), each possessing a non-zero topological charge (see Fig.~\ref{fig:Checkerboard}d), similar to the results presented in Fig.~\ref{fig:AFM_theory}. For MSH ribbon geometries with either ferromagnetic (see Fig.~\ref{fig:Checkerboard}e) or antiferromagnetic (see Fig.~\ref{fig:Checkerboard}f) edges, the resulting electronic dispersion as a function of momentum along the edges is shown in Figs.~\ref{fig:Checkerboard}g and ~\ref{fig:Checkerboard}h, respectively. The edge modes connecting the topological nodes are shown in red: they are chiral and only very weakly dispersing for the ferromagnetic edge, but very strongly dispersing for the antiferromagnetic edge. The reason for this qualitative difference is revealed when plotting the momentum-resolved spectral function at the two edges, as shown in Figs.~\ref{fig:Checkerboard}i and ~\ref{fig:Checkerboard}j, where the upper (lower) panel corresponds to the left (right) edge. For the FM edge, the two chiral modes are spatially separated at the left and right edges, and thus cannot hybridize, resulting in the weak dispersion. On the other hand, for the AFM edge, the two modes are located at the same edge, allowing for a strong hybridization, thus pushing the modes close to the bulk states. As a result, the LDOS at the FM edge shows a pronounced peak at zero energy (see Fig.~\ref{fig:Checkerboard}k), while the signature of the edge modes for the AFM edge occurs at much higher energies (see black arrows in Fig.~\ref{fig:Checkerboard}l). Correspondingly, the zero-energy LDOS of a finite-size island with both FM and AFM edges show large spectral weight at the FM edges, and vanishing spectral weight at the AFM edges, as shown in Fig.~\ref{fig:Checkerboard}m. It was found that these features persist in all of the gapless regions shown in Fig.~\ref{fig:Checkerboard}b, and thus also for systems with a non-vanishing ferromagnetic moment $JS \not = 0$. 

\begin{figure}[H]
\centering
\includegraphics[width=0.9\textwidth]{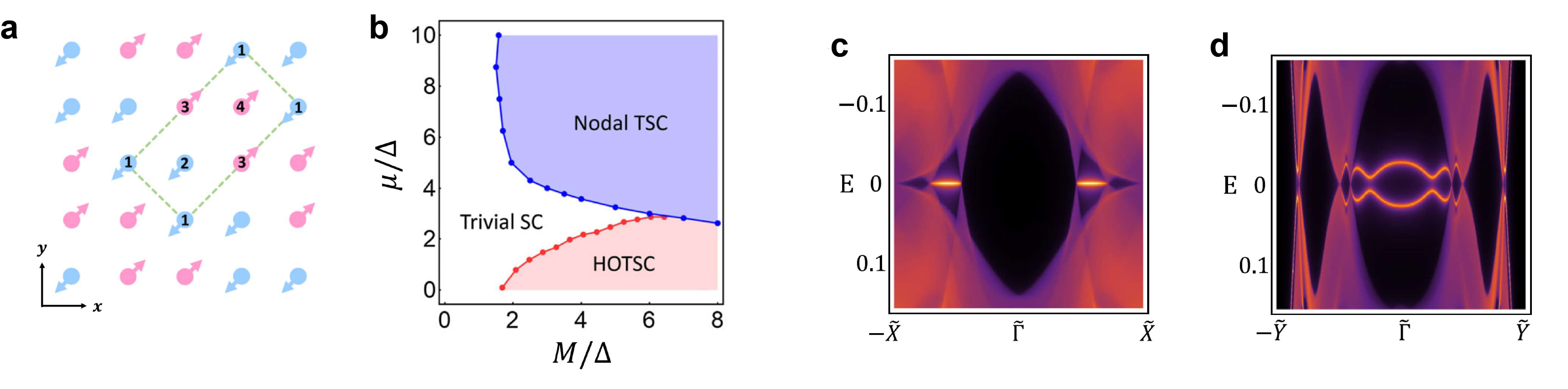}
\caption{\textbf{a} Schematic plot of bi-collinear antiferromagnetic
order in FeTe. The circle and its arrow represent the Fe atom
and its magnetic moment. \textbf{b} Topological phase diagram with respect to $M$ (representing the magnetic exchange strength) and $\mu$ at a fixed $\Delta$. \textbf{c} and \textbf{d} show the dispersions of the AFM edge and the FM edge for the nodal SC phase, respectively. Figure adapted from Ref.~\cite{Zhang2019}.}
\label{fig:bilinearAFM}
\end{figure}

Moreover, it was argued that a similar MSH system can be made by depositing a layer of FeTe, which exhibits a bi-colinear AFM structure, on the surface of FeSe$_{1-x}$Te$_x$ \cite{Zhang2019}, as schematically shown in Fig.\ref{fig:bilinearAFM}a.
Using the Bernevig-Hughes-Zhang model, it was shown that the phase diagram of such a system (see Fig.\ref{fig:bilinearAFM}b) exhibits both topological nodal-point superconducting phases, as well as higher-order topological superconducting phases (see also the discussion in Sec.~\ref{sec:theory-2D-stacked}). In the topological nodal-point superconducting phase, there again exists a significant difference in structure of the edge mode for a ribbon geometry system: while the AFM edge possesses a flat (dispersionless) edge mode, the edge mode along the FM edge shows a significant dispersion. This qualitative difference in the nature of the edge modes in FeTe(ML)/FeSe$_{1-x}$Te$_x$, with regards to the AFM and FM edge, in comparison to the results discussed in Figs.~\ref{fig:AFM_theory} and \ref{fig:Checkerboard} arises from the larger magnetic unit cell of FeTe, and the associated reduced magnetic BZ that possesses fewer nodal points.

\begin{figure}[H]
\centering
\includegraphics[width=0.9\textwidth]{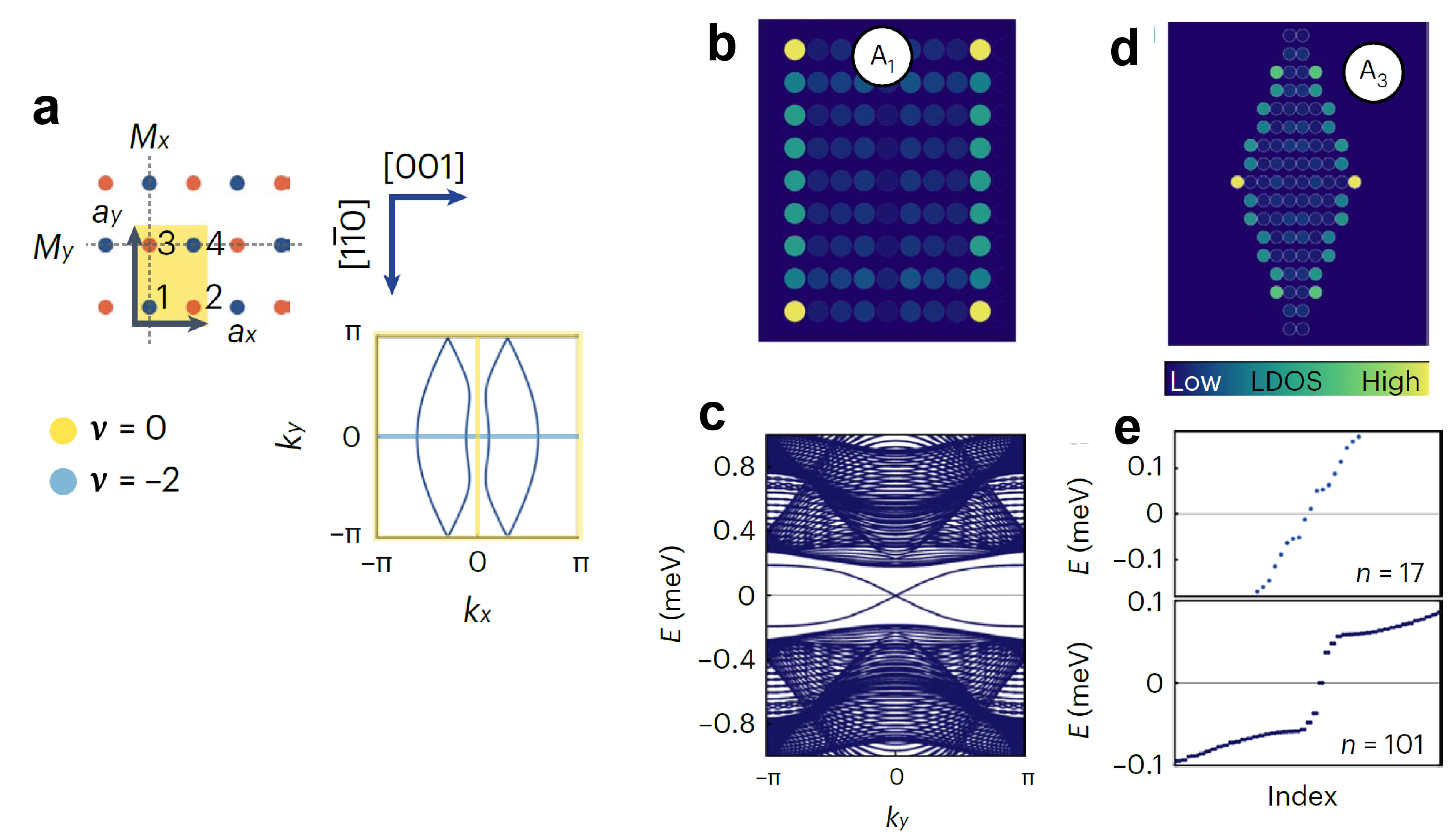}
\caption{\textbf{a} Lattice structure, antiferromagnetic ordering is indicated by red and blue dots. Shown is also the normal state Fermi surface. {\bf b} Zero-energy LDOS, and {\bf c} electronic dispersion of a rectangular island.  {\bf d} Zero-energy LDOS, and {\bf e} electronic structure of a rhombus-like island. Figure adapted from Ref.\cite{SoldiniNAT-PHYS2023}.}
\label{fig:crystalline}
\end{figure}

Finally, it was argued that an antiferromagnetic structure on a rectangular lattice (see Fig.~\ref{fig:crystalline}a) can give rise to a topological superconductor that is protected by crystalline symmetry \cite{SoldiniNAT-PHYS2023}.
In particular, it was shown that the real space structure of a magnetic island strongly determines the type of edge mode that the system exhibits. For a rectangular island, a low-energy edge modes appears only along one type of edge (see Fig.~\ref{fig:crystalline}b) whose electronic band is not connected to the bulk states  (see Fig.~\ref{fig:crystalline}c), similar to the spectroscopic signatures of a weak topological phase discussed in Ref.~\cite{Wong2023}. In contrast, the rhombus like island shown in Fig.~\ref{fig:crystalline}d has large spectral weight of the zero-energy LDOS in two of the four corners, associated with the existence of two zero energy states (see Fig.~\ref{fig:crystalline}e), which were attributed to a higher-order topological phases, similar to those found in other MSH systems \cite{Wong2023}.

\subsubsection{Two-dimensional non-collinear magnetic structures}\label{theory-2D-NC}

Another important class of magnetic structures that have been shown to give rise to topological phases in MSH systems are two-dimensional non-collinear magnetic textures, such as spiral \cite{Nakosai2013}, skyrmionic \cite{Mascot2021a}, or 3${\bf Q}$ \cite{Bedow2020} magnetic structures. These magnetic structures are of particular interest as their non-collinear form induces an additional Rashba spin-orbit interaction, such that even systems with very weak or vanishing conventional Rashba spin-orbit interaction (arising from a broken inversion symmetry on the surface) can exhibit stable topological phases.

\begin{figure}[H]
\centering
\includegraphics[width=\textwidth]{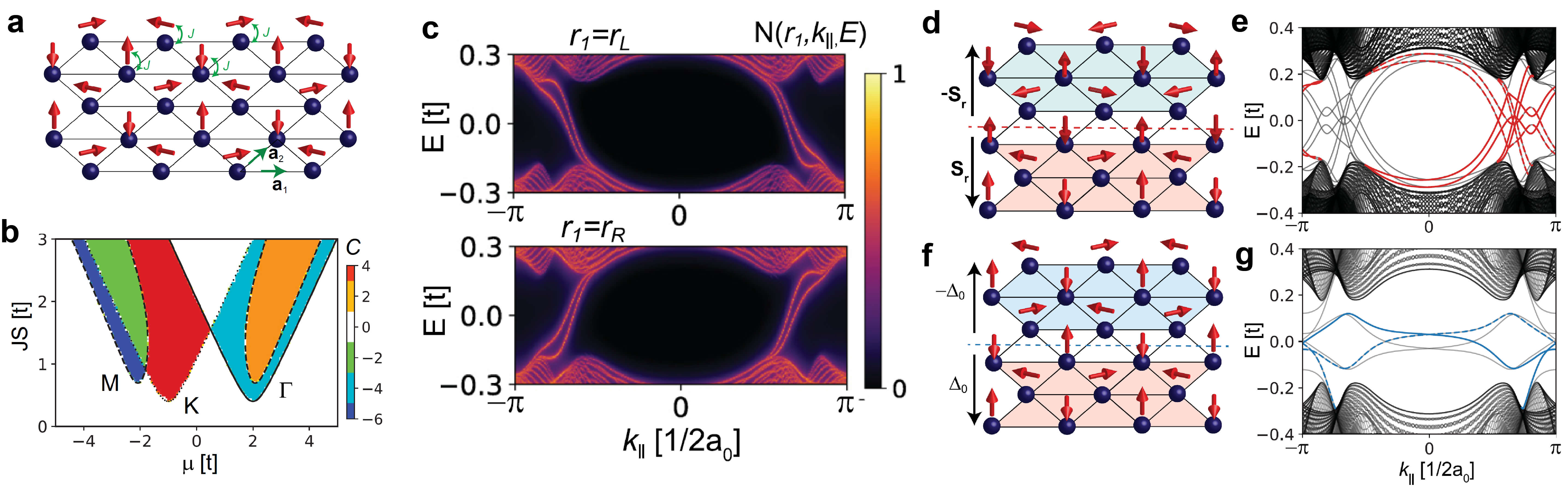}
\caption{\textbf{a} Schematic picture of the 3Q magnetic layer (red arrows) placed on the surface of an $s$-wave superconductor (blue spheres). \textbf{b} Topological phase diagram representing the Chern number in the $(\mu,JS)$-plane. The solid, dashed and dotted lines denote gap closings at the $\Gamma$, $M$, and $(K,K^\prime)$ points, respectively. \textbf{c} Spectral function at the left (upper panel) and right edges (lower panel) of an MSH ribbon. Schematic picture of {\bf d} a spin domain wall where ${\bf S}_{\bf r} \rightarrow -{\bf S}_{\bf r}$ (case I), and {\bf e} a $\pi$-phase domain wall where the superconducting order parameter $\Delta \rightarrow - \Delta$ (case II). Electronic structure as a function of momentum $k_\parallel$ along the domain wall for {\bf f} case I (Majorana modes are shown as red lines), and {\bf g} case II (trivial in-gap states shown in blue). Figure adapted from Ref.~\cite{Bedow2020}. }
\label{fig:3Q}
\end{figure}

An MSH system with a $3{\bf Q}$-magnetic structure \cite{Bedow2020}, whose spatial form is shown in Fig.~\ref{fig:3Q}a, possesses strong topological phases with even Chern numbers only (see Fig.~\ref{fig:3Q}b), due to the double-degeneracy of the electronic structure.  
Despite the complex magnetic structure, one finds that for a ribbon geometry, the Majorana edge modes are still chiral in nature, as follows from a plot of the momentum-resolved spectral function shown in Fig.~\ref{fig:3Q}b. It was suggested that the topological nature of such an MSH system can be identified by considering the electronic structure near domain walls. In particular, at a domain wall at which the magnetic moments are inverted, ${\bf S}_{\bf r} \rightarrow - {\bf S}_{\bf r}$ (see Fig.~\ref{fig:3Q}d), Majorana edge modes emerge (see Fig.~\ref{fig:3Q}e), as required by the bulk-boundary correspondence, since the Chern number between the two regions separated by the domain wall changes from $C \rightarrow -C$.
In contrast, at a $\pi$-phase domain wall where the superconducting order parameter undergoes a sign change, $\Delta \rightarrow -\Delta$ (see Fig.~\ref{fig:3Q}f) only trivial in-gap states emerge (see Fig.~\ref{fig:3Q}g). Thus, domain walls could in general be employed to gain insight into the topological nature of MSH systems.

\begin{figure}[H]
\centering
\includegraphics[width=0.9\textwidth]{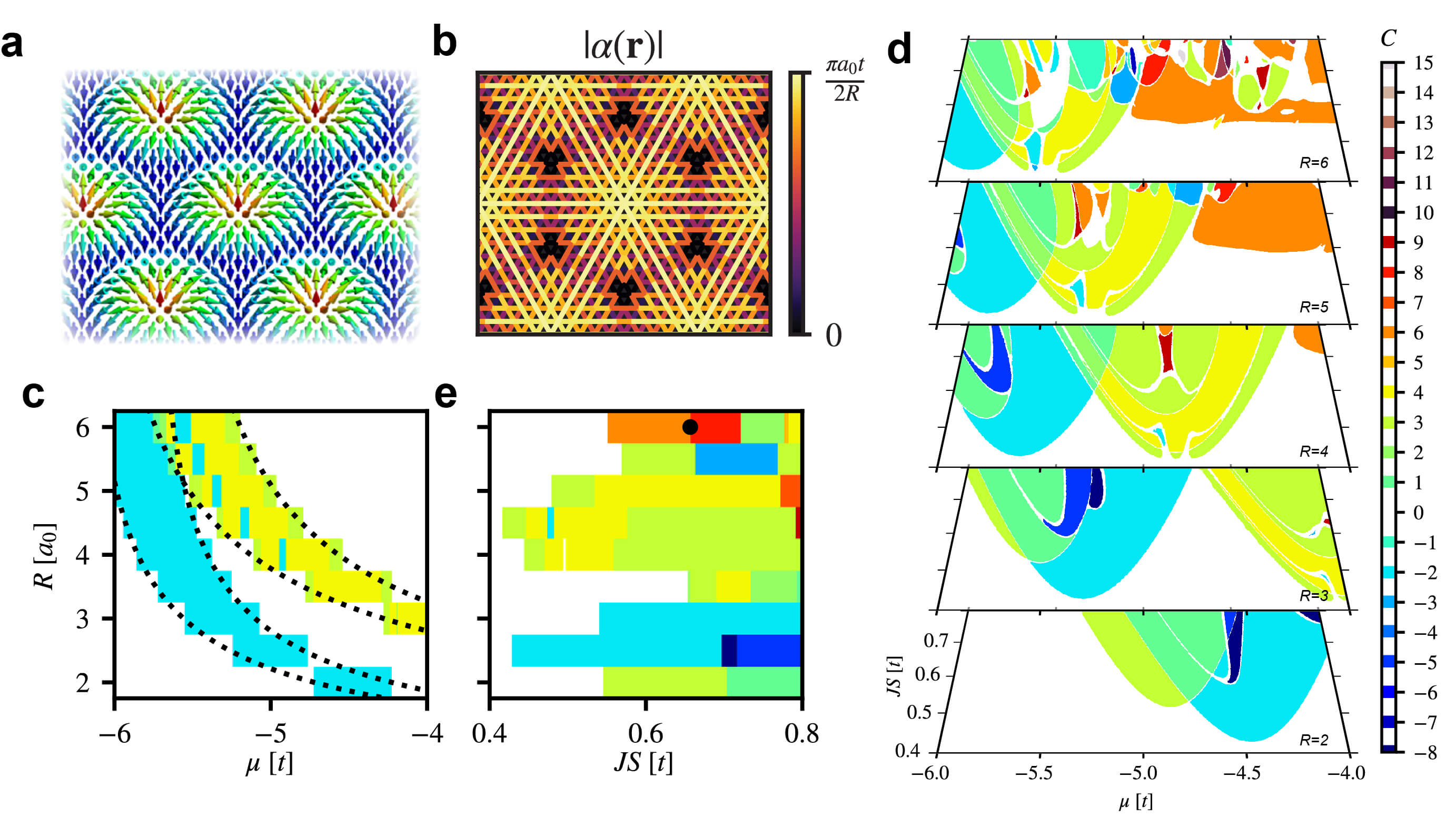}
\caption{\textbf{a} Schematic plot of the skyrmionic magnetic structure. \textbf{b} Spatial plot of the induced Rashba-spin orbit interaction $\alpha({\bf r})$. Topological phase diagram in the \textbf{c}  $(\mu, JS)$-plane for different values of the skyrmion radius $R$, \textbf{d} $(\mu, R)$-plane for fixed $JS$, and \textbf{e} $(JS, R)$-plane for fixed $\mu$. Figure adapted from Ref.~\cite{Mascot2021a}.}
\label{fig:skyrmion}
\end{figure}

A skyrmionic magnetic structure is schematically shown in Fig.~\ref{fig:skyrmion}a, with the induced and spatially varying Rashba spin-orbit interaction $\alpha({\bf r})$ shown in Fig.~\ref{fig:skyrmion}b. Note that the maximum induced Rashba-spin orbit interaction, $\alpha_{max}$, occurs in the center of the skyrmions, and scales inversely with the skyrmion radius $R$, $\alpha_{max} \sim 1/R$.
The resulting topological phase diagram in the $(\mu,R)$-plane (see Fig.~\ref{fig:skyrmion}c) shows that the topological phase of the skyrmion MSH system can be tuned by changing the 
skyrmion radius $R$; the latter can be achieved experimentally through an external magnetic field \cite{Romming2013}. The ability to change between different topological phases of the MSH system is a direct consequence of the magnitude of the induced $\alpha({\bf r})$ varying with the skyrmion radius $R$. The results in Fig.~\ref{fig:skyrmion}c sdemonstrate that the phase transition lines in the $(\mu,R)$-plane obey the relation $\mu = A_i+B_i/R^2$ (see dotted
lines) with constants $A_i,B_i$. Since the maximum Rashba spin-orbit interaction (induced in the center of a skyrmion) is given by $\alpha_{max} \sim 1/R$, the above equation describing the phase transition lines suggests
that the induced RSO interaction leads to an effective renormalization
of the chemical potential, thus allowing one to drive the system through a topological phase transition by changing the skyrmion radius. This dependence of the phase transition lines on $R$ is further exemplified when considering the phase diagram in the $(\mu, JS)$-plane for different skyrmion radii, as shown in Fig.~\ref{fig:skyrmion}d. While these phase diagrams for different values of $R$ show a very similar topological structure, the topological phases move to lower values of $\mu$ with increasing $R$. This result also reveals that the topological phase that can be accessed by varying $R$ depends on the strength of $JS$. In particular, for sufficiently large $JS$, every change in the
skyrmion radius by half a lattice constant changes the system's Chern number (see Fig.~\ref{fig:skyrmion}e). Thus, the ability to change the skyrmion radius externally through a magnetic field opens the venue for exploring a rich topological phase diagram.

\begin{figure}[H]
\centering
\includegraphics[width=0.9\textwidth]{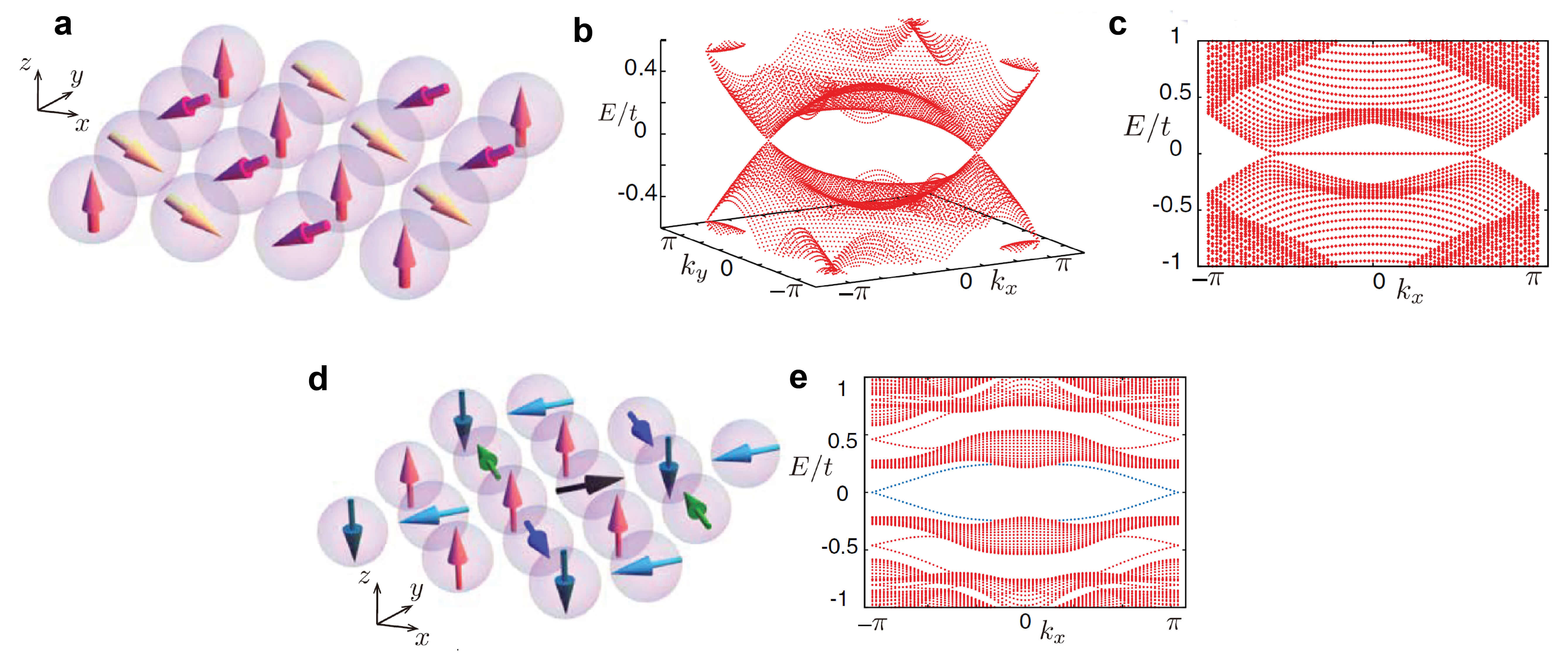}
\caption{\textbf{a} Schematic of the spiral magnetic structure with wave-vector ${\bf Q}=(2\pi/3,2\pi/3)$. \textbf{b} Resulting electronic structure in the Brillouin zone, exhibiting two nodal points. \textbf{c} Electronic structure along the boundary of the system, revealing a zero-energy Andreev state connecting the nodal points.  {\bf d} Schematic picture of a non-collinear magnetic structure that gives rise to a fully gapped state. {\bf e} Resulting electronic structure along an edge, showing an edge mode. Figures are adapted from Ref.~\cite{Nakosai2013}}.
\label{fig:Nakosai}
\end{figure}

Moreover, it was shown \cite{Nakosai2013} that two-dimensional spiral magnetic structures with a wave-vector of ${\bf Q}=(2\pi/3,2\pi/3)$ (see Fig.\ref{fig:Nakosai}a) can give rise to an electronic structure that exhibits nodal points, as shown in Fig.~\ref{fig:Nakosai}b. 
For a system with open boundaries, a dispersionless Andreev state connects the two nodal points, as shown in Fig.~\ref{fig:Nakosai}c. This case was analysed in more detail in Ref. \cite{Chatterjee2024b}. More complex collinear magnetic structures, such as the one shown in Fig.~\ref{fig:Nakosai}d can lead to fully gapped phases. The emergence of edge modes for systems with open boundaries (see Fig.\ref{fig:Nakosai}e) that traverse the superconducting gap, and possess a linear dispersion near zero energy, provide strong evidence for this gapped phase being a strong topological superconducting phase with a non-zero Chern number.

\begin{figure}[H]
\centering
\includegraphics[width=0.8\textwidth]{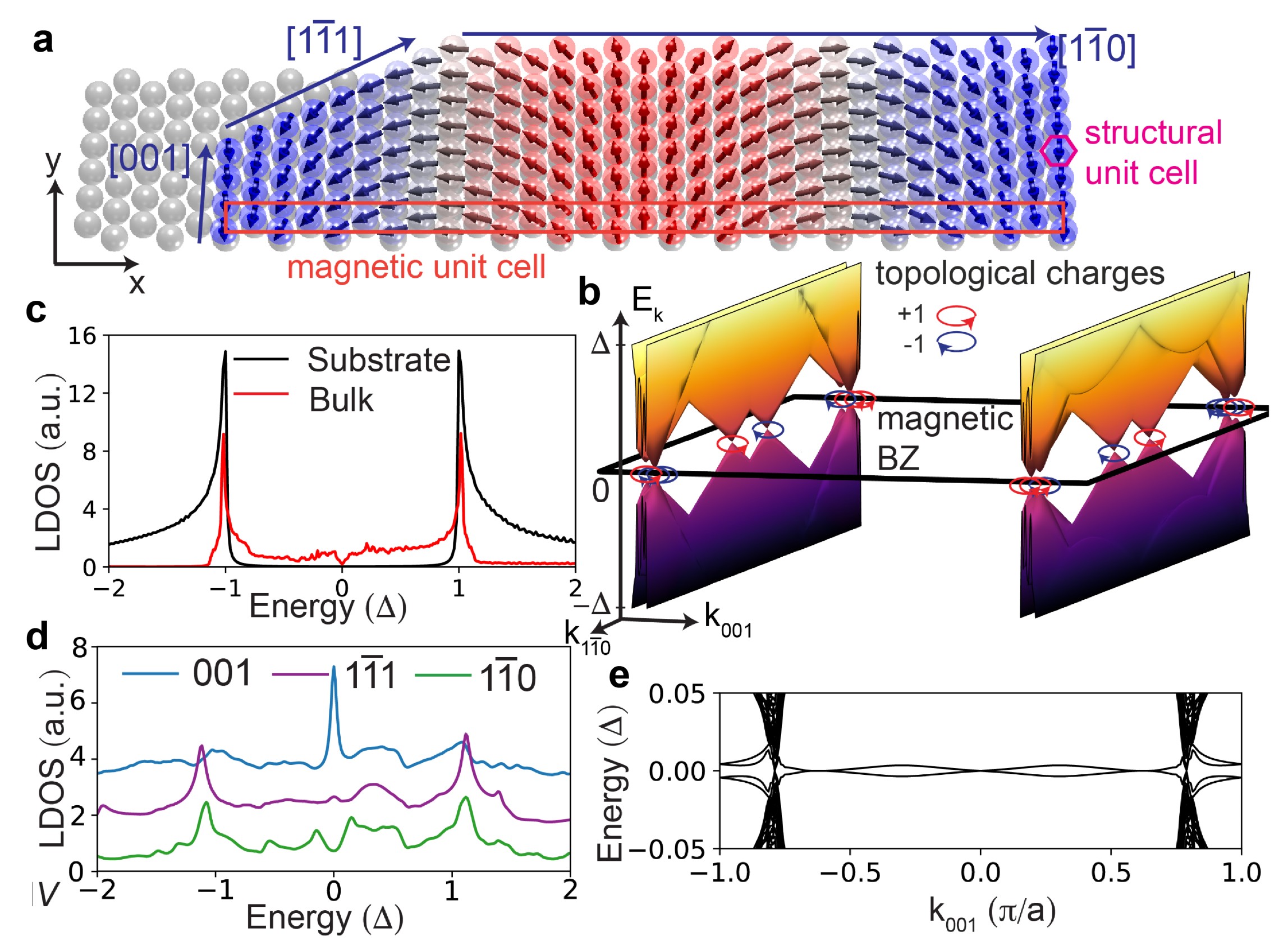}
\caption{\textbf{a} Sketch of the structural and magnetic unit cell for the spiral magnetic structure. \textbf{b}  Electronic structure in the magnetic Brillouin zone exhibiting nodal points with non-zero topological charge. \textbf{c} Theoretical LDOS on the Ta(110) surface (black) and for the Fe/Ta(110) MSH system (red). \textbf{d} Theoretical LDOS at a [001]-edge (blue), [1$\bar{1}$1]-edge (purple) or [1$\bar{1}$0]-edge (green). \textbf{e} Electronic band structure as a function of momentum along the [001]-edges of a ribbon system. The spin spiral terminates with an angle $\theta = 0^{\circ}$ between the spin direction and the normal to the plane at the ribbon's left and right edges. Figure adapted from Ref.\cite{BrüningARXIV2024}. }
\label{fig:FeTa}
\end{figure}

The observation of a spin spiral in the MSH system Fe(ML)/Ta(110)  \cite{BrüningARXIV2024} (see Fig.~\ref{fig:FeTa}a) stimulated interest in the question of whether it can give rise to a topological superconducting phase. 
The electronic band structure of the MSH system (see Fig.\ref{fig:FeTa}b), arising from such a spiral magnetic structure, was shown to possess several nodal points in the magnetic BZ (the corresponding magnetic unit cell in real space is shown in Fig.~\ref{fig:FeTa}a). The quantized topological charge, $q=\pm 1$, of these nodal points implies that the system is in a topological nodal point superconducting phase. As a result, the LDOS of the bulk Fe(ML)/Ta(110) system exhibits a characteristic $V$-shape around zero energy, directly reflecting the existence of these nodal points, in contrast to the uncovered Ta surface, where the LDOS reflects the existence of a hard $s$-wave superconducting gap (see Fig.~\ref{fig:FeTa}c). Since the quasiparticles' velocity in the nodal cones is small (i.e., the bands are dispersing only weakly), the $V$-shape in the LDOS is visible only over a small energy window. 

A unique feature of the TNPSC phase is that it exhibits zero energy modes only along certain real space edges, which are determined by the interplay of the nodal point position in momentum space, their projection onto the edge direction, and the magnetic structure of the edge \cite{BazarnikNAT-COMM2023,Zhang2019,Kieu2023}. Fe/Ta(110) islands in general realize three different types of edges along the $[001]$, $[1{\bar 1}1]$, and $[1{\bar 1}0]$ directions (see Fig.\ref{fig:FeTa}a). A comparison of the LDOS for these three edges (see Fig.~\ref{fig:FeTa}d) reveals that a zero-energy peak occurs only along the [001] edge, but is absent for the other two edges. This peak arises from a low-energy, weakly dispersing chiral edge mode that connects nodal points of opposite topological charge, as follows from a plot of the electronic band structure of a ribbon system shown in  Fig.\ref{fig:FeTa}{\bf e}. As discussed in Sec.~\ref{experiment_2D-NC}, this finding explains the experimentally observed differential tunneling conductance along different edges of Fe/Ta(110) islands.

Finally, it was theoretically predicted that in more complex heterostructures, where a two-dimensional quantum spin Hall insulator is sandwiched between a magnetic spin spiral layer and an $s$-wave superconductor, the spiral structure of the magnetic layer gives rise to a second-order topological superconductor \cite{Chatterjee2024a}. 

\subsubsection{Two-dimensional stacked magnetic structures}\label{sec:theory-2D-stacked}

\begin{figure}[H]
\centering
\includegraphics[width=1.0\textwidth]{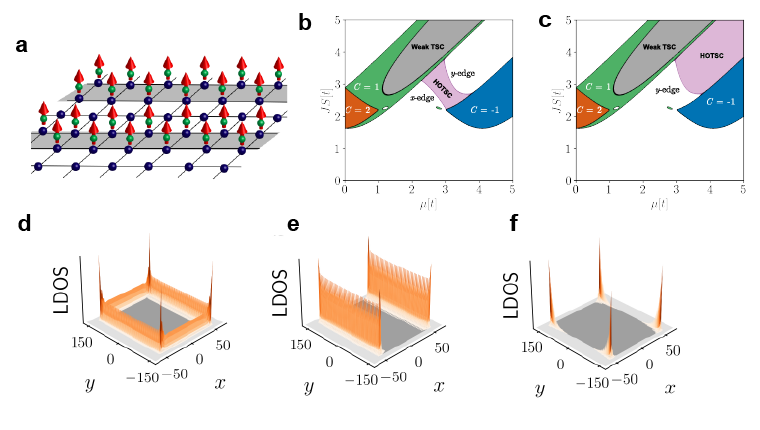}
\caption{\textbf{a} Schematic of an MSH system with a stacked magnetic structure with a double-chain termination. Topological phase diagram for an MSH system with \textbf{b} a single-chain and \textbf{c} a double-chain termination. Zero-energy LDOS for a rectangular MSH island on the surface of an s-wave superconductor in the \textbf{d} strong topological, \textbf{e} weak topological, and \textbf{f} higher order topological superconducting phase. 
Figure adapted from Ref.~\cite{Wong2023}.}
\label{fig:MSH_stacked}
\end{figure}

Another important type of magnetic structure which was recently investigated theoretically are so-called stacked magnetic structures~\cite{Wong2023}, as schematically shown in Fig.~\ref{fig:MSH_stacked}a. 
Here, chains of magnetic adatoms extending along the $x$-axis are stacked in a repeating order along the $y$-direction: for the case shown in Fig.\ref{fig:MSH_stacked}a, two chains of magnetic adatoms are followed by an empty row along the $y$-direction.

If the system is finite along the $y$-direction and ends with two adjacent chains (a single chain) of magnetic adatoms, one refers to it as an MSH system with a double-chain (single-chain) termination. Representative topological phase diagrams for MSH systems with a single-chain and a double-chain termination are shown in Figs.~\ref{fig:MSH_stacked}b and \ref{fig:MSH_stacked}c, respectively. In addition to strong topological phases characterized by non-zero Chern numbers, the system also exhibits weak and higher-order topological superconducting phases. As the phase transition into the strong topological phases are accompanied by gap closings at high symmetry points in the Brillouin zone, i.e., by bulk gap closings, they are insensitive to the termination of the MSH system. Any finite size MSH island in one of these strong topological phases possesses a chiral Majorana edge mode, leading to a large spectral weight of the zero energy LDOS at all edges of the island, as shown in Fig.~\ref{fig:MSH_stacked}d. 

To understand the emergence of the weak topological and higher order topological phases in such MSH systems, it is important to realize that their fundamental building block is the single magnetic chain.
In the region of the phase diagram where the weak topological phase is realized, the two adjacent MSH chains are coupled such that they form a single, effective topological Kitaev chain, exhibiting a single Majorana Zero Mode (MZM) at each end. These double-chain MZMs are uniformly coupled along the $y$-edge, thus generating a dispersive Majorana edge band along the $y$-direction, accompanied by a large spectral weight along the $y$-edge of a finite-size island, as shown in Fig.~\ref{fig:MSH_stacked}e. Since the coupled adjacent magnetic chains are topological, their bulk (which extends along the $x$-axis) is gapped, and thus no spectral weight appears in the zero-energy LDOS along the $x$-edge of the MSH island (see Fig.~\ref{fig:MSH_stacked}e).

In contrast, the HOTSC phase occurs when each adatom chain is  individually topological. In this case, the relative strength of the coupling of MZMs between two adjacent chains (referred to as the intra-pair coupling), or between chains across an empty row 
(referred to as inter-pair coupling) determines whether the HOTSC phase emerges for a single-chain or for a double-chain termination.  
If the intra-pair coupling is stronger than the inter-pair coupling, then a single unpaired MZM remains in the corner of the island for the single-chain termination, thus forming a Majorana corner mode and giving rise to the large spectral weight of the LDOS in the corners of a finite-size island, as shown in Fig.~\ref{fig:MSH_stacked}f. At the same time, for a double-chain termination, all MZMs hybridize and the $y$-edge is completely gapped.  The situation is reversed if the inter-pair coupling is stronger than the intra-pair coupling, implying that the HOTSC phases for singe-chain and double-chain termination are complementary in the phase diagram, as confirmed by the results shown in Figs.~\ref{fig:MSH_stacked}b and \ref{fig:MSH_stacked}c. 

This picture also allows us to understand the two different types of edge transitions that separate the trivial from the HOTSC phases in such stacked MSH systems. An $x$-edge transition occurs when the individual chains themselves undergo a trivial to topological transition, while the $y$-edge transition arises when the strengths of the intra-pair and inter-pair couplings become equal, thus reversing their relative strength between the trivial and HOTSC phases. This also implies that the phase transitions into the HOTSC phases for the single-chain and double-chain terminations share the same $y$-edge transition line, as confirmed by the results shown in Figs.~\ref{fig:MSH_stacked}b and \ref{fig:MSH_stacked}c. Moreover, the complementary pattern of HOTSC phases also implies that by changing the edge termination of the MSH system between single and double chain termination using atomic manipulation techniques, allows one to tune the MSH system between trivial and HOTSC phases. 

We finally note that since the HOTSC phases are separated from the trivial phases by surface (i.e., edge) gap closings, rather than bulk gap closings (as in the case of the strong topological phases), these HOTSC phases emerge as boundary-obstructed phases~\cite{benalcazar2017science,benalcazar2017prb,Geier2018,Khalaf2021} and thus possess an extrinsic higher order topology.  

\section{Experimental investigation of single magnetic impurities, dimers and spin chains on superconducting substrates}\label{experiment_1}
The main focus of this chapter is a review of the experimental investigation of one-dimensional spin chains coupled to superconducting substrates (Sec.~\ref{experiment_1-1D}). It compares the different spin chain systems, ordered by their magnetic structure (see Table~\ref{table:spinchains}), focusing mostly on systems fabricated by STM-tip-based atom manipulation which were not already extensively discussed in previous reviews~\cite{JaeckNAT-REV-PHYS2021, YazdaniSCI2023}. The bulk of experimental work has been done for chains of $3d$ transition metal atoms, while a few works also dealt with chains from $4f$ rare earth atoms. Except for dimers, chains of molecular spins have not been experimentally investigated yet. Moreover, review articles about the theoretical concepts~\cite{Balatsky2006} and experimental work on single atomic and molecular spins and pairs~\cite{Heinrich2018} of such already entered the literature. Therefore, in the first and second sections of this chapter, we only briefly review the experimental work on single atomic spins coupled to superconducting substrates (Sec.~\ref{experiment_0D}) and dimers thereof (Sec.~\ref{experiment_dim}), as far as it strongly relates to the work on atomic spin chains discussed in Sec.~\ref{experiment_1-1D}. In addition, Sec.~\ref{experiment_shibatip} contains a short review of the work related to spins attached to superconducting STM tips.

\subsection{Magnetic single atoms on superconducting surfaces}\label{experiment_0D}
The effects of impurities in superconductors have been first used to study unconventional superconductivity at the atomic scale by STS~\cite{Balatsky2006,Hudson1999Atomic,YazdaniPRL1999Impurity,Pan2000Imaging}. Since the pioneering work of A. Yazdani~\cite{Yazdani1997}, YSR states induced by magnetic impurities in or at the surface of $s$-wave superconductors have been studied in great detail~\cite{Heinrich2018} using atomic~\cite{Cornils2017,Schneider2019,Wang2021,Kamlapure2021,ji2008high,ji2010application,menard2015coherent,ruby2015tunneling,ruby2016orbital,randeria2016scanning,choi2017mapping,senkpiel2019robustness,liebhaber2019yu,odobesko2020observation,huang2020quantum,perrin2020unveiling,trahms2023diode,Wang2024Quantum} or molecular magnetic defects~\cite{hatter2015magnetic,franke2011competition,bauer2013microscopic,heinrich2013protection,hatter2017scaling,farinacci2018tuning,malavolti2018tunable,kezilebieke2019observation,farinacci2020interfering,rubio2021coupled,xia2022spin,vaxevani2022extending,trivini2023cooper,cortes2021observation}. Though the latter studies of YSR states induced by molecular spins contributed in large part to our current understanding of the physics of spins on $s$-wave superconductors~\cite{Heinrich2018}, in particular pertaining to the competition of Kondo coupling and Cooper pairing, here we shall focus largely on those atomic spins which have been used as building blocks of MSH systems (Table~\ref{table:spinchains}).

\begin{figure}[H]
\centering
\includegraphics[width=1\textwidth]{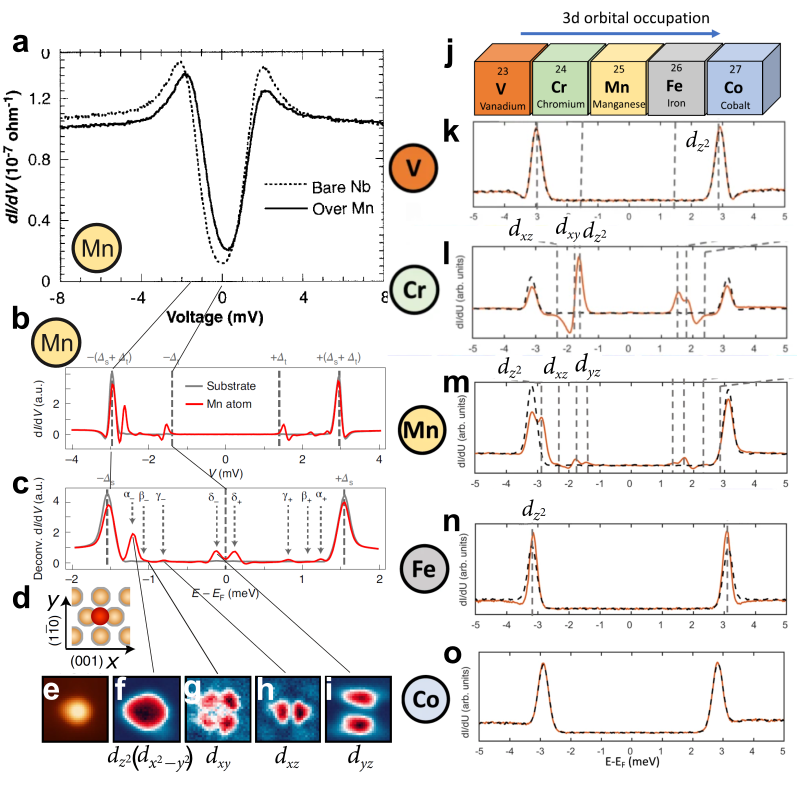}
\caption{YSR states of $3d$ transition metal atoms on Nb(110). \textbf{a} $\didv$ spectra taken on a Mn atom on Nb(110) and on the bare substrate using a normal metal tip at $T\approx\SI{4}{\kelvin}$~\cite{Yazdani1997}. \textbf{b} Same as panel a but measured using a SC tip at $T\approx\SI{0.3}{\kelvin}$~\cite{Schneider2021b}. The bias voltages according to the tip ($\Delta_\textrm{t}$) and sample gaps ($\Delta_\textrm{s}$) are indicated by vertical dashed lines. \textbf{c} Deconvolution of spectra in b in order to remove the effect of the SC tip DOS~\cite{Schneider2021b} with indicated particle hole pairs of YSR states. \textbf{d} Top view sketch of the uppermost layer of the Nb(110) surface with Nb atoms (brown) and a Mn atom (red) as well as crystallographic directions. \textbf{e} to \textbf{i} $\didv$ maps (f to i) around a single Mn atom shown in the STM image in e evaluated at bias voltages indicated in c according to the YSR state energies. The panels each have dimensions of $2 \times 2$ nm$^2$. \textbf{j} $3d$ transition metal elements investigated in this figure. \textbf{k} to \textbf{o} $\didv$ spectra taken with a SC tip at $T=\SI{1.9}{\kelvin}$ on the atom on Nb(110) indicated on the left (colored spectra) and on the substrate (dashed spectra). The dashed horizontal lines indicate the energies of the $d_{z^2}$, $d_{xz}$, $d_{yz}$, and $d_{xy}$ YSR states~\cite{Kuester2021}. Panel a from~\cite{Yazdani1997}. Reprinted with permission from AAAS. Panels b to i adapted from~\cite{Schneider2021b}. Panels j to o adapted from~\cite{Kuester2021}.}\label{fig:YSRatomsNb}
\end{figure}

Fig.~\ref{fig:YSRatomsNb}a shows the $\didv$ spectrum measured, using a normal metal tip, over a single Mn atom on Nb(110) at a temperature of $\approx\SI{4}{\kelvin}$~\cite{Yazdani1997}. In comparison to a spectrum measured on the bare surface, there is an enhancement of the $\didv$ signal inside the gap of the Nb. This enhancement is only detectable up to about $\SI{1}{\nano \meter}$ distance from the impurity. In the same work, Gd and Ag atoms were studied as well~\cite{Yazdani1997}. Gd atoms also revealed in-gap states, though with an oscillatory decay on a similar length scale, and an inverted particle-hole asymmetry. The presumably nonmagnetic Ag atoms did not show any significant enhancement of the in-gap signal. The in--gap signal was interpreted as strong indication of the YSR bound state induced by the magnetic impurity in the superconductor. Note, that in a more recent work on Gd atoms on Nb(110)~\cite{Wang2023}, the spectra over the Gd did not show any in-gap states and were indistinguishable from the spectra taken on clean Nb(110). It was, therefore, speculated, that the substrate used in the pioneering work~\cite{Yazdani1997} shown in Fig.~\ref{fig:YSRatomsNb}a might have been, at least partially, oxygen-reconstructed~\cite{Odobesko2019Preparation}.

Pioneered by the group of Q.-K. Xue~\cite{ji2008high,ji2010application}, the energy resolution in such experiments was largely enhanced by making use of superconductive STM tips from Nb utilizing the resulting energetically narrow coherence peak in the tip DOS to detect the YSR states of single magnetic atoms adsorbed on superconductors (Pb thin films were used in those works). As a result, the energy resolution of STS spectra can reach $\approx \SI{100}{\micro \electronvolt}$ at temperatures around $\SI{300}{\milli \kelvin}$, if the capacitance of the tip is large enough, which is typically achieved by large tip wire diameters with appropriate opening angles~\cite{Ast2016Sensing}. A drawback of this method is, though, that $\didv$ spectra are no longer directly linked to the samples LDOS, but need to be deconvoluted to extract the latter, which, as the tip DOS is usually unknown, can be quite demanding depending on the desired accuracy. To first order approximation, as a result of the tip gap $\Delta_\textrm{t}$, all spectral features are shifted away from zero bias by $\Delta_\textrm{t}/e$, such that the sample's Fermi energy $E_\textrm{F}$ ends up at a bias voltage of $\Delta_\textrm{t}/e$. Fig.~\ref{fig:YSRatomsNb}b shows a $\didv$ spectrum measured at $\SI{300}{\milli \kelvin}$ over a Mn atom on clean, non-oxygen-reconstructed Nb(110) (see STM image and sketch of the atomic positions in Fig.~\ref{fig:YSRatomsNb}d,e) using a superconducting tip, which was prepared by indentation of a W tip into the Nb substrate to a depth of several nanometres so that a superconducting cluster was formed on the tip apex~\cite{Schneider2021b}. In comparison to the substrate spectrum (grey curve) and in contrast to Fig.~\ref{fig:YSRatomsNb}a, it got obvious that the Mn atom induces at least four particle-hole pairs of YSR states in the gap which were labelled with decreasing energy by $\alpha_{+,-}$, $\beta_{+,-}$, $\gamma_{+,-}$, and $\delta_{+,-}$ (see the deconvoluted spectrum in Fig.~\ref{fig:YSRatomsNb}b). Focusing on the most intense two YSR states, the intensities of the close-to gap edge $\alpha_{+,-}$ YSR state were shown to be strongly particle-hole asymmetric, while those of the close-to $E_\textrm{F}$ $\delta_{+,-}$ YSR state are particle-hole symmetric. $\didv$ maps taken at the bias voltages of all four YSR peaks around the Mn atom (Fig.~\ref{fig:YSRatomsNb}f-i) revealed that the symmetries and spatial appearances of these states, in short-range distance from the atom, are akin to those of the 3$d$ orbitals as viewed from above. This leads to the conclusion that the $\alpha_{+,-}$, $\beta_{+,-}$, $\gamma_{+,-}$, and $\delta_{+,-}$ YSR states are induced by the $d_{z^2}$ (possibly with an additional faint $d_{x^2-y^2}$ character), $d_{xy}$, $d_{xz}$, and $d_{yz}$ orbitals, respectively~\cite{Beck2021}.

Similar measurements on Mn at $T=\SI{1.9}{\kelvin}$ using SC tips were reported in Ref. ~\cite{Kuester2021} (Fig.~\ref{fig:YSRatomsNb}m). At this measurement temperature, the spectra showed additional thermal replica of the $d_{yz}$ YSR state and the weakest intensity $d_{xy}$ YSR state was not detected (note, that the $x-$ and $y-$ axes are interchanged in Ref.~\cite{Kuester2021} with respect to the nomenclature in Ref.~\cite{Schneider2021b}). By a systematic analysis of $\didv$ spectra of all five elements from the 3$d$ transition metal series, i.e., V, Cr, Mn, Fe, and Co, measured in the SC state (Figs.~\ref{fig:YSRatomsNb}k-o) and normal metallic state, as well as Josephson spectroscopy (both not shown), and comparison to time-dependent DFT and many-body
perturbation theory, the following general trends were concluded. With increasing 3$d$ orbital occupation, the multi-orbital YSR states start to split off from the coherence peak for V (Fig.~\ref{fig:YSRatomsNb}k, $d_{z^2}$ very close to coherence peak), have just crossed $E_\textrm{F}$ for Cr (Fig.~\ref{fig:YSRatomsNb}l), then move towards the coherence peak on the other side of the gap for Mn and Fe (Figs.~\ref{fig:YSRatomsNb}m, n, for the latter $d_{z^2}$ is again very close to the coherence peak), and finally disappear for Co (Fig.~\ref{fig:YSRatomsNb}o). The Josephson spectroscopy enabled to directly measure the local variation of the SC pairing amplitude for the different atoms (not shown here). V revealed only a tiny pair breaking effect, Cr showed the strongest effect, and it got subsequently weaker for Mn and Fe, and finally vanished completely for Co. The $\didv$ spectra around $E_{\textrm F}$ in the metallic state of the substrate (not shown here) additionally unveiled a relatively featureless signal for V and Co, while there were close-to zero bias resonances with increasing broadening and decreasing intensity for Cr, Mn, and Fe. These effects were correlated with the following trends observed in the \emph{ab-initio} calculations. The calculated total magnetic moment of the adatom-polarized-substrate complex is small for V ($\SI{1.4}{}\mu_{\textrm B}$), largest for Cr ($\SI{3.3}{}\mu_{\textrm B}$), and then decreases for Mn ($\SI{3.0}{}\mu_{\textrm B}$) and Fe ($\SI{1.5}{}\mu_{\textrm B}$), until it is completely quenched for Co ($\SI{0}{}\mu_{\textrm B}$). All five elements experience a relatively strong hybridization with the substrate which successively decreases from V to Co by a factor of about 1/2. In addition the \emph{ab-initio} calculations revealed an in-plane magnetic anisotropy for V and Cr, while Mn and Fe prefer an out-of-plane magnetic orientation. Overall, there is a complex interplay of magnetic moments, anisotropy and hybridization-induced LDOS effects which results in the experimentally observed trends discussed above, in the energies of YSR states shifting through the gap, the pair breaking effects and the successively decreasing lifetimes (increasing broadening) of spin excitations from Cr over Mn to Fe.

\begin{figure}[H]
\centering
\includegraphics[width=\textwidth]{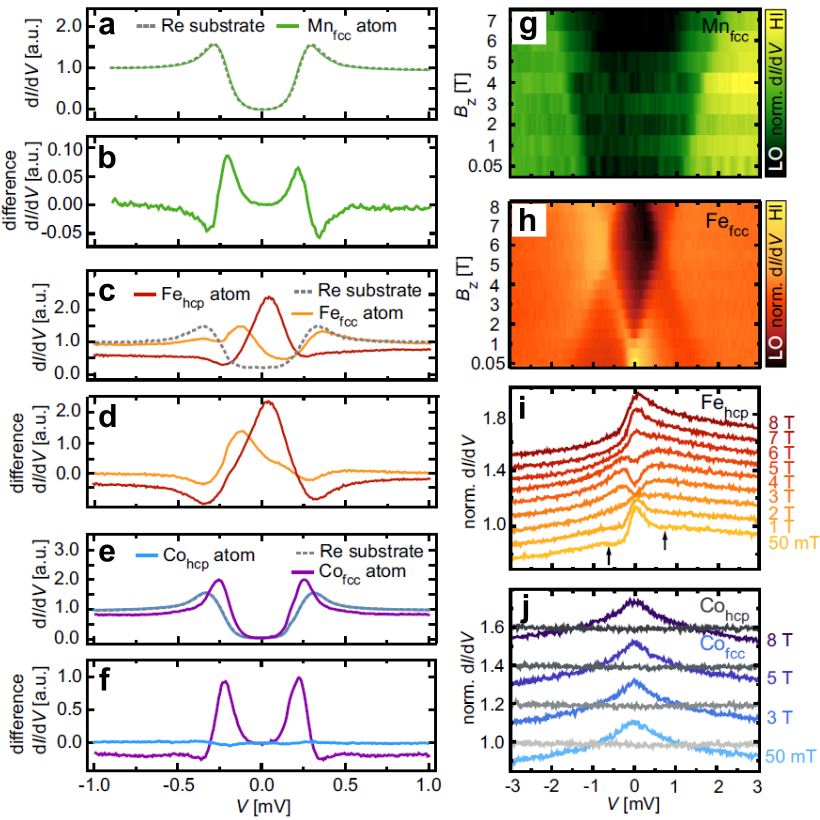}
\caption{YSR states and spin excitations of $3d$ transition metal atoms on Re(0001). \textbf{a} to \textbf{f} $\didv$ spectra of a Mn atom on the fcc adsorption site, and of Fe and Co atoms on fcc and hcp adsorption sites on Re(0001) in comparison to spectra taken on the substrate, as indicated. In panels b, d, and f, the substrate spectra were subtracted from the spectra taken on the atoms. \textbf{g} to \textbf{j} Magnetic field dependent $\didv$ spectra taken on the different atomic species. The spectra were normalized by division of a substrate spectrum taken with the same microtip. Figure adapted from~\cite{Schneider2019}.}\label{fig:YSRatomsRe}
\end{figure}

Similarly, the orbital-occupation dependent YSR state properties were experimentally investigated and compared to \emph{ab-initio} calculations for Mn, Fe, and Co adsorbed on Re(0001)~\cite{Schneider2019} as reviewed in the following (Fig.~\ref{fig:YSRatomsRe}). Because of the approximately five times smaller gap compared to Nb ($\Delta_\textrm{Nb}\approx\SI{1.5}{\milli\electronvolt}$, $\Delta_\textrm{Re}\approx\SI{0.28}{\milli\electronvolt}$) it was not possible to resolve the multi-orbital character, neither energetically nor spatially, of the observed YSR states at the measurement temperature of $T=\SI{0.3}{K}$. The two possible three-fold coordinated hollow adsorption sites fcc and hcp on the hcp(0001) surface were observed for Fe and Co, while Mn only occupies the fcc site. Due to the different substrate-hybridizations expected for the two different adsorption sites, the energetical positions of the YSR states also strongly depend on the latter, and not only on the orbital occupation (Figs.~\ref{fig:YSRatomsRe}a-f). The generally observed trend was, that, for each adatom species, the hcp adsorption site revealed a stronger coupling to the substrate electrons compared to the fcc site. In contrast to the Nb case discussed above, however, the coupling also increases going through the 3d series from Mn over Fe to Co. The interplay of magnetic moments, anisotropy and hybridization-induced LDOS effects lead to a closest energetical position of the dominant YSR states to $E_{\textrm F}$ for the hcp Fe, while the YSR states of fcc Mn and Co are barely split off from the coherence peak, and finally, hcp Co is again nonmagnetic without showing any YSR states (Figs.~\ref{fig:YSRatomsRe}a-f). Also differently from the Nb case, while Mn still features an out-of plane magnetic anisotropy, Fe has an easy-plane magnetic anisotropy. This was nicely visible in the magnetic-field dependent $\didv$ spectra around $E_{\textrm F}$ in the metallic state of the substrate (Figs.~\ref{fig:YSRatomsRe}g and h) which were shown to be consistent with the tunnel-current driven spin excitations of a Mn atom with spin $S=2$ and a relatively strong out-of plane magnetic anisotropy, and of an fcc Fe atom with $S=3/2$ and a relatively strong easy-plane magnetic anisotropy with weak Kondo screening in the zero field case, respectively. The hcp Co (Figs.~\ref{fig:YSRatomsRe}j) does not show any resonance around $E_{\textrm F}$ which is also consistent with the theoretically predicted quenching of its magnetic moment. For the hcp Fe and the fcc Co atoms (Figs.~\ref{fig:YSRatomsRe}i and j), the conclusions which can be drawn from the experimental spectra were less straight forward. For the fcc Co, there is a broad peak which does not change in a magnetic field up to $B=\SI{8}{\tesla}$, which would be consistent with a strong Kondo-screening based explanation. However, for the hcp Fe, there is a complex mix of a broadened step, a relatively broad and two narrow Frota function like resonances, where the two energetically narrow features shift with the magnetic field consistent with spin excitations of a $S=3/2$ impurity, while the two broader features do not change considerably. It is thus still an open question, whether the latter $\didv$ data in the metallic regime are better described within a (weakly) Kondo coupled scheme, or within a spinaron picture~\cite{Bouaziz2020Spinaron,Brinker2022Anomalous,Friedrich2024Evidence}. Overall, the results for both, the Nb and Re substrate, showed that Cr, Mn, and Fe with close to half 3$d$ filling reveal signatures of large spins subject to considerable magnetic anisotropies. For such impurities Hund's coupling~\cite{Khajetoorians2015Hund} and/or strong substrate coupling~\cite{Gauyacq2015Decoherence,Delgado2017SpinDecoherence} tend to quench strong Kondo effects and, thus, these systems might be well-described by a quasi-classical magnetic moment acting on the superconductor.

\begin{figure}[H]
\centering
\includegraphics[width=0.9\textwidth]{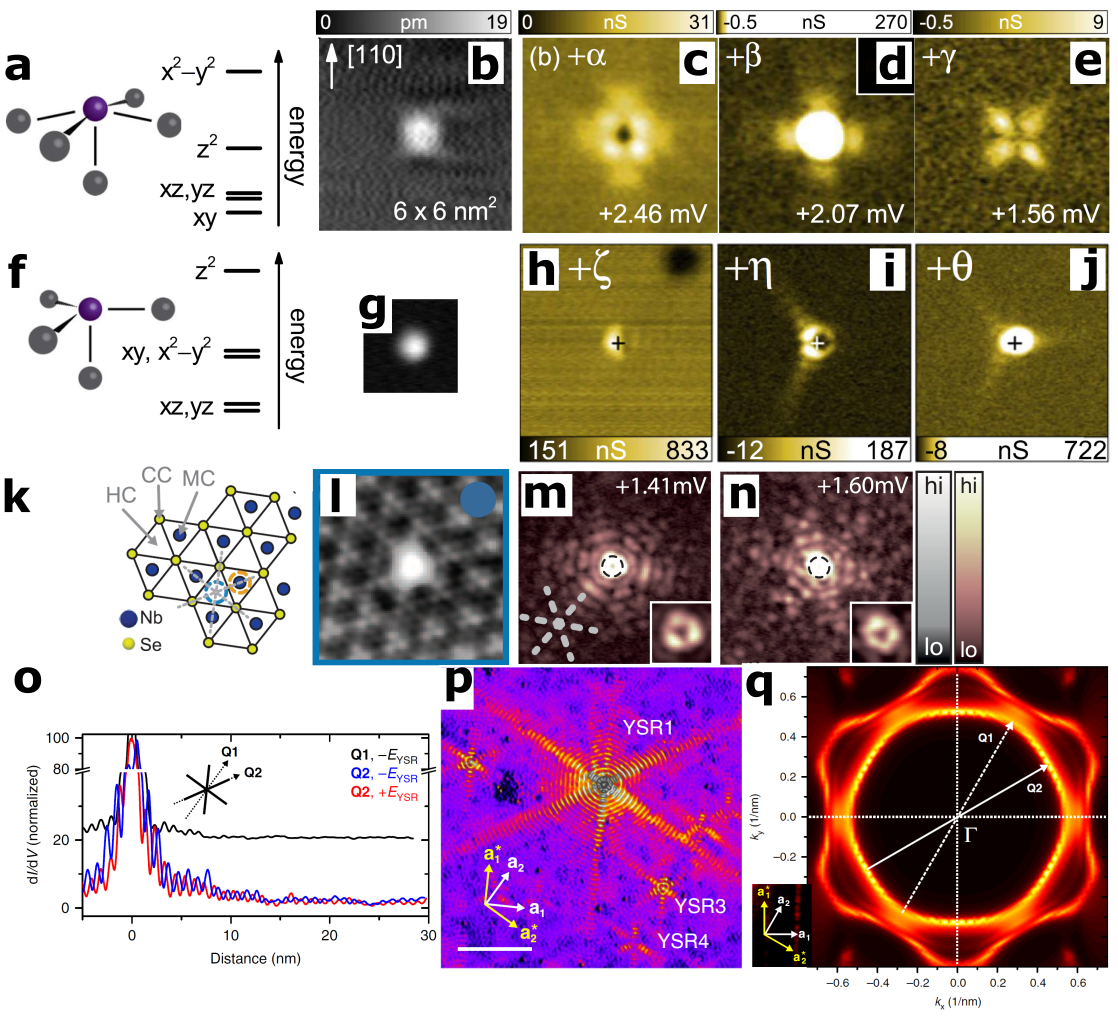}
\caption{Short- and long-range wavefunctions of YSR states of $3d$ transition metal atoms at the surfaces of different SCs. \textbf{a} to \textbf{j} Crystal field splittings of 3$d$ levels (panels a,f)~\cite{ruby2016orbital}, constant-current STM images (panels b, g~\cite{ruby2016orbital}, size of g is 2.6 x 2.6 nm$^2$), and $\didv$ maps (panels c-e and h-j~\cite{ruby2016orbital}) at the YSR state energies of Mn atoms on Pb(001) and Pb(111), respectively. \textbf{k} to \textbf{n} Top
view indicating the hollow-centered (HC) adsorption lattice site (panel k), constant-current STM image (panel l), and $\didv$ maps (panels m,n) at the $\alpha$ (m) and $\beta$ (n) YSR state energies of a HC Fe atom on $2H$-NbSe${_2}$~\cite{liebhaber2019yu}. Insets in m, n show a $2 \times 2$~nm$^2$ close-up view around the
center of the atoms. \textbf{o} and \textbf{p} $\didv$ line-profile along YSR long-range streak (panel o) and $\didv$ map (panel p) of the YSR states of magnetic impurities below the (0001) surface of La~\cite{Kim2020Longrange}. Scale bar, $\SI{15}{\nano\meter}$. \textbf{q} Calculated Bloch spectral function at the Fermi energy for the surface atomic layer of La(0001), highlighting the Fermi surface formed by the surface bands~\cite{Kim2020Longrange}. The white arrow labelled $Q_2$ represents the quasiparticle focus direction leading to the YSR streaks observed in panel p. Panels a to j adapted with permission from Ref.~\cite{ruby2016orbital}. Copyrighted by the American Physical Society. Panels k to n adapted with permission from~\cite{liebhaber2019yu}. Copyright 2024 American Chemical Society. Panels o to q adapted from~\cite{Kim2020Longrange}.}\label{fig:YSRatomsother}
\end{figure}

In addition to the orbital occupation and adsorption-site dependent trends~\cite{odobesko2020observation} discussed above, the dependencies of YSR state energies and short-range wavefunctions on various other effects have been studied experimentally: 1) on the orbital-substrate-coupling, e.g. by changing the substrate DOS using a charge density wave~\cite{liebhaber2019yu} or by affecting the adatom substrate hybridization by variation of the STM tip-adatom separation~\cite{farinacci2018tuning,malavolti2018tunable,huang2020quantum}; 2) on the magnetic anisotropy~\cite{hatter2015magnetic}; 3) on the exchange interaction to neighboring spin assemblies~\cite{Kamlapure2018}. In the following, we will review the effects of the adsorption site symmetry and substrate electronic structure on the short-range spatial appearance and on the long-range decay, respectively, of the YSR state. In two contemporaneous works, these effects were first studied experimentally for Cr atoms on Pb(111)~\cite{choi2017mapping} and Mn atoms on Pb(001) and Pb(111)~\cite{ruby2016orbital} (Fig.~\ref{fig:YSRatomsother}a-j), and were compared to density functional theory calculations in the former case. As discussed above, the short-range YSR state wave functions inherit their shapes from the atom's $d$ orbital wave functions. These are split by interaction with the surrounding anisotropic crystal field such that the potential and exchange coupling of the conduction electrons with the impurity get orbital-dependent. E.g., for Mn on a hollow adsorption site of Pb(001), the nearest neighbors form a square pyramidal coordination symmetry (Fig.~\ref{fig:YSRatomsother}a),
which removes the degeneracy of the five 3$d$ levels~\cite{ruby2016orbital}. Due to the resulting crystal field, the $d_{x^2-y^2}$-orbital is highest in energy, followed by the $d_{z^2}$ orbital, the degenerate $d_{xz}$ and $d_{yz}$ orbitals, and the energetically very close $d_{xy}$ orbital. As a result, there are only three pairs of YSR states with a four-fold symmetric spatial shape, as illustrated in Figs.~\ref{fig:YSRatomsother}c-e. In contrast, for Mn adsorbed on the hollow site of Pb(111), which is subject to a trigonal pyramidal crystal field (Fig.~\ref{fig:YSRatomsother}f), the $d_{z^2}$ orbital has highest energy, followed by the degenerate $d_{x^2-y^2}$ and $d_{xy}$ orbitals, and finally the degenerate $d_{xz}$ and $d_{yz}$ orbitals are lowest in energy~\cite{ruby2016orbital}. This again results in three YSR states (Fig.~\ref{fig:YSRatomsother}h-j). However, in this case, it was concluded~\cite{ruby2016orbital}, that the $C_{3v}$ ligand field symmetry polarizes the 3$d$ orbitals due to hybridization with the $p$
orbitals, such that the resulting shape of the YSR states is twofold symmetric. For the Cr atom on Pb(111)~\cite{choi2017mapping}, a strong relaxation results in a pseudo-fcc adsorption site between the topmost and second subsurface layer, and consequently a distorted octahedral crystal field symmetry.  Here, all degeneracies between the $3d$ states are again lifted, and, as a result, there are five pairs of YSR states.

While, as discussed above, the short-range spatial appearance of the YSR state wave function is largely governed by the crystal-field-splitted 3$d$ impurity orbitals, there is an additional long-range tail of the YSR state wave functions which is governed by the anisotropy of the Fermi surface of the superconducting substrate (see, e.g. the diagonal streaks in Figs.~\ref{fig:YSRatomsother}i, j for Mn atoms on Pb(111) or the long-range wavy pattern in Figs.~\ref{fig:YSRatomsother}m, n for Fe atoms on $2H$-NbSe${_2}$). This long-range wave function shows $2k_\textrm{F}$ periodic oscillations with a scattering-phase shift between the particle and hole parts of the wave function~\cite{menard2015coherent,Kim2020Longrange} (see, e.g., Fig.~\ref{fig:YSRatomsother}o). Note, that the closer the YSR states are to the gap edge, the smaller is the phase shift between the positive and negative YSR component~\cite{ruby2016orbital}. The long-range part gets more pronounced and intense, both, by the quasi-two-dimensional character of the SC in case of $2H$-NbSe${_2}$~\cite{liebhaber2019yu,Liebhaber2022} (Figs.~\ref{fig:YSRatomsother}m,n), and due to quasiparticle-focusing effects along directions perpendicular to low-curvature regions of the Fermi surface (e.g. along $Q_2$ in Fig.~\ref{fig:YSRatomsother}q) as shown for impurities below the (0001) surface of La (Fig.~\ref{fig:YSRatomsother}p)~\cite{Kim2020Longrange}. Liebhaber \emph{et al.}~\cite{liebhaber2019yu,Liebhaber2022} furthermore investigated the effect of the position of the magnetic impurity with respect to the charge density wave maxima for Fe atoms on $2H$-NbSe${_2}$. For most of the possible adsorption sites of the Fe atoms with respect to the charge density wave, the symmetries of the short- and long-range parts of the single atom YSR wavefunction were already reduced from $D_3$ (three-fold symmetric adsorption site) to a $D_1$ symmetry due to the symmetry reduction by the charge density wave. They also found that, in their system, the YSR energies of the two deepest-in-the-gap pairs of YSR states $\alpha$ and $\beta$ (Figs.~\ref{fig:YSRatomsother}m,n) are (anti)correlated with the charge density wave potential and the local density of states for the situation after (before) the YSR state had crossed $E_\textrm{F}$ (strong or weak coupling, respectively).

\subsection{Magnetic dimers on superconducting surfaces}\label{experiment_dim}

An adequate hybridization of the single atom or molecule YSR state wave functions discussed above after binding two atoms or molecules to a dimer is a prerequisite for the formation of YSR bands in MSH systems, and finally for the emergence of TSC and Majorana bound states (MBSs). The YSR state wave 
 function hybridization was, therefore, experimentally investigated thoroughly for atomic and molecular dimers. The early work~\cite{ji2008high}, which investigated the YSR states in dimers of Mn and Cr atoms on Pb(111), already reported on the splitting of the YSR states after the formation of Mn dimers and on the fact that their spatial shapes somewhat resembled antisymmetric linear combinations of the single atom YSR state wave function. These observations were tentatively assigned to a ferromagnetic coupling in the Mn dimer.

\begin{figure}[H]
\centering
\includegraphics[width=0.85\textwidth]{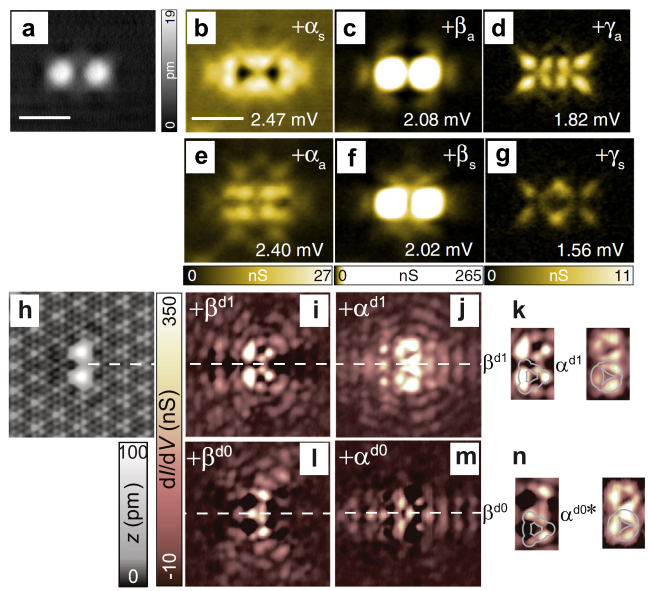}
\caption{YSR states of dimers of 3$d$ transition metal atoms on different substrates. \textbf{a} to \textbf{g} STM image (a) and $\didv$ maps (b to g) of a dimer of Mn atoms on Pb(001) taken at the indicated bias voltages showing the symmetric (subscript s) and antisymmetric (subsctript a) linear combinations of the given single atom YSR wave functions (cf. Figs.~\ref{fig:YSRatomsother}c to e)~\cite{ruby2018wave}. Scale bar, $\SI{2}{\nano\meter}$. \textbf{h} to \textbf{n} STM image (h) and $\didv$ maps (i to n) of a dimer of HC Fe atoms on $2H$-NbSe${_2}$ at different bias voltages showing the symmetric (superscript d0, zero nodal lines) and antisymmetric (superscript d1, one nodal line) linear combinations of the given single atom YSR wave functions (c.f. Figs.~\ref{fig:YSRatomsother}m, n~\cite{Liebhaber2022}. k and n are close-ups around the dimer atoms' centers. Panels a to g adapted with permission from Ref.~\cite{ruby2018wave}. Copyrighted by the American Physical Society. Panels h to n adapted from~\cite{Liebhaber2022}.}\label{fig:YSRdimersother}
\end{figure}

Later, there were four contemporaneous works which studied these effects more systematically for dimers of Cr atoms on $\beta-$Bi$_2$Pd(001)~\cite{choi2018influence}, dimers of Mn atoms on Pb(001)~\cite{ruby2018wave}, dimers of Fe atoms on Re(0001)~\cite{Kim2018} and dimers of Cobalt phtalocyanine molecules on NbSe$_2$~\cite{Kezilebieke2018coupled}. For a dimer of Mn atoms on Pb(001)~\cite{ruby2018wave} (Fig.~\ref{fig:YSRdimersother}a) the distinct shapes of the splitted YSR state wave functions and their interpretations as symmetric (Figs.~\ref{fig:YSRdimersother}b, f, g) and antisymmetric (Figs.~\ref{fig:YSRdimersother}e, c, d) linear combinations of the single atom YSR wave functions (cf. Figs.~\ref{fig:YSRatomsother}c to e) were revealed very clearly. From these results it was deduced that the atom spins in the Mn dimer cannot be aligned antiferromagnetically. In the same publication, the authors also systematically analyzed how the splittings (up to $\approx\SI{300}{\micro\electronvolt}$) and shifts of the YSR states in dimers decay with the distances between the atoms in the dimer, but also depend crucially on the orientations of the atoms in the dimer with respect to the substrate lattice. The latter result was rationalized by the strongly anisotropic shapes of the single atom YSR wave functions (Figs.~\ref{fig:YSRatomsother}c to e). However, because the range of interatomic distances in the dimer over which the splittings were observed was only slightly larger than half of the substrate's Fermi wavelength $\lambda_\textrm{F}$, it was not possible to test whether the experiment also shows the theoretically expected $\lambda_\textrm{F}/2$-periodic oscillatory behavior of the YSR splittings due to the long-range decay of the YSR state wave function. From the result of the formation of symmetric and antisymmetric linear combinations of the single atom YSR wave functions, the authors furthermore concluded, that a direct exchange or superexchange via a single substrate atom can be neglected. Instead, the size of the splittings is comparable to what would be expected for a RKKY coupling~\cite{ruby2018wave}.

Dimers of Cr atoms with different separations on $\beta-$Bi$_2$Pd(001)~\cite{choi2018influence} also revealed either shifts in the energy of a single YSR state from the individual Cr atom, or a splitting of $\approx\SI{250}{\micro\electronvolt}$. In the latter case, the spatial shape of one of the two YSR states also resembled a symmetric linear combination of the single atom YSR state wavefunction, while the other resembled an antisymmetric linear combination with a nodal line in between the two atoms of the dimer. With the help of DFT and model calculations, these observations were rationalized by an antiferromagnetic coupling in the dimer for the former, and a ferromagnetic coupling for the latter case~\cite{choi2018influence}.

Splittings up to $\approx\SI{500}{\micro\electronvolt}$ of a single molecular YSR state after molecular dimer formation have been also observed for Cobalt phtalocyanine molecules on NbSe$_2$~\cite{Kezilebieke2018coupled}. In that work, an oscillatory behaviour of the splittings as a function of molecular separation was indeed observed. However, the period was not consistent with $\lambda_\textrm{F}/2$, which was explained by the authors by an atomic-scale variation of the wave function overlap of the two impurity states following from their modelling. The effects of the oscillatory long-range interaction between magnetic atoms coupled to superconductors were investigated experimentally and theoretically in detail for Cr dimers on Nb(110)~\cite{Küster2021longrange} and for Gd dimers on Bi(110) ultrathin films grown on Nb(110)~\cite{Ding2021tuning}.

Liebhaber \emph{et al.}~\cite{Liebhaber2022} investigated the YSR wavefunction hybridization for dimers of Fe atoms adsorbed on hollow-centered adsorption sites on $2H-$NbSe$_2$ (Figs.~\ref{fig:YSRdimersother}h to n). They focused on the effect of the hybridization on the two deepest-in-the-gap pairs of YSR states $\alpha$ and $\beta$ (Figs.~\ref{fig:YSRatomsother}m,n). For each of these single atom YSR states, upon dimer formation, two states (splitting up to $\approx\SI{200}{\micro\electronvolt}$) appear in a similar energy range, which are mirror symmetric with respect to the normal to the dimer's axis (Figs.~\ref{fig:YSRdimersother}h to n). These pairs of states were again interpreted as symmetric ($+\alpha^{d0}$, $+\beta^{d0}$, Figs.~\ref{fig:YSRdimersother}l to n) and antisymmetric ($+\alpha^{d1}$, $+\beta^{d1}$, Figs.~\ref{fig:YSRdimersother}i to k) linear combinations of the single atom YSR wavefunctions. Importantly, these symmetries and the nodal lines for the antisymmetric case can be observed in both the short- as well as the long-range parts of the respective wavefunctions. In conclusiuon, the authors argued that the experimentally observed large YSR state shifts, which are in opposite directions for the two different YSR states, can only be explained within a quantum spin model reported in detail in their publication~\cite{Liebhaber2022}. They also showed, that the $\alpha$ YSR state, which is deepest in the gap, undergoes a quantum phase transition due to the RKKY interaction between the atoms in the dimer, which is another strong indication for the quantum spin nature of the Fe atoms.

\begin{figure}[H]
\centering
\includegraphics[width=0.85\textwidth]{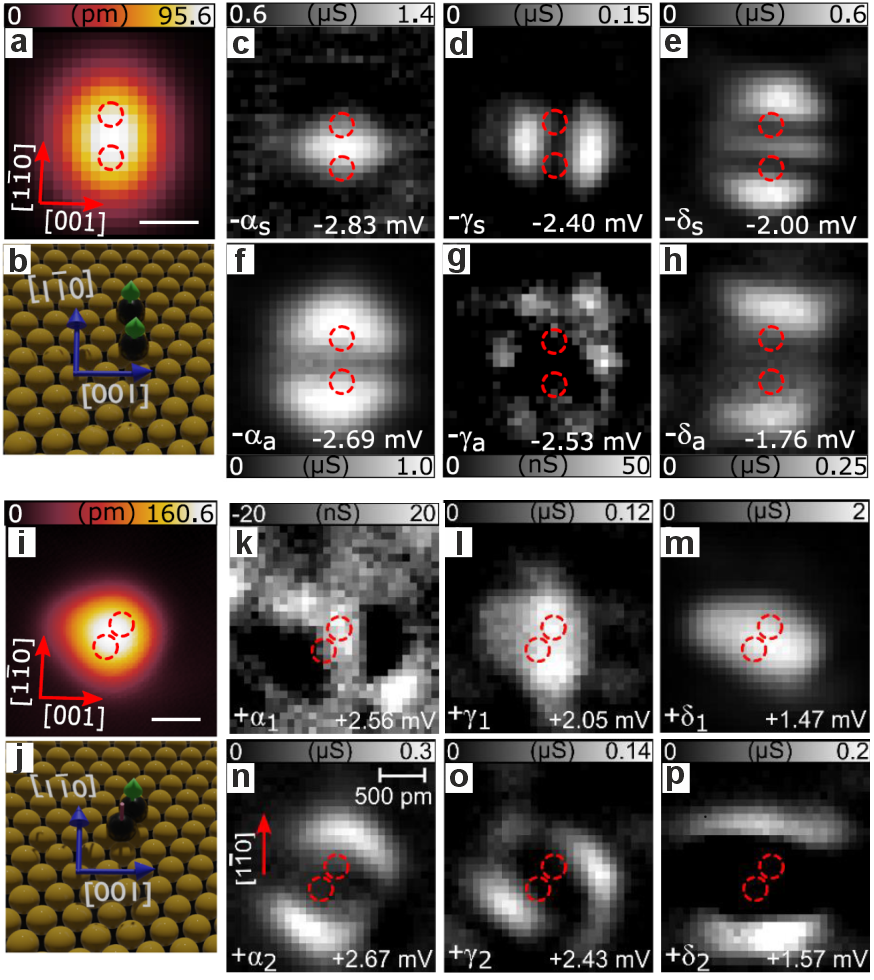}
\caption{YSR states of ferromagnetic and antiferromagnetic Mn dimers on Nb(110). \textbf{a} STM image and \textbf{b} sketch of a ferromagnetically coupled dimer of Mn atoms on Nb(110) oriented along the $[1\bar 10]$ direction as indicated in panel b. \textbf{c} to \textbf{h} $\didv$ maps of the dimer in panel a taken at the indicated bias voltages showing the symmetric (subscript s) and antisymmetric (subsctript a) linear combinations of the given single atom YSR wavefunctions (c.f. Figs.~\ref{fig:YSRatomsNb}f to i). \textbf{i} STM image and \textbf{j} sketch of an antiferromagnetically coupled dimer of Mn atoms on Nb(110) oriented along the $[1\bar 11]$ direction as indicated in panel j. \textbf{k} to \textbf{p} $\didv$ maps of the dimer in panel i taken at the indicated bias voltages showing the (approximately) symmetric (subscript 1) and antisymmetric (subscript 2) linear combinations of the given single atom YSR wavefunctions. Scale bars in panel a and i, $\SI{500}{\pico\meter}$. Red circles in panels a, c to h, i and k to p denote the locations of the Mn atoms in the dimers. Figure adapted from~\cite{Beck2021}.}\label{fig:YSRdimersNb}
\end{figure}

In the publications discussed so far, it was generally assumed that a splitting of the YSR states in dimers can only be rationalized by a ferromagnetic coupling in the pair, while strictly antiferromagnetic couplings can only lead to shifts in the single atom YSR states after dimer formation. Beck \emph{et al.}~\cite{Beck2021}
 showed that this is no longer the case, if there is considerable spin-orbit coupling (Figs.~\ref{fig:YSRdimersNb}i to p). Figs.~\ref{fig:YSRdimersNb}a to h illustrate the $\didv$ maps of the YSR wavefuncions of a dimer of Mn atoms with a separation of $\sqrt{2}a$ along the $[1\bar 10]$ direction on Nb(110). Using spin-resolved measurements on chains of atoms with the same separation having the same orientation as the dimer (see Section~\ref{experiment_shibatip}), a ferromagnetic coupling between the two atoms in the pair was experimentally revealed (as shown in Fig.~\ref{fig:YSRdimersNb}b). Similar to the examples discussed above, the corresponding $\didv$ maps showed that the $d_{z^2}$ (Figs.~\ref{fig:YSRdimersNb}c, f), $d_{xz}$ (Figs.~\ref{fig:YSRdimersNb}d, g), and $d_{yz}$ (Figs.~\ref{fig:YSRdimersNb}e, h) YSR state wave functions of the single atom (Figs.~\ref{fig:YSRatomsNb}f to i) formed symmetric (top panels of Figs.~\ref{fig:YSRdimersNb}c to e) and antisymmetric (bottom panels) linear combinations upon dimer formation. The splittings were on the order of $\approx\SI{200}{\micro\electronvolt}$. However, such splittings were also observed for the dimer of Mn atoms with a separation of $\sqrt{3}a/2$ along the $[1\bar 11]$ direction (Fig.~\ref{fig:YSRdimersNb}i), which was proven experimentally to have antiferromagnetic coupling (as schematically shown in Fig.~\ref{fig:YSRdimersNb}j). The according $\didv$ maps of the splitted YSR states are shown in Figs.~\ref{fig:YSRdimersNb}k-p, and still somewhat resemble symmetric (top panels) and antisymmetric (bottom panels) linear combinations of the single atom's YSR wavefunctions. These splittings were shown to be of similar strength compared to the ferromagnetic case. With the help of numerical simulations based on \emph{ab initio} calculated parameters, the authors were further able to show~\cite{Beck2021}, that the splitting of the YSR states in the antiferromagnetic dimer is not a result of a slight non-collinear orientation in the dimer, but must be induced by the spin-orbit coupling. The simulations also showed, that for the general antiferromagnetic case including spin-orbit coupling, the spatial distributions of the splitted YSR wavenfunctions no longer resemble pure symmetric and antisymmetric linear combinations of the original single atom YSR wavefunctions. Finally, it was shown that the wavefunction hybridization and splittings are also induced in dimers with a non-collinear spin alignment, even if there is no spin-orbit coupling.

The formation of symmetric and antisymmetric linear combinations of single atom YSR state wave functions in dimers with splittings on the order of several hundreds of $\SI{}{\micro\electronvolt}$ was further explored for the tailoring of YSR bands in chains of atoms as will be discussed in Section~\ref{experiment_1-1D}. First, we will discuss in the next section, how YSR states on SC tips can be used, amongst others, for SPSTS measurements to reveal the spin couplings and their orders in chains and lattices of magnetic atoms on SCs.

\subsection{Magnetic atoms on superconducting tips}\label{experiment_shibatip}

The physics of the YSR state has been furthermore studied in detail experimentally by attaching magnetic atoms to the apex of superconducting STM tips, which adds additional tunability of the parameters. In this Section we will give a very concise summary of the main results, before we discuss in more detail how STM tips hosting YSR states can be used for SP-STS of MSH systems. H. Huang \emph{et al.} used magnetic impurities attached to the apex of superconducting V tips. With such experimental platform, they were able to study the lifetimes in the transition between sequential and coherent tunneling regimes from the YSR state of the magnetic atom at the tip to a YSR state of an intrinsic impurity on a superconducting V(100) sample~\cite{huang2020tunnelling}, as well as the spin dynamics in the tunneling process by experimentally extracting the angle between the spins~\cite{huang2021spin}. The same group also made use of changing the distance between the YSR-functionalized tip and the superconducting substrate to tune the impurity-substrate coupling via the changing atomic forces between tip and sample. This enabled them to realize a 0 to $\pi$ transition of the Josephson junction through the YSR state~\cite{karan2022superconducting} as well as a quantitative analysis of the scaling between the YSR state energy and the Kondo temperature~\cite{Huang2023}. They, furthermore, showed the first evidence of an excitation of the YSR states on the tip by microwave irradiation~\cite{siebrecht2023microwave}. Finally, Odobesko \emph{et al.}~\cite{Odobesko2023} attached a $CO$ molecule to a superconducting STM tip and were able to, thereby, maximize both the spatial and energy resolution of the STM tip.

\begin{figure}[H]
\centering
\includegraphics[width=0.8\textwidth]{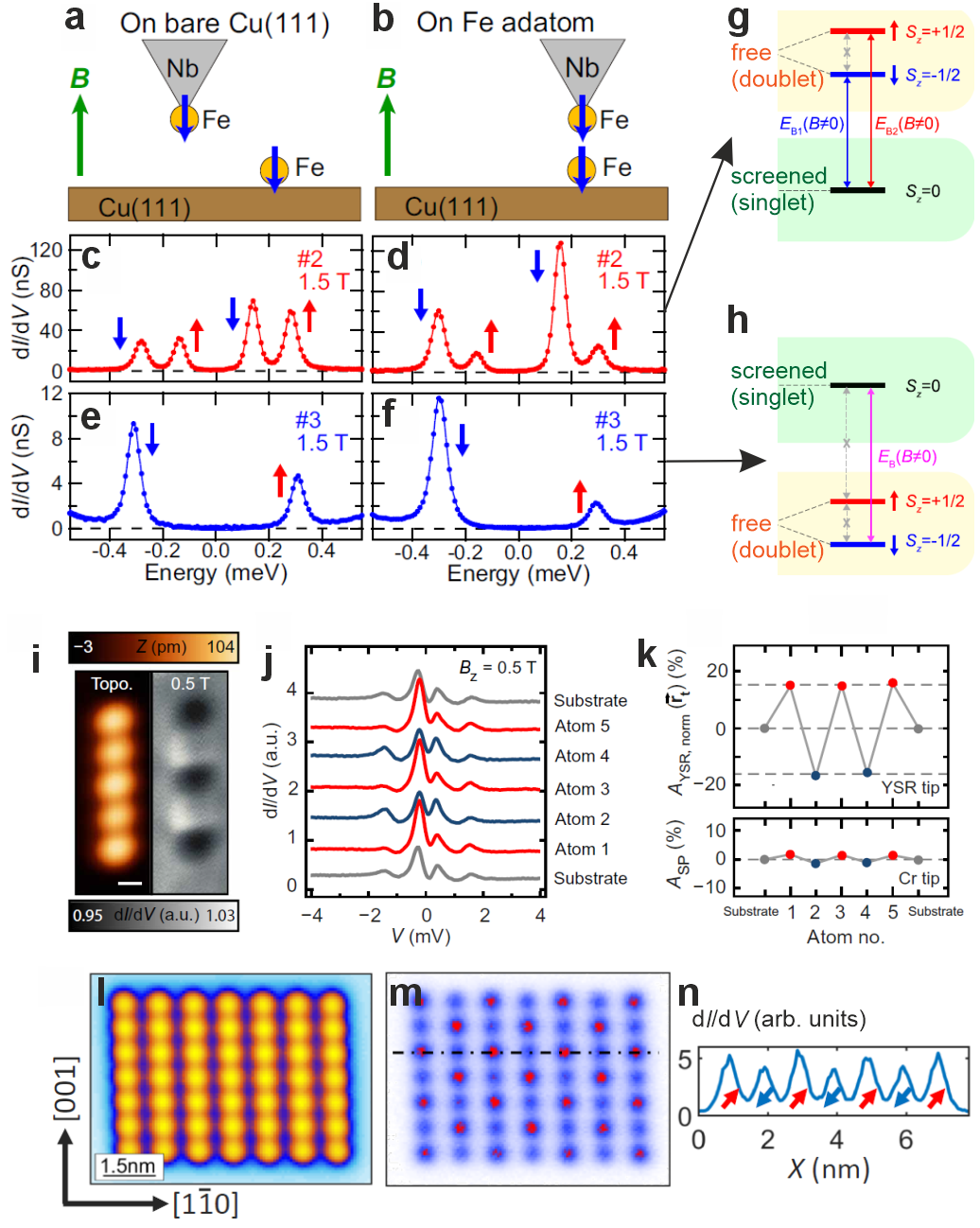}
\caption{Spin-resolved STS (SPSTS) using superconducting tips functionalized by YSR states. \textbf{a}, \textbf{b} Schematic illustrations of the experimental configurations used for the magnetic field ($B$) dependent YSR-state based SPSTS, e.g. on Fe atoms on Cu(111). Arrows on the Fe atoms represent the directions of the majority spin of the Fe atoms, which are antiparallel to the magnetic
moments. \textbf{c} to \textbf{f} $\didv$ spectra taken at $B=\SI{1.5}{\tesla}$ perpendicular to the surface with two different micro-tips (\#2 and \#3 as indicated) on the bare Cu(111) (panels c, e) and on the Fe atom (panels d, f). Arrows indicate the spin orientations of the YSR states. \textbf{g}, \textbf{h} Many-body state energy diagrams of a spin $1/2$ magnetic impurity on the tip in a magnetic field assuming screened (singlet, panel g) or free (doublet, panel h) spin ground states, which reflect the experimental results shown in panels c, d, or panels e, f, respectively ~\cite{Machida2022Zeeman}. \textbf{i} STM image (left) and $\didv$ map (right) of an antiferromagnetic five-atom Mn chain on Nb(110) measured with a SC tip with an attached Fe$_2$ cluster at an out-of-plane field $B=\SI{0.5}{\tesla}$. \textbf{j} $\didv$ spectra taken under the same conditions using this tip on the substrate and on each atom of the chain, as indicated. \textbf{k} Particle-hole intensity asymmetry of the YSR peaks in panel j along the chain (top panel) normalized to the asymmetry of the substrate spectrum compared to the spin asymmetry obtained with a conventional Cr tip (bottom panel)~\cite{Schneider2021}. \textbf{l} STM image and \textbf{m} $\didv$ map of a lattice of Cr atoms on Nb(110) taken using a SC tip with an attached Cr atom in a tilted magnetic field ($B=\SI{0.7}{\tesla}$ in-plane, $B=\SI{1}{\tesla}$ out-of-plane). \textbf{n} Profile of the $\didv$ map in panel m along the dashed line where the arrows indicate the suggested tilted spin order in the lattice~\cite{Küster2023}. Panels a to h adapted from~\cite{Machida2022Zeeman}. Panels i to k from~\cite{Schneider2021}. Reprinted with permission from AAAS. Panels l to n adapted from~\cite{Küster2023}.}\label{fig:Shibatip}
\end{figure}

Schneider \emph{et al.}~\cite{Schneider2021} first demonstrated that the spin-polarization of the YSR states of magnetic atoms attached to the STM tip can be used for SP-STS investigations of MSH systems (Figs.~\ref{fig:Shibatip}i-k, see discussion below). The underlying physics was thereafter experimentally studied in detail by tunneling between the YSR states of Fe atoms attached to a superconducting Nb tip and Fe atoms on a normal metallic Cu(111) surface in an out-of-plane magnetic field~\cite{Machida2022Zeeman} (Figs.~\ref{fig:Shibatip}a,b). Fe atoms were attached to the Nb tip by stabilizing the tip above an Fe atom on Cu(111), then opening the STM feedback loop and applying a positive voltage pulse to the sample. To avoid complications deriving from multiple orbitals, magnetic anisotropy, or multichannel Kondo screening, the authors furthermore selected only such tips, where a single particle-hole pair of YSR states dominated
the $\didv$ spectrum. For some of the tips prepared as such, e.g. tip \#2 (Figs.~\ref{fig:Shibatip}c,d), the $\didv$ spectra taken on the bare Cu(111) in an out-of-plane magnetic field of $B=\SI{1.5}{\tesla}$ revealed a splitting of the single pair of YSR states (Fig.~\ref{fig:Shibatip}c) with an increased intensity asymmetry if measured on an Fe/Cu(111) atom (Fig.~\ref{fig:Shibatip}d). For other tips, e.g. tip \#3 (Figs.~\ref{fig:Shibatip}e,f), the single pair of YSR states was just shifted in energy (Fig.~\ref{fig:Shibatip}e) by the application of the magnetic field, and showed an increase in the intensity asymmetry between the electron and hole part of the YSR state if measured on the Fe/Cu(111) atom (Fig.~\ref{fig:Shibatip}f). The data was explained by the two different possible combinations of ground and excited states of a magnetic atom attached to the superconducting tip. For strong coupling between the magnetic atom spin and the substrate conduction electrons, the atom-tip system is in the screened, i.e., singlet, ground state and its lowest possible excited state is the doublet spin state (Fig.~\ref{fig:Shibatip}g). In the magnetic field the doublet will Zeeman split leading to two (particle-hole) spin-polarized pairs of excitations found with tip \#2 (Fig.~\ref{fig:Shibatip}d). For weak coupling the magnetic atom on the SC tip is in the free, doublet, ground state and the lowest possible excited state is the singlet spin state (Fig.~\ref{fig:Shibatip}h). In the magnetic field the doublet will again Zeeman split, but only the lower energy state is occupied, leading to a single particle-hole pair of excitations which is also spin-polarized, as the one found with tip \#3 (Fig.~\ref{fig:Shibatip}f). While this reasoning on the base of Figs.~\ref{fig:Shibatip}g,h is based on a spin 1/2 atom, it was argued in~\cite{Machida2022Zeeman} that the same behaviour in the spectra is also expected for an atom with larger spin. The arguments furthermore rely on a Zeeman splitting which is strong compared to the thermal energy, thus requiring sufficiently strong magnetic fields and/or low temperatures. Under these conditions, the YSR peaks are fully spin-polarized. Combined with the enhanced energy resolution provided by superconducting tips (see Section~\ref{experiment_0D}), such YSR-functionalized tips constitute an ideal choice for high-energy resolution SP-STS measurements with the possibility to measure the atomically resolved absolute values of the sample's spin-polarization~\cite{Machida2022Zeeman,Schneider2021}.

The possibility to quantify the spin-polarization of samples via YSR-tips was demonstrated by Schneider \emph{et al.}~\cite{Schneider2021} for ferromagnetic and antiferromagnetic Mn chains assembled by STM-based atom manipulation on Nb(110), as illustrated in Figs.~\ref{fig:Shibatip}i-k. The $\didv$ map of the five-atom Mn chain on Nb(110) (left panel of Fig.~\ref{fig:Shibatip}i) measured using a superconducting tip with two attached Fe atoms, i.e., a Fe$_2$ cluster, in a magnetic field of $B=\SI{0.5}{\tesla}$ is shown in the right panel of Fig.~\ref{fig:Shibatip}i. It revealed a strong alternating contrast between every other atom due to the antiferromagnetic out-of-plane oriented spin order of this Mn chain. Accordingly, the $\didv$ spectra taken on each of the Mn atoms along the chain displayed an alternating intensity asymmetry between the particle and hole parts of the YSR peaks (Fig.~\ref{fig:Shibatip}j), very similar to what was found in Fig.~\ref{fig:Shibatip}f. This indicates that the used YSR state of the Fe$_2$ cluster on this tip was in the weak coupling regime. The particle-hole asymmetries of the $\didv$ spectra along the chain normalized to the asymmetry in the spectrum taken on the substrate shown in the top panel of Fig.~\ref{fig:Shibatip}k in fact demonstrated an extraordinarily strong spin contrast, when compared to the typical asymmetry of SP-STS maps taken with a conventional metallic Cr tip (bottom panel of Fig.~\ref{fig:Shibatip}k).

Similar YSR-functionalized tips were also used by Küster \emph{et al.}~\cite{Küster2023} to demonstrate that they can be used as well for sensing in-plane and tilted magnetization directions, imaging the spin orders in two-dimensional lattices of magnetic atoms assembled by STM-based atom manipulation on superconductors, and magnetic field driven transitions in such spin orders. As an example, Fig.~\ref{fig:Shibatip}m shows the $\didv$ map, recorded with a superconducting tip with an attached Cr atom, of a rectangular lattice of Cr atoms on Nb(110) (Fig.~\ref{fig:Shibatip}l) in a tilted magnetic field ($B=\SI{0.7}{\tesla}$ in-plane, $B=\SI{1}{\tesla}$ out-of-plane). It clearly revealed a tilted antiferromagnetic spin order (see the line profile in Fig.~\ref{fig:Shibatip}n along the dashed line in Fig.~\ref{fig:Shibatip}m). For another lattice of rhombic structure, the antiferromagntic spin order was finally driven into a ferromagnetic spin order when further increasing the out-of-plane magnetic field~\cite{Küster2023}.

\newpage

\subsection{Spin-chains on superconducting substrates}\label{experiment_1-1D}
In this chapter, we will largely focus on reviewing the experimental work on linear chains with equidistant interatomic separations, which were build by STM tip-based atom manipulation (am; see Table~\ref{table:spinchains}). However, interesting effects were also observed in dimerized linear chains~\cite{Küster2022}, which might be related to topologically distinct phases~\cite{Drost2017topological}. We will start by reviewing the methodology that was used to study experimentally accessible evidences for TSC and MBSs in such chains (Section~\ref{Sec:ExpMeth}). We will also introduce a comprehensive table which overviews the experimental work on 1D MSHs that was done up to date, and indicates which of these systems fulfil which of the experimentally accessible criteria for TSC and MBSs (Table~\ref{table:spinchains}). Afterwards, we will review the experimental works on ferromagnetic (Section~\ref{experiment_1-1D-FM}), antiferromagnetic (Section~\ref{experiment_1-1D-AFM}), and non-collinear (Section~\ref{experiment_1-1D-NC}) spin chains.

\subsubsection{Experimentally accessible criteria to verify 1D TSC and MBSs}
\label{Sec:ExpMeth}
There are mainly the following eight  experimentally accessible criteria which have been used up to date in order to experimentally investigate, whether a 1D MSH system could be a TSC with MBSs (see Table~\ref{table:spinchains}). (i) It has to have a suitable spin order (referred to as \emph{spin order}). This is categorized into collinear ferromagnetic (\emph{FM}) or antiferromagnetic (\emph{AFM}) spin orders oriented out-of-plane (\emph{op}) or in-plane (\emph{ip}), and non-collinear spin-spirals (\emph{SS}). (ii) It has to have a zero bias DOS localized at the chains end (\emph{loc. DOS}). (iii) This localized DOS has to be associated with an energetically narrow peak located exactly at zero bias (\emph{ZBP}). (iv) The zero bias peak has a spin-polarization which is distinct from that of the YSR bands (\emph{ZBP SP}). (v) The zero bias peak resides in an induced minigap of the YSR bands of size $\Delta_\textrm{ind}$. (vi) The YSR bands, in which the minigap opens up, cross $E_\textrm{F}$ an odd number of times as can be revealed by quasi-particle interference (QPI) measurements (\emph{odd $E_\textrm{F}$  QPI}). (vii) If the chain end is disturbed by a local defect imposing a magnetic or charge perturbation, the localized zero bias DOS has to behave consistently with what is expected for a MBS (\emph{ZBP pert.}). Finally (viii) the criteria (ii) to (iv) need to be checked on both ends of the same defect-free chain (\emph{both ends}). It should be noted, that other scanning tunnel spectroscopy based techniques like tip-sample distance dependent STS~\cite{Zhu2020nearly} and shot noise spectroscopy~\cite{Ge2023Singleelectron} can be used to define additional experimental criteria to verify MBSs, but haven't been applied so far to the 1D MSH systems we discuss here. In the following, we will illustrate, referring to particular examples from the literature, how the different criteria above can be experimentally verified or falsified.

\begin{figure}[H]
\centering
\includegraphics[width=\textwidth]{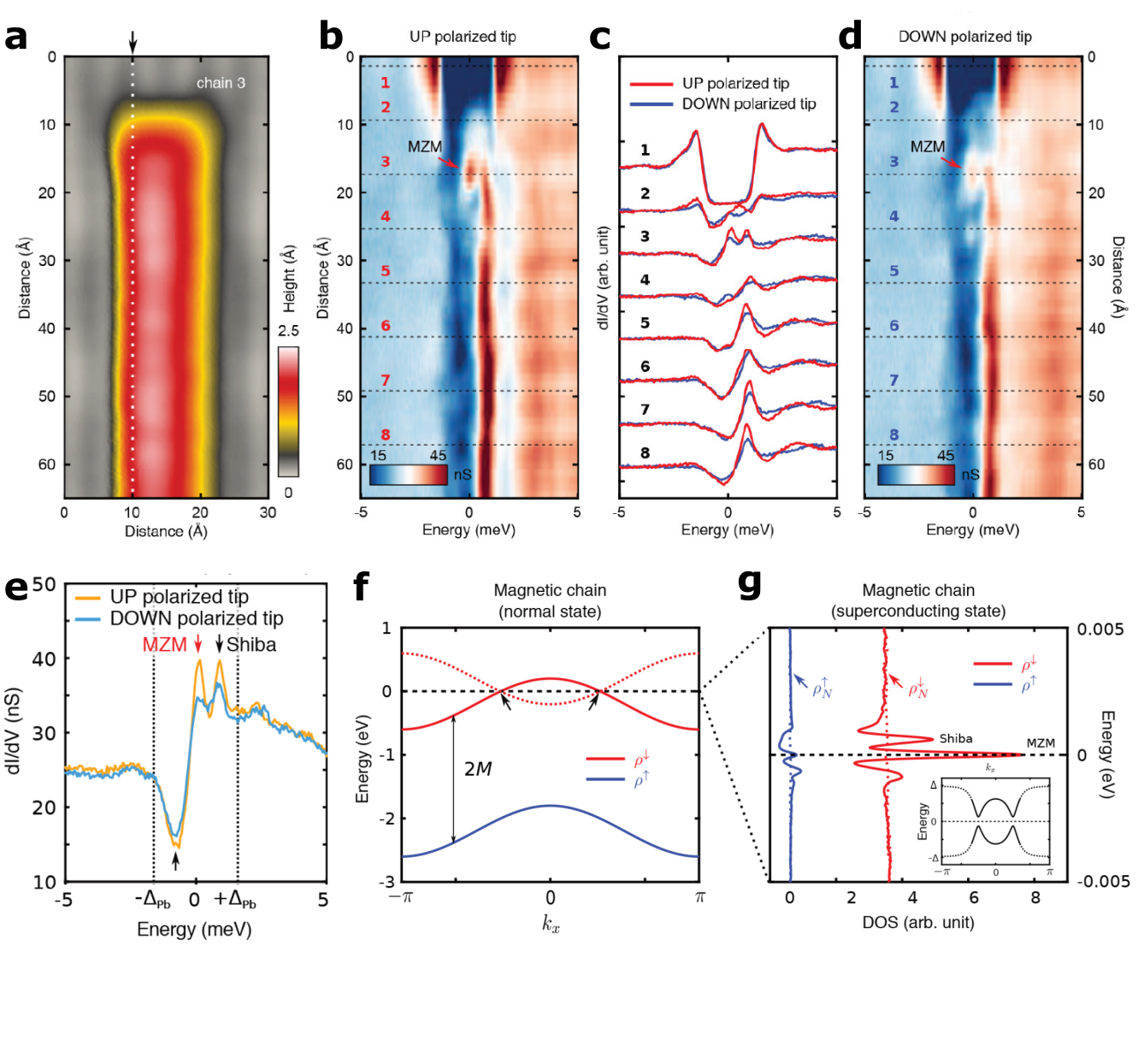}\caption{SPSTS of an edge mode for the verification of criterium \emph{ZBP SP}. \textbf{a} STM image of part of a Fe chain on Pb(110). \textbf{b}, \textbf{d} $\didv$ line profiles taken in zero magnetic field along the dashed line in panel a with upwards (panel b) and downwards (panel d) spin-polarized tips. \textbf{c} $\didv$ spectra taken from panels b (red curves) and d (blue curves) at the indicated locations. The edge mode appears at spectrum indicated by "MZM". \textbf{e} $\didv$ spectra taken in zero magnetic field at the end of the chain with upwards and downwards spin-polarized tips as indicated. The red arrow marks the zero-energy edge mode and the black arrows mark the van Hove singularities of the YSR bands. \textbf{f} Model calculation for a magnetic chain embedded in a SC. Solid and dashed curves in panel f are the dispersions of original d-orbital bands and hole-copy, respectively (red is minority, blue is majority band), which are energetically splitted by the exchange energy 2M. Black arrows mark Fermi level crossing of minority band. \textbf{g} Spin-resolved DOS calculated at the end of chain show MBSs (indicated by "MZM") and van Hove singularities of YSR bands. Red and blue dashed lines are normal state spin-resolved DOSs of minority and majority bands, respectively. The inset shows the dispersion of the YSR bands. From~\cite{Jeon2017}. Reprinted with permission from AAAS.}\label{fig:SPSTS}
\end{figure}

Figure~\ref{fig:SPSTS} shows spin-resolved STS performed on the first 1D MSH system where indications of MBSs were found~\cite{NadjPerge2014,Feldman2017,Jeon2017}: an Fe chain on Pb(110) fabricated by molecular beam epitaxy (\emph{fab. MBE}) of Fe and annealing (Fig.~\ref{fig:SPSTS}a). The $\didv$ spectra taken along the dashed line in Fig.~\ref{fig:SPSTS}a on the chain (Figs.~\ref{fig:SPSTS}b-e) indeed reveal an energetically narrow resonance at zero bias (Fig.~\ref{fig:SPSTS}e) (criterion \emph{ZBP}) localized at the chain end which decays to negligible intensity over a length of $\SI{1}{\nano\meter}$ to $\SI{2}{\nano\meter}$ (criterion \emph{loc. DOS}). Note, that the spectra were taken slightly off the longitudinal axis of the chain (dashed line in Fig.~\ref{fig:SPSTS}a) as this zero bias state has a reduced intensity along the longitudinal chain axis (dubbed \emph{double-eye}) which was related to the suppression of SC along the FM chain and effects due to the STM tip trajectory~\cite{Feldman2017,Crawford2022}. In addition, the chain has been investigated by SP-STS, which revealed that the chain is in a ferromagnetic state with an out-of-plane orientation. This magnetic state, in conjunction with the orientation of the SOC in the MSH system, is favorable for the development of TSC~\cite{NadjPerge2014} (criterion \emph{FM (op)}). Moreover, SP-STS was also performed on the zero bias peak itself, and was compared to that of the YSR bands and the overall FM spin-polarization of the chain (Figs.~\ref{fig:SPSTS}b-e)~\cite{Jeon2017}. In this particular system, the chain's magnetization is stable at $B=\SI{0}{\tesla}$, and up to very large magnetic fields the chain's magnetization is not reoriented (DFT has estimated a coercive field of $\SI{1.4}{\milli\electronvolt}$ per Fe atom). Therefore, SP-STS measurements in a remanent chain magnetization state and in the SC state of the substrate were feasible. Using Fe-coated Cr STM tips, which can be polarized up or down in smaller magnetic fields of $B=\pm\SI{1}{\tesla}$, respectively, and then going to  $B=\SI{0}{\tesla}$, SP-STS of the YSR bands and of the zero bias peak were measured in a constant magnetic state of the chain, either up or down (Figs.~\ref{fig:SPSTS}b-e). It is important to note that in these measurements, the chain's average spin polarization was compensated by adjusting the tip height for the two tip magnetization cases using a particular stabilization bias in the $\didv$ spectra. It turned out that the electron and hole components of the YSR bands have negative and positive spin polarizations (Figs.~\ref{fig:SPSTS}c,e), respectively, as expected from the overall FM magnetization direction of the chain. This was experimentally confirmed around the same time for the YSR states of single atoms~\cite{Cornils2017,Wang2021} as well as for Co chains on Pb(111)~\cite{Ruby2017}. However, while a YSR state which is accidentally at zero bias would show zero spin-polarization, the spin polarization of the zero bias peak in Figs.~\ref{fig:SPSTS}c,e is strong and positive. Such a spin polarization was indeed expected for a MBS, which directly originates from the minority $d$-orbital band crossing the Fermi level, as shown within model calculations (Figs.~\ref{fig:SPSTS}f,g). The system, therefore, also satisfies criterion \emph{ZBP SP}. Concerning the other criteria of $\Delta_\textrm{ind}$ and \emph{both ends}, which have been tested as well, we will come back to this MSH system in Section~\ref{experiment_1-1D-FM}.

In general, the verification of the criterion \emph{ZBP SP} is very challenging for spin chains that do not have a remanent spin-ordered state at zero magnetic field, which is the case for many of the investigated 1D MSH systems in Table~\ref{table:spinchains}. For those systems the chain's spin-polarization can only be studied by SP-STS in non-zero magnetic field which typically quenches the SC phase of the substrate, and thus the related YSR bands or potential MBSs. Under these conditions, the spin order of the chain itself can still be extracted in the metallic phase as described in Section~\ref{experiment_shibatip}. Since the interatomic exchange constants in the chain are typically much larger than the SC gap and the applied magnetic field strengths, except for very dilute chains in the RKKY coupled regime, the resulting spin order measured at small magnetic field can often be extrapolated into the SC phase. However, since the chain's magnetic anisotropy energy is usually smaller and might be on the order of the substrates SC energy gap and the applied magnetic fields, the direction of the magnetic anisotropy was inferred from DFT calculations in many cases (see footnotes in column \emph{spin order (op/ip)} in Table~\ref{table:spinchains}). It might be possible to overcome these experimental difficulties in future experiments using thin film SCs and in-plane oriented magnetic fields~\cite{Weerdenburg2023Extreme} and/or stabilization of the chain's spin order by RKKY coupling to stable magnetic islands~\cite{SteinbrecherNAT-COMM2018}.

\begin{figure}[H]
\centering
\includegraphics[width=\textwidth]{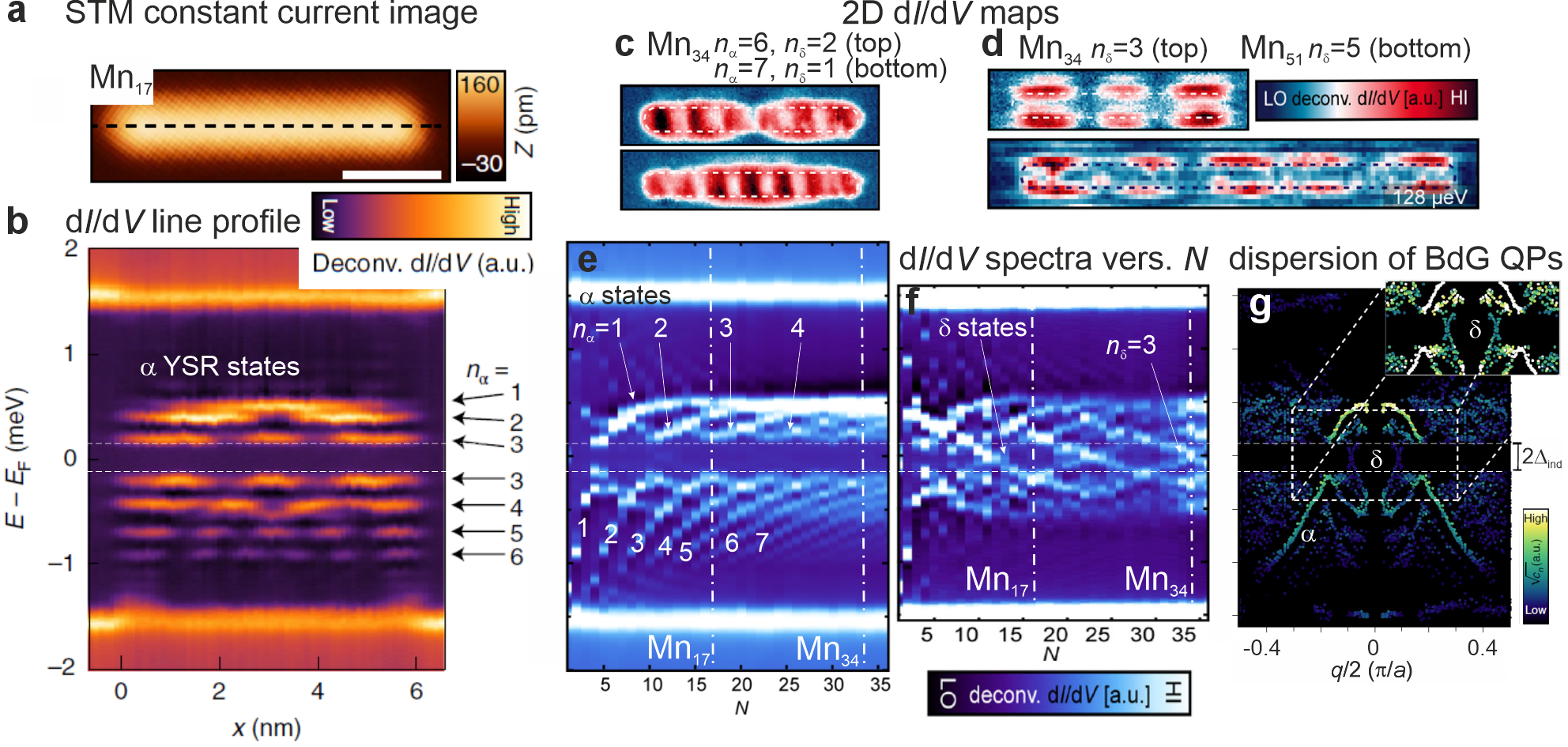}
\caption{STS methodology for the verification of criteria $\Delta_\textrm{ind}$ and \emph{odd $E_\textrm{F}$ QPI}. \textbf{a} STM constant current image of a closed packed Mn$_{17}$ chain along the $[001]$ direction constructed from $N=17$ Mn atoms, atom by atom, using STM-tip based atom manipulation on the Nb(110) surface. Scale bar, $\SI{2}{\nano \metre}$. $\textbf{b}$ $\didv$ line profile along the central longitudinal axis of the chain indicated by the dashed line in a. The arrows indicate the BdG quasiparticles from the $\alpha$ YSR band with the given quantum numbers $n_\alpha$. $\textbf{c}, \textbf{d}$ 2D $\didv$ maps of the confined BdG quasiparticles $n_\alpha=6$, $n_\alpha=7$ of the $\alpha$ YSR band and $n_\delta=2$, $n_\alpha=1$ of the other orbital, $\delta$, YSR band in a Mn$_{34}$ chain (panel c), and of the $n_\delta=3$ and $n_\delta=5$ BdG quasiparticles in a Mn$_{34}$, respectively, Mn$_{51}$ chain (panel d). $\textbf{e}, \textbf{f}$ $\didv$ spectra averaged across part of the Mn$_{N}$ chains plotted against $N$. In panel e the $\didv$ signal is averaged along the longitudinal axis, such that mainly the signal of the $\alpha$ BdG quasiparticles is measured, in panel f the $\didv$ signal is measured on the side of the chains' one end, such that mainly the $\delta$ BdG quasiparticle's intensity is picked up. The dash-dotted vertical lines indicate the Mn$_{17}$ and Mn$_{34}$ chains of (a,b) and (c,d), respectively. The arrows indicate the signal due to the confined $\alpha$ and $\delta$ BdG quasiparticles of given quantum numbers $n_\alpha$ and $n_\delta$, respectively. $\textbf{g}$ Dispersion of $\alpha$ and $\delta$ BdG quasiparticle scattering vectors extracted from similar data as in panels b to d of Mn$_{N}$ chains with $N=17$ to $N=36$. The branches assigned to the $\alpha$ and $\delta$ YSR band are indicated, as well as the minigap of width $\Delta_\textrm{ind}=\SI{180}{\micro \eV}$ in the $\alpha$ YSR band. Panels a to c, d (top part), e, g adapted from~\cite{Schneider2021b}. Panel d (bottom part), f adapted from~\cite{Crawford2022}.}\label{fig:1DExpMethods}
\end{figure}

\begin{figure}[H]
\centering
\includegraphics[width=\textwidth]{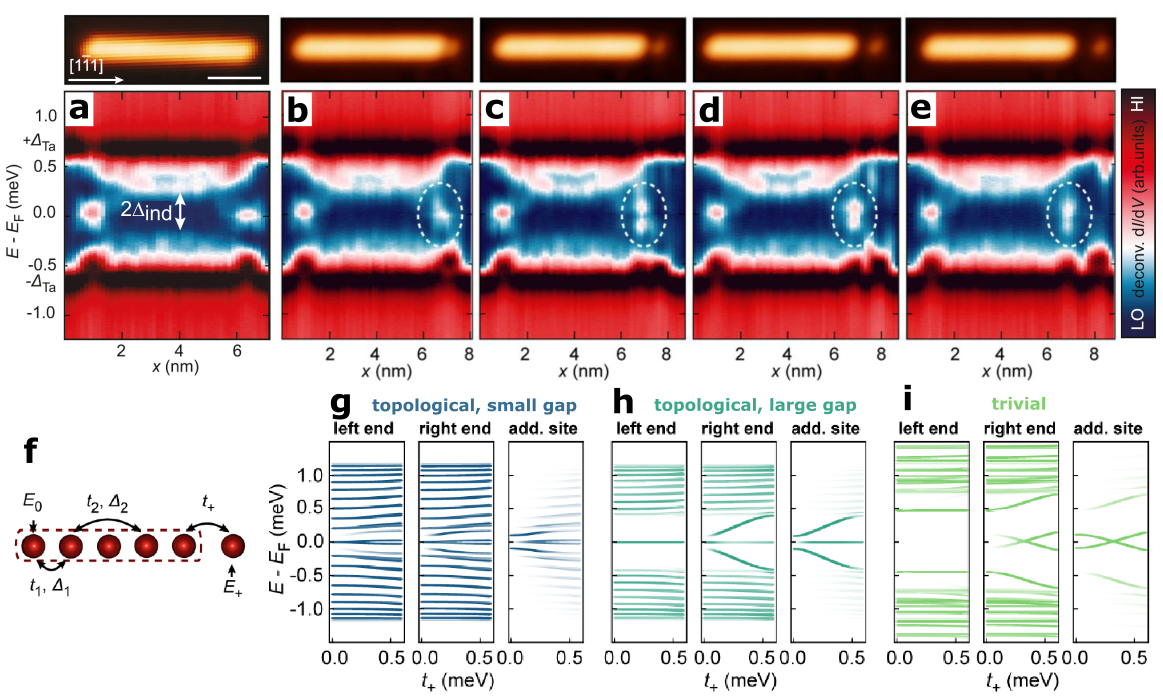}
\caption{Falsification of criterium \emph{ZBP pert.} by a perturbation experiment. \textbf{a} STM image of an antiferromagnetic Mn$_{22}$ chain assembled with interatomic distance of $\sqrt{3}a/2$ along the $[1\bar 11]$ direction on Ta(110) (top panel) and deconvoluted $\didv$ line profile taken along the longitudinal central axis of a similar Mn$_{20}$ chain (bottom panel). The arrow indicates the induced minigap $\Delta_\textrm{ind}$. \textbf{b} to \textbf{e} STM images (top panels) and deconvoluted $\didv$ line profiles taken along the longitudinal central axis of a Mn$_{22}$ chain and a single Mn atom which is subsequently moved to positions two (panel b), three (panel c), four (panel d), and five (panel e) lattice sites away from the right chain end. The lateral position of the line profile is aligned with the STM image above. The dashed white ellipses mark the edge modes which are subject to change as the single Mn atom is moved. \textbf{f} Sketch of the tight binding model to simulate the antiferromagnetic chain with hoppings $t_1$ and $t_2$ and pairings $\Delta_1$ and $\Delta_2$ between nearest and next nearest neighbouring sites, respectively, and on-site YSR state at energy $E_0$. The chain is perturbed with an additional site with onsite YSR state at energy $E_+$ and via a hopping of strength $t_+$ on the right end. \textbf{g} to \textbf{i} Spectral weight versus $t_+$ calculated from the tight-binding model illustrated in panel f in the small-gap topologically non-trivial (panel g), in the large-gap topologically non-trivial (panel h), and in the topologically trivial regime (panel i) evaluated on the left and right chain ends, and on the additional site. Model parameters, see ~\cite{Schneider2023}. Figure adapted from~\cite{Schneider2023}.}\label{fig:Pert}
\end{figure}

The verification of the further criteria got facilitated by a fabrication of the MSH systems atom by atom using STM tip-based atom manipulation (\emph{fab. am}). By this fabrication method, an electronic disorder in the YSR bands is largely inhibited, and only limited by the substrate's electronic inhomogeneity which can be minimized by a proper substrate preparation and choice of fabrication location on the substrate's surface. By minimizing the degree of disorder, it became possible to observe the successive formation of the multi-orbital YSR bands, starting from the multi-orbital single atom YSR states described in Section~\ref{experiment_0D} (e.g. Fig.~\ref{fig:YSRatomsNb}) via the hybridized YSR states in pairs of atoms described in Section~\ref{experiment_dim} (e.g. Fig.~\ref{fig:YSRdimersNb}) to the bands in the chains containing different numbers $N$ of atoms (e.g. Fig.~\ref{fig:1DExpMethods}). Also, the QPI patterns of the BdG quasiparticles confined inside the chain get very clear and can be analyzed to extract the dispersion of the YSR bands, numbers of $E_\textrm{F}$ crossings (criterion \emph{odd $E_\textrm{F}$ QPI}), and thereby reveal, whether a potential minigap ($\Delta_\textrm{ind}$) is topological or trivial. Furthermore, it enables to check all the criteria discussed above related to the local appearance of features on the chain's ends for both ends of the same defect-free chains (criterion \emph{both ends}). Finally, perturbation experiments can be performed to check, whether zero-bias features are topologically trivial or non-trivial (\emph{ZBP pert.}, see Fig.~\ref{fig:Pert} which will be discussed further below).

As an example for the extraction of the multi-orbital YSR band dispersion by QPI and it's minigaps, we illustrate in Fig.~\ref{fig:1DExpMethods} the work on Mn chains assembled along the $[001]$ direction on Nb(110) with an interatomic separation of $1a$~\cite{Schneider2021b,Crawford2022}. $\didv$ spectra can be taken along a line, e.g. along the longitudinal axis of the Mn$_{17}$ chain containing $N={17}$ Mn atoms (dashed line in Fig.~\ref{fig:1DExpMethods}a) resulting in the so-called $\didv$ line profile shown in Fig.~\ref{fig:1DExpMethods}b. It reveals confined BdG quasiparticle states of the YSR bands inside the substrate SC gap, with different wavelengths resulting in the given numbers of maxima $n_\alpha$. Already this data shows that there is a well-defined minigap $\Delta_\textrm{ind}\approx\SI{180}{\micro \eV}$. These confined BdG quasiparticle states are also visible in maps of the $\didv$ signal taken at the according bias voltage across a whole image area around a chain (e.g. Fig.~\ref{fig:1DExpMethods}a), so called 2D $\didv$ maps shown in Fig.~\ref{fig:1DExpMethods}c,d for a Mn$_{34}$ and a Mn$_{51}$ chain. However, in contrast to the $\didv$ line profile which mostly reflects the BdG states of YSR bands which have a strong signal on top of the chain (called $\alpha$ YSR band), the $\didv$ maps also reveal the confined BdG quasiparticle states originating from a YSR band which has a nodal line along the chain's longitudinal axis and a strong signal on both sides along the chain (called $\delta$ YSR band). The numbers of maxima of both of these states, the ones with strong intensity along the chain's top and the ones with strong intensity along the chain's side are given by $n_\alpha$ and $n_\delta$, respectively, in Fig.~\ref{fig:1DExpMethods}c,d. Furthermore, the multi-orbital YSR states, that these two bands stem from, can be deduced in two ways: (i) by following the successive emergence of these YSR bands from the single atom YSR states ($N=1$) over the hybridized YSR states ($N=2$) into longer chains ($N=5-10$), as shown in Figs.~\ref{fig:1DExpMethods}e,f; and (ii) by comparing the wavefunctions of the confined BdG quasiparticles from the 2D $\didv$ maps as in Figs.~\ref{fig:1DExpMethods}c, d to the ones from the single atoms (Figs.~\ref{fig:YSRatomsNb}f to i) and pairs with the same interatomic seperation and orientation as that of the chain (similar to Fig.~\ref{fig:YSRdimersNb}). Fig.~\ref{fig:1DExpMethods}e,f show the $\didv$ spectra averaged across part of the Mn$_{N}$ chains plotted against $N$. In Fig.~\ref{fig:1DExpMethods}e the $\didv$ signal is averaged along the longitudinal axis, such that mainly the signal of the $\alpha$ BdG quasiparticles is measured. In Fig.~\ref{fig:1DExpMethods}f the $\didv$ signal is measured on the side of the chains' one end, such that mainly the $\delta$ BdG quasiparticle's intensity is picked up. By this detailed comparison~\cite{Schneider2021b}, it was concluded, that the $\alpha$ BdG quasiparticles originate from the hybridized $d_{z^2}$ YSR states, and the $\delta$ BdG quasiparticles from the hybridized $d_{yz}$ YSR states (see Fig.~\ref{fig:YSRatomsNb}f to i). 

Finally, by analysing the wavelengths of the two types of confined BdG quasiparticles as a function of their energy for chains of many different lengths $N=17$ to $N=36$, it was possible to extract the dispersion of scattering vectors $q$ from both of these YSR bands, given in Fig.~\ref{fig:1DExpMethods}g. This analysis leads to the following conclusions. The $d_{z^2}$ ($\alpha$) YSR band passes $E_\textrm{F}$ only once (mirror symmetric around $q=0$, criterion \emph{odd $E_\textrm{F}$ QPI}), but, instead of a crossing, there is a minigap of size $\Delta_\textrm{ind}=\SI{180}{\micro \eV}$. This minigap is, thus, most probably induced by SOC and of $p$-wave nature~\cite{Schneider2021b}. Note, however, that there were no indications of a MBS associated with this minigap~\cite{Schneider2021b}. On the other hand, the $d_{yz}$ ($\delta$) YSR band crosses this minigap also once (see inset in Fig.~\ref{fig:1DExpMethods}g) without an experimentally resolvable additional minigap in this band. Additionally, other potential YSR bands stemming from hybridizations between the other three single atom YSR states (Figs.~\ref{fig:YSRatomsNb}f to i) were not detected, possibly because of their weak intensity. This means, that experimentally, it can only be concluded that \emph{one of the multi-orbital YSR bands} ($\alpha$) has a gap with an odd number of $E_\textrm{F}$ crossings. That is why we introduce the additional criterion \emph{[1 orb.]} in the columns of Table~\ref{table:spinchains} giving the minigap and odd $E_\textrm{F}$ crossing criteria. Experimentally, it also wasn't possible to exclude, though, that the $d_{yz}$ YSR band (and other undetected YSR bands) have a minigap of a width below the experimental energy resolution of about $\Delta_\textrm{ind}=\SI{100}{\micro \eV}$, such that the overall system might still be a TSC with strongly hybridizing MBSs from both ends of the chain which would have a similar signature as the confined states from the $d_{yz}$ YSR band discussed above~\cite{Crawford2022}.

An additional advantage of 1D MSH systems fabricated by atomic manipulation is that they enable one to perform perturbation experiments of the zero bias features on the ends of the chains, in order to verify or falsify their MBS nature, as discussed in the following on the example of an antiferromagnetic (AFM) Mn chain assembled along the $[1\bar{1} 1]$ direction on Ta(110) with an interatomic separation of $(\sqrt{3}a/2)$~\cite{Schneider2023} (Fig.~\ref{fig:Pert}). $\didv$ line profiles, e.g. along the central longitudinal axis of a Mn$_{20}$ chain shown in Fig.~\ref{fig:Pert}a, revealed a substantial minigap $\Delta_\textrm{ind}=\SI{200}{\micro \eV}$ which hosts zero bias modes localized on both chain ends which are clearly energetically separated from the YSR bands outside $\Delta_\textrm{ind}$. This means all the criteria \emph{loc. DOS}, \emph{ZBP}, \emph{both ends}, and \emph{$\Delta_\textrm{ind}$} are fulfilled, and the AFM spin order in conjunction with strong SOC is also principally able to realize TSC~\cite{Schneider2023}. However, chain perturbation experiments were able to falsify a MBS origin of the zero bias mode and the TSC in this case. Figs.~\ref{fig:Pert}b-e show $\didv$ line-profiles along the central longitudinal axis of a Mn$_{22}$ chain which was perturbed on the right end by manipulating a Mn atom to adsorption sites at different distances from that chain's end (see the top panels of Figs.~\ref{fig:Pert}b-e). The $\didv$ line-profiles also extend over the perturbing atom whose YSR state is visible at about $\pm\SI{0.4}{\milli\electronvolt}$ around $x\approx\SI{8}{\nano\meter}$ on the $\didv$ line-profiles in Figs.~\ref{fig:Pert}c-e. While the $\didv$ signal on the unperturbed chain's end ($x\approx\SI{1}{\nano\meter}$) always stayed unaffected in spatial location and at zero bias, the $\didv$ signal on the perturbed end ($x\approx\SI{7}{\nano\meter}$) splitted off from $E_\textrm{F}$ for the farther atom positions in Figs.~\ref{fig:Pert}c-e without any noticeable appearance of a zero bias signal at other locations, neither shifted further inside the chain, nor on the perturbing atom. Only for very close separations, the perturbing Mn atom got part of the chain itself, the $\didv$ signal shifts back to $E_\textrm{F}$ and extends throughout the perturbing atom (Fig.~\ref{fig:Pert}b). These experimental signatures safely ruled out a MBS origin of the observed zero energy end state, as argued below using a minimal next-nearest-neighbor tight binding model (Figs.~\ref{fig:Pert}f-i), i.e., the criterion \emph{ZBP pert.} has a negative result. Therefore, the observed minigap also most probably has a topologically trivial character. Using a minimal next-nearest-neighbor tight binding model to simulate the AFM chain, with a perturbing site with energy levels $E_\pm=\pm\SI{0.1}{\milli\electronvolt}$ coupled via the hopping parameter $t_+$ to the right chain's end (Fig.~\ref{fig:Pert}f), the LDOSs on the left and right chain's ends and on the perturbing site were calculated as a function of perturbation strength $t_+$ for different topological properties of the model (Figs.~\ref{fig:Pert}g-i)~\cite{Schneider2023}. In the situation of Fig.~\ref{fig:Pert}g the chain is in the topologically non-trivial phase, but has a relatively narrow minigap such that the close to zero bias MBS has a large localization length and the two parts from both chain ends still weakly hybridize (so-called precursor MBSs~\cite{Schneider2022}). In that case, if the perturbation is switched on, the zero bias end states at both sides of the chain were shown to split off from $E_\textrm{F}$ (Fig.~\ref{fig:Pert}g), in contrast to what is observed experimentally in Figs.~\ref{fig:Pert}b to e. In the situation of Fig.~\ref{fig:Pert}h the chain is deep in the topologically non-trivial phase and has a large minigap $\Delta_\textrm{ind}\gg E_\pm$ hosting MBS at the chain's ends. In that case, if the perturbation is switched on, the zero bias end state on the left chain's end is completely unaffected, while that on the perturbed right end stays at zero bias, but laterally shifts onto the perturbing site, while simultaneously, the energy levels of the perturbing site change in energy and leak into the right chain's end (Fig.~\ref{fig:Pert}h). The observation of such an experimental signature in a 1D MSH (\emph{ZBP pert.}) would be a strong indication for a 1D TSC hosting a MBS. However, this signature was also not observed experimentally in Figs.~\ref{fig:Pert}b-e. Finally, in the situation of Fig.~\ref{fig:Pert}i the chain is the topologically trivial phase and also has a large minigap. Obviously, within the minigap, an end state at finite energy has emerged (criterion \emph{off-ZBP} in Table~\ref{table:spinchains}) which is a particular trivial end state occuring in AFM chains as will be discussed in Section~\ref{experiment_1-1D-AFM}. When the perturbation is switched on, this off-zero bias end state is affected only on the right chain's end, i.e., shifts to larger energy, and leaks into the perturbing site, while simultaneously, the energy levels of the perturbing site change in energy and leak into the right chain's end (Fig.~\ref{fig:Pert}i). Since this end state in an AFM chain can also be, accidentally, located at zero bias, it was concluded in~\cite{Schneider2023}, that the experiments in Figs.~\ref{fig:Pert}b-e strongly suggest a topologically trivial phase of the AFM $[1\bar{1} 1]$Mn chains on Ta$(110)$.

After having introduced the experimentally accessible criteria for the verification, or falsification, of TSC with MBSs in 1D MSHs, we present in Table~\ref{table:spinchains} an extensive compilation of the experimentally investigated 1D MSH systems. The used substrates, given in bold letters in the leftmost column, have been ordered from top to bottom according to their increasing atomic number $Z$. Where appropriate, the numbers of atoms $N$ in the longest investigated chains are given. Orientation and spacing refer to the chain's orientation along the given substrate's crystallographic directions and the interatomic spacing within the chain in units of surface unit cell lattice constant $a$, respectively. The second to fourth column give the according references, the fabrication method, and the spin orders and magnetic orientations, respectively, of all investigated 1D MSH systems. If the spin order or magnetic anisotropy is coming from another publication, e.g. using SPSTS or DFT calculations, the according source is given by a footnote. We added a question mark behind the spin order or magnetic anisotropy, if the result was discussed controversially. In the remaining columns, checkmarks ($\checkmark$) indicate which of the experimental criteria introduced above have been investigated for the corresponding MSHs. A value of $\Delta_\textrm{ind}$ or a checkmark for the criterion \emph{odd $E_\textrm{F}$ QPI} in brackets ($\left[...\right]$) indicates that the corresponding criterion was only found for one of the multi-orbital YSR bands. A cross ($\times$) instead of a checkmark for the criterion \emph{ZBP pert.} indicates that the perturbation experiment was performed, but disfavours a MBS origin of the observed edge state.

Overall, we can see in Table~\ref{table:spinchains}, that so far in none of the investigated MSH systems all possible experimental criteria were checked alltogether. From those systems where all of the criteria \emph{investigated for the particular system} were in favor of TSC with a MBS, two systems fulfilled most of the criteria (in total 6 out of 8). Firstly, for the ferromagnetic Mn$[1\bar{1} 0]$ chains on Nb(110)~\cite{Schneider2022} only the experimental proof of a finite $\Delta_\textrm{ind}$ and SP-STS measurements of the zero bias peak are missing. We will discuss this system in Section~\ref{experiment_1-1D-FM}. Secondly, for the system of Fe clusters attached to Bi(111) hinge states on Nb(110)~\cite{Jaeck2019}, with the largest $\Delta_\textrm{ind}=\SI{1500}{\micro\electronvolt}$ of all MSHs observed to date, the perturbation experiment has not been performed, and the part of the Majorana wavefunction on the other side of the Fe cluster wasn't detected (missing criterion \emph{both ends}). This system and the EuS islands coupled to the proximitized and confined surface state at the (111) surface of Au grown on V~\cite{Manna2020}, are rather special cases not merely consisting of spin chains coupled to a superconducting substrate, as all the other investigated MSH systems in Table~\ref{table:spinchains}. As they were already included in a recent review~\cite{JaeckNAT-REV-PHYS2021}, we will not discuss them here. It is also noteworthy, that for all investigated AFM MSH systems given in Table~\ref{table:spinchains}, the edge states are of trivial origin, as they are either off zero energy, or failed the perturbation experiment (see Section~\ref{experiment_1-1D-AFM}). Finally, a spin spiral ground state could only be unambiguously found for one MSH system, i.e., the Fe$[1\bar 10]$ chains on Re(0001)~\cite{Kim2018,Schneider2020}. This system will be reviewed in Section~\ref{experiment_1-1D-NC}.

For the following MSH systems in Table~\ref{table:spinchains}, a conclusion on the potential for TSC is difficult, since the spin order is completely unknown. Therefore, we review them only briefly here. The Fe$[100]$ chain on Ta(100)-$(3\times3)$O~\cite{Kamlapure2018} system was one of the first two fabricated by STM-based atom manipulation, together with the Fe$[1\bar 10]$ chain on Re(0001) system~\cite{Kim2018,Schneider2020} (Section~\ref{experiment_1-1D-NC}). For the former substrate, a chain built by Fe atoms on the nearest neighbouring centers of cross-shaped plaquettes of the $(3\times3)$O-reconstruction resulted in a negligible hybridization of the close to gap edge YSR states. In order to increase the hybridization as well as shift the YSR states farther into the gap, subsurface interstitial Fe atoms were inserted in between the chain atoms by STM-based atom manipulation~\cite{Kamlapure2018,Kamlapure2021}. However, the resulting interatomic couplings were still too weak and short-range, and the disorder of YSR state energies too large due to substrate inhomogeneities~\cite{Mozara2019atomically}, such that the formation of a YSR band was not achieved. Gd chains assembled on Nb(110) in two different crystallographic orientations~\cite{Wang2023} were the second investigated MSH systems using 4$f$ transition metal atoms as chain material, after Gd triplets assembled on Bi(110) grown on Nb(110)~\cite{Ding2021tuning} (see Section~\ref{experiment_1-1D-NC}), with the hope to increase the effective spin-orbit interaction within the YSR bands. In the Gd chains oriented along $[1\bar 10]$ on Nb(110), there is a relatively large minigap. In order to explain the observed close-to-edge changes in the YSR bands it was argued that strain effects in that system lead to a smooth transition in the YSR band energy between inside of the chain and the edge, but no real edge states were observed. In contrast, in the Gd chains oriented along $[001]$ on the same substrate, there is zero bias edge state fulfilling the criteria \emph{ZBP}, \emph{loc. DOS}, and \emph{both ends}. However, a hard minigap inside the chain is not really resolved. For the Mn$[001](2a)$ chains on Ta(110)~\cite{Beck2023} none of the criteria are fulfilled, except for \emph{both ends}. Finally, Mn chains on Re(0001)~\cite{Schneider2020} (which are not straight but zig-zag shaped with Mn atoms occupying alternating fcc and hcp adsorption sites) have a considerable minigap with a slightly changing width at the chain ends suggesting an off-zero bias edge state.

In the following Sections~\ref{experiment_1-1D-FM}, \ref{experiment_1-1D-AFM}, and \ref{experiment_1-1D-NC} we will discuss some of the remaining MSH systems from Table~\ref{table:spinchains}, starting with the FM chains, then the AFM chains, and finally the non-collinear spin chains.

\begin{table}[!htbp]
\caption{Experimentally accessible evidences of TSC with MBSs in 1D MSHs. The table summarizes the experimentally studied chain on substrate (/substrate) systems (sorted by increasing atom number $Z$ of substrate atoms) with given maximum number of studied atoms $N$, chain orientation ([orient.]), and interatomic spacing (spacing) fabricated (fab.) by molecular beam epitaxy (MBE) or atom manipulation (am) and with ferromagnetic (FM), spin spiral (SS), or antiferromagnetic (AFM) spin orders with out-of-plane (op) or in-plane (ip) orientation, or zero magnetic moment ($\times$). Compliances with the following experimental criteria for TSC with isolated MBSs are indicated: close-to zero bias DOS localized on the chain's ends (loc. DOS); resolved zero bias peak at chain's ends (ZBP); ZBP spin polarization measured and consistent with MBSs (ZBP SP); features measured on both ends of the same defect-free chain (both ends); resolved mini-gap [in one multi-orbital YSR band] ($\Delta_\textrm{ind}$ [1 orb.]); QPI shows odd Fermi level crossings of the YSR bands [in one multi-orbital band] (odd $E_\mathrm{F}$ [1 orb.] QPI); when perturbed, the ZBP behaves consistently with a (precursor) MBS (ZBP pert., $\times$: experiment performed, but negative result); off-ZBP on chains ends (off-ZBP, counter indicative of TSC).}\label{tab1}%

\begin{tabular}{@{}lm{39pt}m{20pt}m{49pt}m{10pt}m{10pt}m{10pt}m{10pt}m{40pt}m{28pt}m{10pt}m{15pt}@{}}
\toprule
\textbf{/substrate} & cit. & fab. & spin &  loc. & ZBP & ZBP & both & $\Delta_\textrm{ind}$ & odd $E_\mathrm{F}$ & ZBP & off- \\
chain$_N$[orient.] &  & (MBE/ & order & DOS & & SP & ends & [1 orb.] & [1 orb.] & pert. & ZPB\\
(spacing) & & am)& (op/ip) &   & & & & & QPI & \\
\midrule
\textbf{/2$H$-NbSe}$_2$/Fe$_{51}(3a)$ & \cite{Liebhaber2022}& am& FM or SS\footnotemark[1]& & & & $\checkmark$&  & [$\checkmark$]& &\\
\textbf{/Nb(110)} & & & & &  &  & & & & &\\
Cr$_{15}[1\bar{1}3](\sqrt{11}a/2)$ & \cite{Küster2022}& am& FM (ip)& $\checkmark$& $\checkmark$& & $\checkmark$&$\SI{100}{\micro \eV}$\footnotemark[5] & & &\\
Cr$_{15}[1\bar{1} 1](\sqrt{3}a)$ & \cite{Küster2022}& am& FM (ip)& $\checkmark$& $\checkmark$& & $\checkmark$& & & &\\
Cr$_{5}[1\bar{1} 1](3\sqrt{3}a/2)$ & \cite{Küster2022}& am& AFM\footnotemark[2] (ip\footnotemark[3])& & & & $\checkmark$& & & & $\checkmark$\\
Cr$_{5}[1\bar{1} 1](2\sqrt{3}a)$ & \cite{Küster2022}& am& FM (ip\footnotemark[3])& & & & $\checkmark$& & & & $\checkmark$\\
Cr$_{5}[1\bar{1} 1](5\sqrt{3}a/2)$ & \cite{Küster2022}& am& SS (ip)\footnotemark[3]& & & & $\checkmark$& & & & $\checkmark$\\
Cr$_{15}[001]2a$ & \cite{Küster2022}& am& AFM (ip)& & & & $\checkmark$& $\SI{350}{\micro \eV}$\footnotemark[5] & & & $\checkmark$\\
Cr$_{15}[1\bar{1} 0](2\sqrt{2}a)$ & \cite{Küster2022}& am& AFM (ip)& & & & $\checkmark$& $\SI{150}{\micro \eV}$\footnotemark[5] & & &$\checkmark$\\
Mn$_{40}[1\bar{1} 1](\sqrt{3}a/2)$ & \cite{Schneider2023} & am & AFM\footnotemark[2] (op\footnotemark[3]) &  &  &  & $\checkmark$ & $\SI{700}{\micro \eV}$& & $\times$ & $\checkmark$\\
Mn$_{36}[001](1a)$ & \cite{Schneider2021b}&  am & FM\footnotemark[2] (op\footnotemark[3]) & &  &  & $\checkmark$ & [$\SI{180}{\micro \eV}$] & [$\checkmark$] & \\
Mn$_{45}[1\bar{1} 0](\sqrt{2}a)$ & \cite{Schneider2022} & am & FM (op\footnotemark[3]) & $\checkmark$ & $\checkmark$ &  & $\checkmark$ & $\lesssim
\SI{50}{\micro \eV}$& [$\checkmark$] & $\checkmark$\\
Fe$_4[1\bar{1} 1](\sqrt{3}a/2)$ & \cite{Friedrich2021} & am &  FM (op\footnotemark[3])& &  &  &\checkmark  & & & \\
Fe$_{16}[001](1a)$ & \cite{Crawford2022}&  am & FM (op\footnotemark[3]) & \checkmark&  &  & $\checkmark$ &  &  & \\
Gd$_{7}[001](1a)$ & \cite{Wang2023}& am & & $\checkmark$& $\checkmark$ &  & $\checkmark$& & & \\
Gd$_{23}[1\bar{1} 0](\sqrt{2}a)$ & \cite{Wang2023} & am &  &  &  &  & $\checkmark$ & $\approx\SI{250}{\micro \eV}$&  & & \\
\textbf{/Ta(110)} & & & & & &  & & & & \\
Mn$_{22}[1\bar{1} 1](\sqrt{3}a/2)$ & \cite{Schneider2023} & am & AFM & $\checkmark$ & $\checkmark$ &  & $\checkmark$ & $\SI{200}{\micro \eV}$& & $\times$ & \\
Mn$_{41}[001](1a)$ & \cite{Beck2022} & am & FM (ip\footnotemark[4])& & & & $\checkmark$ & [$\SI{146}{\micro \eV}$]& & \\
Mn$_{20}[1\bar{1} 0](\sqrt{2}a)$ & \cite{Beck2023} & am & FM? &  &  &  & $\checkmark$ & &  & & $\checkmark$\\
Mn$_{18}[001](2a)$ & \cite{Beck2023}&  am &  & &  &  & $\checkmark$ &  &  & \\
\textbf{/Ta(100)-$(3\times3)$O}    &    &  & & & & & & & \\
Fe$_{63}[100]3a$    &  \cite{Kamlapure2018} & am &    &  & & & $\checkmark$ & & \\
\textbf{/Re(0001)}    &    &  & & & & & & & \\
Mn$_{101}[1\bar{1} 0](a)$    &  \cite{Schneider2020}  & am &    &  & & & $\checkmark$  & $\approx\SI{100}{\micro \eV}$ & & & $\checkmark$ \\
Fe$_{40}[1\bar{1} 0](a)$    &  \cite{Kim2018,Schneider2020}  & am &  SS  & $\checkmark$ & & & $\checkmark$ & & \\
Co$_{84}[1\bar{1} 0](a)$    &  \cite{Schneider2020}  & am &  $\times$  &  & & & $\checkmark$ & & \\
\textbf{/Au(111)/V} &  &  & & & &  & & & & \\
EuS island & \cite{Manna2020}& MBE & FM (ip)& $\checkmark$& $\checkmark$& & $\checkmark$& & & \\
\textbf{/Au(110)/Nb(110)} &  &  & & & &  & & & & \\
Fe$_{16}[001](2a)$ & \cite{Beck2023b}& am& FM (op)& & & & $\checkmark$& & [$\checkmark$]& \\
\textbf{/Pb(110)}    &   &  & &  & & & & & & \\
Fe    &  \cite{NadjPerge2014,Jeon2017,Feldman2017,Ruby2015,Pawlak2016}   & MBE & FM (op) & $\checkmark$  
 & $\checkmark$ & $\checkmark$ & & $\lesssim
\SI{80}{\micro \eV}$& & \\
  Co    &  \cite{Ruby2017}  & MBE & FM (op) & & & &$\checkmark$ & & & & $\checkmark$\\
\textbf{/Bi(111)/Nb(110)} &  &  & & & &  & & & & \\
Fe cluster & \cite{Jaeck2019}& MBE & FM (ip \& op)& $\checkmark$& $\checkmark$& $\checkmark$& & $\SI{1500}{\micro \eV}$& $\checkmark$\footnotemark[6]& &\\
\textbf{/Bi(110)/Nb(110)} &  &  & & & &  & & & & \\
Gd$_3(2a)$ & \cite{Ding2021tuning}& am & SS? (ip?)& & & & $\checkmark$& & & \\
\textbf{/$\beta-$Bi$_2$Pd(001)} &  &  & & & &  & & & & \\
Cr$_4[110](\sqrt{2}a)$ & \cite{Mier2021}& am & FM (op) & & & & $\checkmark$& & & \\
Cr$_{12}[100](2a)$ & \cite{Mier2021}& am & SS?& & & & $\checkmark$& $\SI{1000}{\micro \eV}$& & \\
\botrule
\end{tabular}
\footnotetext[1]{Suggested by YSR band formation.}
\footnotetext[2]{from SP-STS~\cite{Schneider2021,Küster2023}} \footnotetext[3]{from DFT~\cite{Laszloffy2021,Crawford2022,Küster2021longrange}}
\footnotetext[4]{from DFT, B. \'{U}jfalussy, private communication.}
\footnotetext[5]{P. Sessi, private communication.}
\footnotetext[6]{\cite{Drozdov2014,Jaeck2020Observation}}
\label{table:spinchains}
\end{table}

\subsubsection{Ferromagnetic spin chains}\label{experiment_1-1D-FM}

As already introduced in the previous section, the first investigated MSH system showing indications for MBSs were the Fe chains grown by MBE at elevated substrate temperatures on Pb(110)(Fig.~\ref{fig:SPSTS})~\cite{NadjPerge2014,Feldman2017,Jeon2017}, fulfilling the criteria \emph{loc. DOS}, \emph{ZBP}, and \emph{ZBP SP}. Ruby \emph{et al.}~\cite{Ruby2015} and Pawlak \emph{et al.}~\cite{Pawlak2016} emphasized that there are two types of chains, one with a protrusion at the end, the other without. According to~\cite{Pawlak2016} zero bias end states were only present in the chains which show a protrusion in the STM constant current images. However, even in some of those chains, the zero bias end state was absent~\cite{Ruby2015}. Ruby \emph{et al.} noticed in their measurements performed at $T=\SI{1.1}{\kelvin}$ using SC tips~\cite{Ruby2015}, that the zero bias peaks appeared asymmetric, indicating the possibility of an overlap with a state of the YSR band. It was later shown by Feldmann~\emph{et al.}~\cite{Feldman2017} using STS also with SC tips at lower temperatures ($T=\SI{20}{\milli\kelvin}$) but at an energy resolution corresponding to an electronic temperature of $T=\SI{250}{\milli\kelvin}$, that one can find many chains where the zero bias peak at the chain's end is indeed symmetric, indicating its MBS origin. However, it was also noticed in this work that, in some of the investigated chains, the zero bias peak is asymmetric. Concerning the criterion of the minigap, it was shown that the limited energy resolution in both experiments only enabled to fix an upper bound of $\Delta_\textrm{ind}\lesssim\SI{80}{\micro\electronvolt}$~\cite{Ruby2015, Feldman2017}. Unfortunately, the growth of the chains out of Fe clusters at elevated substrate temperatures for this system prevented to check all these criteria of the zero bias end state for two structurally identical ends of the same defect free chain. So the criterion \emph{both ends} is not fulfilled and perturbation experiments have been too challenging to perform because of the impossibility of carrying out STM-based atom manipulation on Pb(110).

Co chains grown on the same substrate Pb(110)~\cite{Ruby2017} showed zero bias states delocalized along the chain, so no indications for a MBS. There is only an off-zero energy edge state (\emph{off-ZBP}). Using tight-binding model calculations, it was argued that in this system, an even number of four bands cross $E_\textrm{F}$, suppressing the formation of TSC. 
It is worth noting that, in this MSH system, a stable chain magnetization with a coercivity of at least $\SI{3}{\tesla}$ enabled zero magnetic field SP-STS measurements in order to establish the FM spin order.

As discussed in the previous section, Mn chains assembled along the $[001]$ direction on Nb(110) with an interatomic separation of $1a$~\cite{Schneider2021b} (Fig.~\ref{fig:1DExpMethods}) revealed a minigap of considerable width in the $d_{z^2}$ YSR band. In order to investigate the influence of spin orbit coupling on the width of the minigap in the $d_{z^2}$ YSR band, Beck \emph{et al.} investigated structurally identical chains on Ta(110)~\cite{Beck2022}. The lattice constants of Ta and Nb are only $0.3$\%
apart, their work functions differ by only $1.5$\%,
they have almost identical Fermi surfaces and
both have an occupied $d_{z^2}$-like surface state with similar
effective masses and binding energies. The only major difference is the strength of spin orbit coupling which is increased by a factor of $\approx3$ for Ta(110)
with respect to Nb(110)~\cite{Beck2022}. Therefore, the authors expected similar YSR states and YSR bands of the structurally identical Mn chains on the two substrates. Indeed, the chains on Ta(110) also revealed a minigap in the YSR bands of width $\Delta_\textrm{ind}=\SI{146}{\micro\electronvolt}$. If normalized to the SC gaps of the two materials, it turned out that the minigap for the Ta system was enhanced by a factor of about 2. However, since the SC gap of Ta is about a factor of 2 smaller than that of Nb, the experimental energy resolution was considerably worse and the assignment of the YSR bands to the different orbitals was not as straightforward as for the Nb case shown in Fig.~\ref{fig:1DExpMethods}, which hinders a very clear direct comparison. In addition, later investigations revealed that the ferromagnetic spin order for the Ta-based MSH system has an in-plane orientation, while it is out-of-plane for the Nb based MSH system (see Table~\ref{table:spinchains}) which additionally impedes to disentangle the effects of SOC and spin-orientation on the width of the minigap.

\begin{figure}[H]
\centering
\includegraphics[width=\textwidth]{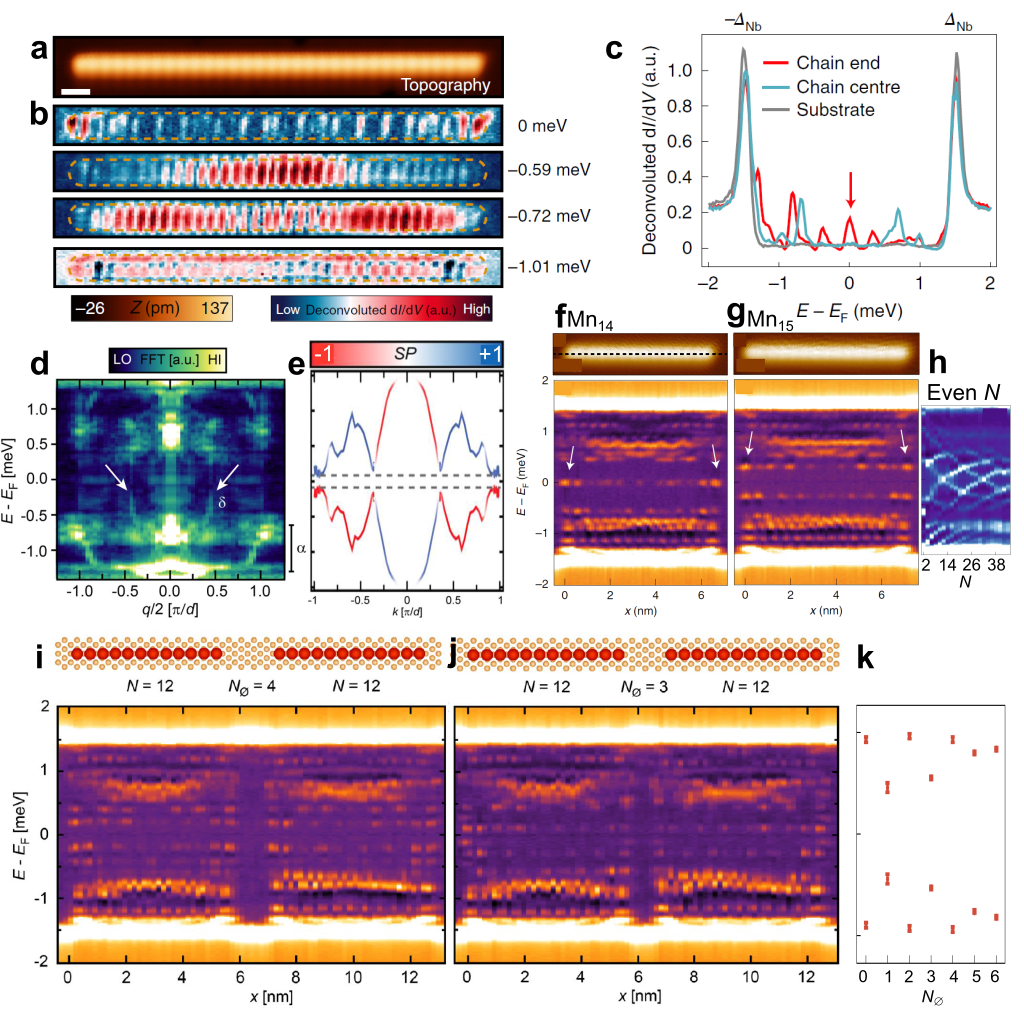}
\caption{Finite size effects and hybridization of precursors of MBSs. \textbf{a} STM image of a ferromagnetic Mn$_{32}$ chain assembled with interatomic distance of $\sqrt{2}a$ along the $[1\bar 10]$ direction on Nb(110). \textbf{a} Deconvoluted $\didv$ maps over the chain in panel a taken at the indicated bias voltages. \textbf{c} Deconvoluted $\didv$ spectra taken on the chain's end, in the chain centre and on the substrate, as indicated. The arrow highlights the spectroscopic signature of the precursor MBS. \textbf{d} QPI extracted from similar Mn chains of
length $N = 10$ to $N = 32$ where $q$ is the scattering wavevector. The arrows indicate the position where the $\delta$ YSR band is crossing $E_\textrm{F}$. The range of the $\alpha$ YSR band is indicated on the right. \textbf{e} YSR band structure calculated from a first-principles based model with topological gap $\Delta_\textrm{ind}$ indicated by the dashed lines. The spin polarization is given by the color of the band. \textbf{f}, \textbf{g} STM images (top panels) and deconvoluted $\didv$ line profiles (bottom panels) taken along the longitudinal central axis as marked by the dashed line in the top panel f of Mn$_{14}$ (panel f) and Mn$_{15}$ (panel g) chains. \textbf{h} Sequence of $\didv$ spectra taken on the ends of Mn$_N$ chains with even numbers $N$ of atoms as a function of $N$. \textbf{i}, \textbf{j} Sketch of geometric chain structure (top panels) and deconvoluted $\didv$ line profiles (bottom panels) taken along the longitudinal central axis of two Mn$_{12}$ chains separated by $N_\emptyset=4$ (panel i) and $N_\emptyset=3$ (panel j) empty lattice sites as indicated in the sketches above. \textbf{k} Energies of the lowest-energy state measured from similar data as in panels i and j as a function of $N_\emptyset$. Figure adapted from~\cite{Schneider2022}.}\label{fig:Precursor}
\end{figure}

Schneider \emph{et al.}~\cite{Schneider2022} also constructed Mn chains along $[1\bar 10]$ on Nb(110) (Fig.~\ref{fig:Precursor}a) leading to a $\sqrt{2}$ times larger interatomic separation (i.e., along $y$, see Fig.~\ref{fig:YSRatomsNb}d). Therefore, on the one hand, the hybridization of the close to gap edge $d_{z^2}$ YSR states and, thus, the bandwidth of their according $\alpha$ YSR band was so much reduced, that it did no longer cross $E_\textrm{F}$ (Fig.~\ref{fig:Precursor}d). On the other hand, due to their lobes along the chain direction $y$, the close-to mid gap $d_{yz}$ YSR states still hybridized sufficiently strong such that their $\delta$ YSR band was of sufficient width to cross $E_\textrm{F}$ (Fig.~\ref{fig:Precursor}d). The measured band structure fitted quite well to the band structure in Fig.~\ref{fig:Precursor}e from a single-orbital tight-binding model using the bulk Nb dispersion, Rashba-type spin-orbit coupling parameter and scattering potentials from \emph{ab initio} calculations, as well as adatom-substrate coupling parameters from comparison to the experimental data of the single atom $d_{yz}$ YSR state~\cite{Beck2021}. From these calculations~\cite{Schneider2022}, it turned out that there is indeed a topologically non-trivial minigap in the $\delta$ YSR band of $\Delta_\textrm{min}\approx\SI{50}{\micro\electronvolt}$, which would lead to overall TSC assuming that there are no other, experimentally undetectable, YSR bands crossing $E_\textrm{F}$ (\emph{odd $E_\textrm{F}$ QPI}). This width is, unfortunately, beyond the experimental energy resolution in~\cite{Schneider2022}. Nevertheless, separated from the confined states within the off-zero energy, trivial $\alpha$ YSR band, the zero bias state of the $\delta$ YSR band localized on both chain's ends was clearly visible (Fig.~\ref{fig:Precursor}b and ~\ref{fig:Precursor}c, criteria \emph{ZBP}, \emph{loc. DOS}, \emph{both ends} fulfilled). The state shows a peculiar even-odd-$N$ oscillation where the energy is jumping between close to zero energy and off zero energy as a function of $N$ (Fig.~\ref{fig:Precursor}f-h). The authors of~\cite{Schneider2022} assigned this to a significant overlap of the MBS wave functions from both ends of the chain for the corresponding chain lengths, indicating that the hybridization is changing alternately between weak and strong. The reason for the even-odd character of this oscillation is the steep slope of the corresponding YSR band which opens up its minigap close to the particular value $q=\pi/d$, where $d$ is the interatomic distance. Only for very particular lengths $N$, presumably where the hybridization is tuned to almost zero, the end state was found at zero energy within the experimental energy resolution. Using this MSH system, it was also possible to perform a slightly different perturbation experiment compared to that introduced in the last section (Fig.~\ref{fig:Pert}). Two structurally identical Mn$_{12}$ chains with the close to zero energy edge state were built such that their ends are separated by a small number of $N_\emptyset$ lattice sites (Fig.~\ref{fig:Precursor}i-k). From Fig.~\ref{fig:Precursor}i to Fig.~\ref{fig:Precursor}j, this number has been decreased by 1. As it can be seen, the edge states are slightly shifting in energy for both chains, indicating a considerable hybridization of the two edge state's wavefunctions from both chains. This hybridization primarily results from the two ends of the chains which are facing towards each other. However, both edge states are shifting all over the chain. It was thus argued that this proves that there is one quantum state on each chain, which is delocalized along the whole chain~\cite{Schneider2022}. The hybridization between the edge states on the two chains as measured by the energy shift (Fig.~\ref{fig:Precursor}k), moreover, oscillates as a function of void sites between the chains and finally dies out for $N_\emptyset>5$. Such a behaviour is consistent with a MBS of a chain which is still so short that the parts of the MBS wavefunction from the two chain's ends hybridize with each other (a so called precursor MBS). Together with all the other criteria discussed above (\emph{ZBP}, \emph{loc. DOS}, \emph{both ends}, $\Delta_\textrm{min}\lesssim\SI{50}{\micro\electronvolt}$, and \emph{odd $E_\textrm{F}$ QPI}) the authors concluded that this MSH system realizes a small-gap TSC with a precursor MBS. With the help of tight-binding model calculations, they also concluded that the precursor would evolve into a localized, unhybridized, MBS for chain lengths of more that 70 atoms corresponding to $\SI{35}{\nano\meter}$ chain lengths. Unfortunately, such chain lengths could not be realized until now due to the residual oxygen contamination of the Nb(110) surface.

\begin{figure}[H]
\centering
\includegraphics[width=0.995\textwidth]{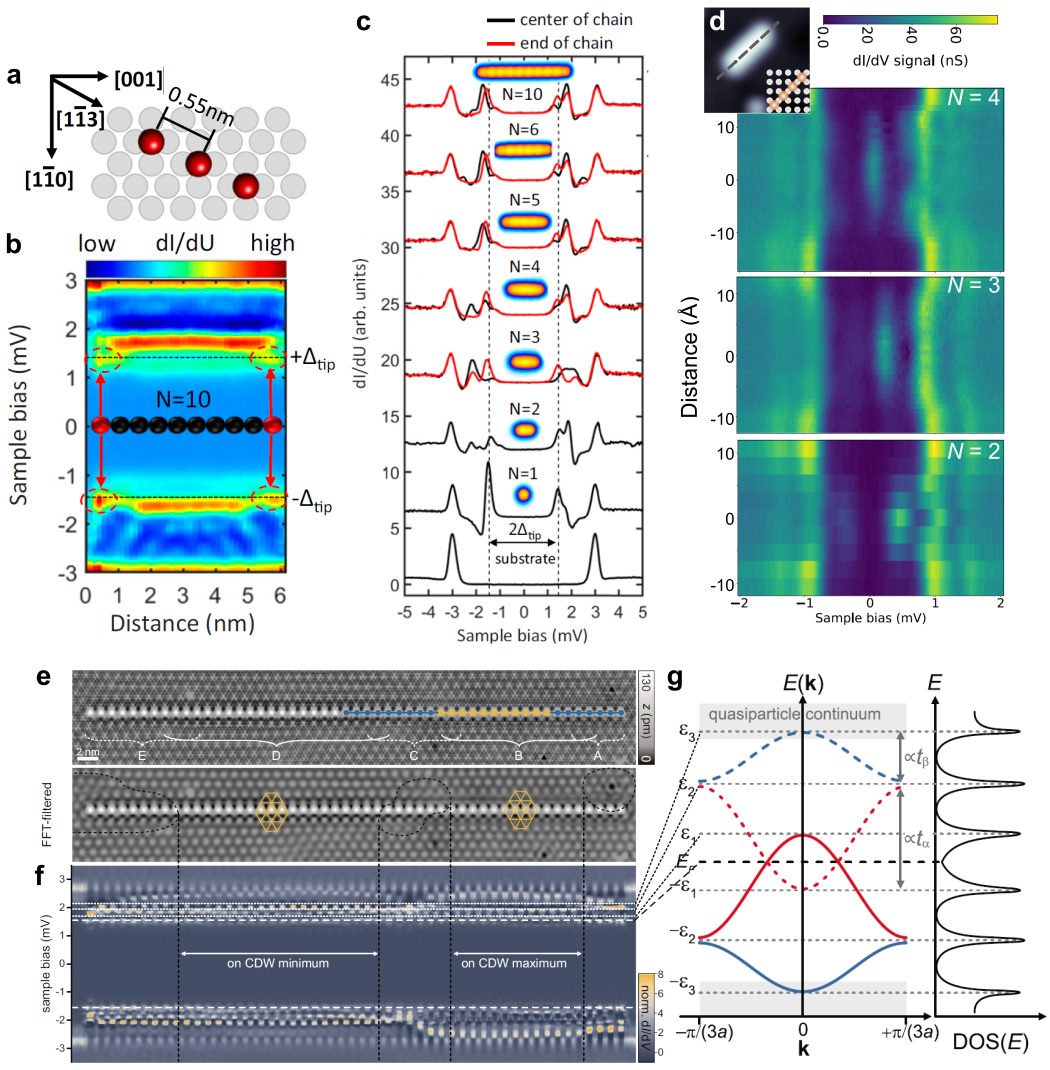}
\caption{Ferromagnetic spin chains on different SC substrates. \textbf{a} to \textbf{c} Sketch of atomic positions (panel a), $\didv$ line profiles along the central longitudinal axis (panel b), and $\didv$ spectra at ends and centers of Cr$_N$ chains with different numbers $N$ of Cr atoms as indicated, assembled with an interatomic distance of $\sqrt{11}a/2$ along the $[1\bar 13]$ direction on Nb(110)~\cite{Küster2022}. The red arrows with dashed ellipses in panel b indicate the edge states. The insets in panel c show STM images of the chains from which the spectra were taken. \textbf{d} $\didv$ line profiles along the central longitudinal axis of Cr$_N$ chains (e.g. dashed line in inset) with different numbers $N$ of Cr atoms as indicated, assembled with an interatomic distance of $\sqrt{2}a$ along the $[110]$ direction on $\beta$-Bi$_2$Pd(001)~\cite{Mier2021}. The inset shows an STM image of a Cr$_4$ chain with the sketch of the atomic positions on the lattice of substrate atoms. \textbf{e}, \textbf{f} Original (top panel in e) and FFT-filtered (bottom panel in e) STM images and $\didv$ line profile (panel f) along the central longitudinal axis of a Fe$_{51}$ chain assembled on HC adsorption sites on $2H$-NbSe$_2$ with interatomic distances of $3a$~\cite{Liebhaber2022}. The dashed ellipses encircle the most pronounced charge density wave (CDW) distortions. The yellow grids indicate the positions of atoms with respect to the CDW maxima (crossings of lines and upwards pointing triangles mark CDW maxima and minima, respectively). In panel f, the dashed lines indicate the tip gap $\Delta_\textrm{t}$, and the dotted lines refer to the van Hove singularities of YSR bands (see lines guding to panel g). \textbf{g} Schematic illustration of the YSR bands (left part) and DOS (right part). The solid (dashed) lines indicate the electron- (hole-) parts of the $\alpha$ (red) and $\beta$ (blue) YSR bands. Panels a to c adapted from~\cite{Küster2022}. Panel d adapted with permission from Ref.~\cite{Mier2021}. Copyrighted by the American Physical Society. Panels e to g adapted from~\cite{Liebhaber2022}.}\label{fig:FMspinchain}
\end{figure}

On the same substrate, Cr chains have been assembled by STM based atom manipulation along many different crystallographic directions and were studied by STS~\cite{Küster2022} (see Table~\ref{table:spinchains} and lattice orientations in Fig.~\ref{fig:FMspinchain}a). It was argued, that due to the relatively large used interatomic spacings $>\sqrt{11}a/2$, the direct hopping between the $3d$ levels is already so weak~\cite{Küster2021longrange}, that such chains are best described by the so called dilute spin chain case. In that case, the YSR bands are formed by hybridizing YSR states of the individual atoms in the chain. Note, that this was also assumed for some of the tight-binding models used to simulate the Mn chains on Nb(110) and Ta(110) described above~\cite{Schneider2021b,Schneider2022,Schneider2023}. It is worth noting here that for the Cr atoms the $d_{z^2}$ YSR state is already close to $E_\textrm{F}$ (Fig.~\ref{fig:YSRatomsNb}l) and a relatively weak hybridization can result in a sufficient width of the resulting YSR band such that it crosses $E_\textrm{F}$. This is one of the necessary requirements to form a TSC. An example of one particular orientation of a Cr chain along $[1\bar{1} 3]$ resulting in an in-plane ferromagnetic spin order is shown in Fig.~\ref{fig:FMspinchain}a-c. The development of the corresponding $\didv$ spectra taken at the center and end of the chains from $N=1$ to $N=10$ atoms is shown in Fig.~\ref{fig:FMspinchain}c and the $\didv$ line-profile along the central longitudinal axis in Fig.~\ref{fig:FMspinchain}b. There is indeed a YSR band formed close to $E_\textrm{F}$ which has a minigap of width $\Delta_\textrm{min}\approx\SI{100}{\micro\electronvolt}$ and hosts a zero bias end state detected on both ends, so the criteria \emph{loc. DOS}, \emph{ZBP}, \emph{both ends}, and \emph{$\Delta_\textrm{min}$} are fulfilled. However, the authors did not analyse their data by FFT as in Fig.~\ref{fig:1DExpMethods} such that it is hard to judge, whether the gapped YSR band has crossed $E_\textrm{F}$, and the minigap could, therefore, be topologically non-trivial. Instead, by comparison to many other chain orientations with different interatomic distances, which all show zero or close-to zero bias edge states, as well as with a minimal tight-binding model~\cite{Küster2022}, the authors concluded, that the observed edge states are most probably topologically trivial YSR states. The other in-plane ferromagnetic examples of Cr chains assembled along $[1\bar{1} 1]$ with separations of $\sqrt{3}a$ or $2\sqrt{3}a$ fulfilled the criteria \emph{loc. DOS}, \emph{ZBP}, and \emph{both ends}, however, without detectable minigap, or with an off-zero bias end state and also no minigap, respectively. The other Cr chains which were studied in that work have antiferromagnetic or spin spiral order, and will be discussed in Sections~\ref{experiment_1-1D-AFM} and \ref{experiment_1-1D-NC}. The tight-binding model without spin-orbit coupling studied in the same work~\cite{Küster2022} showed that the energy of the lowest YSR band state strongly oscillates as a function of the chain length. It was argued that, if all the multiple lowest energy states of the YSR band are added up, which are detected due to the finite energy resolution in the experiment, the resulting measured $\didv$ line-profile will have a peak at the chain's ends with a very similar appearance as in the experiment. It was finally noted in this work, that even in the metallic regime, i.e., if the substrate's superconductivity is quenched, there are edge states which make the end of the chain distinct with respect to the center, and which have a single atom localization length.

Another investigated MSH system having out-of-plane ferromagnetic spin order, is the Cr chain assembled by STM based atom manipulation along the $[110]$ direction on the $(001)$ surface of the two-compound SC $\beta-$Bi$_2$Pd~\cite{Mier2021} (Fig.~\ref{fig:FMspinchain}d).  The experiments, which were performed up to chain lengths of only $N=4$, showed the formation of a minigapped YSR band moving closer to $E_\textrm{F}$, and the minigap was almost closed for $N=4$. The simulations performed in the same work predicted, that for $N=8$, a topological minigap would have reopened hosting a MBS. But unfortunately, up to the date of finishing this review article, experimental studies of chains of that length have not been published.

Figures~\ref{fig:FMspinchain}e to g illustrate another MSH system which is based on the two-compound SC $2H-$NbSe$_2$ used as a substrate and chains from Fe atoms assembled by STM based atom manipulation~\cite{Liebhaber2022}. The substrate features a charge density wave which imposes additional difficulties for this MSH system. In order to maximize hybridization between the YSR states, the authors used as identical as possible adsorption sites for the Fe atoms within the chain, i.e., the hollow-centered adsorption site which was already described in Figs.~\ref{fig:YSRatomsother}k to m and \ref{fig:YSRdimersother}h to n. After chain formation, $\didv$ line-profiles reveal that the hybridized YSR states form bands (Figs.~\ref{fig:FMspinchain}e,f). The authors argue, that the band formation as such already suggests a ferromagnetic or a weak spin spiral order. However, a considerable spin-orbit interaction in the system challenges this conclusion~\cite{Beck2021}. The three most intense features in the YSR band visible in Fig.~\ref{fig:FMspinchain}f were interpreted to originate from four van-Hove singularities of the two bands within the quantum model which has been used (Fig.~\ref{fig:FMspinchain}g), where two of these singularities are almost degenerate. If this interpretation is correct, the experimental data would suggest a single Fermi level crossing (criterion \emph{odd $E_\textrm{F}$}). However, the chains show no signatures of zero bias edge states. As already discussed for the according dimers in Section~\ref{experiment_dim} there were strong experimental evidences, that the Fe atom spins in this chain system have to be described as spin 1/2 states and treated fully quantum mechanically including the Kondo effect. The authors of~\cite{Liebhaber2022} argued, that this possibly reduces the parameter range for topological superconductivity. As also visible in the data shown in Fig.~\ref{fig:FMspinchain}f, the incommensurate nature of the charge density wave induces a smoothly varying potential onto the YSR bands. In other words there is a band bending varying locally along the chain. This band bending induces strong disorder which is visible in the form of apparent domains in the YSR bands.

\begin{figure}[H]
\centering
\includegraphics[width=\textwidth]{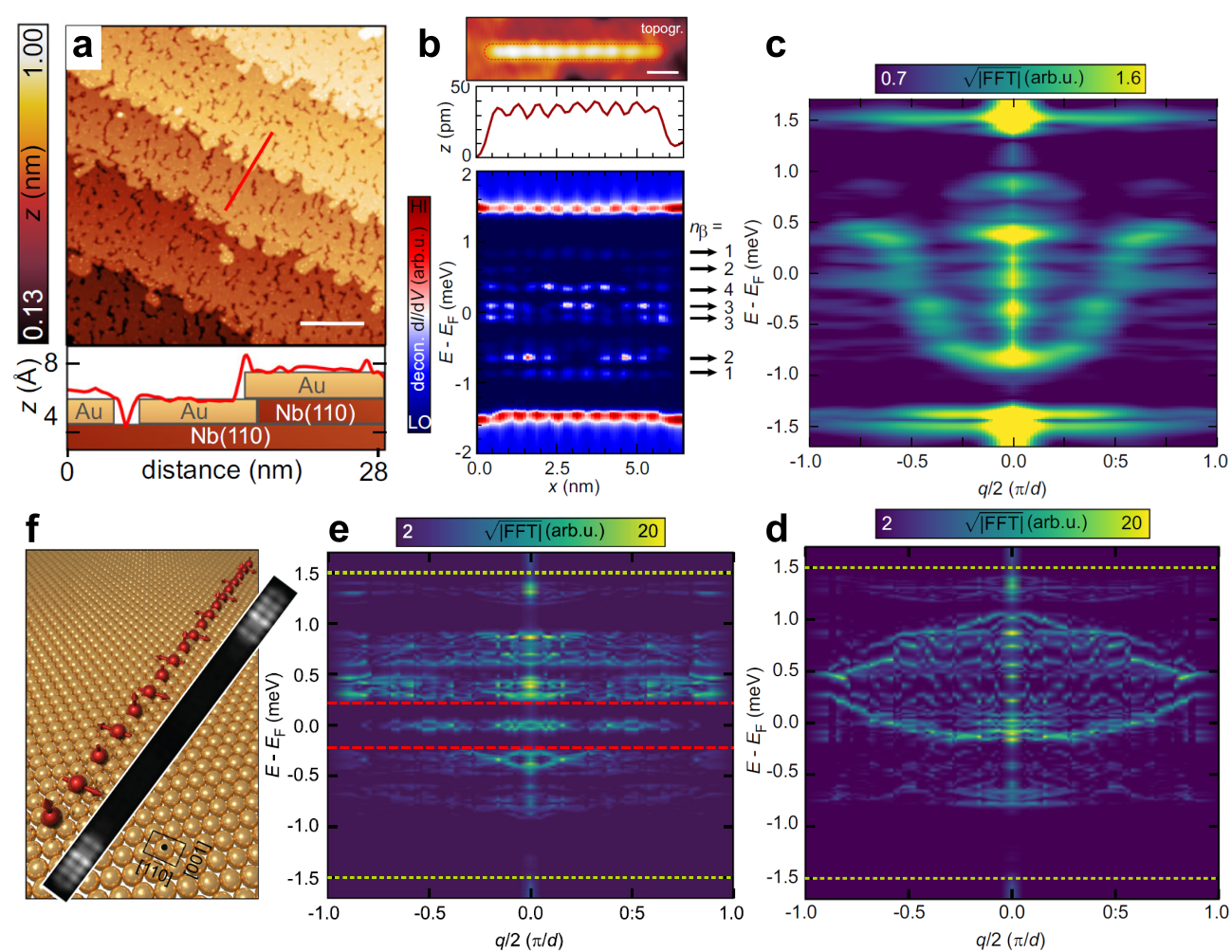}
\caption{Spin chains on proximity-SC thin layers of high-$Z$ metals. \textbf{a} STM image of a monolayer of Au on Nb(110) decorated with single Fe atoms (top panel). The bottom panels shows the height profile along the red line in the top panel with the surface composition sketched beneath the profile. Scale bar, $\SI{20}{\nano\meter}$. \textbf{b} STM image (top panel), height profile (middle panel), and deconvoluted  $\didv$ line profile (bottom panel) along the central longitudinal axis of a Fe$_{9}$ chain assembled with interatomic distances of $2a$ along the $[001]$ direction on the Au(110)/Nb(110) surface shown in panel a. Scale bar in the top panel, $\SI{1}{\nano\meter}$. The black arrows in the bottom panel indicate the confined BdG quasiparticle states of the $\beta$ YSR band with the given numbers of maxima $n_\beta$. \textbf{c} Averaged energy-wise 1D-FFT extracted from similar data as in the bottom panel of b from the ferromagnetic Fe$_N$ chains with lengths $N=7$ to $N=14$. \textbf{d}, \textbf{e} Calculated dispersions of scattering wave vectors obtained by 1D-FFT from the calculated LDOSs of ferromagnetic (panel d) and helical spin spiral (panel e) Fe$_N$ chains averaged for different lengths between $N=9$ to $N=19$. The green and red dashed lines indicate the energy gap of the SC substrate and the induced minigap $\Delta_\textrm{ind}$, respectively. \textbf{f} Sketch of the $90^\circ$ helical spin spiral (red arrows) in the chain of Fe atoms (red spheres) on the lattice of Au atoms (golden spheres) assumed in the calculation shown in panel e (note, that this is not the experimental spin order, which is ferromagnetic). The inset shows the spatial distribution of the zero energy LDOS evaluated in the first vacuum layer above the spin spiral Fe$_{19}$ chain. Figure adapted from~\cite{Beck2023b}.}\label{fig:FeAu}
\end{figure}

As visible from the compilation of all the properties of the above discussed ferromagnetic MSHs in Table~\ref{table:spinchains}, those systems which are merely comprised of a spin chain coupled to a SC and are indicating at least some of the criteria for TSC have all minigaps which are on the order of $\SI{100}{\micro\electronvolt}$. In TSCs based on ferromagnetic MSHs the size of the topological minigap is expected to scale with the size of the proximity-induced SC gap and with the strength of the SOC in the relevant YSR bands. It was proposed that one way to increase this SOC is using a heavier substrate with a larger atomic number $Z$, as the atomic spin-orbit coupling roughly scales with $Z$. In order to increase the proximity-induced SC gap, however, a substrate with a large SC gap to start with is essential. One way to satisfy these two requirements at the same time is making use of the proximity effect to induce superconductivity into thin layers of high-$Z$ materials grown onto a superconductor with a large $s$-wave gap and then utilize these layers as the substrate for the assembly of spin chains. Beck~\emph{et al.} investigated this concept for monolayers of Au grown on Nb(110)~\cite{Beck2023b} (Fig.~\ref{fig:FeAu}a). Indeed it was possible to proximity-induce a SC gap of almost the same size as that on Nb(110) into the Au monolayer (Fig.~\ref{fig:FeAu}b, bottom panel, $\didv$-spectra at $x=\SI{0}{\nano\meter}$ and $x=\SI{6.5}{\nano\meter}$). It was then possible to realize a YSR band in chains from Fe atoms assembled by STM-based atom manipulation along the $[001]$ direction with an interatomic separation of $2a$ which turned out to have a ferromagnetic spin order with out-of-plane orientation (Fig.~\ref{fig:FeAu}b, top panels). The bottom panel of Fig.~\ref{fig:FeAu}b illustrates the $\didv$ line profile which was taken along the Fe$_9$ chain in the top panel. It shows the confined BdG quasiparticle states of the resulting YSR band. While this YSR band turned out to cross $E_\textrm{F}$ only once (Fig.~\ref{fig:FeAu}c), it does not host a detectable minigap and also no zero bias edge states. DFT calculations reported in the same publication indeed reproduced the ungapped YSR bands (Fig.~\ref{fig:FeAu}d). They, furthermore, showed that the Dzyaloshinskii-Moriya term in the exchange interaction between the Fe atoms (in a dimer) is only 10\% as big as the Heisenberg exchange interaction. Although this is not particularly weak, the spin-orbit coupling in the Au layer also induces a very strong out-of-plane on-site anisotropy. Overall, these energies prevent the formation of a spin-spiral in the Fe chain, but instead stabilize an out-of-plane ferromagnetic spin order with negligible minigap. The calculations also showed, that if the system is forced into a spin spiral order (Fig.~\ref{fig:FeAu}f) a considerable topological minigap would open up in the $d_{x^2-y^2}$ YSR band (Fig.~\ref{fig:FeAu}e), which hosts a MBS (inset of Fig.~\ref{fig:FeAu}f).

\subsubsection{Antiferromagnetic spin chains}\label{experiment_1-1D-AFM}
\begin{figure}[H]
\centering
\includegraphics[width=0.99\textwidth]{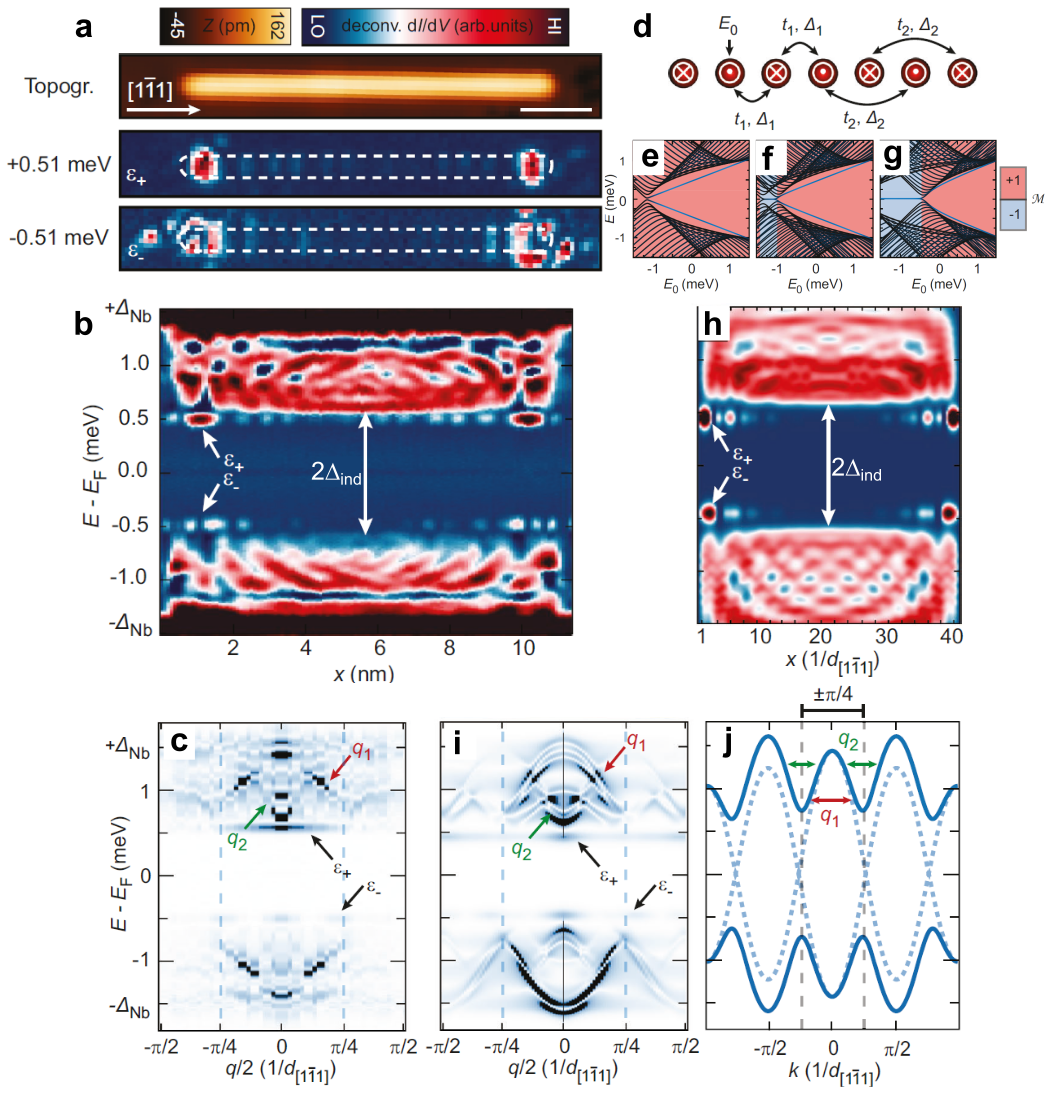}
\caption{Off-zero bias edge mode in antiferromagnetic Mn chain on Nb(110). \textbf{a} STM image (top panel) and deconvoluted $\didv$ maps (bottom panels) taken at the indicated off-zero bias voltages (energies $\varepsilon_{+,-}$) of the edge mode of an antiferromagnetic Mn$_{40}$ chain assembled with interatomic distance of $\sqrt{3}a/2$ along the $[1\bar 11]$ direction on Nb(110). Scale bar, $\SI{2}{\nano\meter}$. \textbf{b} Deconvoluted $\didv$ line profile taken along the longitudinal central axis of the Mn$_{40}$ chain with marked off-zero bias energy edge mode at $\varepsilon_{+,-}$ in the induced gap of width $2\Delta_\textrm{ind}$. \textbf{c} Absolute values of the line-wise FFT of the data in panel b. Dispersive scattering vectors $q_1$, $q_2$, and edge state energies $\varepsilon_{+,-}$ are indicated. \textbf{d} Sketch of the
same tight-binding model used in Fig.~\ref{fig:Pert}f to i to simulate the antiferromagnetic chain. \textbf{e} to \textbf{g} Calculated eigenenergies from the tight-binding model in panel d of a chain with $N=40$ sites as a function of on-site YSR state energy $E_0$ and with pairing and hopping parameters $\Delta_1=\SI{0.5}{\milli\electronvolt}$, $\Delta_2=\SI{0.0}{\milli\electronvolt}$, $t_2=\SI{0.6}{\milli\electronvolt}$, $t_1=\SI{0.0}{\milli\electronvolt}$ (panel e), \textbf{f} $t_1=\SI{0.1}{\milli\electronvolt}$ (panel f), and unrealistically large $t_1=\SI{0.4}{\milli\electronvolt}$ (panel g). The Majorana number $\mathcal{M}$ is trivial ($+1$) in the red regions and non-trivial ($-1$) in the blue regions. \textbf{h} to \textbf{j} Calculated LDOS along a chain of $N = 40$ sites (panel h), absolute values of the line-wise FFT of a similar calculated LDOS as in panel h but with $N = 100$ sites (panel i), and YSR band structure (panel j), using the parameters from panel f ($t_1=\SI{0.1}{\milli\electronvolt}$). Dashed lines in panel j additionally show the case $\Delta_1=\SI{0.0}{\milli\electronvolt}$. The induced minigap $2\Delta_\textrm{ind}$, the edge state energies $\varepsilon_{+,-}$, and the scattering vectors $q_1$ and $q_2$ corresponding to the experimental result in panels b and c are indicated. Figure adapted from~\cite{Schneider2023}. }\label{fig:AFMspinchains}
\end{figure}

It is striking, that four out of the five experimentally investigated MSH systems having antiferromagnetic spin orders have comparatively large minigaps (see Table~\ref{table:spinchains}), but at the same time they show either no indications for zero bias edge states, or a negative outcome of the perturbation experiment, ruling out a topologically nontrivial character of the edge states and minigaps. It is also conspicuous, that four out of the five systems host off-zero bias edge states. The largest minigap of all these systems ($\Delta_\textrm{ind}=\SI{700}{\micro\electronvolt}$) was observed for close-packed Mn chains assembled along the $[1\bar 11]$ orientation on Nb(110)~\cite{Schneider2023} (Fig.~\ref{fig:AFMspinchains}). In the following, we will review that work and discuss the possible reason for the large topologically trivial minigap commonly hosting off-zero bias edge states in antiferromagnetic MSH systems. Fig.~\ref{fig:AFMspinchains}a shows an STM constant-current image (top panel) and $\didv$ maps over the same area at the biases of the off-zero bias edge states localized on both chain ends (bottom panels). The $\didv$ line-profile taken along the central longitudinal axis of the chain (Fig.~\ref{fig:AFMspinchains}b) reveals that the edge states are located at energies $\varepsilon_{+,-}$ inside the large minigap of the YSR bands energetically quite close to the gap edges. After a local perturbation of one side of the chain, similar to the one in Fig.~\ref{fig:Pert}, it was shown that the left and the right end state have slightly different energies, proving the local nature of these states in clear contrast to a non-local, spatially correlated MBS. The line-wise FFT of the $\didv$ line-profile, shown in Fig.~\ref{fig:AFMspinchains}c, uncovers two prominent scattering vectors $q_1$ and $q_2$ within the YSR band, one with a downwards ($q_1$) and one with an upwards ($q_2$) dispersion, respectively. These results were rationalized within a minimal next-nearest-neighbor tight-binding model Hamiltonian (Fig.~\ref{fig:AFMspinchains}d) including an onsite YSR state at energy $E_0$, nearest neighbor (NN) and next-nearest neighbor (NNN) hopping ($t_1$ and $t_2$) and pairing ($\Delta_1$ and $\Delta_2$), as well as Rashba spin-orbit coupling. The authors further argue, that, without spin-orbit coupling, hopping is mainly allowed between the sites with the same spin, i.e., between NNNs, while the effective superconducting pairing is most effective between sites with opposite spins, i.e., NNs. For non-zero spin-orbit coupling of reasonable strengths these restrictions are only softly lifted, i.e., the model should still be restricted to the cases $t_1\ll t_2$ and $\Delta_1\gg \Delta_2$. Since the YSR band arose from a YSR state close to $E_\textrm{F}$ in this system, the authors chose $E_0\approx0$ in their model. Within these reasonable parameter ranges, the resulting calculated eigenenergies and Majorana number for different values of $t_1$ (Fig.~\ref{fig:AFMspinchains}e-g) indeed showed, that a minigap is easily opened by $\Delta_1$, and then, the system always has non-zero energy edge states close to the minigap edge in the trivial phase ($\mathcal{M}=+1$), one on each side. Moreover, the topologically non-trivial phase with zero-energy MBSs ($\mathcal{M}=-1$), for realistic parameters reflecting reasonable values of the spin-orbit coupling, takes up only a very small proportion of the phase space (Fig.~\ref{fig:AFMspinchains}f). Finally, the model parameters were adapted to reproduce the experimental scattering vectors (c.f. Fig.~\ref{fig:AFMspinchains}c and Figs.~\ref{fig:AFMspinchains}i,j). Using these parameters, the calculated local density of states along a chain of $N = 40$ sites (Fig.~\ref{fig:AFMspinchains}h) closely resembled the experimental result (Fig.~\ref{fig:AFMspinchains}b).
 
\begin{figure}[H]
\centering
\includegraphics[width=\textwidth]{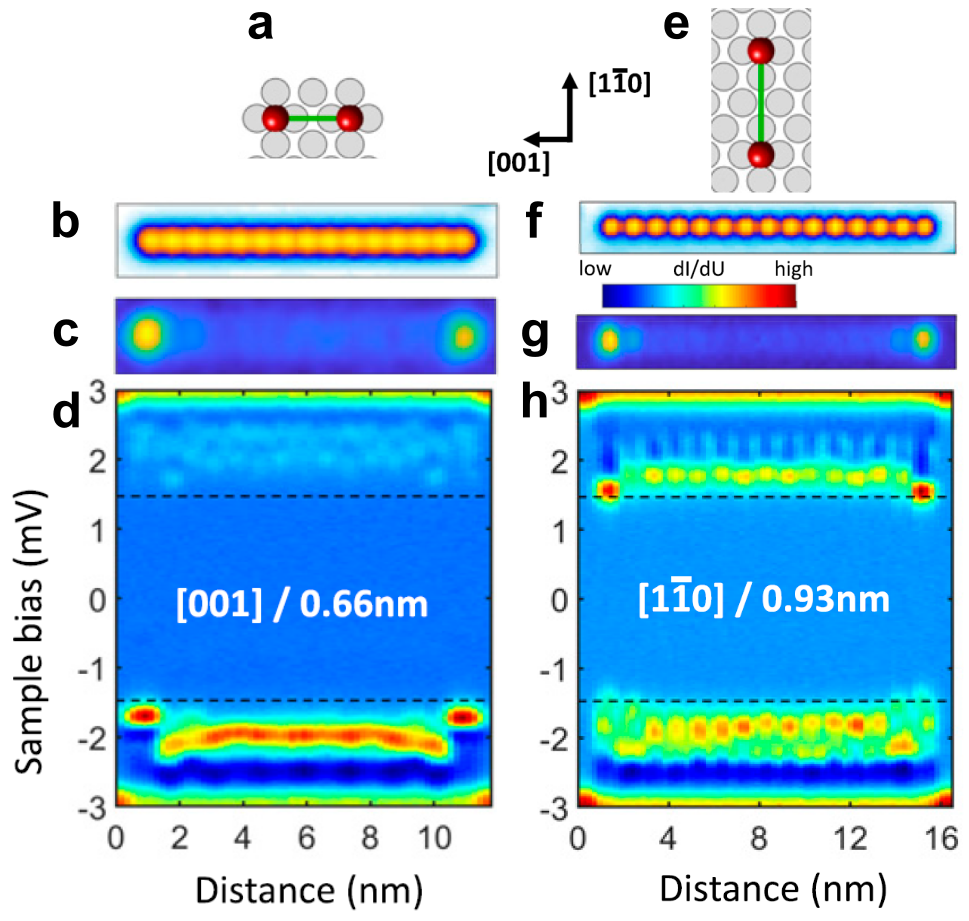}
\caption{Off-zero bias edge modes in antiferromagnetic Cr chains on Nb(110). \textbf{a} to \textbf{d} Illustration of the Cr atoms (red) on the Nb(110) surface lattice (grey, panel a), STM image (panel b), sum of $\didv$ maps taken at the bias voltages $\pm\Delta_\textrm{tip}$ corresponding to $E_\textrm{F}$, and $\didv$ line profile taken along the longitudinal central axis of a Cr$_{15}$ chain assembled with interatomic distance of $2a$ along the $[001]$ direction on Nb(110). \textbf{e} to \textbf{h} Same as panels a to d, but for an antiferromagnetic Cr$_{15}$ chain assembled with interatomic distance of $2\sqrt{2}a$ along the $[1\bar10]$ direction on Nb(110). Note, that the dashed lines in panels d and h indicate the biases of $\pm\Delta_\textrm{tip}$ showing that the signal imaged in panels c and g at $E_\textrm{F}$ stems from the off-zero edge modes in panels d and h. Figure adapted from~\cite{Küster2022}. }\label{fig:AFMspinchainsCr}
\end{figure}

The general validity of the model is substantiated by the other experimental results on antiferromagnetic MSHs described in Section~\ref{Sec:ExpMeth} (Fig.~\ref{fig:Pert}) and illustrated in Fig.~\ref{fig:AFMspinchainsCr}. The Mn chains on Ta(110) (Fig.~\ref{fig:Pert})~\cite{Schneider2023} which are structurally identical to the ones on Nb(110) discussed above, also revealed a large minigap hosting topologically trivial, in that case accidentally close to zero energy, edge states. Antiferromagnetic Cr chains assembled by STM-based atom manipulation on Nb(110) along two different orientations (Fig.~\ref{fig:AFMspinchainsCr})~\cite{Küster2022} as well manifest considerable minigaps which host off-zero energy edge modes.

In summary, it can be concluded, that, in contrast to ferromagnetic chains, the antiferromagnetic spin order facilitates the formation of a large minigap in the YSR band even without SOC. However, the formation of the topologically non-trivial phase then is a threshold effect, as the SOC has to compete with the pairing potential, such that unrealistically large values of SOC are needed in order to enable topological superconductivity~\cite{Schneider2023}. 

\subsubsection{Non-collinear spin chains}\label{experiment_1-1D-NC}

\begin{figure}[H]
\centering
\includegraphics[width=\textwidth]{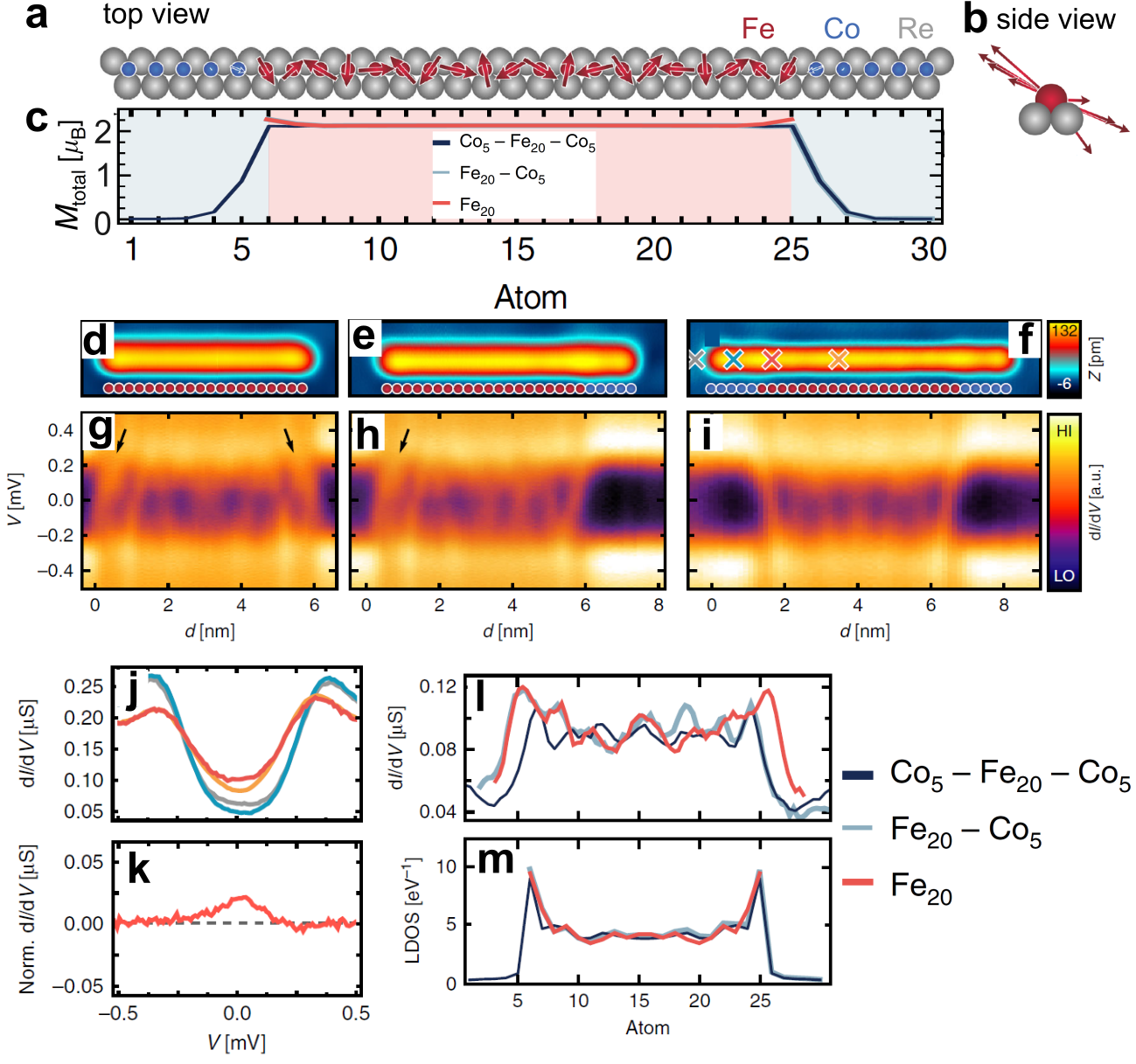}
\caption{Zero-bias edge modes in a spin spiral realized by an Fe chain on Re(0001). \textbf{a} to \textbf{c} Top (panel a) and side views (panel b) of the calculated spin structure, and calculated total magnetic moment of the atoms (panel c, black curve) along a Co$_5-$Fe$_{20}-$Co$_5$ 
chain assembled with interatomic distance of $a$ along the $[1\bar10]$ direction on Re(0001). Blue, red, and grey spheres correspond to Co, Fe, and surface Re atoms, respectively. The length of the arrows in panels a and b is proportional to the size of the magnetic moment at each particular site (for panel a the in-plane component is given). The red and blue curves in panel c additionally show the calculated magnetic moments of the atoms along a Fe$_{20}$ and a  Fe$_{20}-$Co$_5$ chain, respectively. \textbf{d} to \textbf{f} and \textbf{g} to \textbf{i} STM images (panels d to f) and according $\didv$ line profiles (panels g to i) taken along the longitudinal central axis of an Fe$_{20}$ chain (panels d, g), a Fe$_{20}-$Co$_5$ chain (panels e, h), and a Co$_5-$Fe$_{20}-$Co$_5$ chain (panels f, i) as illustrated by the spheres in the STM images. The arrows in panels g and h indicate the additional off-zero bias edge 
modes. \textbf{j} $\didv$ spectra of the Co$_5-$Fe$_{20}-$Co$_5$ chain taken at the positions marked by the accordingly colored crosses
in panel f. \textbf{k} $\didv$ spectrum from panel j taken at the Fe/Co transition and normalized by subtraction of a spectrum averaged along the chain’s interior. \textbf{l} Zero-bias $\didv$ signal along the three different chains as indicated on the right extracted from the data in panels g to i. \textbf{m} Spectral weight at $E_\textrm{F}$ along the three different chains as indicated on the right, calculated from the tight-binding model. Figure adapted from~\cite{Schneider2020}. }\label{fig:Spinspiralspinchains}
\end{figure}

Considering the results from the previous two Sections, the most promising path towards large topologically non-trivial minigaps lies in the realization of YSR bands in MSHs with non-collinear spin order. Unfortunately, there is up to date only one MSH system, for which a non-collinear spin order has been confirmed directly using SP-STS measurements, i.e., Fe chains on Re(0001)~\cite{Kim2018,Schneider2020}. For the four other systems given in Table~\ref{table:spinchains}, the spin spiral order was either a result of DFT calculations~\cite{Küster2022,Küster2021longrange}, or was concluded rather indirectly from the experimentally detected splittings of YSR states~\cite{Ding2021tuning} or from the shape of the dispersion of the experimentally measured YSR bands~\cite{Liebhaber2022,Mier2021}. Moreover, those four systems also showed no indications for TSC or MBSs. We, therefore, focus here on the system of chains of close-packed hcp-site Fe atoms on Re(0001) (Fig.~\ref{fig:Spinspiralspinchains}, see Fig.~\ref{fig:YSRatomsRe} for the properties if single atoms), which revealed a spin-spiral order in SP-STS measurements~\cite{Kim2018,Schneider2020}. \emph{Ab initio} calculations using the Korringa–Kohn–Rostoker Green’s function method, together with an effective spin model, revealed strongly antiferromagnetic exchange interactions between the nearest and next-nearest neighbors, leading to spin frustration~\cite{Schneider2020}. The frustration is resolved by a cycloidal spin spiral formation with a wavelength of three to four lattice spacings (Fig.~\ref{fig:Spinspiralspinchains}a). This result is in agreement with the SP-STS measurements reported in the earlier publication~\cite{Kim2018}.  While there is also considerable Dzyaloshinskii–Moriya interaction between the nearest neighbors, this only determines the plane of rotation of the spin spiral at an angle of $30^\circ$ to the surface plane and the rotational sense (Fig.~\ref{fig:Spinspiralspinchains}b). $\didv$ maps of chains of increasing numbers of Fe atoms $N$ showed an enhanced zero bias $\didv$ intensity at both chains ends, decaying in an oscillatory fashion in intensity toward the center of the chain (criteria \emph{loc DOS.} and \emph{both ends}), and emerging around $N\approx8$ which was fully developed for $N\approx12$~\cite{Kim2018}. The enhanced zero bias $\didv$ intensity at both chains ends was later reproduced~\cite{Schneider2020} (see $\didv$ line-profile in Figs.~\ref{fig:Spinspiralspinchains}d,g as well as zero bias $\didv$ intensity along the chain in the red line in Fig.~\ref{fig:Spinspiralspinchains}l). However, it was also noted in the $\didv$ line-profile, that the YSR bands produce additional off-zero bias states with an enhanced intensity at the chains ends (see arrows in Fig.~\ref{fig:Spinspiralspinchains}g). The authors showed that this close-to-chain-end YSR band intensity can be moved out of the gap by attaching chains of non-magnetic hcp-site Co atoms to the Fe chain's ends (Fig.~\ref{fig:Spinspiralspinchains}e, f, h, i, see Fig.~\ref{fig:YSRatomsRe} for the properties of the single atoms). As the \emph{ab initio} calculations proved, this had a negligible effect on the spin-structure of the Fe chain (Fig.~\ref{fig:Spinspiralspinchains}a-c). The effect of the attached Co chains on the Fe chain's edge states is illustrated in Figs.~\ref{fig:Spinspiralspinchains}e, h, l first for the left Fe chain's end and subsequently in Figs.~\ref{fig:Spinspiralspinchains}f, i, l also for the right Fe chain's end. It can be seen that the zero energy $\didv$ weight maximum is persistent at the Co terminated Fe chains, but its position is slightly shifted toward
the interior of the Fe part of the hybrid chain (Figs.~\ref{fig:Spinspiralspinchains}h, i, l). As a result, the $\didv$ spectra taken at the location of the transition between spin-spiral Fe chain and nonmagnetic Co chain revealed more clearly that the edge state has its maximum intensity at zero bias (Fig.~\ref{fig:Spinspiralspinchains}j,k). Finally, tight-binding model calculations using parameters extracted from the \emph{ab initio} calculations~\cite{Schneider2020}, for an appropriately tuned superconducting energy gap, reasonably reproduced the experimentally observed changes to the zero-energy spectral weight localized at the spin-spiral ends (Fig.~\ref{fig:Spinspiralspinchains}m). Furthermore, using the same superconducting energy gap parameter, the tight-binding model resulted in a topologically superconducting phase for an infinite Fe chain. Unfortunately, due to the small SC energy gap of Re ($\Delta_\textrm{Re}=\SI{280}{\micro\electronvolt}$), a topologically nontrivial minigap would probably be too small to be experimentally detectable at the used experimental temperatures of $T\approx\SI{300}{\milli\kelvin}$, calling for measurements at even lower temperatures.

In conclusion, we emphazise that the effort to induce non-collinear spin order in one-dimensional MSH systems by high-$Z$ substrate materials is very promising. However, it can be hampered by the fact that a large spin orbit interaction usually not only increases the Dzyaloshinskii–Moriya interaction, but also enhances the magnetic anisotropy within the magnetic chain. This can prevent the desired formation of a spin spiral. As the example of the atomic Fe chains on Re(0001) discussed above shows, it is worth to perform experiments at mK temperatures or to proximitize ultrathin layers of a heavy-element material by a superconducting Nb(110) substrate in order to increase the superconducting gap size, thereby facilitating the experimental observation of the in-gap band structure. 

\section{Experimental investigation of two-dimensional spin textures on superconducting substrates}\label{experiment_2D}
Apart from the investigation of YSR states associated with single magnetic impurities on superconducting surfaces and the emerging YSR bands in 1D atomic chains, great interest has emerged in understanding the interaction between superconductors and spin textures in 2D magnetic systems. Here, the most interesting aspect is to understand the relationship between the topological properties of the emergent bands and edge modes and the magnetic ground state of the hybrid system. As will be discussed in this section, the experimental investigations originally focused on the study of 2D MSHs with a ferromagnetic (FM) ground state, with the aim of observing a topological phase with a gapped bulk band structure and chiral edge modes, whose number is defined by the topological invariant \textit{C}, as discussed in detail in Sec.~\ref{theory-2D-FM}. More recently, the attention moved towards 2D MSHs hosting non-ferromagnetic ground states. In particular, a few recent works investigated the nature of superconductivity in hybrid systems hosting collinear antiferromagnetic (AFM) order~\cite{LoContePRB2022,BazarnikNAT-COMM2023,SoldiniNAT-PHYS2023}. In those studies, both gapped and gapless superconducting phases were observed, together with edge and/or corner modes. Here, the antiferromagnetic ground state and the crystallographic symmetry of the hybrid system are at the core of the observed unconventional superconductivity. Very recently there was the report of a first experimental study of a 2D MSH hosting a non-collinear spin texture. The non-collinearity of the magnetic ground state is understood to be key for the emergence of the observed edge modes, an experimental fingerprint of the topological superconducting phase hosted by the hybrid system~\cite{BrüningARXIV2024}. 

A summary of the experimentally studied 2D MSH systems discussed in this Section is reported in Table \ref{tab_2DMSHs-exp}. Analogously to what was done for the 1D MSH systems in Section \ref{experiment_1}, all the most important physical properties and experimentally accessible criteria for topological superconductivity of the listed systems are indicated in Table \ref{tab_2DMSHs-exp}.

\begin{table}[!htbp]
\caption{Experimentally accessible evidences of topological superconductivity (TSC) with edge modes in 2D MSHs. The table summarizes the experimentally studied 2D magnetic layers on substrate (/substrate) systems (sorted by increasing atom number $Z$ of substrate atoms), fabricated (fab) via molecular beam epitaxy (MBE) or atom manipulation (am), with ferromagnetic (FM), antiferromagnetic (AFM) and spin spiral (SpS) spin orders with out-of-plane (op) or in-plane (ip) orientation. Compliance with the following experimental criteria for TSC with edge modes are indicated: zero-bias DOS localized on the 2D islands edges (ZBEM); edge modes' spin polarization measured (ZBEM SP); uniform (uni) and non-uniform (non-uni) edge modes observed; decay length of edge mode (decay length); resolved mini-gap ($\Delta_\textrm{ind}$). In the various columns, checkmarks ($\checkmark$) indicate which of the experimental criteria introduced above have been investigated/observed for the corresponding MSHs.}\label{tab_2DMSHs-exp}

 \begin{tabular}{@{}m{80pt}m{27pt}m{15pt}m{15pt}m{20pt}m{20pt}m{40pt}m{50pt}m{50pt}@{}}
\toprule
\textbf{/substrate} & cit. & fab. & spin & ZBEM & ZBEM & ZBEM & ZBEM & $\Delta_\textrm{ind}$\\
magnetic material &  & (MBE & order & & SP & (uni, & decay & \\
 &  & /am) & (op/ip) & & & non-uni) & length & \\
 
\midrule
\textbf{/NbSe\textsubscript{2}} & & & & & & & & \\
CrBr\textsubscript{3}(ML) & \cite{KezilebiekeNAT2020}& MBE & FM (op)\footnotemark[1] & $\checkmark$ & & non-uni & $\sim2.5$ nm & 350 $\mu$eV\\

\textbf{/Nb(110)} & & & & & & & & \\
Mn(ML) & \cite{LoContePRB2022,BazarnikNAT-COMM2023}& MBE & AFM (op) & $\checkmark$ & $\checkmark$\footnotemark[2] & non-uni & $\sim1.5$ nm & 0 $\mu$eV\footnotemark[3]\\
Cr(lattice) & \cite{SoldiniNAT-PHYS2023}& am & AFM (op) & $\checkmark$ & & non-uni & $\sim0.5-1$ nm & 0 $\mu$eV\footnotemark[3]\\
 & &  &  & & & & & 200-300 $\mu$eV\footnotemark[4]\\
Fe(ML) & \cite{Goedecke2022}& MBE & FM (op) & & & & & 0 $\mu$eV\\

\textbf{/Ta(110)} & & & & & & & & \\
Fe(ML) & \cite{BrüningARXIV2024}& MBE & SpS & $\checkmark$ & & non-uni & $\sim2$ nm & 0 $\mu$eV\footnotemark[3]\\

\textbf{/Re(0001)-O(2$\times$1)} & & & & & & & & \\
Fe(ML) & \cite{Palacio-MoralesSCI-ADV2019}& MBE & FM (op)\footnotemark[5] & $\checkmark$ & & uni & $\sim3$ nm\footnotemark[6] & 240 $\mu$eV\\

\textbf{Pb(ML)/.../Si(111)} & & & & & & & & \\
Co-Si & \cite{MenardNAT-COMM2017}& MBE & FM (op)\footnotemark[7] & $\checkmark$ & & uni & $\sim0.5$ nm & 300 $\mu$eV\\
Co-Si & \cite{MenardNAT-COMM2019}& MBE & FM (op)\footnotemark[8] & & & & & 300 $\mu$eV\\

\botrule
\end{tabular}
\footnotetext[1]{from DFT~\cite{KezilebiekeNAT2020} and MOKE measurements~\cite{KezilebiekeADV-MAT2021}.}
\footnotetext[2]{From tight-binding calculations, unpublished.}
\footnotetext[3]{Nodal-point TSC.}
\footnotetext[4]{Values from tight-binding model calculations~\cite{SoldiniNAT-PHYS2023}.}
\footnotetext[5]{Out-of-plane magnetization is assumed in the tight-binding model since it was the magnetic ground state that could best reproduce the experimental results~\cite{Palacio-MoralesSCI-ADV2019}.}
\footnotetext[6]{From private communication with the authors of~\cite{Palacio-MoralesSCI-ADV2019}.}
\footnotetext[7]{from DFT~\cite{JoJMMM2006} and MOKE measurements~\cite{ChangJAP2015}.}
\footnotetext[8]{The authors in~\cite{MenardNAT-COMM2019} speculate that there is the possibility that the Co island hosts a non-collinear magnetic texture instead. However, there is no experimental evidence of such a magnetic state.}
\end{table}

\subsection{Two-dimensional ferromagnets in proximity to superconductors}\label{experiment_2D-FM}
The early experimental investigations on 2D MSHs focused primarily on ferromagnetic nano-islands interacting with superconducting substrates and thin films. The main goal of those studies was to observe the emergence of a topological superconducting phase with a gapped bulk band structure – a topological mini-gap – and chiral edge modes propagating along the rim of the 2D ferromagnetic island. This is the topological superconducting state predicted to emerge in 2D hybrid systems combining a FM ground state with an out-of-plane anisotropy, spin-orbit coupling (Rashba and/or atomic), and conventional $s$-wave superconducting pairing, whose topological state is defined by the integer topological invariant \textit{C} (Chern number, see Sec.~\ref{theory-2D-FM}).
An early study was reported by G. C. M\'{e}nard et al.~\cite{MenardNAT-COMM2017} in 2017, where ferromagnetic 2D Co-Si islands grown on Si(111) were covered by a superconducting Pb monolayer (ML). The buried magnetic islands are of the order of 5 to 10 nm in diameter. A large Rashba-SOC is expected to occur at the interface between the Pb ML and the Si(111) surface. This, together with the $s$-wave superconducting phase of the Pb ML and the time-reversal symmetry breaking arising from the ferromagnetic Co islands satisfies all necessary requirements for the emergence of topological superconductivity.

\begin{figure}
    \centering
    \includegraphics[width=0.75\linewidth]{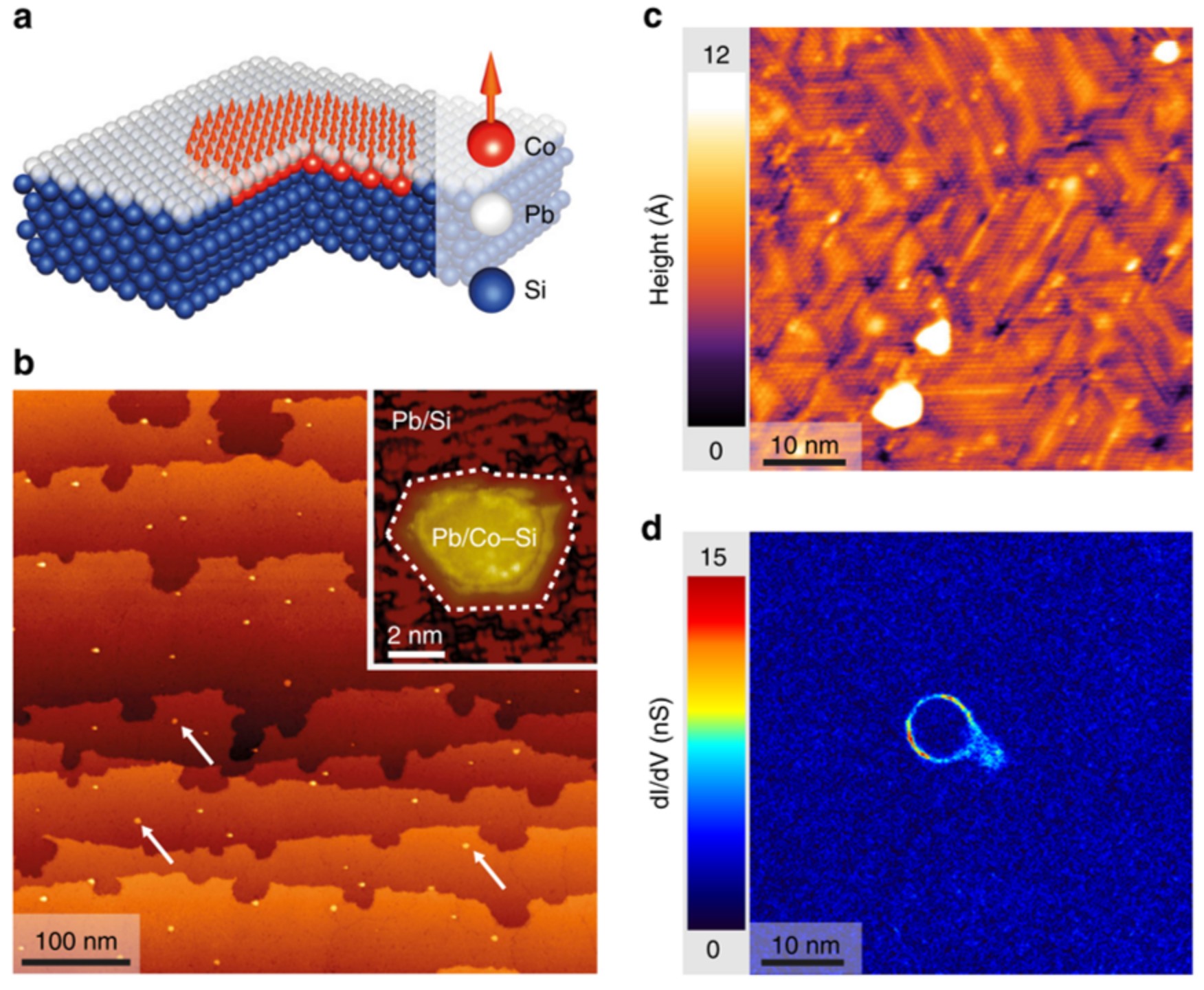}
    \caption{Magnet-superconductor hybrid Pb(ML)/Co-Si/Si(111). \textbf{a} Schematic of the hybrid system, with red arrows indicating the magnetic moments. \textbf{b} Large field of view constant-current STM image of the prepared sample, showing the presence of Co-Si islands underneath the Pb ML on Si(111). \textbf{c} and \textbf{d} constant-current STM image 
    and $\didv$ map at \textit{V} = 1.32 mV of the same sample region, showing no topographic evidence of the buried magnetic island but a spectroscopic feature resembling a contour of a buried 2D Co island. The d\textit{I}/d\textit{V} map was acquired with a superconducting Pb tip (\(\Delta\)\textsubscript{tip} \(\approx\) 1.30 meV) at \textit{T} = 300 mK. Accordingly, the differential tunneling conductance spectra are a convolution of the sample LDOS with the BCS superconducting gap of the Pb tip (see Section~\ref{experiment_0D}). This means that the $\didv$ map shown here is representative of the LDOS of the sample close to the Fermi level. Figure adapted from \cite{MenardNAT-COMM2017}.}
    \label{fig:Co-Pb_Si111_topo}
\end{figure}

\begin{figure}
    \centering
    \includegraphics[width=1\linewidth]{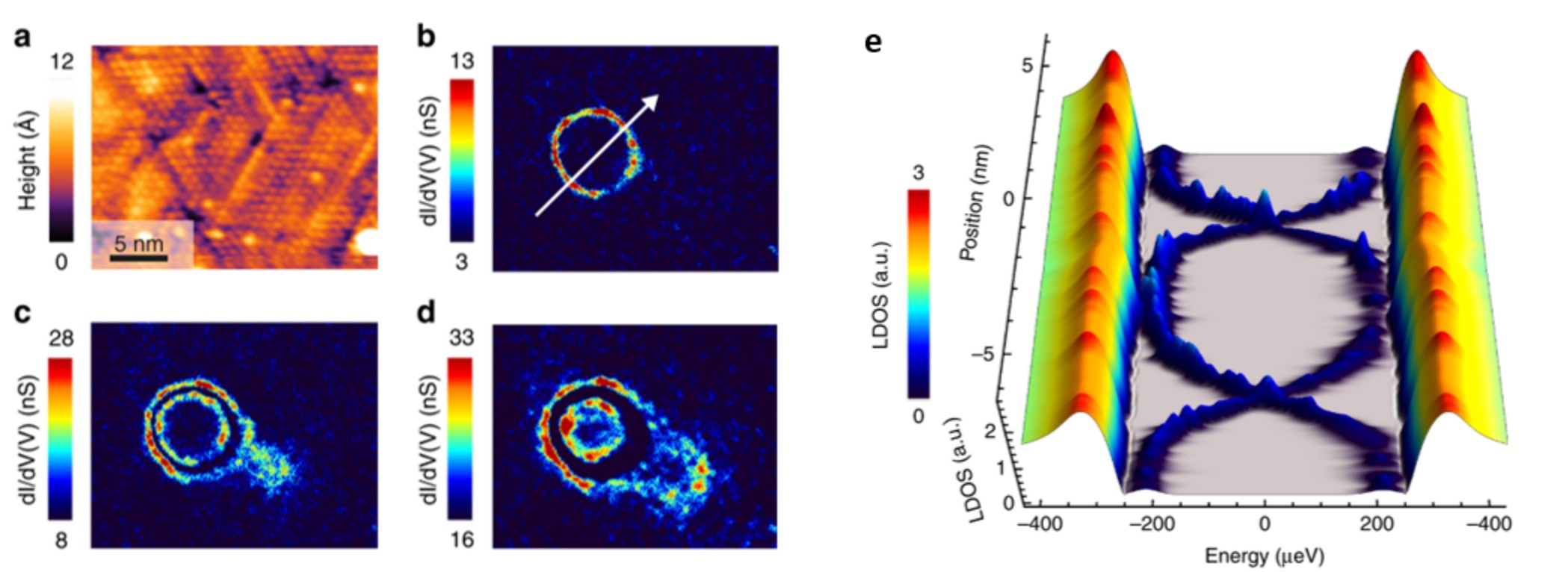}
    \caption{Edge states in Pb(ML)/Co-Si/Si(111). \textbf{a} Constant-current STM image 
    of a different region of the sample. \textbf{b}-\textbf{d }$\didv$maps of the same sample region in panel a, acquired at \textit{V} = 1.30 mV, 1.43 mV, and 1.50 mV, respectively, showing the energy evolution of the edge states appearing on the magnetic island. The $\didv$ map was acquired with a superconducting Pb tip (\(\Delta\)\textsubscript{tip} \(\approx\) 1.30 meV) at \textit{T} = 300 mK. \textbf{e} Deconvoluted $\didv$ line profile of the magnetic island taken along the direction indicated by the white arrow in panel b, showing the spatial dispersion of the in-gap energy modes across the island. Figure adapted from~\cite{MenardNAT-COMM2017}.} 
    \label{fig:Co-Pb_Si111_spec}
\end{figure}

The out-of-plane ferromagnetic ground state of the Co-Si islands (Fig. \ref{fig:Co-Pb_Si111_topo}a and \ref{fig:Co-Pb_Si111_topo}b) is not directly observed during the STM investigation. However, it was previously predicted by DFT calculations~\cite{JoJMMM2006} and experimentally observed via magneto-optic Kerr effect (MOKE) measurements performed by a different research group~\cite{ChangJAP2015}. The observation of in-gap  states is interpreted by M\'{e}nard and colleagues as a confirmation of the out-of-plane ferromagnetic state of the Co-Si islands investigated via STM. The differential conductance map in Fig. \ref{fig:Co-Pb_Si111_topo}d shows the presence of in-gap low-energy spectroscopic features resembling the shape of a Co-Si island. This contour-like low-energy spectroscopic feature seems to indicate the presence of an edge mode. More evidence regarding the nature of these in-gap features is shown in Fig. \ref{fig:Co-Pb_Si111_spec}a-d. Interestingly, $\didv$ maps of the Pb/Co-Si island acquired at different in-gap biases show the emergence of split features. The appearance of those split features becomes more apparent in the line-spectroscopy acquired along the diameter of the Co-Si island. Two in-gap features show a dispersion in real space along the diameter of the magnetic island, with two crossings symmetrically positioned with respect to the center of the island. 
The absence of an anti-crossing observed for the in-gap states has been regarded as evidence for a topological phase established at the location of the Pb/Co-Si islands, with chiral topological dispersive states. The touching/crossing of the in-gap dispersive states shown in  Fig.~\ref{fig:Co-Pb_Si111_spec}e is also proposed as evidence for a transition from a topologically trivial superconducting phase (outside the Fermi level LDOS circle) to a topologically non-trivial one (inside the Fermi level LDOS circle). The local Zeeman field generated by the magnetic island is considered key in establising the chiral topological superconducting phase.
Finally, one interesting observation made by M\'{e}nard and colleagues is the extremely short lateral decay length of the observed in-gap electronic features. The lateral width is much smaller than that of the two most relevant physical quantities, namely the superconducting coherence length and the mean free path. For the system in question, the former is on the order of 50 nm~\cite{BrunNAT-PHYS2014}, and the latter is on the order of a few nm. Both of them are much larger than the lateral confinement of the in-gap modes observed in the Pb/Co-Si islands, which is on the order of 0.5 nm.

\begin{figure}
    \centering
    \includegraphics[width=1\linewidth]{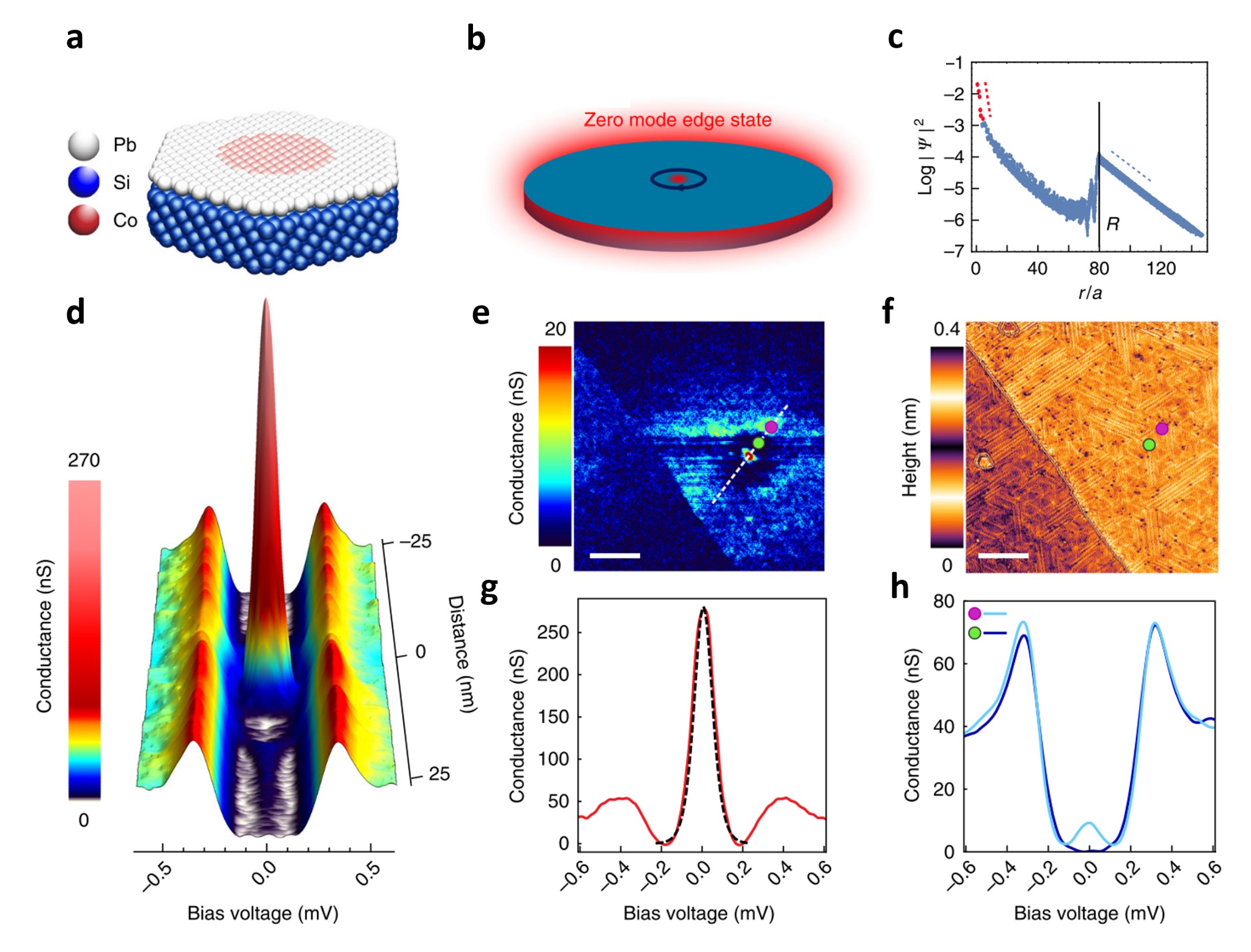}
    \caption{Isolated Majorana zero-energy modes in a disk-shaped Pb(ML)/Co-Si/Si(111) hybrid system. \textbf{a} Schematic view of the hybrid system. \textbf{b} Schematic view of the Majorana zero-energy bound state-pair, with one at the center and one at the rim of the disk-shaped topological superconductor. \textbf{c} Majorana states’ lateral distribution as calculated by the spin-orbit vortex model. \textbf{d} $\didv$ line profile in the magnetic island taken along the direction indicated by the white arrow in panel e, showing the spatial distribution of the zero-energy modes in the island. \textbf{e} zero-bias $\didv$ map and \textbf{f} constant-current STM image of one of the buried magnetic islands, showing the spatial distribution of the two zero-energy states. Scale bars are 20 nm. \textbf{g}, \textbf{h} $\didv$ spectra acquired at the location indicated in panel e and f, showing the amplitude of the LDOS at the location of the two zero-energy states (g, center state; h, edge state). The second spectrum in h was acquired outside the magnetic island. Data acquired with a Pt tip at T = 320 mK. Figure adapted from~\cite{MenardNAT-COMM2019}.}
    \label{fig:Co-Pb_Si111_Vortex}
\end{figure}

In a follow up work, the same research team also explors the possibility to stabilize Majorana zero-energy bound state-pairs in the same materials platform discussed above~\cite{MenardNAT-COMM2019}. They explore new samples where the buried Co-Si islands has a larger diameter (D $\thicksim15-20$ nm) than the one of the islands discussed in the previous work (D $\thicksim5-10$ nm)~\cite{MenardNAT-COMM2017}. As shown in Fig.~\ref{fig:Co-Pb_Si111_Vortex}, in those larger islands two zero-bias spectral features are observed, one at the center and one at the rim of the buried magnetic islands, energetically well isolated from all the other electronic states by a hard gap similar to the one of the superconducting Pb monolayer. In particular, the zero-bias feature at the center of the disk (Fig.~\ref{fig:Co-Pb_Si111_Vortex}d and Fig.~\ref{fig:Co-Pb_Si111_Vortex}g) shows a stronger spectral weight than the one at the perimeter of the island (Fig.~\ref{fig:Co-Pb_Si111_Vortex}h). This is in agreement with the theoretical situation where a pair of Majorana zero-energy bound states is stabilized in a topological superconducting disk~\cite{IvanovPRL2001,GurariePRB2007}, as schematically illustrated in Fig.~\ref{fig:Co-Pb_Si111_Vortex}b. Given the fact that the two Majorana zero-energy bound states should account for the same integrated (in space) DOS, the Majorana bound state localized at the center of the island is expected to show a larger LDOS than the other Majorana bound state distributed over the entire edge of the island. In addition, even in this case the spatial decay lengths of the bound state’s spectral weight away from its location (the center and the rim of the island) are extremely short, on the order of 1 nm. In particular, this short decay length inside the magnetic island is in stark contrast with the longer decay length of the edge bound state outside the magnetic island ($\thicksim15-20$ nm) - as shown in the zero-bias $\didv$ map in Fig.~\ref{fig:Co-Pb_Si111_Vortex}e - where it is much closer to the coherence length expected for the superconducting Pb(ML)~\cite{BrunNAT-PHYS2014}. 

These experimental observations are interpreted within the theoretical framework of a spin-orbit vortex model. In this model, the Hamiltonian describing the hybrid system contains a superconducting pairing term, a uniform exchange interaction term and a SOC vortex term, with a spin-momentum locking angle which is a function of the spatial coordinate \textit{\textbf{r}}. Apart from the stabilization of the Majorana zero-energy state-pair, the spin-orbit vortex model is also able to give an explanation for the different decay lengths inside and outside the magnetic island, as shown in Fig.~\ref{fig:Co-Pb_Si111_Vortex}c. While the decay length inside the magnetic island is defined by a magnetic length scale (which is inversely proportional to the magnetic exchange strength), the decay-length outside the magnetic island is defined by the superconducting coherence length. Unfortunately, direct experimental evidence of such a spin-orbit vortex situation is not available. However, the disordered Pb layer is expected to be characterized by a spatially-disordered spin-momentum locking. Interestingly, from a theoretical point of view, a similar situation is expected to emerge in a system with a uniform Rashba spin-orbit coupling and a spatially-varying exchange field texture~\cite{MenardNAT-COMM2019}. This consideration demonstrates the importance of a direct characterization of the magnetic state of the hybrid systems for the correct understanding of the origin of experimentally observed low-energy states and modes.\newline

\begin{figure}
    \centering
    \includegraphics[width=0.5\linewidth]{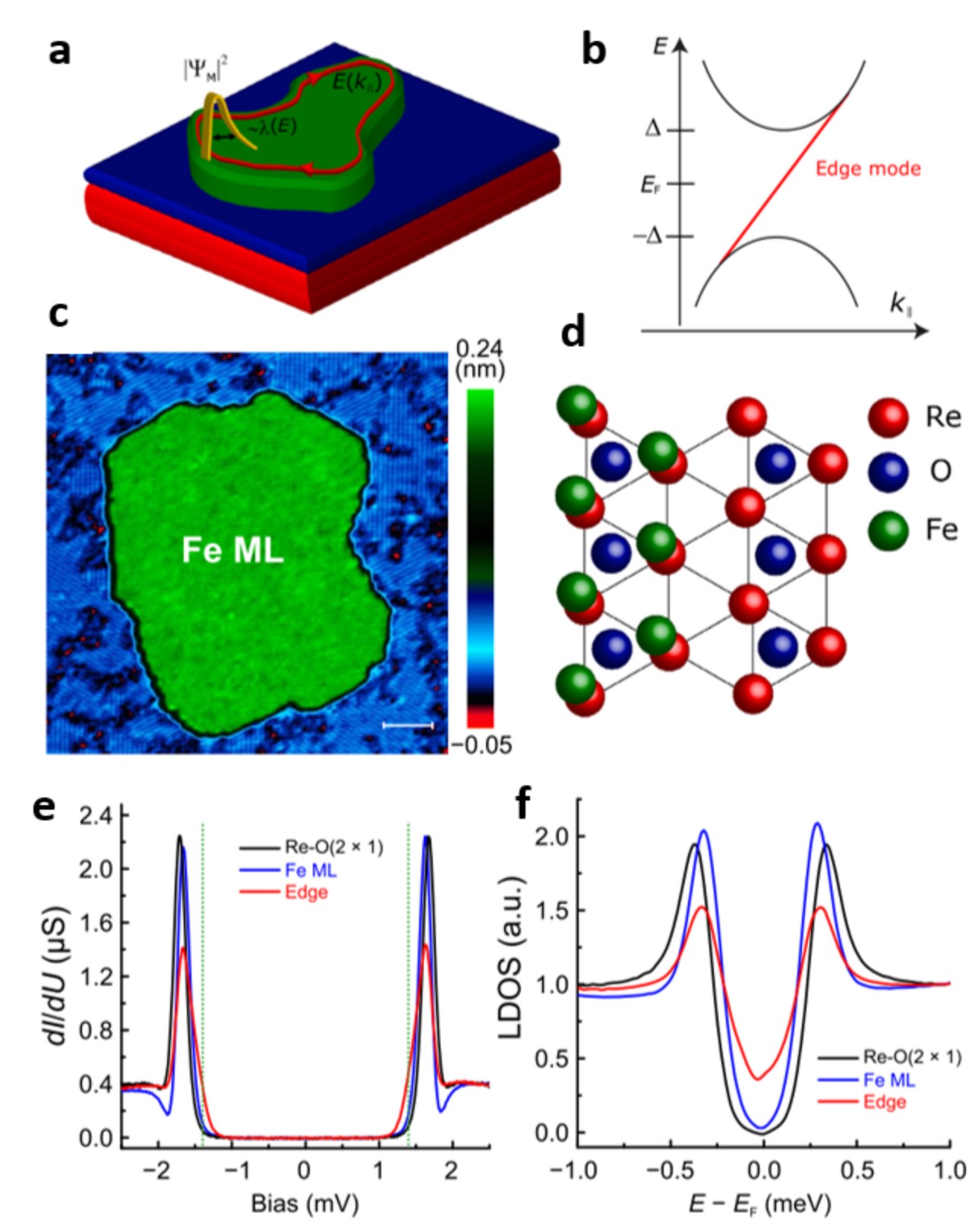}
    \caption{Magnet-superconductor hybrid Fe(ML)/Re(0001)-O(2x1). \textbf{a} Schematic of the hybrid system, with a ferromagnetic island (green) on the oxidized surface (blue) of the superconducting substrate (red), showing the presence of a propagating chiral edge mode (red contour) localized at the edge (yellow profile) \textbf{b} Single orbital band structure showing a gapped bulk band with a dispersive edge mode (red) going across the entire topological superconducting gap. \textbf{c} Constant-current STM image of the prepared sample, showing a Fe ML island on top of a Re(0001)-O(2x1) substrate. 
    \textbf{d} Sketch of the positions of the atoms in the three top-most layers. \textbf{e} Measured $\didv$ spectra on the Re(0001)-O(2x1) substrate (black curve), on the Fe ML island (blue curve) and at the edge of the Fe island (red curve). Measurements performed with a superconducting Pb tip. Vertical dotted lines indicate the tip superconducting gap ($\pm\Delta\textsubscript{tip}$). All STM measurements performed at \textit{T} = 360 mK. \textbf{f} Deconvoluted $\didv$ spectra from panel e. Figure adapted from~\cite{Palacio-MoralesSCI-ADV2019}. © The Authors, some rights reserved; exclusive licensee AAAS. Distributed under a CC BY-NC 4.0 license http://creativecommons.org/licenses/by-nc/4.0/”. Reprinted with permission from AAAS.}
    \label{fig:Fe-Re0001_structure}
\end{figure}

\begin{figure}
    \centering
    \includegraphics[width=1\linewidth]{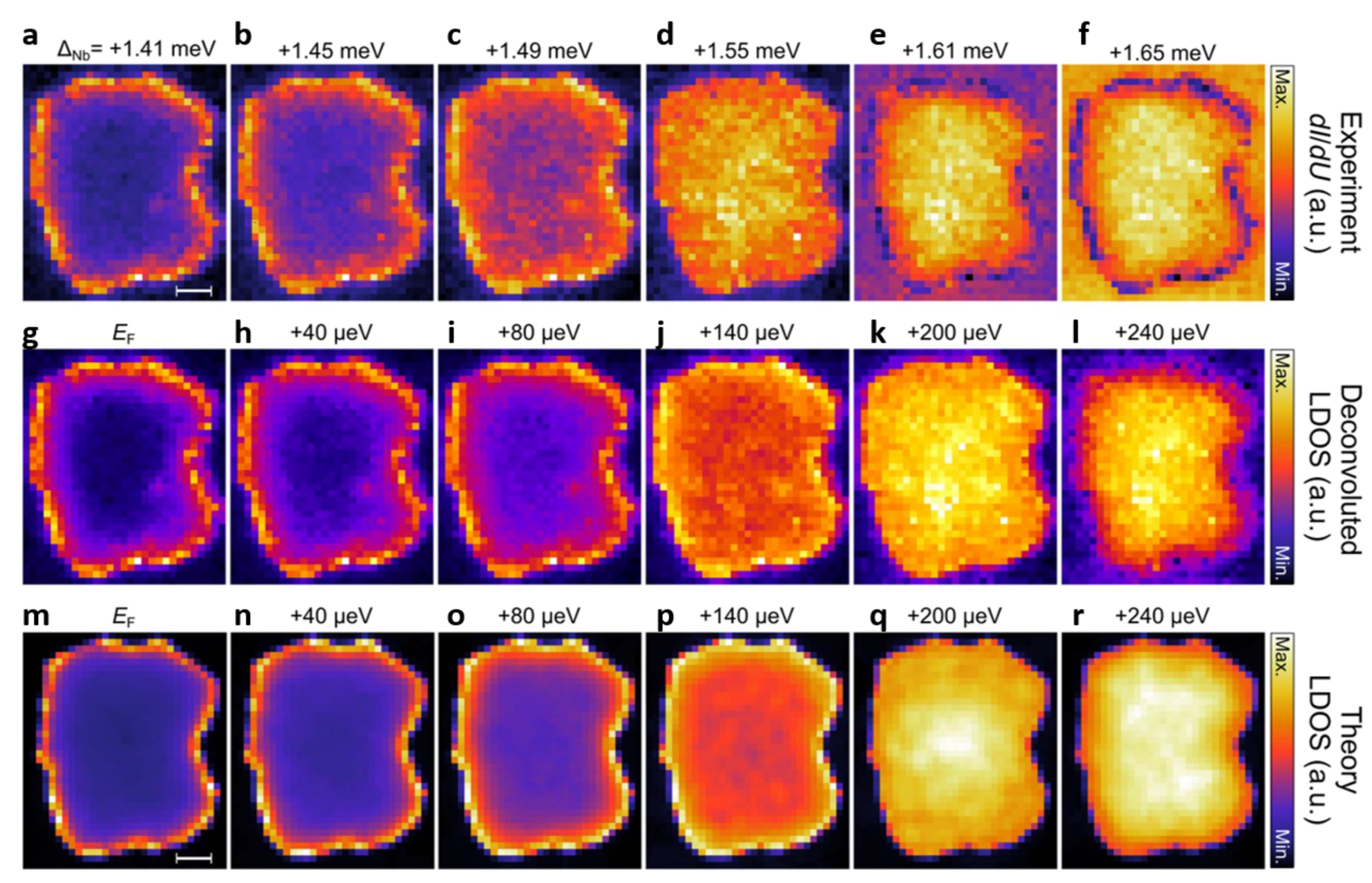}
    \caption{Visualization of low-energy edge mode in Fe(ML)/Re(0001)-O(2x1). \textbf{a-f} Measured $\didv$ maps at different biases, as indicated, with a superconducting STM tip ($\Delta\textsubscript{tip}=1.41$ meV). $\didv$ maps acquired at \textit{T} = 360 mK. \textbf{g-l} Deconvoluted $\didv$ maps from panels a-f with the indicated energies of the maps. \textbf{m-r} LDOS maps at the same energies as in panels g-l obtained from tight-binding model calculations. Figure adapted from~\cite{Palacio-MoralesSCI-ADV2019}. © The Authors, some rights reserved; exclusive licensee AAAS. Distributed under a CC BY-NC 4.0 license http://creativecommons.org/licenses/by-nc/4.0/”. Reprinted with permission from AAAS.}
    \label{fig:Fe-Re0001_specs}
\end{figure}

The first observation of low-energy edge modes localized at the rim of a ferromagnetic-superconducting island which is fully consistent with the predicted properties of a 2D TSC was reported by A. Palacio-Morales et al.~\cite{Palacio-MoralesSCI-ADV2019}. The material system is a Fe ML island deposited on top of a superconducting Re(0001)-O(2x1) substrate (see Fig. \ref{fig:Fe-Re0001_structure}). Even though no direct experimental imaging of the magnetic ground state of the Fe island is obtained, it is concluded to be an out-of-plane ferromagnetic order, due to the fact that tight-binding calculations where such magnetic state is assumed are able to best reproduce the experimental observations. The proximity-induced superconductivity observed in the Fe island is significant, with a measured superconducting gap of approximately 240 µeV, compared to a gap of 280 µeV measured for the superconducting substrate. The large proximity-induced gap in the interior of the Fe island is accompanied by a clear increase in LDOS at the rim of the island for energies smaller than the induced gap, with a maximum at the Fermi level (see Fig \ref{fig:Fe-Re0001_structure}e and \ref{fig:Fe-Re0001_structure}f, and Fig. \ref{fig:Fe-Re0001_specs}). In addition, the observed low-energy LDOS at the Fe island’s rim has a lateral decay length on the order of 3 nm, which is small compared with the superconducting coherence length of 24 nm for Re.

Based on a tight-binding model of the experimentally investigated Fe/Re(0001)-O(2x1) system~\cite{Palacio-MoralesSCI-ADV2019}, the observed in-gap LDOS at the rim of the Fe island are understood as evidence of an emergent topological superconducting phase, where a non-zero Chern number defines the number of chiral edge modes propagating along the rim of the Fe island (as discussed in the theory Sec. \ref{theory-2D-FM}). The calculated zero-energy (i.e., Fermi level) LDOS maps are shown in Fig.~\ref{fig:Fe-Re0001_specs}m-r. 
Interestingly, the oxygen layer at the interface between Fe and Re(0001) is found to play a key role establishing the topological superconducting phase. Indeed, in Fe(ML)/Re(0001) a rather delocalized LDOS at the Fermi level is observed, in stark contrast to the localized rim state observed for the hybrid system with the oxygen interlayer shown in Fig.~\ref{fig:Fe-Re0001_specs}. Again, this experimental observation is explained by tight-binding calculations, which highlight the importance of the stacking of the Fe atoms over the Re(0001) surface for establishing the topological phase. The presence/absence of the oxygen interlayer results in a different stacking of the Fe atoms with respect to the Re(0001) top atomic layer. In addition, the magnetic ground state of Fe ML on Re(0001) is expected to be a 120$^{\circ}$ antiferromagnetic N\'{e}el state~\cite{OuaziSURF-SCI2014,Palacio-MoralesNANO-LETT2016}, and not a ferromagnetic one as assumed for Fe ML on Re(0001)-O(2x1). Accordingly, the work by A. Palacio-Morales et al. shows how both the actual positions of atoms at the magnet-superconductor interphase and the magnetic ground state of the hybrid system can play a crucial role for determining the topological nature of the emergent band structure.\newline

\begin{figure}
    \centering
    \includegraphics[width=1\linewidth]{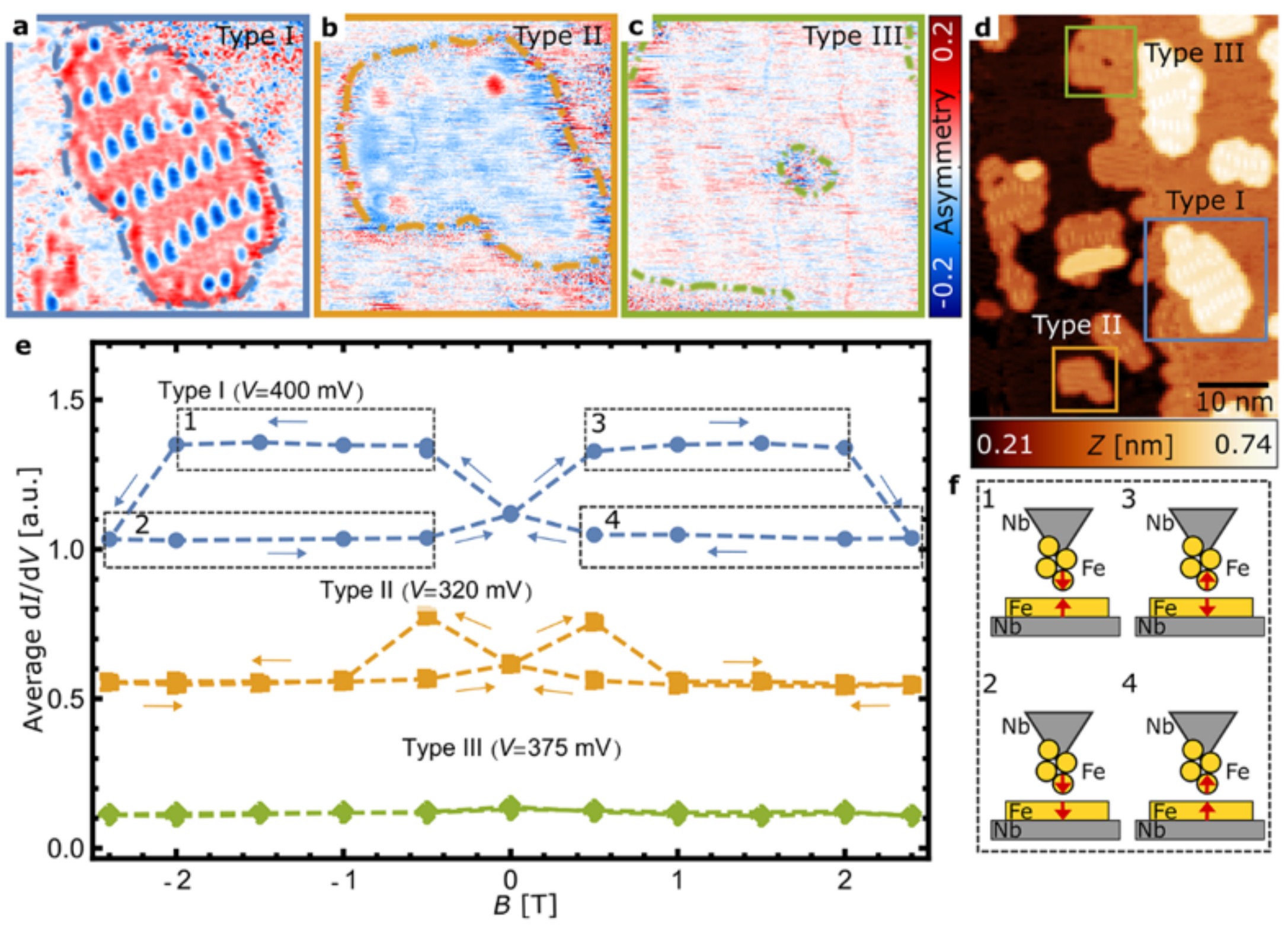}
    \caption{Magnet-superconductor hybrid Fe(ML)/Nb(110). \textbf{a-c} Spin-asymmetry maps (\textit{B}\textsubscript{z }= -0.5 T, \textit{B}\textsubscript{z} = +0.5 T) of the three types of Fe(ML) reconstructions shown in the constant-current STM image in \textbf{d}, acquired with the same spin-polarized STM tip. \textbf{e} Magnetic hysteresis loops obtained from the spin-resolved $\didv$ maps of the three different Fe(ML) reconstructions as a function of the external out-of-plane field \textbf{B}\textsubscript{z}. \textbf{f} Schematics showing the field-dependent orientations of the Fe island and tip magnetization with numbers referring to panel e. Figure adapted from~\cite{Goedecke2022}. https://pubs.acs.org/doi/10.1021/acsnano.2c03965. Further permissions related to the material excerpted should be directed to the ACS.}
    \label{fig:Fe-Nb110_mag}
\end{figure}

With the intent of exploring magnet-superconductor hybrids with a larger critical temperature and superconducting energy gap, Goedecke et al.~\cite{Goedecke2022} studied the correlation between magnetism and in-gap bands formation in Fe ML islands on a superconducting Nb(110) substrate. Nb is the elemental superconductor with the highest critical temperature ($T_c = 9.25$ K) and a large energy gap ($\Delta = 1.5$ meV), which makes it very attractive for the design of hybrid systems hosting potentially more stable topological superconducting phases, which can be experimentally more easily identified. The Fe ML deposited on top of the Nb(110) surface via physical vapor deposition shows three different reconstructions, as visible in Fig.~\ref{fig:Fe-Nb110_mag}a-d. Those three different reconstructed Fe ML islands exhibit out-of-plane ferromagnetic ground states, as shown by the $\didv$ maps acquired with a SP-STM tip (see Figs.~\ref{fig:Fe-Nb110_mag}a-c and the hysteresis loops in Fig.~\ref{fig:Fe-Nb110_mag}e). The field-dependent $\didv$ measurements show different magnetic coercivity for the three types of Fe reconstructions, which indicates different exchange and anisotropy energies. 

\begin{figure}
    \centering
    \includegraphics[width=1\linewidth]{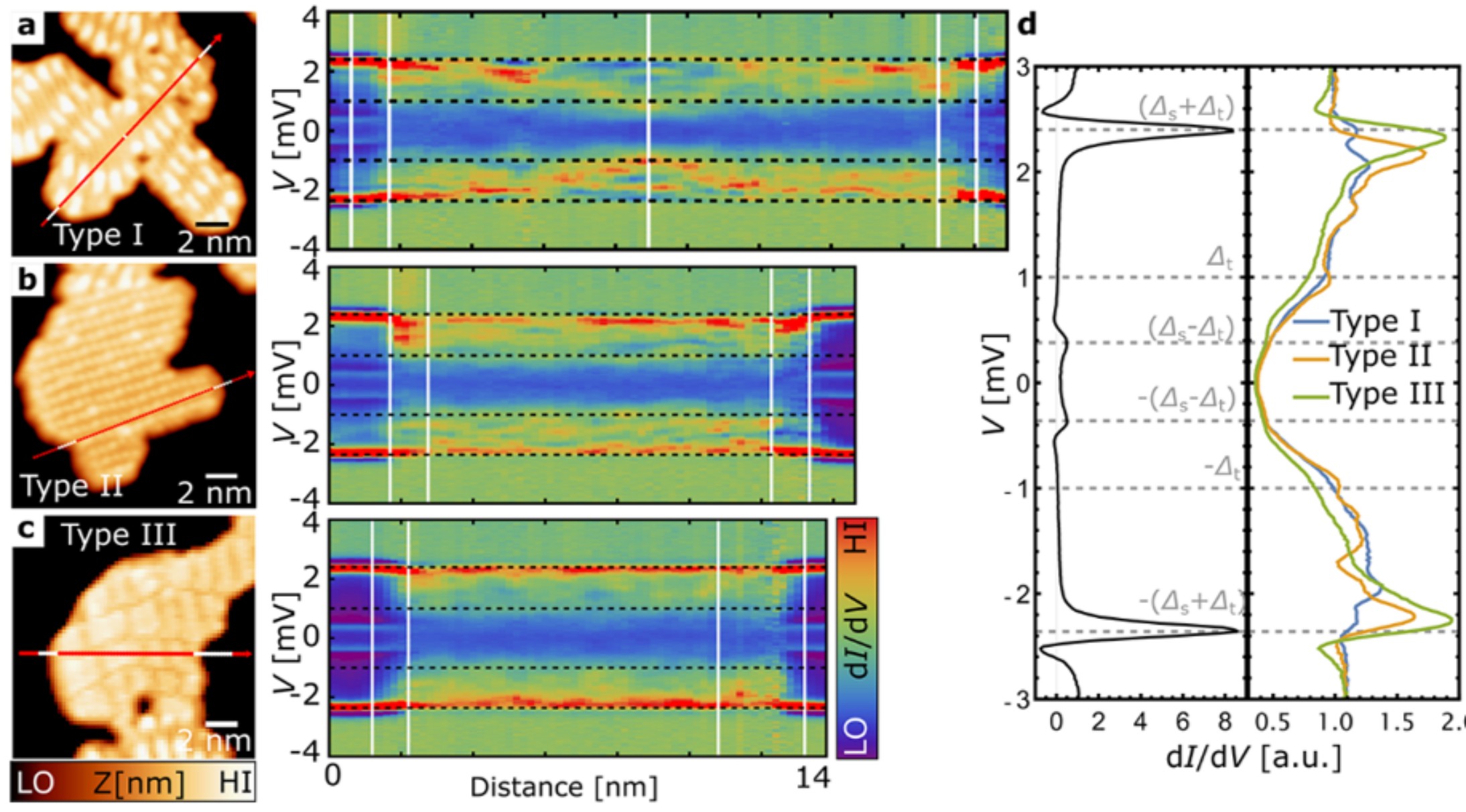}
    \caption{Visualization of dispersive and disordered YSR bands in Fe(ML)/Nb(110) islands. \textbf{a-c} Constant-current STM images (left panels) 
    of the three types of Fe ML islands on Nb(110), and the corresponding dispersive YSR bands observed via $\didv$ line profiles (right panels) measured along the red lines in the topographic images. \textbf{d} Measured $\didv$ spectra on the superconducting Nb(110) (left) and on the three different Fe ML islands (right). Gray dashed horizontal lines in the spectra are at \textit{e}\textit{V} = $\pm$(\(\Delta\)\textsubscript{t} - \(\Delta\)\textsubscript{s}), \textit{eV} = $\pm$\(\Delta\)\textsubscript{t}, and \textit{eV} = $\pm$(\(\Delta\)\textsubscript{t} + \(\Delta\)\textsubscript{s}). All data acquired with a superconducting Nb tip at \textit{T} = 4.5 K. Figure adapted from~\cite{Goedecke2022}. https://pubs.acs.org/doi/10.1021/acsnano.2c03965. Further permissions related to the material excerpted should be directed to the ACS.}
    \label{fig:Fe-Nb110_specs}
\end{figure}

The differential tunneling conductance spectroscopy investigation reveals also the presence of a spatially inhomogeneous in-gap LDOS, for all the three types of Fe(ML) islands (Fig. \ref{fig:Fe-Nb110_specs}). A strong spatial variation of the spectral weight of the in-gap states is observed over a length scale comparable to the reconstruction period of the three types of observed Fe islands. Furthermore, neither a superconducting mini-gap nor an enhanced low-energy LDOS at the rim is observed for any of the Fe islands. All those observations are interpreted as evidence of the trivial nature of the observed in-gap bands, suggesting that structural disorder is detrimental to the formation of 2D topological superconductivity. It is important mentioning, however, that an alternative explanation for the non-topological state of the Fe/Nb(110) system could be that such a material system is simply in a parameter regime where its superconducting state is not topological.\newline

Efforts have also been made to study the potential emergence of topological superconductivity in hybrid systems consisting of van der Waals (vdW) heterostructures. The first major result in this direction was obtained by S. Kezilebieke et al.~\cite{KezilebiekeNAT2020}, who reported the observation of a topological superconducting state in CrBr\textsubscript{3}(ML) islands deposited on NbSe\textsubscript{2}. CrBr\textsubscript{3} is a semiconducting transition metal trihalide, which in the monolayer form is an out-of-plane ferromagnet~\cite{ChenSCIENCE2019}. Those magnetic properties are conserved when CrBr\textsubscript{3} is deposited on the NbSe\textsubscript{2} substrate, as confirmed via MOKE measurements~\cite{KezilebiekeADV-MAT2021} and density functional theory calculations~\cite{KezilebiekeNAT2020}. NbSe\textsubscript{2} is a superconducting transition metal dichalcogenide, with a hard double-gap~\cite{HuangPRB2007,BorisenkoPRL2009}. 

\begin{figure}
    \centering
    \includegraphics[width=\linewidth]{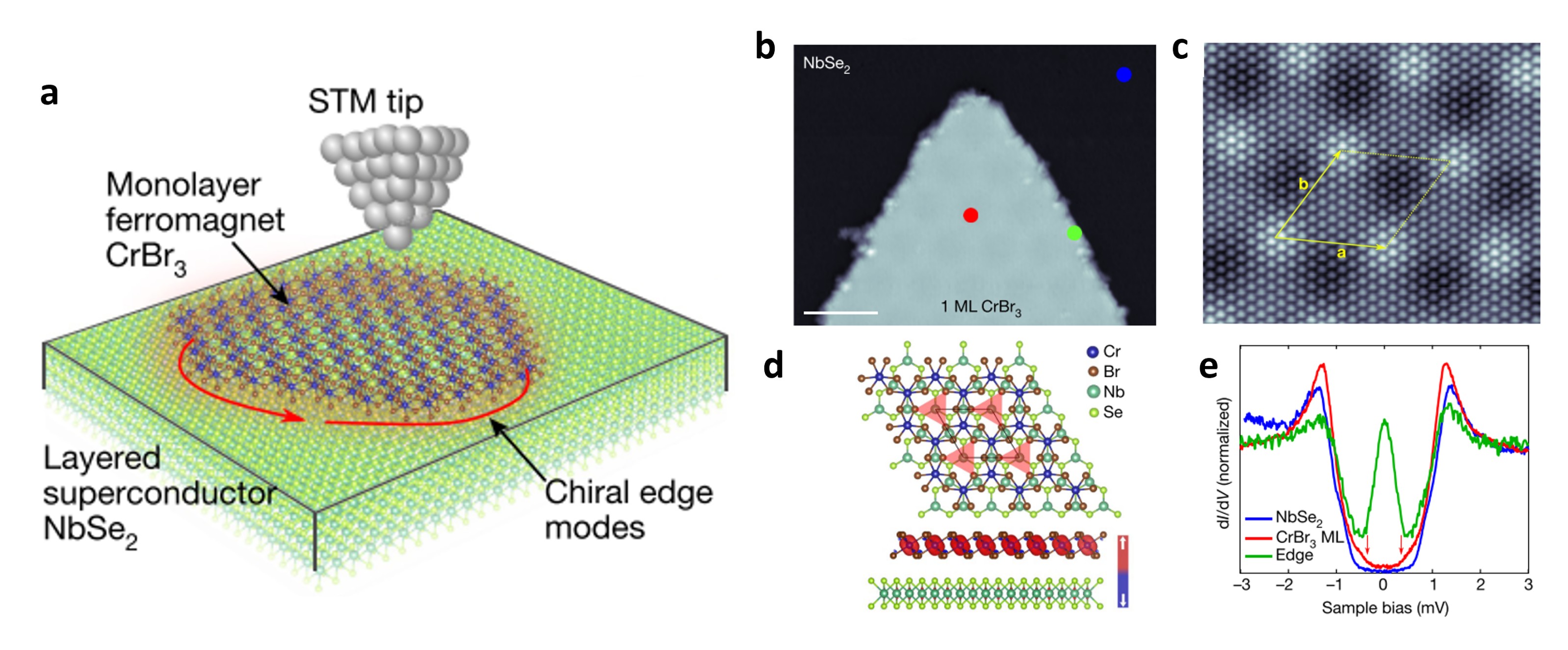}
    \caption{The van der Waals heterostructure CrBr\textsubscript{3}(ML)/NbSe\textsubscript{2}. \textbf{a} Sketch of the experimental set-up including the magnet-superconductor hybrid structure, the STM tip, and the emergent topological edge mode. \textbf{b} Constant-current STM image of the prepared sample, showing part of a CrBr\textsubscript{3}(ML) island on top of the NbSe\textsubscript{2} substrate. Scale bar is 10 nm. \textbf{c} Constant-current STM image of the Moiré pattern with a 6.3 nm periodicity, formed due to the lattice mismatch between the magnetic monolayer and the superconducting substrate. \textbf{d} Calculated structure and the induced spin-polarization from density-functional theory calculations. \textbf{e} Measured $\didv$ spectra at the three locations indicated in panel b. Measurements performed at \textit{T} = 350 mK. Figure adapted from~\cite{KezilebiekeNAT2020}.}
    \label{fig:vdW_topo}
\end{figure}

A sketch of the experimental set-up is shown in Fig. \ref{fig:vdW_topo}a. Fig. \ref{fig:vdW_topo}b shows the topographic image of a CrBr\textsubscript{3} ML island on the NbSe\textsubscript{2} substrate, while the details of the Moir\'{e} pattern emerging from the lattice mismatch between magnetic monolayer and superconducting substrate are reported in Fig.~\ref{fig:vdW_topo}c. Fig. \ref{fig:vdW_topo}e reports three d\textit{I}/d\textit{V} point spectra acquired on the NbSe\textsubscript{2} substrate (blue), the middle of the CrBr\textsubscript{3} ML island (red), and one of the edges of the magnetic island (green), acquired with a normal conducting tip at $T=350$ mK. Those spectra confirm the double-gap nature of NbSe\textsubscript{2} (shoulder at 1.0 mV and coherence peak at 1.5 mV), still detectable through the monolayer of the magnetic semiconductor. In particular a mini-gap of about 0.35 meV is observed for the CrBr\textsubscript{3}(ML)/NbSe\textsubscript{2} hybrid, which is interpreted as an induced topological mini-gap, $\Delta\textsubscript{ind}$. Finally, the spectrum over the ferromagnetic island’s edge shows a peaked differential tunneling conductance around zero-bias, which is interpreted as evidence of a chiral edge mode propagating along the island rim and dispersing across the topological mini-gap.

\begin{figure}
    \centering
    \includegraphics[width=0.75\linewidth]{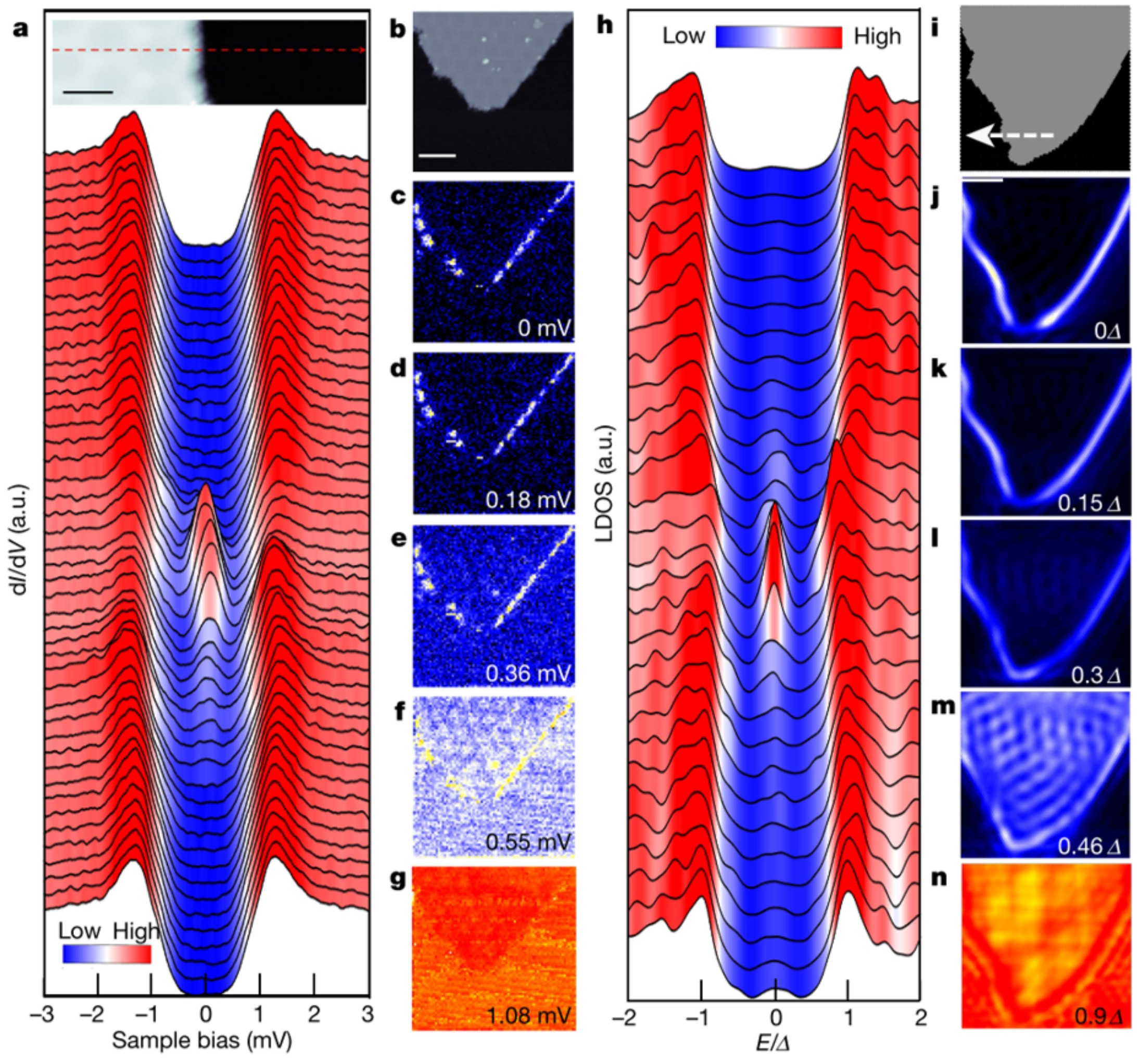}
    \caption{Visualization of low-energy edge modes at the rim of a CrBr\textsubscript{3}(ML) island on top of the NbSe\textsubscript{2} substrate. \textbf{a} $\didv$ line profiles across a CrBr\textsubscript{3}(ML) island edge taken along the dashed horizontal line in the inset STM image. Scale bar is 4 nm. \textbf{b} Constant-current STM image of part of a CrBr\textsubscript{3}(ML) island on top of the NbSe\textsubscript{2} substrate. Scale bar is 12 nm. \textbf{c-g} d\textit{I}/d\textit{V} maps in the same sample region shown in panel b, acquired at different in-gap biases. Measurements performed at \textit{T} = 350 mK. \textbf{h-n} Tight-binding model calculation results, reproducing the experimental observations in panels a to g. Figure adapted from~\cite{KezilebiekeNAT2020}.}
    \label{fig:vdW_specs}
\end{figure}

Further experimental evidence for the topological nature of the spectroscopic features observed in this vdW heterostructure is obtained from the mapping of the spatial distribution of the in-gap spectral features. These results are shown in Fig. \ref{fig:vdW_specs}, where $\didv$ line profiles across the magnetic island edge (Fig. \ref{fig:vdW_specs}a), and d\textit{I}/d\textit{V} maps at different in-gap energies (Fig. \ref{fig:vdW_specs}c-g) consistently reveal the presence of a mini-gap in the bulk of the CrBr\textsubscript{3} island and of a strong spectral weight on the edge for energies around the Fermi level. Some of those experimental evidences are reproduced via tight-binding model calculations (Fig. \ref{fig:vdW_specs}h-n). The calculated LDOS is similar to what is observed experimentally. In particular, the lateral confinement of the edge mode (c.a. 2.5 nm) is well reproduced by the model. However, there are also some discrepancies between the model results and the experimental evidences, the most evident being the non-homogeneous intensity of the edge mode observed at the rim of the CrBr\textsubscript{3} islands.

\begin{figure}
    \centering
    \includegraphics[width=0.75\linewidth]{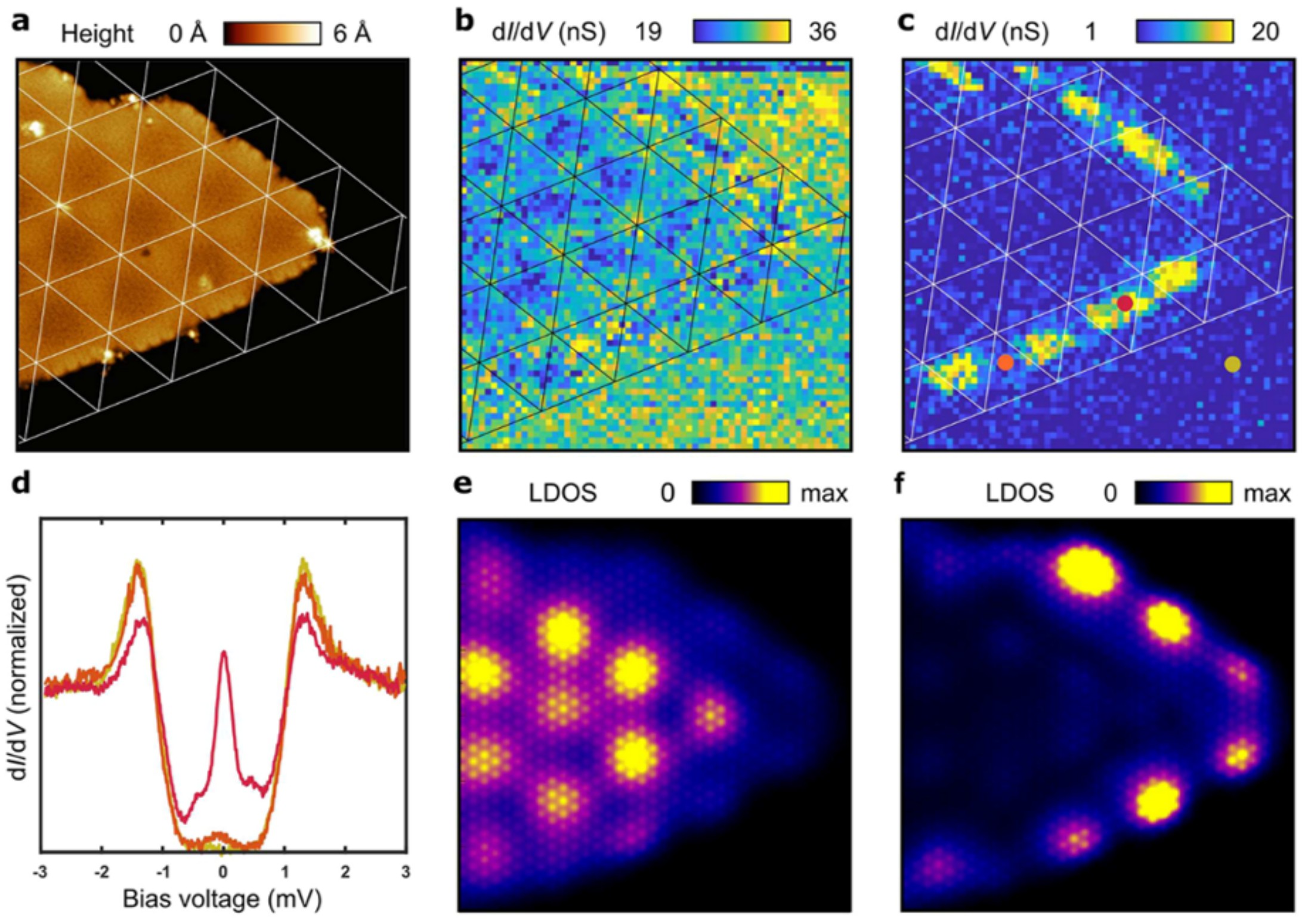}
    \caption{Edge states of the topological superconducting van der Waals heterostructure with a Moir\'{e} pattern. \textbf{a} Constant-current STM image of CrBr\textsubscript{3} island on NbSe\textsubscript{2}. Image size is 33 x 33 nm\textsuperscript{2}. \textbf{b} d\textit{I}/d\textit{V} map at the in-gap energy \textit{eV}= 0.8 meV. A Moir\'{e} pattern appears in the bulk of the CrBr\textsubscript{3} island. \textbf{c} d\textit{I}/d\textit{V }map at \textit{eV} = 0 meV showing the edge modes. \textbf{d }d\textit{I}/d\textit{V} spectra acquired at the positions indicated in panel c (red: strong edge mode intensity; orange: weak edge mode intensity; yellow: NbSe\textsubscript{2}). \textbf{e} and \textbf{f} Theoretically computed LDOS in the presence of a moir\'{e}-modulated exchange, as shown in Fig. \ref{fig:Moire}, at an energy in the middle of the in-gap bands and at zero energy, respectively. Experimental measurements performed at \textit{T} = 350 mK. Figure adapted from~\cite{KezilebiekeNANOLETT2022}. https://pubs.acs.org/doi/10.1021/acs.nanolett.1c03856.}
    \label{fig:vdW_moire-theory_2}
\end{figure}

This peculiar feature of the observed edge modes could only be explained in a follow up work by Kezilebieke et al.~\cite{KezilebiekeNANOLETT2022}, where the Moir\'{e} pattern emerging in the vdW heterostructure is also taken into consideration. In particular, it is shown that by assuming that the spatial Moir\'{e} pattern originating from the lattice mismatch between CrBr$_3$ and NbSe$_2$ generates a similar spatial modulation in the magnetic exchange and chemical potential, it is possible to explain the spatial variation of the edge mode's spectral weight, as discussed in Sec. \ref{theory-2D-FM} and shown in Fig. \ref{fig:Moire}.
The strongly modulated electronic potential and exchange coupling in the Moir\'{e} structure gives rise to modulated 
in-gap bands, which facilitates the emergence of a topological superconducting phase and generates spatially modulated edge modes. A direct comparison between the experimental observations and the outcome of the tight-binding model calculations is shown in Fig. \ref{fig:vdW_moire-theory_2}.

\subsection{Two-dimensional antiferromagnets on superconducting substrates}\label{experiment_2D-AFM}
Interesting new superconducting phases can emerge from the combination of antiferromagnetic spin textures and superconductivity. This was proposed in a theoretical work by Zhang et al.~\cite{Zhang2019}, who predicted that a bilinear AFM structure in combination with superconductivity can establish a topological superconducting phase. The theoretical prediction indicated the possible emergence of both gapless nodal-point superconductivity as well as higher-order topological superconducting phases.  In the case of nodal-point superconductivity, the emergent in-gap band structure possesses nodal-points along specific directions of the 2D Brillouin zone.  This is similar to the situation of non-superconducting Weyl semimetals~\cite{Yan-Felser2017}. In those topological systems the conduction band and the valence band touch each other in particular points of the Brillouin zone, which result to be sinks or sources of Berry curvature and are characterized by opposite winding numbers ($\pm$1), as discussed in Sec.~\ref{theory-2D-AFM}. This motivated the expansion of the search for topological superconductivity towards hybrid systems which host an antiferromagnetic ground state.
\newline

\begin{figure}
    \centering
    \includegraphics[width=1\linewidth]{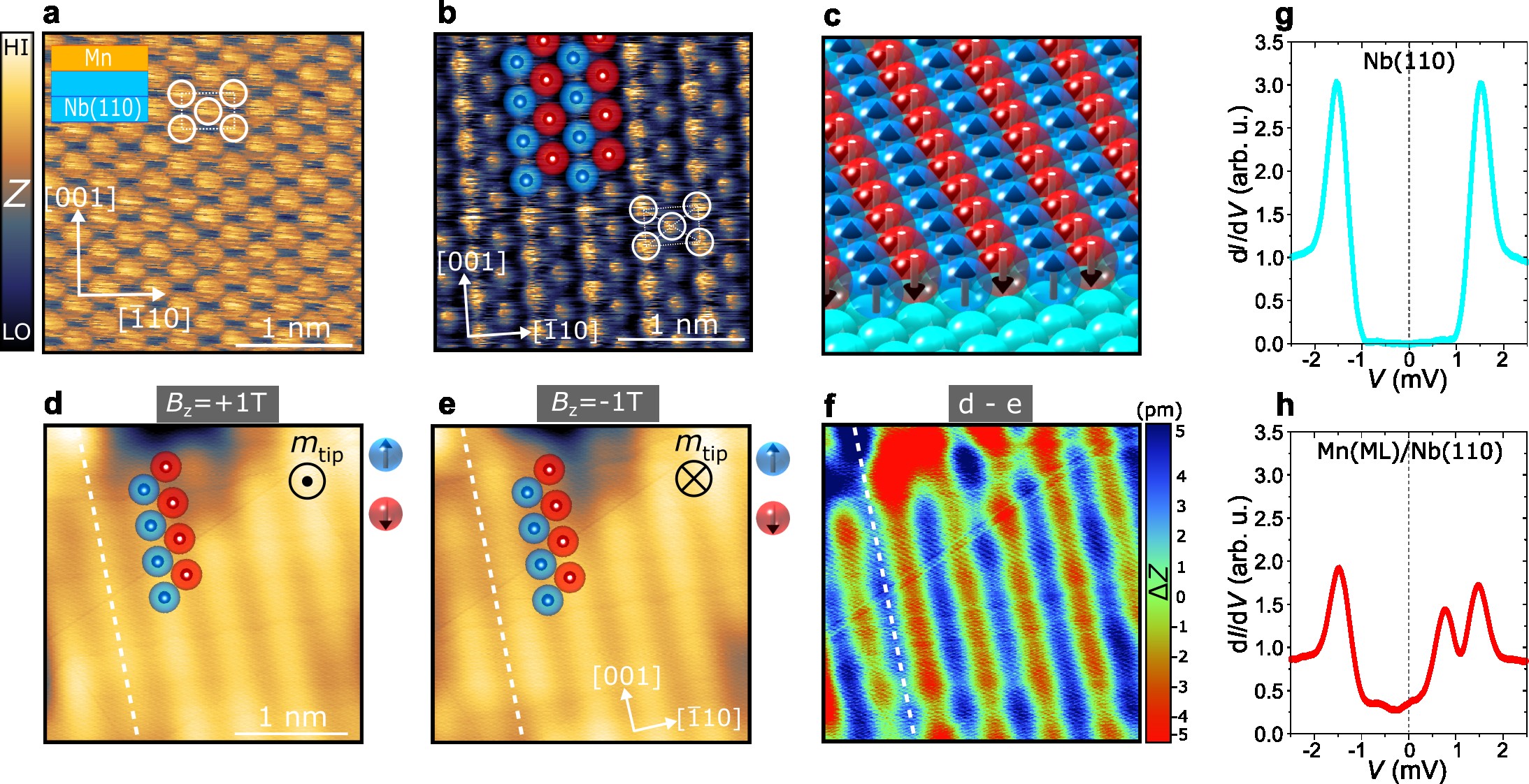}
    \caption{Magnet-superconductor hybrid Mn(ML)/Nb(110). \textbf{a} Constant-current STM image reporting the atomic configuration of the Mn(ML), showing the bcc(110) surface symmetry (atomic unit cell in white). \textbf{b} SP-STM image of the Mn monolayer revealing the presence of the $c(2\times2)$ AFM ground state. \textbf{c} Sketch of the $c(2\times2)$ AFM ground state of the Mn(ML) on the Nb(110) surface (blue sphere = spin-up; red sphere = spin-down). \textbf{d}-\textbf{e} SP-STM image of Mn(ML) obtained with a soft magnetic tip in a +1 T/-1 T out-of-plane magnetic field. \textbf{f} Computed difference image [d - e], revealing the out-of-plane magnetic contrast in the Mn(ML). \textbf{g}-\textbf{h} Low bias $\didv$spectra acquired on the bare Nb surface (panel g) and Mn(ML) (panel h). Spectra acquired at \textit{T} = 1.8 K with a Nb-coated superconducting tip. Each curve shows the point spectrum after numerical deconvolution of the superconducting tip DOS. Part of the figure adapted from~\cite{LoContePRB2022}.}
    \label{fig:Mn(ML)-Nb110_mag}
\end{figure}

The first attempt in this new direction was performed by R. Lo Conte et al.~\cite{LoContePRB2022}, who investigated the interaction between a collinear antiferromagnetic ground state and s-wave superconductivity in a monolayer of Mn deposited on top of a Nb(110) substrate. The Mn(ML) was found to grow pseudomorphically on the Nb(110) surface (Fig. \ref{fig:Mn(ML)-Nb110_mag}a) and to host a $c(2\times2)$ antiferromagnetic ground state (Figs. \ref{fig:Mn(ML)-Nb110_mag}b, c). The easy-axis magnetic anisotropy along the out-of-plane direction was resolved via SP-STM imaging in magnetic field with a soft magnetic tip (Figs. \ref{fig:Mn(ML)-Nb110_mag}d to f). A schematic of the magnetic ground state of the system is shown in Fig. \ref{fig:Mn(ML)-Nb110_mag}c. Finally, STS measurements verified the presence of proximity-induced superconductivity in the Mn(ML)/Nb(110) hybrid system, together with in-gap features in the LDOS testifying for the formation of 
new bands (see Figs. \ref{fig:Mn(ML)-Nb110_mag}g, h).

\begin{figure}
    \centering
    \includegraphics[width=1\linewidth]{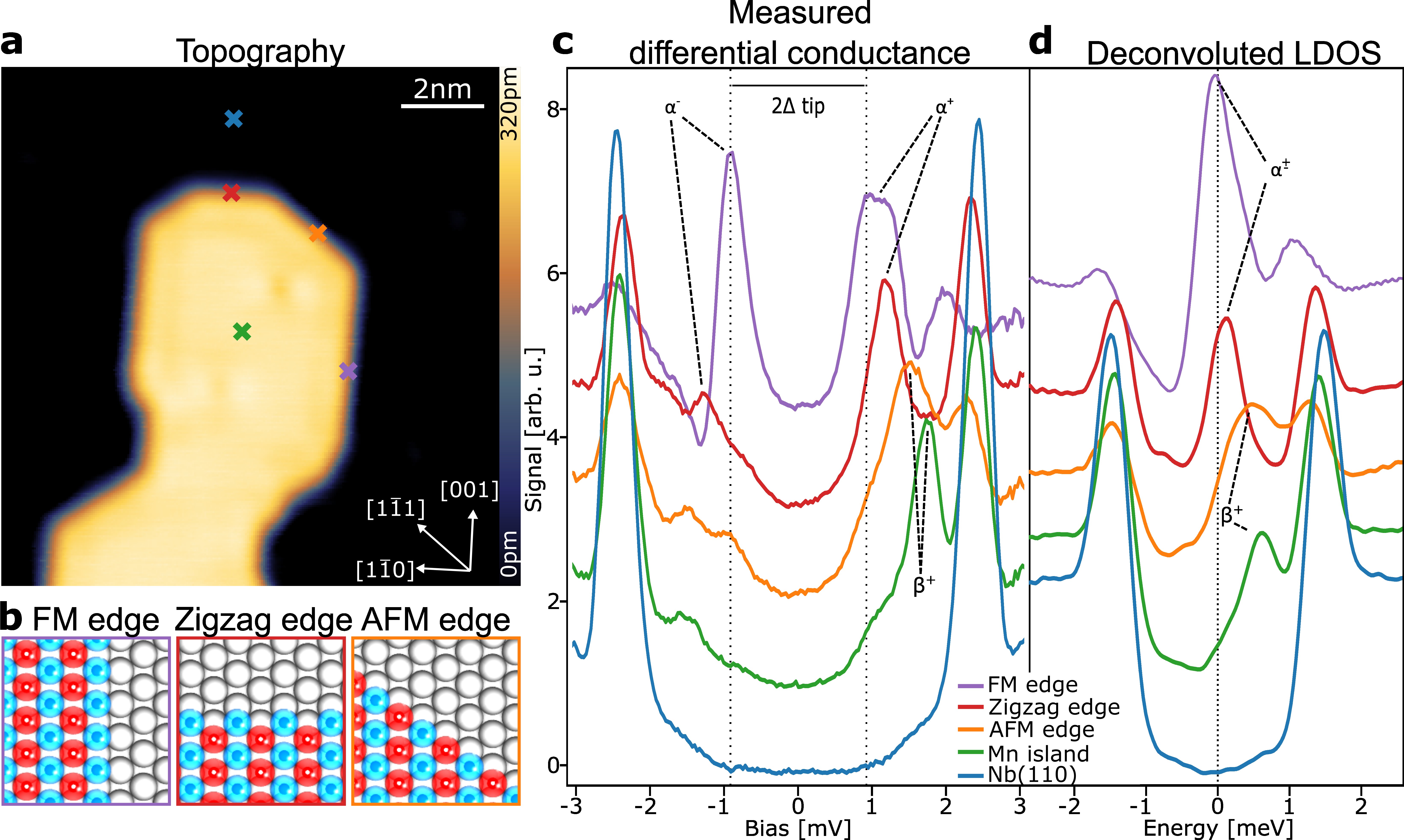}
    \caption{Measured LDOS of Mn ML island on Nb(110). \textbf{a} Constant-current STM-image of a Mn island with well-developed edges along the main crystallographic directions. \textbf{b} Schematics of the atomic and spin structure of a ferromagnetic (FM), zig-zag (ZZ), and antiferromagnetic (AFM) edge, respectively. \textbf{c} $\didv$ spectra acquired with a superconducting Nb-coated Cr tip at the positions indicated in panel a. \textbf{d} Deconvoluted $\didv$ spectra obtained from the spectra displayed in panel c. All d\textit{I}/d\textit{V} measurements were performed at \textit{T} = 1.3 K. The position of the Fermi level in panel \textbf{c} and \textbf{d} is indicated by the dotted lines. Figure adapted from~\cite{BazarnikNAT-COMM2023}. }
    \label{fig:Mn(ML)-Nb110_specs}
\end{figure}

A follow up paper by the same research team shed more light on the nature of the observed in-gap electronic states~\cite{BazarnikNAT-COMM2023}. New samples with Mn(ML) islands on Nb(110) were prepared (see Fig. \ref{fig:Mn(ML)-Nb110_specs}a), which allowed for a detailed investigation of the in-gap LDOS in the bulk as well as at the different edges of the antiferromagnetic islands: the ferromagnetic (FM), the zig-zag (ZZ) and the antiferromagnetic (AFM) edges oriented along the [001], [1$\bar{1}$0] and [1$\bar{1}$1] crystallographic directions on the Nb(110) surface, respectively (see Fig. \ref{fig:Mn(ML)-Nb110_specs}b). $\didv$ spectra (Fig. \ref{fig:Mn(ML)-Nb110_specs}c) acquired on the bare Nb (blue), the bulk Mn (green) and the three different edges of the Mn island (purple, red and orange), and the numerically deconvoluted $\didv$ spectra (Fig. \ref{fig:Mn(ML)-Nb110_specs}d) show both the presence of in-gap 
bands as well as the different electronic nature of the three edges. In particular, it is worth noting that from the obtained $\didv$ spectra, the AFM edge appears to be fundamentally different from the FM and the ZZ edge. The spectra acquired on the FM and ZZ edges show a clear peak (or peaks) close to zero-bias ($\alpha\textsuperscript{±}$ features in Figs.~\ref{fig:Mn(ML)-Nb110_specs}c, d), while the main spectroscopic feature for the AFM edge is at a finite bias in the superconducting gap of the Nb substrate ($\beta\textsuperscript{±}$ feature in Figs.~\ref{fig:Mn(ML)-Nb110_specs}c, d), very similar to what is observed for the bulk Mn(ML) on Nb(110) (see Fig~\ref{fig:Mn(ML)-Nb110_mag}h).

\begin{figure}
    \centering
    \includegraphics[width=1\linewidth]{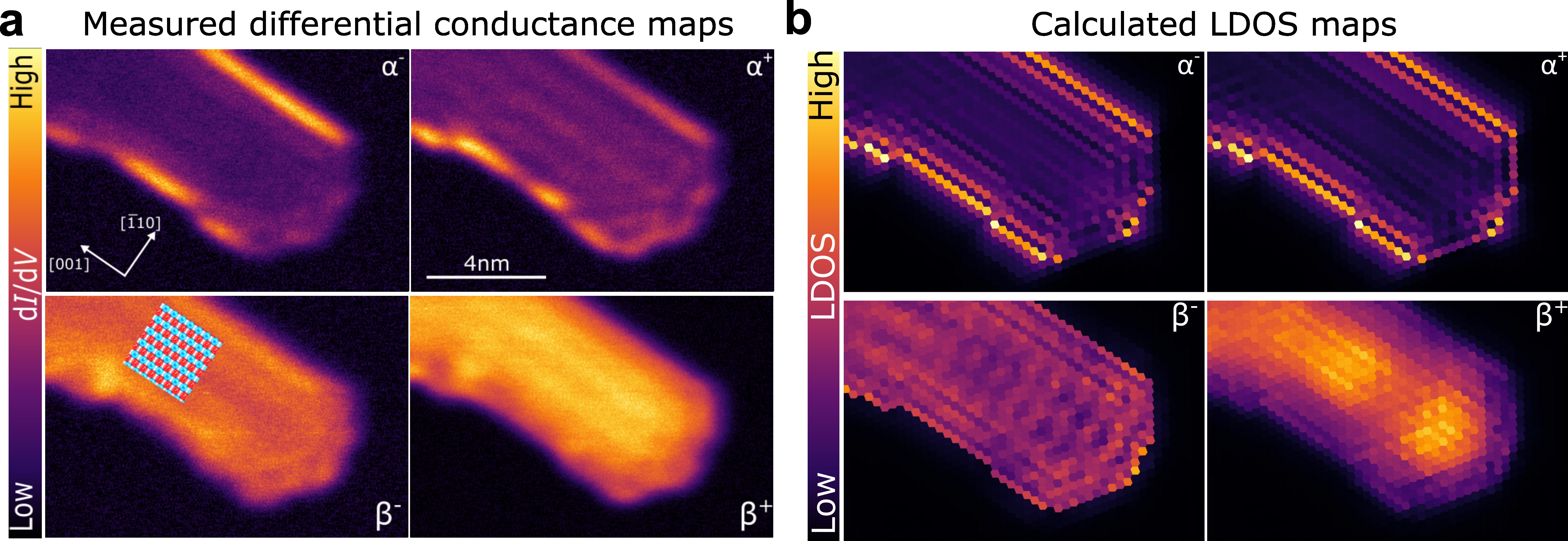}
    \caption{Measured and calculated LDOS maps of a Mn ML island on Nb(110). \textbf{a} $\didv$ maps acquired on a Mn(ML) island with a superconducting Nb-coated Cr tip at characteristic energies as indicated for $\alpha\textsuperscript{±}$ and $\beta\textsuperscript{±}$. Inset shows a sketch of the $c(2\times2)$ AFM spin texture in the Mn(ML) island. Measurements were performed at \textit{T} = 1.3 K. \textbf{b} Calculated LDOS maps for the same spectroscopic features as in Fig. \ref{fig:Mn(ML)-Nb110_specs}c. Figure adapted from~\cite{BazarnikNAT-COMM2023}.}
    \label{fig:Mn(ML)-Nb110_LDOS-maps}
\end{figure}

The different low-energy LDOS observed at the three different edges is further confirmed by d\textit{I}/d\textit{V} maps acquired at biases corresponding to the $\alpha\textsuperscript{±}$ and $\beta\textsuperscript{±}$ spectral features (Fig. \ref{fig:Mn(ML)-Nb110_LDOS-maps}a). Those d\textit{I}/d\textit{V} maps allow to establish the following two points: 1) the $\alpha\textsuperscript{±}$ spectral features originated from an edge mode; 2) the $\beta\textsuperscript{±}$ spectral features originate from the emergent band structure in the interior of the Mn monolayer on Nb(110). This was clear evidence that, in this material system, the spectroscopic character of the AFM edge is different from that of the FM and ZZ edges, which is in stark contrast to what is usually observed for ferromagnetic islands (see Sec.~\ref{experiment_2D-FM}). For FM systems a uniform LDOS was observed on the rim of the islands, except for the vdW heterostucture, where the Moir\'{e} pattern is actually inducing a lateral modulation in the LDOS of the system. However, in the work by M. Bazarnik et al.~\cite{BazarnikNAT-COMM2023}, the Mn ML is perfectly registered to the surface underneath and free from any structural reconstruction, which excludes the possibility that the observed inhomogeneous LDOS at the edges originates from some kind of nm-scale modulation of the electronic landscape.

An understanding of the puzzling experimental observations described above was provided by effective low-energy tight-binding model calculations, as explained in detail in Sec.~\ref{theory-2D-AFM}. The model includes all the main characteristics of the hybrid system, and shows that the interplay of the antiferromagnetic ground state with superconductivity gives rise to a quasiparticle band structure which has eight nodal points located on the border of the magnetic Brillouin zone (Fig.~\ref{fig:AFM_theory}). 
The calculated band structures for the three different edges show the emergence of weakly dispersive edge modes connecting nodal points with opposite topological charge. In particular, the calculated spectral functions show a large low-energy spectral weight for the ferromagnetic ([001]-like) and the zig-zag ([1$\bar{1}$0]-like) edges, but not for the antiferromagnetic ([1$\bar{1}$1]-like) edges, in agreement with the experimental observations in Figs.~\ref{fig:Mn(ML)-Nb110_specs}c and~\ref{fig:Mn(ML)-Nb110_specs}d. Using the same parameters of the model which predict the emergence of such a topological superconducting phase, it is possible to reproduce also the lateral distribution of the low-energy edge modes observed experimentally, as shown in Fig. \ref{fig:Mn(ML)-Nb110_LDOS-maps}b. The comparison between experimentally measured low-energy differential tunneling conductance maps and calculated LDOS maps provides a powerful way to identify the topological nature of the experimentally investigated AFM-MSH system. \newline

\begin{figure}
    \centering
    \includegraphics[width=1\linewidth]{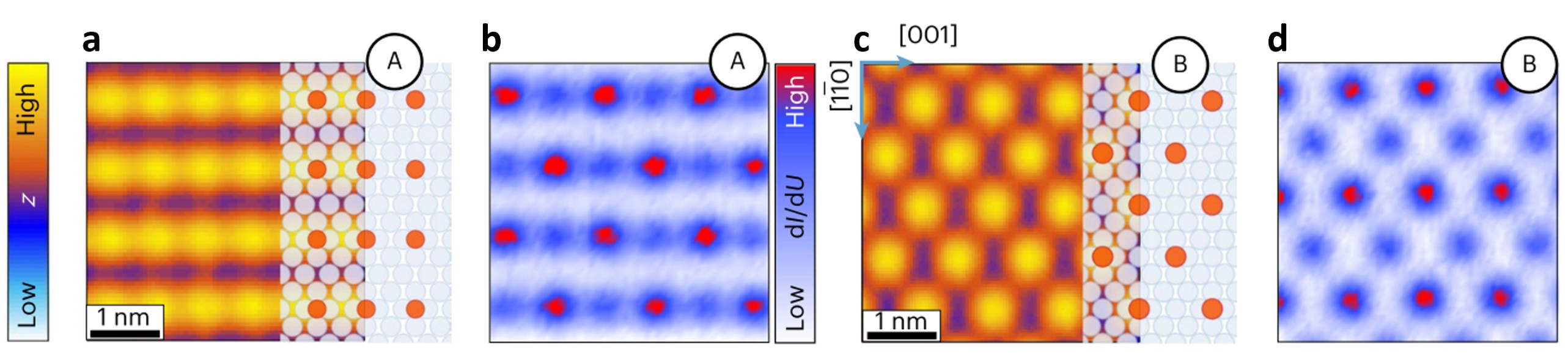}
    \caption{2D Cr spin lattices on Nb(110). \textbf{a} Constant-current STM image of the type-A (simple rectangular) spin lattice, with additional sketch of the lattice structure. \textbf{b} Spin-polarized d\textit{I}/d\textit{V} map of type-A lattice, showing a checkerboard antiferromagnetic spin order (blue = spin-up, red = spin-down). \textbf{c} Constant-current STM image of the type-B (centered rectangular) spin lattice, with additional sketch of the lattice structure. \textbf{d} Spin-polarized d\textit{I}/d\textit{V} map of type-B lattice, showing a $c(2\times2)$ antiferromagnetic spin order (blue = spin-up, red = spin-down). Figure adapted from~\cite{SoldiniNAT-PHYS2023}.}
    \label{fig:Cr(lattice)-Nb110_mag}
\end{figure}

\begin{figure}
    \centering
    \includegraphics[width=1\linewidth]{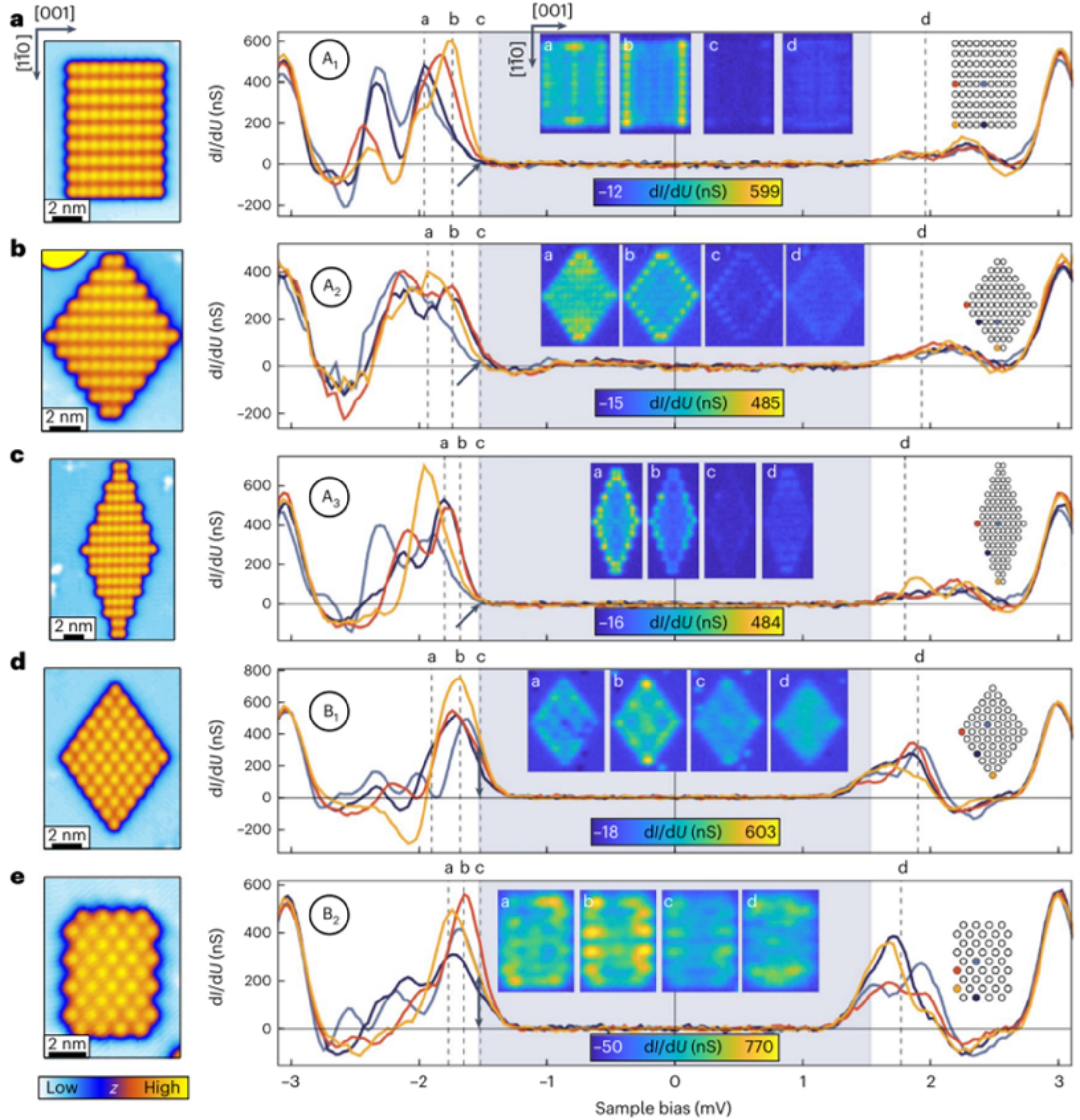}
    \caption{Visualization of edge modes in 2D Cr spin lattices on Nb(110). \textbf{a–e} Topography of the 2D Cr spin lattice terminations (left) and $\didv$ spectra (right) acquired with a superconducting tip at four representative lattices sites, given by the colored dots in the schematic lattice in the insets on the right, for system A${_1}$ (panel a), A${_2}$ (panel b), A${_3}$ (panel c), B${_1}$ (panel d) and B${_2}$ (panel e). Here A and B refer to the two types of 2D Cr spin lattices described in Fig.~\ref{fig:Cr(lattice)-Nb110_mag}, and the numbers 1, 2, 3 indicate islands of lattice A and B with edges along different crystallographic directions. The insets in the center show the $dI/dV$ maps of the prepared structures acquired at specific energies marked in the main plots, labelled by the letters a–d. The grey area corresponds to the tip superconducting gap $[-\Delta\textsubscript{tip}, +\Delta\textsubscript{tip}]$. $\didv$ map parameters: $I = 500$\,pA; $V = -5$\,mV; $V_{\rm mod} = 50\,\mu$V. Figure adapted from~\cite{SoldiniNAT-PHYS2023}.
    }
    \label{fig:Cr(lattice)-Nb110_specs}
\end{figure}

Another work combining a 2D antiferromagnetic spin lattice and the superconducting Nb(110) substrate was reported by M. O. Soldini et al.~\cite{SoldiniNAT-PHYS2023}. The investigated hybrid system consists of 2D Cr spin lattices prepared via atom-manipulation on the superconducting Nb(110) surface (see Fig. \ref{fig:Cr(lattice)-Nb110_mag}). Cr single atoms evaporated on the cold Nb(110) surface show typical in-gap YSR bound states (see Fig. \ref{fig:YSRatomsNb}). Atom-manipulation is used to build two different lattices of the hybridizing Cr YSR states with different symmetries: a simple rectangular lattice (Fig. \ref{fig:Cr(lattice)-Nb110_mag}a) and a centered rectangular lattice (Fig. \ref{fig:Cr(lattice)-Nb110_mag}c). Both lattices are found to host a collinear antiferromagnetic ground state: checkerboard and $c(2\times2)$ antiferromagnetic order for the simple rectangular and the centered rectangular lattice, respectively. This is shown by the spin-resolved d\textit{I}/d\textit{V} maps reported in Figs. \ref{fig:Cr(lattice)-Nb110_mag}b, d (see also Section~\ref{experiment_shibatip} and Fig.~\ref{fig:Shibatip}).

$\didv$ spectra and maps revealed the presence of a proximity-induced superconducting state and the emergence of in-gap bands for both spin lattices (see Fig. \ref{fig:Cr(lattice)-Nb110_specs}). For each lattice type, 2D spin lattice islands with different shapes - and so with different types of edges - were built via atom-manipulation (see Figs. \ref{fig:Cr(lattice)-Nb110_specs}a-e). $\didv$ maps (insets in the d\textit{I}/d\textit{V} spectra in Fig. \ref{fig:Cr(lattice)-Nb110_specs}) show a very inhomogeneous spectral weight close to the Fermi level on the edges of the constructed spin islands. Tight-binding model calculations explain those experimental observations as evidence for crystalline topological superconductivity (see discussion in Sec. \ref{theory-2D-AFM} and Fig. \ref{fig:crystalline}). In such framework, the non-homogeneous intensity of the low-energy LDOS on the edges is interpreted as the potential manifestation of higher order topological superconductivity, where the crystal symmetry of the constructed spin lattices is a key ingredient for the emergence of the observed topological phase.

\subsection{Two-dimesional non-collinear spin textures on superconducting substrates}\label{experiment_2D-NC}
Apart from the use of the collinear magnetic states discussed in Sec. \ref{experiment_2D-FM} and Sec. \ref{experiment_2D-AFM}, an additional powerful approach to the stabilization of topological superconductivity in 2D MSH systems is the use of non-collinear spin textures such as spin spirals, skyrmions and multi-q states. Some attempts in this direction have been made in recent years. For example, non-collinear and non-coplanar spin textures were observed in magnetic thin film islands on a Ru~\cite{HerveNAT-COMM2018} and a Re substrate~\cite{KubetzkaPRM2020}, which become superconducting only below 0.5~K and 1.7~K, respectively. However, the low critical temperature of those material systems did not allow for the exploration of their superconducting state due to experimental limitations. It was only very recently that Br\"uning et al.~\cite{BrüningARXIV2024} were able to stabilize a spin spiral in a Fe monolayer on top of a superconducting Ta(110) substrate and investigate the hybrid system in its superconducting state. This investigation revealed a very intriguing interplay between superconductivity, non-collinear spin texture and spin-orbit coupling that was not observed in any of the previously studied collinear hybrid systems.

\begin{figure}
    \centering
    \includegraphics[width=0.75\linewidth]{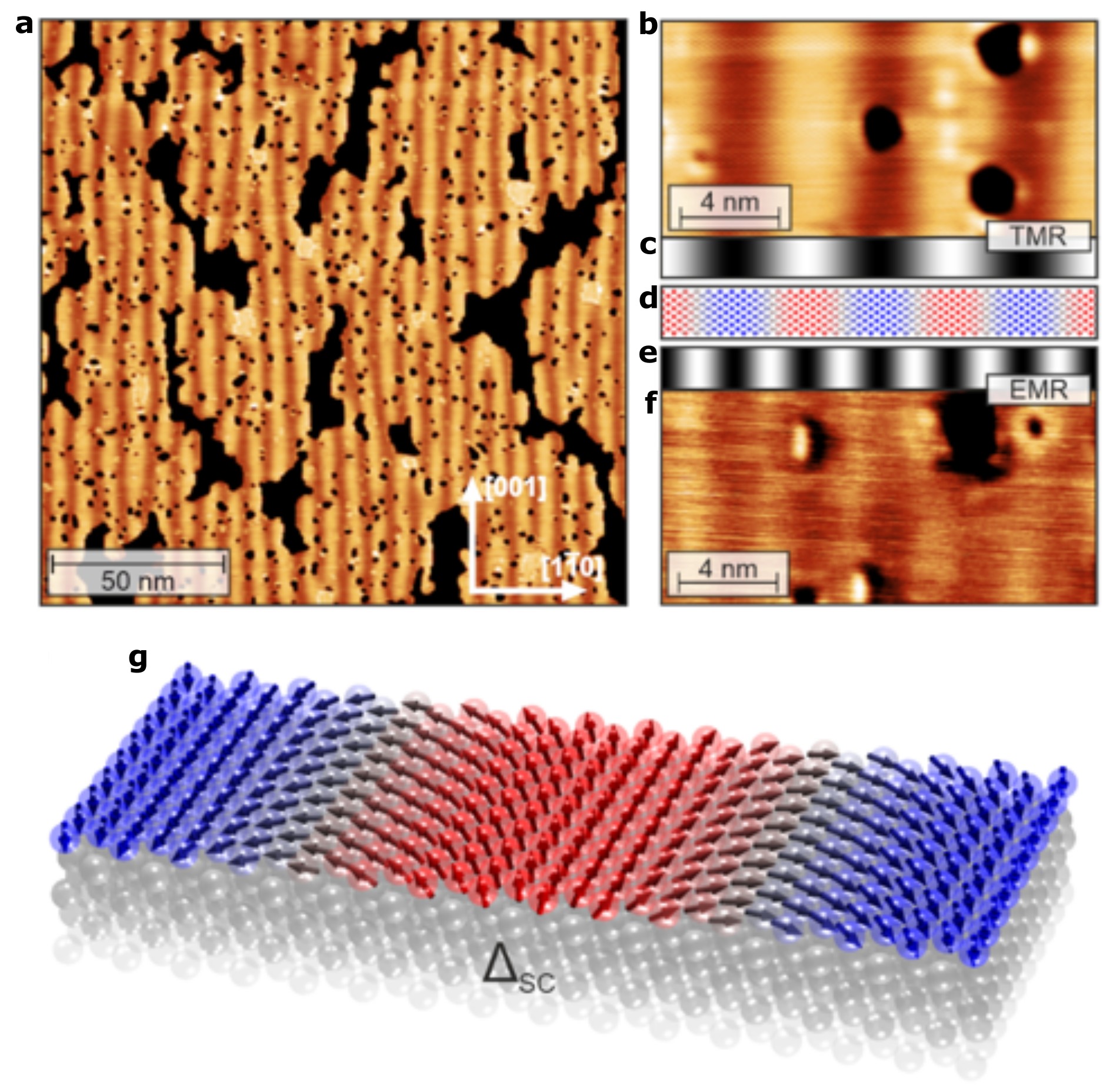}
    \caption{Spin spirals in Fe(ML)-island on Ta(110). \textbf{a} SP-STM constant-current image of 0.8 atomic layers of Fe on Ta(110), showing a spin spiral ground state. \textbf{b} Zoom-in SP-STM image of the spin spiral in the Fe(ML), revealing a spin spiral wavelength of circa 6 nm. \textbf{c} Expected SP-STM signal for the spin spiral due to the tunneling magnetoresistance (TMR) effect. \textbf{d} Sketch of a homogeneous spin spiral; red and blue indicate up and down magnetization directions, respectively. \textbf{e} Expected STM signal for the spin spiral with a non-magnetic tip due to the electronic magnetoresistance (EMR) effect. \textbf{f} STM constant-current image acquired with a non-magnetic tip, showing a pattern with half of the spin spiral wavelength, in agreement with EMR contrast. \textbf{g} Visualization of the magnet-superconductor hybrid system. Figure adapted from~\cite{BrüningARXIV2024}.}
    \label{fig:Fe(ML)-Ta(110)_Mag}
\end{figure}

\begin{figure}
    \centering
    \includegraphics[width=1\linewidth]{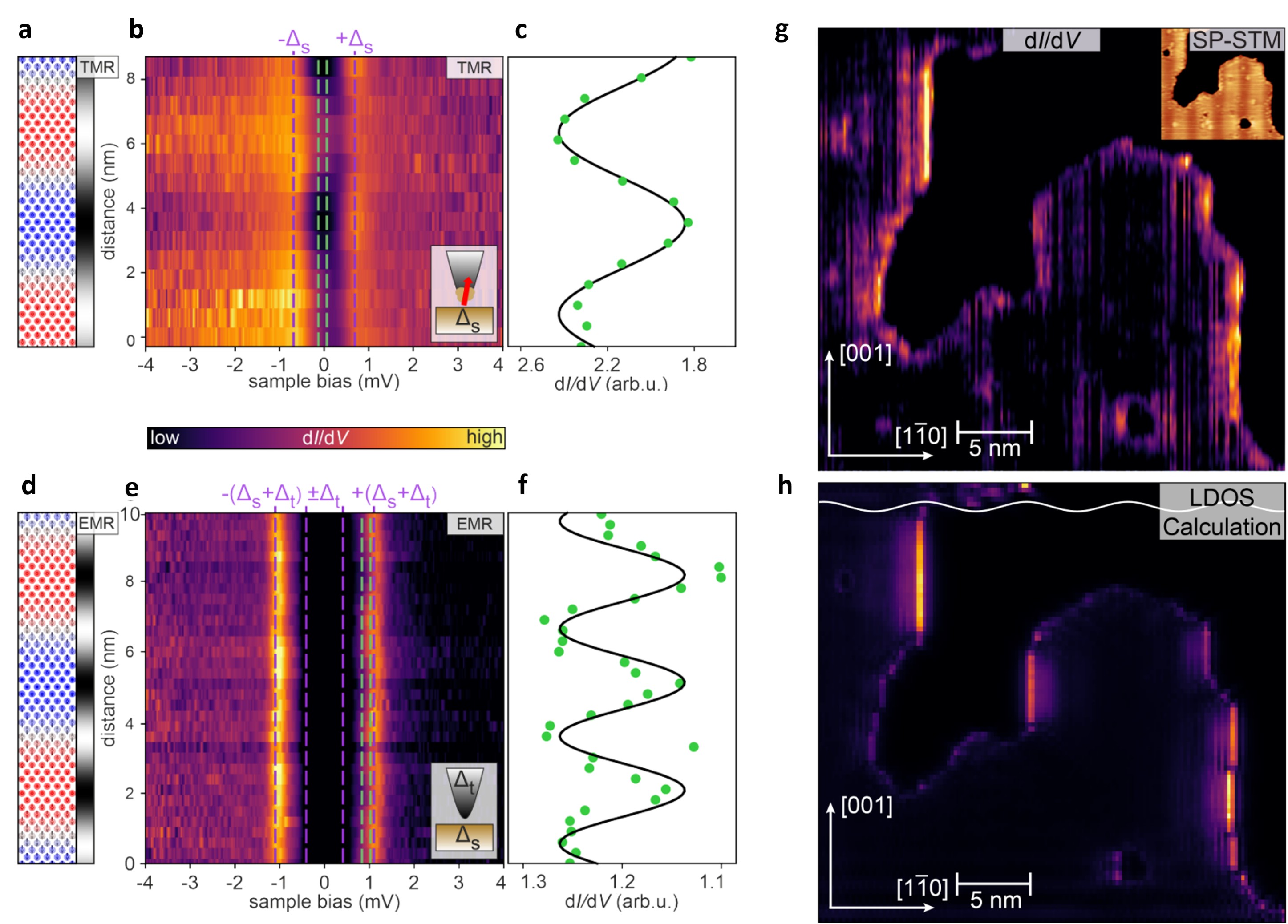}
    \caption{Spin spiral-driven topological superconductivity in Fe(ML)-island on Ta(110). \textbf{a} Sketched spin configuration of the magnetic spin spiral and the expected SP-STM contrast due to the TMR effect. \textbf{b} $\didv$ line profile acquired with a spin-polarized tip along the spin spiral propagation direction. \textbf{c} d\textit{I}/d\textit{V} intensities along the spin spiral averaged over an energy range in the superconducting gap (see green lines in panel b); the solid line is a cosine function with the spin spiral periodicity. \textbf{d} Sketched spin configuration of the magnetic spin spiral and the expected imaging contrast due to the EMR effect. $\didv$ line profile obtained with a superconducting tip along the spin spiral propagation direction. \textbf{f} d\textit{I}/d\textit{V} intensities along the spin spiral averaged over an energy range in the superconducting gap (see green lines in panel e); the solid line is a cosine function with half the spin spiral periodicity. $\Delta$\textsubscript{s} and $\Delta$\textsubscript{t} in panels b and e indicate the superconducting gap of the sample and the tip, respectively. \textbf{g} Experimental spin-averaged zero-bias d\textit{I}/d\textit{V} map of an irregularly shaped Fe(ML) island; the inset shows the spin spiral ground state of the same island imaged via SP-STM. Edges along the [001] direction are ferromagnetic. \textbf{h} Calculated spin-averaged zero-energy LDOS for a Fe island of the same size and shape as that shown in panel g. The white line at the top represents the m\textsubscript{z}-component of the spin spiral along the propagation direction of the spiral. Figure adapted from~\cite{BrüningARXIV2024}.}
    \label{fig:Fe(ML)-Ta(110)_Spec}
\end{figure}

The magnetic ground state observed via SP-STM is shown in Fig.~\ref{fig:Fe(ML)-Ta(110)_Mag}a. Figure~\ref{fig:Fe(ML)-Ta(110)_Mag}b shows a zoom-in of the spin spiral as detected via SP-STM. The observed pattern is in agreement with the tunneling magnetoresistance (TMR) effect, as explained in Fig. ~\ref{fig:Fe(ML)-Ta(110)_Mag}c and Fig.~\ref{fig:Fe(ML)-Ta(110)_Mag}d. These experimental observations were a confirmation of the prediction of previously performed first principles DFT calculations~\cite{RozsaPRB2015}, which indeed predicted a spin spiral with a wavelength of about 6 nm as one of the potential magnetic ground states of a monolayer of Fe on Ta(110). The stabilization of the spin spiral state is driven by the presence of a non-negligible Dzialoshinskii-Moriya interaction. In addition, the spin spiral state was also detected with spin-unpolarized tips. In this case, constant-current STM-images show a periodic pattern with a periodicity that is half the wavelength of the spins spiral (see Fig.~\ref{fig:Fe(ML)-Ta(110)_Mag}f). Such a periodic pattern with half the spatial wavelength is understood as originating from an electronic magnetoresistance (EMR) effect, which can have two different contributions~\cite{HannekenNAT-NANOTECH2015}: anisotropic magnetoresistance (AMR); or non-collinear magnetoresistance (NCMR). Fundamentally, both AMR and NCMR originate from spin-orbit coupling. The direct observation of the spin spiral ground state via EMR underlines the importance of SOC for this particular hybrid system. 

The role of SOC coupling becomes even more evident in the investigation of the spectroscopic properties of the Fe(ML)/Ta(110) system. Figures~\ref{fig:Fe(ML)-Ta(110)_Spec}a-c show the dependence of the spin-polarized low-energy LDOS as a function of the position along the spin spiral direction of propagation. In particular, the d\textit{I}/d\textit{V} line spectroscopy data acquired with a spin-polarized tip and reported in Fig.~\ref{fig:Fe(ML)-Ta(110)_Spec}b show that the amplitude of the coherence peaks of the superconducting LDOS of Fe(ML)/Ta(110) is modulated with the same periodicity as the spin spiral. Figures~\ref{fig:Fe(ML)-Ta(110)_Spec}d-f show the results of another d\textit{I}/d\textit{V} line profile recorded with a non-spin-polarized tip (more precisely, a superconducting tip; note the enlarged energy range around zero bias in Fig.~\ref{fig:Fe(ML)-Ta(110)_Spec}e due to the convolution of the sample LDOS with the superconducting LDOS of the tip). In this case, a new type of modulation of the coherence peaks' amplitude is observed along the spin spiral propagation direction, with a spatial frequency which is two times that of the spin spiral, similarly to what is observed in the constant-current STM images shown in Fig.~\ref{fig:Fe(ML)-Ta(110)_Mag}f. Br\"uning and colleagues interpreted this modulation of the intensity at the superconducting gap edge as a direct result of the spatial variation of the SOC in the spin spiral-hosting MSH system.

The topological nature of the superconducting state in the Fe(ML) on Ta(110) system was identified via the observation of edge modes. Figure~\ref{fig:Fe(ML)-Ta(110)_Spec}g shows a zero-bias d\textit{I}/d\textit{V} map acquired on a Fe(ML) island on Ta(110). The measured d\textit{I}/d\textit{V} map clearly shows the presence of edge modes, which seem to be present only on certain edges (the ferromagnetic edges along the [001] crystallographic direction). This experimental observation was successfully reproduced by tight-binding model calculations (see Sec.~\ref{theory-2D-NC}). The model predicts the emergence of a nodal-point topological superconducting state, where pairs of nodal-points with opposite winding number are located inside the magnetic Brillouin zone deriving from the large magnetic unit cell of the spin spiral (see Fig.~\ref{fig:FeTa} for more details). In particular, the calculated band structure for the different edges (Fig.~\ref{fig:Fe(ML)-Ta(110)_Spec}h) shows the presence of an edge mode at the edges parallel to the [001] crystallographic direction. The predicted edge modes are very weakly dispersing, resulting in a large DOS close to zero energy. This translates, for the calculated LDOS map at zero energy (see Fig.~\ref{fig:Fe(ML)-Ta(110)_Spec}h), into the emergence of a strong spectral feature at these specific ferromagnetic edges, which reproduces well the experimentally observed edge modes (compare Fig.~\ref{fig:Fe(ML)-Ta(110)_Spec}g with Fig.~\ref{fig:Fe(ML)-Ta(110)_Spec}h). Finally, it is worth mentioning that the magnitude of the zero energy LDOS is observed to be different on different ferromagnetic edges. The tight-binding model explains this as a magnetization-dependent edge mode dispersion, where approximately out-of-plane (in-plane) ferromagnetic edges host edge modes with a minimal (maximal) dispersion, resulting in a larger (smaller) zero energy LDOS. This unique feature predicted for MSH systems hosting a spin spiral ground state is well captured in the experimental data, as shown in Fig.~\ref{fig:Fe(ML)-Ta(110)_Spec}g. For more details on this very intriguing effect, we refer to the original work by Br\"uning et al.~\cite{BrüningARXIV2024}.

\section{Summarizing remarks}\label{conclusion}
The atomic lateral resolution and extremely high energy resolution of low-temperature scanning tunneling microscopy and spectroscopy, together with the capability to directly characterize the magnetic ground state at those length-scales by adding spin resolution, make STM the premier experimental technique for the investigation of the emergent physics in magnet-superconductor hybrid systems. Furthermore, the combination of (SP-)STM/STS investigations with theoretical modeling efforts based on first principle calculations and tight-binding model calculations have supported significantly the development of new hybrid systems potentially hosting new topological superconducting phases. As discussed in the present review, the joint effort of both experimentalists and theoreticians in the last decade has resulted in the discovery of new emergent electronic properties and their understanding, going from single paramagnetic impurities on surfaces to 2D magnetic islands in proximity to superconductors. The combination of magnetism and superconductivity has emerged as a very powerful approach to the design of material systems hosting novel topological properties, where the magnetic ground state plays a crucial role for determining the type of topological phase hosted by the hybrid quantum system. Given the large variety of magnetic ground states available, especially for 2D systems, it is likely that new topological phases of matter will continue to be discovered by the combination of those magnetic systems with superconductors. Our intent for this review article is two-fold. On the one hand, we hope it will serve as an introduction to magnet-superconductor hybrids as platforms for the establishment of topological superconductivity to the new generations of physicists, motivating them to join this exciting and fast developing field of modern condensed matter physics. On the other hand, we expect this review paper to stimulate new efforts in the study of magnet-superconductor hybrids which go beyond the current state-of-the-art, allowing us to reach an higher understanding of the fascinating new physics emerging from the interplay of magnetism and superconductivity.

\backmatter

\bmhead{Acknowledgments}
 R.L.C. is grateful to T. Cren, L. Schneider and H. Kim for helpful comments. J.W. is grateful to P. Sessi and B. Jäck for helpful comments and acknowledges editorial help by K. T. That. S.R. acknowledges useful discussions with E. Mascot. J.W. and R.W. gratefully acknowledge funding by the Cluster of Excellence ‘Advanced Imaging of Matter’ (EXC 2056 – project ID 390715994) of the Deutsche Forschungsgemeinschaft (DFG). R.W. acknowledges support by the EU via the ERC Advanced Grant ADMIRE (No. 786020). J.W. acknowledges support by the DFG project WI 3097/4-1 (project No. 543483081). S.R. acknowledges support from the Australian Research Council through Grants No. DP200101118 and DP240100168. D.K.M. acknowledges support by the U. S. Department of Energy, Office of Science, Basic Energy Sciences, under Award No. DE-FG02-05ER46225.

\bmhead{Author contributions}
R.W. wrote Chapter~\ref{intro}. S.R. wrote Chapter~\ref{theory-1D}. D.K.M. wrote Chapter~\ref{theory-2D}. J.W. wrote Chapter~\ref{experiment_1}. R.L.C. wrote Chapter~\ref{experiment_2D} and Chapter~\ref{conclusion}, and coordinated the writing of this review article. All authors provided input on the writing of the whole article.

\bibliography{sn-bibliography}
  
\end{document}